\documentclass[12pt]{book}
\usepackage{amsmath}
\usepackage{amsthm}
\usepackage{amssymb}
\usepackage{here}
\usepackage{psfig}
\usepackage{cite}

\usepackage{fancyheadings}
\begin{document}
\makeatletter
\def\cleardoublepage{\clearpage\if@twoside \ifodd\c@page\else
  \hbox{}
  \vspace*{\fill}
  \begin{center}
  \end{center}
  \vspace{\fill}
  \thispagestyle{empty}
  \newpage
  \if@twocolumn\hbox{}\newpage\fi\fi\fi}
\makeatother

\begin{titlepage}
\pagestyle{empty}
\begin{center}
\vspace{3.5cm}
\begin{large}
{\bf Jagellonian University \\
Institute of Physics \\}
\end{large}
\vspace{3.0cm}
\begin{LARGE}
{\bf Study of the production mechanism of the $\eta$ meson in proton-proton collisions \\
\vspace{1mm}
by means of analysing power measurements} \\
\end{LARGE}
\vspace{2cm}
\begin{large}
{\bf Rafa{\l} Czy{\.z}ykiewicz} \\
\end{large}
\vspace{5.0cm}
\end{center}
\begin{large}
{\bf A doctoral dissertation prepared at the Institute 
of Nuclear Physics of the Jagellonian University
and at the Institute of Nuclear Physics of the Research 
Centre J\"ulich, submitted to the Faculty of Physics,
Astronomy and Applied Computer Science at the Jagellonian University,
conferred by Dr. hab. Pawe{\l} Moskal \\}
\end{large}
\vspace{2.0cm}
\begin{center}
\begin{large}
{\bf Cracow 2006}
\end{large}
\end{center}
\cleardoublepage
\end{titlepage}
\begin{titlepage}
\pagestyle{empty}
\begin{center}
\vspace{3.5cm}
\begin{large}
{\bf Uniwersytet Jagiello{\'n}ski \\
Instytut Fizyki \\}
\end{large}
\vspace{3.0cm}
\begin{LARGE}
{\bf Badanie mechanizmu produkcji \\ mezonu $\eta$ w zderzeniach proton-proton \\
\vspace{1mm}
za pomoc\c{a} pomiar{\'o}w zdolno{\'s}ci analizuj\c{a}cej \\} 
\end{LARGE}
\vspace{2cm}
\begin{large}
{\bf Rafa{\l} Czy{\.z}ykiewicz} \\
\end{large}
\vspace{5.0cm}
\end{center}
\begin{large}
{\bf Praca na stopie{\'n} doktora nauk fizycznych wykonana
w~Zak{\l}adzie Fizyki J\c{a}drowej Instytutu Fizyki 
Uniwersytetu Jagiello{\'n}skiego oraz w Instytucie 
Fizyki J\c{a}drowej Centrum Badawczego J\"ulich
pod kierunkiem dr hab. Paw{\l}a Moskala, przedstawiona 
Radzie Wydzia{\l}u Fizyki, Astronomii i~Informatyki Stosowanej
Uniwersytetu Jagiello{\'n}skiego.
\\}
\end{large}
\vspace{2.0cm}
\begin{center}
\begin{large}
{\bf Krak{\'o}w 2006}
\end{large}
\end{center}
\cleardoublepage
\end{titlepage}
\vspace{-0.5cm}
\pagestyle{empty}

   \begin{center}
   \large
   {\bf Abstract}
   \end{center}
   \vspace{0.3cm}

\normalsize

The analysing power measurements for the $\vec{p}p\to pp\eta$ reaction  
studied in this dissertation are used in the determination 
of the reaction mechanism of the $\eta$ meson production in nucleon-nucleon 
collisions.

Measurements 
have been performed in the close-to-threshold energy region at 
beam momenta of p$_{beam}=2.010$ and 2.085~GeV/c, corresponding 
to the excess energies of Q~=~10 and 36~MeV, respectively. The experiments
were realised by means of a cooler synchrotron and 
storage ring COSY along with a cluster 
jet target. For registration of the reaction products 
the COSY-11 facility has been used. The identification of the $\eta$ meson 
has been performed with the missing mass method. 

The results
for the angular dependence of the analysing power combined with the 
hitherto determined isospin dependence of the total cross section for the 
$\eta$ meson production in the nucleon-nucleon collisions, reveal 
a statistically significant indication that the excitation of the nucleon 
to the S$_{11}$ resonance, the process which intermediates the production 
of the $\eta$ meson, is predominantly due to the exchange of a $\pi$ meson
between the colliding nucleons.

The determined values of the analysing power at both excess energies
are consistent with zero implying that the $\eta$ meson is produced
predominantly in the $s$-wave at both excess energies.

\vspace{0.4cm}

   \begin{center}
   \large
   {\bf Streszczenie}
   \end{center}
   \vspace{0.3cm}

\normalsize

Zaprezentowane w tej pracy pomiary
zdolno{\'s}ci analizuj\c{a}cej dla reakcji
\mbox{$\vec{p}p\to pp\eta$} maj\c{a} na celu wyznaczenie mechanizmu
produkcji mezonu $\eta$ w zderzeniach nukleon{\'o}w.

Pomiary wykonano w przyprogowym obszarze energii.
W eksperymentach wykorzystano wi\c{a}zk\c{e} proton{\'o}w
spolaryzowanych poprzecznie o p\c{e}dach \mbox{p$_{beam}=2.010$ i 2.085~GeV/c},
odpowiadaj\c{a}cych energiom wzbudzenia \mbox{Q~=~10 oraz 36~MeV}
dla reakcji $\vec{p}p\to pp\eta$.
Do bada{\'n} wykorzystano synchrotron COSY, tarcz\c{e} klastrow\c{a},
oraz system detekcyjny COSY-11. W do{\'s}wiadczeniach rejestrowano produkty reakcji
obdarzone {\l}adunkiem elektrycznym, natomiast identyfikacji mezonu $\eta$
dokonano za pomoc\c{a} metody masy brakuj\c{a}cej.

Uzyskane do{\'s}wiadczalnie rozk{\l}ady k\c{a}towe
zdolno{\'s}ci analizuj\c{a}cej, w~po{\l}\c{a}czeniu ze zmierzon\c{a}
uprzednio zale{\.z}no{\'s}ci\c{a} izospinow\c{a} produkcji mezonu $\eta$
w~zderzeniach nukleon-nukleon, pozwalaj\c{a} stwierdzi{\'c}, 
i{\.z} wzbudzenie nukleonu do rezonansu
S$_{11}$(1535) -- proces po{\'s}rednicz\c{a}cy w produkcji mezonu $\eta$ --
nast\c{e}puje  w g{\l}{\'o}wnej mierze poprzez wymian\c{e} mezonu $\pi$
pomi\c{e}dzy zderzaj\c{a}cymi si\c{e} nukleonami.

Warto{\'s}ci zdolno{\'s}ci analizuj\c{a}cej dla
obu energii wzbudzenia s\c{a} r{\'o}wne zeru w granicach b{\l}\c{e}d{\'o}w
statystycznych, co wskazuje, i{\.z} mezon $\eta$ jest w
g{\l}{\'o}wnej mierze produkowany w fali $s$.


            \cleardoublepage
	\pagenumbering{Roman}
        \tableofcontents
            \cleardoublepage
	\pagenumbering{arabic}

\cleardoublepage
\newpage
\clearpage
\chapter{Introduction} 
\pagestyle{fancy}
\label{intro}

The main goal of this dissertation is the closer insight into the production mechanism 
of the $\eta$ meson in the interaction of nucleons.  
Despite the fact that the discovery of this meson -- a member of the 
pseudoscalar meson nonet (see Appendix~\ref{mezony}) -- took place almost half a century ago
\cite{pevsner:61} its production dynamics has remained an open question
for a long time.

From the precise and extensive measurements of the total cross section 
for the $\eta$ meson production in the $pp\to pp\eta$ reaction in the close-to-threshold 
region~\cite{bergdolt:93, chiavassa:94, calen:96, calen:97, hibou:98, smyrski:00, moskal-prc, moskal:02, moskal:02-3}
it was concluded~\cite{nakayama} that this process proceeds through the 
excitation of one of the protons to the S$_{11}$(1535) state which 
subsequently deexcites via the emission of the $\eta$ meson and a proton.
The crucial observations here were the large value of the 
absolute cross section and isotropic distributions of the angle 
of the $\eta$ meson emission in the reactions centre-of-mass 
system~\cite{moskal-prc,abdelbary:02}. However, there are plenty of 
possible scenarios of the excitation of S$_{11}$(1535). 
In fact, any of the $\pi$, $\eta$, $\omega$ or/and $\rho$ mesons may 
contribute to the resonance creation. Taking solely into account 
the total cross section one cannot deduce  
which one out of this rich spectrum of mesons 
plays the most important role 
in the excitation of the intermediate resonance.

Further investigations of the $\eta$ meson production process, 
namely the determination of the isospin dependence of the total cross section 
by the WASA/PROMICE collaboration~\cite{calen_pn}, put some restrictions 
to the abovementioned possibilities of the resonance excitation process.
The ratio \mbox{R$_{\eta}=\sigma(pn\to pn\eta)/\sigma(pp\to pp\eta)$} was 
determined to be about 6.5 in the excess energy range between 16 and 109~MeV, 
which revealed strong isospin dependence of the production process.
The production of the $\eta$ meson with the total isospin I~=~0
exceeds the production with the isospin I~=~1 by a factor of 12, suggesting~\cite{calen_pn} 
that the isovector meson exchange -- the $\pi$ or/and $\rho$ meson exchange --
is the dominant process in the 
excitation of the S$_{11}$ resonance. However, the relative contributions 
of the pseudoscalar $\pi$ meson and vector $\rho$ meson still remained to be 
determined.

Here the measurements of the polarisation observables can assist, 
because the predictions of the one boson exchange 
models~\cite{nakayama,wilkin} with respect to the most basic
polarisation observable as the analysing power
are sensitive to the type of meson being exchanged between the protons in order to excite 
one of them into the resonant state.     
Measurements of the beam analysing power for the 
$\vec{p}p\to pp\eta$ reaction performed by the DISTO collaboration~\cite{balestra}
as well as the tentative experiment of the COSY-11 group~\cite{winter:02-2,winter:02-1-en}
did not bring the univocal solution of this problem. The interpretation 
of results of the DISTO measurements, 
performed in the far-from-threshold region at the excess energies of 
Q~=~324, 412, and 554~MeV 
suffered from the lack of a theoretical prediction for the analysing power. 
This is due to the fact that far from the reaction threshold
the higher partial waves are involved in the reaction process, and so 
the theoretical description becomes more complicated. 
COSY-11 results at the excess energy of Q~=~40~MeV, due to the insufficient 
statistics, could not have been used in order to judge between 
the predictions of pseudoscalar and vector meson exchange models.

This gap is filled by the results of experiments described in this 
dissertation. For the first time ever the COSY-11 group was able 
to perform the measurements of the analysing power 
for the $\vec{p}p\to pp\eta$ reaction in the close-to-threshold region 
that has brought the answer to the open question of the $\eta$ meson production mechanism  
in hadronic collisions.

It is worth mentioning that the precise determination of this 
mechanism -- a process that 
proceeds through the strong interaction -- delivers information
about the nature of the strong forces.  
Thanks to the results 
presented in this dissertation the theoretical models might be revisited 
with new input parameters applied to these models: the coupling 
constants in the description of the production process of the $\eta$ meson, 
the initial and final state interactions and also dimensions of the reaction region.

Experiments presented in this work have been performed
by the COSY-11 collaboration
by means of the COSY-11 facility~\cite{Brauksiepe,klaja,smyr}
at the COoler SYnchrotron and storage ring COSY~\cite{Maier,webcosy}
in the Research Center J\"ulich in Germany. The analysing powers have been measured 
during two runs at different beam momenta: p$_{beam}=2.010$~GeV/c (May 2003) 
and 2.085~GeV/c (September 2002), which for the 
$\vec{p}p\to pp\eta$ reaction correspond to the excess energies 
of Q~=~10 and 36~MeV, respectively.

\vspace{1cm}

This thesis consists of eight chapters. In the following one 
the basic definitions are given, which are important in the understanding of the general methods 
that are applied to the experiments with 
polarised beams. In particular the definitions of the spin of a quantum particle,
the polarisation vector, and the analysing power are
given. We introduce the reference frames used 
in the polarisation experiments, in particular
we show how to construct the so called {\it Madison frame}  
following the Madison convention. This is relevant 
in order to determine the sign of the analysing power. 
 At the end of this part of the thesis we derive the 
formula which will be used to determine the analysing power. 

In Chapter~\ref{theory} an overview of the theoretical 
models of the $\eta$ meson production in hadronic collisions 
is sketched. Superimposed are the phenomenological models 
working within the {\it one boson exchange} frame. 
A special emphasis is put on the pseudoscalar meson 
exchange model of Nakayama et al.~\cite{nakayama} and the vector 
meson exchange model of F\"aldt and Wilkin~\cite{wilkin} as these are the only 
models giving the predictions of the analysing power  
for the $\vec{p}p\to pp\eta$ process in the close-to-threshold 
region. The summary of the so far measured observables for 
the $pp\to pp\eta$ reaction and the conclusions from these 
measurements are given in this chapter.

The general description of the cooler synchrotron COSY
along with the method of the formation and acceleration 
of the polarised proton beam are elucidated in Chapter~\ref{experiment}. 
Also the experimental facility COSY-11 is briefly described.  
The detector setup and the method of analysis
has been previously presented in various dissertations~\cite{moskal4,Wolke,winter-phd}
so here we will only focus on the  description of 
the detectors which are interesting from the point of view of 
the analysis of the \mbox{$\vec{p}p\to pp\eta$} reaction.

In Chapter~\ref{dat_anal} the calibration of the detectors 
needed to register the $\vec{p}p\to pp\eta$ process
is described. Also the methodology of the analysis
is given i.e. the missing mass technique, 
identification of the beam spin mode from the number of 
elastically scattered events, the methods of determination 
of the production yields, relative luminosity, and the beam polarisation. 
At the end of this chapter the final results of both analyses are presented.

Chapter~\ref{interpretation} contains the confrontation of the 
theoretical models with the experimentally determined 
values of the analysing power. Using statistical inference 
we conclude about the production mechanism 
of the $\eta$ meson in hadronic collisions.

The possibilities of extending the experiments presented in this 
work are discussed in Chapter~\ref{perspectives}.

In Chapter~\ref{summary} we summarize the thesis.   
Conclusions from the analyses presented in this dissertation 
can be found in this part.

Three appendices have been added at the end of this dissertation 
in order to elucidate in more detail some issues discussed in the text.
We shall explain the differences between the pseudoscalar and 
vector mesons (Appendix~\ref{mezony}), prove a  
property of the analysing power (Appendix~\ref{prove}) postulated 
and used in the analysis in Chapter~\ref{interpretation}. 
We will also demonstrate that the elastic scattering in the polarisation 
plane does not depend on the degree of polarisation 
(Appendix~\ref{parity}), a property which 
originates from the parity invariance rule and is used 
in Section~\ref{mmm} for the calculation of the relative luminosity. 
  
\newpage
\clearpage
\chapter{Definitions} 
\pagestyle{fancy}

\label{theory}

\section{
Spin and the beam polarisation
} 
\label{definitions}
\vspace{3mm}
{\small The formal quantum mechanical description of spin 1/2 particle is given. 
The polarisation vector and polarisation plane are defined.}
\vspace{5mm}

The {\it spin} of a particle is an internal degree 
of freedom, which is governed by the same equations 
of motion as the angular momentum. Therefore, we can assign to  
observable ``spin" a vectorial operator ${\bf \hat{S}}=[\hat{S_x},\hat{S_y},\hat{S_z}]$, 
where $\hat{S_i}$ (i=x,y,z) are the hermitian operators.   
Subsequently we will only consider the spin-1/2 particles. 
In this case the spin operators can be expressed in terms of 
the Pauli matrices $\sigma_i$:
\begin{equation}
\hat{S_i} = \frac{\hbar}{2} \sigma_i, 
\label{operators}
\end{equation} 
where
\begin{equation} 
\sigma_x = \begin{pmatrix} 0 & 1 \\ 1 & 0 \end{pmatrix} ,
\sigma_y = \begin{pmatrix} 1 & 0 \\ 0 & -1 \end{pmatrix} ,
\sigma_z = \begin{pmatrix} 0 & -i \\ i & 0 \end{pmatrix} .
\label{matrices}
\end{equation}

With respect to the arbitrary quantization axis $Oy$, a
spin 1/2 particle can be found in one of the two eigenstates
of $\hat{S_y}$. These eigenstates may be represented by the 
eigenfunctions  
\begin{equation}
\chi_{+} = \begin{pmatrix} 1 \\ 0 \end{pmatrix},  \chi_{-} = \begin{pmatrix} 0 \\ 1 \end{pmatrix}.
\label{eigenfunctions}
\end{equation}
The $\chi_{\pm}$ are the eigenfunctions 
of $\hat{S_y}$ with the eigenvalues $\pm\hbar/2$.
In what follows we will refer
to the $\chi_{+}$ and $\chi_{-}$ states as to the ``spin up" and ``spin down" states, respectively.

Let us denote the fractions 
of protons in the pure spin up and spin down states by $n_{+}$ and $n_{-}$, respectively. 
Define a {\it polarisation $P_y$} along the $Oy$ quantization axis
as the asymmetry of the populations $n_{+}$ and $n_{-}$:
\begin{equation}
P_y = \frac{n_{+}-n_{-}}{n_{+}+n_{-}}.
\label{polll}
\end{equation}
Analogously we could define the polarisations $P_x$ and $P_z$ with respect to the $Ox$ and $Oz$
axes, but we will assume that these two cancel out in the experiment
described in this thesis.  
From these properties we can construct a {\it polarisation vector} ${\bf\vec{P}}$
defined as follows: 
\begin{equation}
{\bf\vec{P}}\equiv[P_x,P_y,P_z]=[0,P,0].
\label{polar_vector}
\end{equation}  
The modulus of ${\bf\vec{P}}$ shall be called the {\it degree of polarisation}. 
A single spin~1/2 particle in a pure spin state $\chi_{+}$ or $\chi_{-}$ 
is fully polarised, and for such a particle $|{\bf\vec{P}}|=1$, which 
follows from definition~\ref{polll} and the fact that single fermions can be found 
only in one out of two states: $\chi_{+}$ or $\chi_{-}$. 
In reality we 
deal with the systems containing a large number of particles, and for
such systems the degree of polarisation 
is usually smaller than 1.

By the {\it polarisation plane} we will refer to a plane spanned by the beam momentum
vector ${\bf\vec{p}_{beam}}\equiv [0,0,p_z]$ and a polarisation vector
${\bf\vec{P}}=[0,P,0]$.
\label{kosne}

\section{
Reference frames 
}
\label{refi}
\vspace{3mm}
{\small The reference frames 
along with the angles of production are defined.
}
\vspace{5mm}

We will describe the polarisation observables
within the Cartesian coordinates, however in the beginning we have to introduce
the certain reference frames that are convenient to this description.

\begin{figure}[H]
  \unitlength 1.0cm
        \begin{center}
  \begin{picture}(14.5,5.5)
    \put(3.5,0.0){
      \psfig{figure=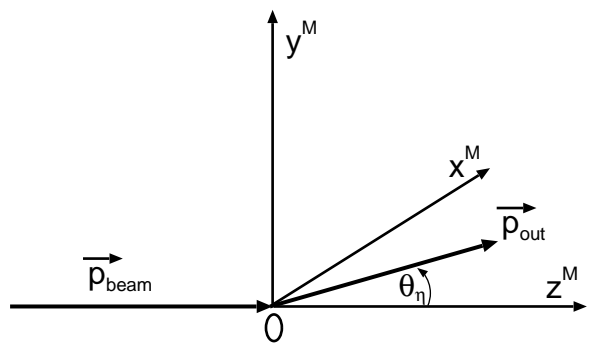,height=4.5cm,angle=0}
    }
  \end{picture}
  \caption{ {\small Definition of the reference system according to the Madison
                convention for the description of the polarisation
                observables. }
 \label{dupa}
  }
        \end{center}
\end{figure}

According to the Madison convention~\cite{madison} we construct the
reference system (${\bf\hat{x}^M},{\bf\hat{y}^M},{\bf\hat{z}^M}$) 
in the following manner: let a unit vector ${\bf\hat{z}^M}$ of the $Oz^M$-axis be parallel to the
beam momentum direction given by the vector ${\bf \vec{p}_{beam}}$,
and the $Oy^M$-axis unit vector ${\bf \hat{y}^M}$ along
the direction of the ${\bf \vec{p}_{beam} \times \vec{p}_{out}}$ vector, where
${\bf \vec{p}_{out}}$ is the momentum vector of the $\eta$ meson.
Both ${\bf \vec{p}_{beam}}$ and ${\bf \vec{p}_{out}}$ vectors
are defined 
in the centre-of-mass (CM) frame. The ${\bf\hat{x}^M}$ -- a
unit vector along the $Ox^M$-axis is defined such that the basis 
(${\bf\hat{x}^M},{\bf\hat{y}^M},{\bf\hat{z}^M}$) forms an
orthonormal, right-handed set. Such defined reference frame is sketched in Fig.~\ref{dupa}.
We will refer to this frame as to the {\it Madison frame}.

The angle $\theta_{\eta}$ will be called the {\it polar 
angle of $\eta$ production}.
This is an angle between 
${\bf \vec{p}_{out}}$ and ${\bf\hat{z}^M}$:
\begin{equation}
\cos(\theta_{\eta}) = \frac{\vec{p}_{out} \cdot \hat{z}^M}{|\vec{p}_{out}|}. 
\label{theta_definition}
\end{equation}

\begin{figure}[H]
  \unitlength 1.0cm
  \begin{picture}(15.0,5.5)
      \put(0.00,1.0){
         \psfig{figure=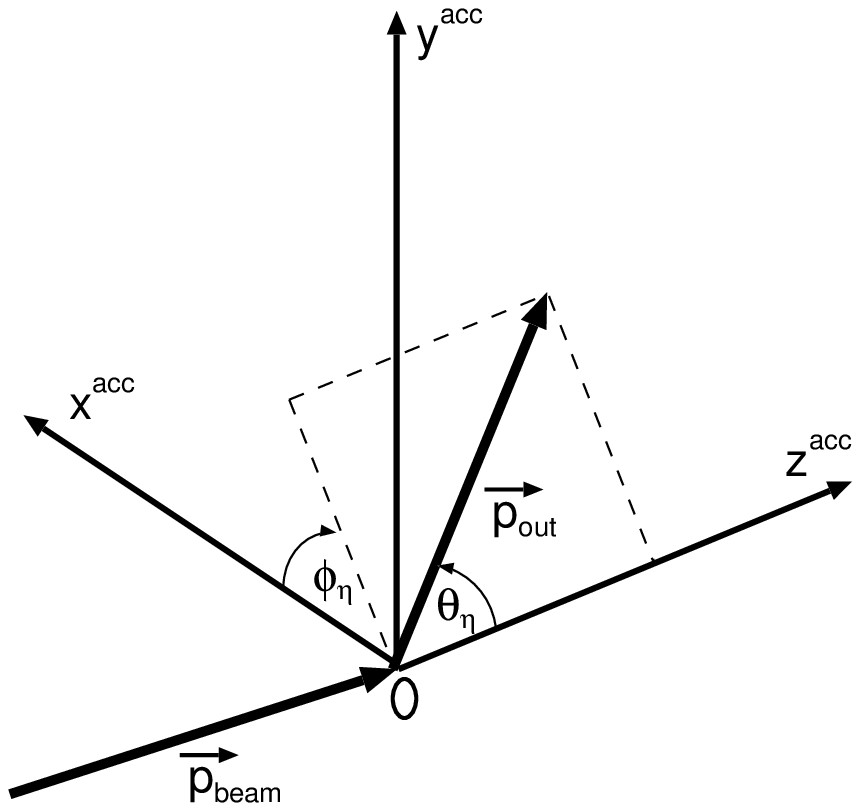,height=4.5cm,angle=0}
      }
      \put(5.00,1.0){
         \psfig{figure=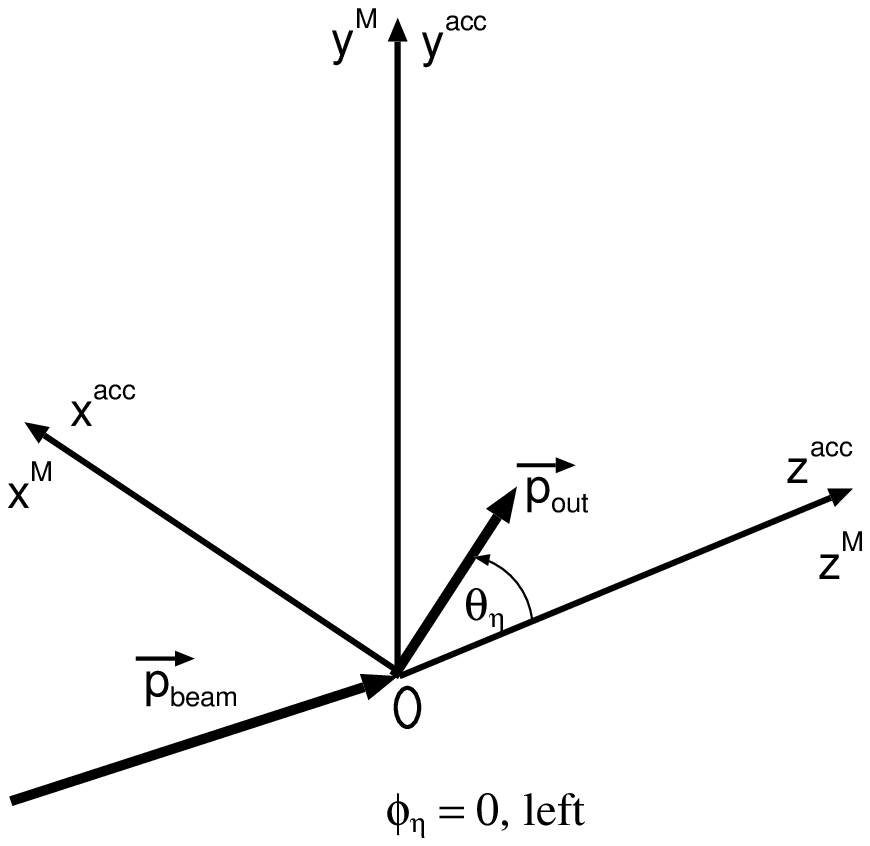,height=4.5cm,angle=0}
      }
      \put(10.0,1.0){
         \psfig{figure=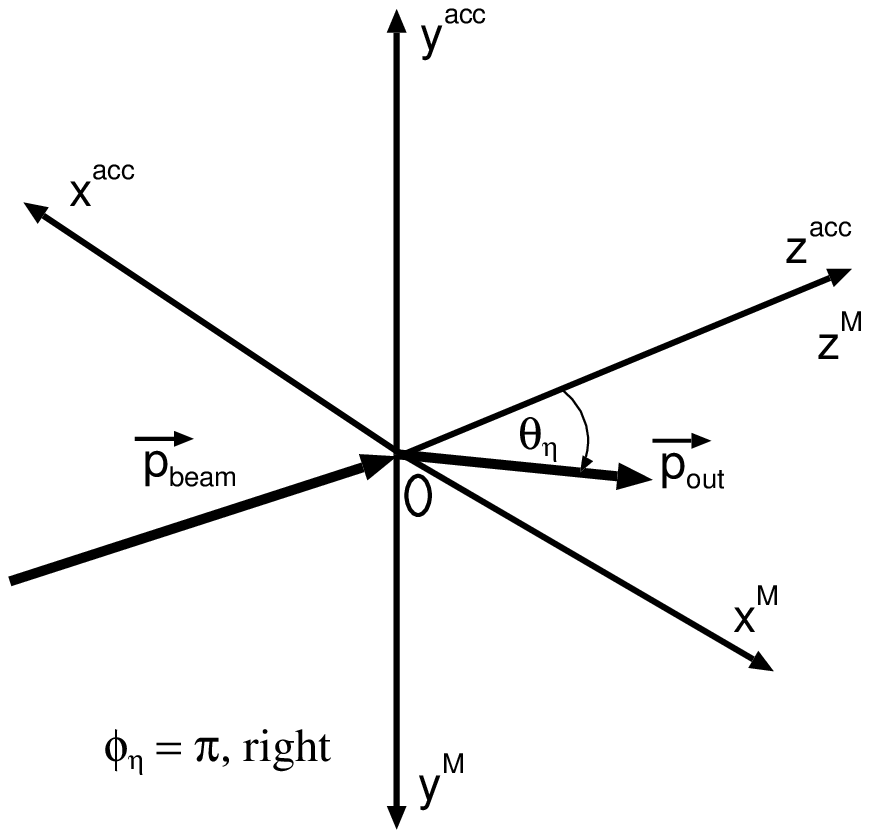,height=4.5cm,angle=0}
      }
      \put(4.0,0.8){{\normalsize {\bf a)}}}
      \put(9.0,0.8){{\normalsize {\bf b)}}}
      \put(14.0,0.8){{\normalsize {\bf c)}}}
  \end{picture}
  \caption{ \small{ (a) Accelerator reference frame.
		    Definition of the scattering to the left (b) and 
		    to the right (c) with respect to the polarisation plane.  } 
  \label{3rrr}
  }
\end{figure}

Now, we introduce the fixed in the centre-of-mass an {\it accelerator coordinate frame}
$(\bf\hat{x}^{acc},\bf\hat{y}^{acc},\bf\hat{z}^{acc})$. 
Define $\bf\hat{z}^{acc}$ --
a unit vector that is parallel
to the beam momentum, 
a unit vector $\bf\hat{y}^{acc}$ perpendicular to the
accelerator plane, pointing up, and
$\bf\hat{x}^{acc}$ which completes the right-handed basis.
The accelerator frame is shown in Figure~\ref{3rrr}.a.

In the case of the accelerator frame 
the polar angle of $\eta$ -- $\theta_{\eta}$ -- is the same as in 
the case of Madison frame, as the $Oz^M$ and $Oz^{acc}$
axes are defined in the same manner. 

By the {\it azimuthal angle of the $\eta$ meson emission} -- $\phi_{\eta}$ -- we will 
understand the angle between the ${\bf\hat{x}^{acc}}$ vector and the projection of ${\bf \vec{p}_{out}}$  
onto the x$^{acc}$-y$^{acc}$ plane.
Both angles 
are presented in Figure~\ref{3rrr}.a.
If the momentum of the $\eta$ meson is lying in the horizontal plane, 
i.e. if $\phi_{\eta}=0$ or $\phi_{\eta}=\pi$, then the 
$Oy^M$ and $Oy^{acc}$ axes (and also $Ox^M$ and $Ox^{acc}$ axes) may be either 
parallel or antiparallel,
depending on whether the $\eta$ meson is emitted to the 
left (Fig.~\ref{3rrr}.b) or to the right side (Fig.~\ref{3rrr}.c) in the accelerator system.  

It is important to distinguish between the Madison and
the accelerator frame. Whereas the accelerator frame
is fixed in the centre-of-mass frame, the Madison coordinate system
may vary in space from event to event. According to the Madison convention  
all the physical observables, like for example
the beam polarisation vector, should be considered in the 
Madison frame in order to avoid the ambiguities in the determination 
of the sign of the analysing power.  
\section{
Analysing power
} 
\label{biecz}
\vspace{3mm}
{\small 
A five-dimensional phase space for the description of the $NN\to NNM$
reactions is defined. The most general formula for 
the cross section for experiments with polarised beam and 
target is given, which contains the definition of 
the analysing power. 
}
\vspace{5mm}

For a nuclear reaction with a given initial channel and
a three body final state 
twelve parameters (four-momenta of three particles) have to be known to fully describe the 
exit channel. Some of these twelve parameters are bound
with the relativistic formula:  
\begin{equation}
E^2_i=m^2_i+({\bf\vec{p}_i})^2, \hspace{1cm} i=1,2,3 ; 
\label{einstein}
\end{equation} 
where $E_i$, $m_i$, ${\bf\vec{p}_i}$ denote
the total energy, mass, and the momentum vector of the i-th particle. 
Equations~\ref{einstein} reduce the number of variables to nine, 
and these variables are dependent on the initial 
state parameters on the basis of the four-momentum conservation (four additional equations).
Therefore, there are eventually five linearly independent variables 
that need to be measured for each reaction in order to
fully describe its kinematics. Here, we follow the references~\cite{moskal-prc,moskal-hab}
in the choice of these kinematical variables.
We will use the invariant masses of the proton-proton system $m_{pp}$ and proton-$\eta$
system $m_{p\eta}$, the polar $\theta_{\eta}$ and azimuthal $\phi_{\eta}$ angles  
of the $\eta$ momentum in the centre-of-mass (CM) frame, as well as the 
angle $\psi$ -- describing the rotation around the direction 
established by the momentum of the $\eta$ meson~\cite{moskal-hab}.
These variables are orthogonal and form a basis in the five-dimensional 
phase space. Let this basis be denoted by 
\mbox{$\zeta=\{m_{pp},m_{p\eta},\phi_{\eta},\theta_{\eta},\psi\}$}.

Let us denote by
\begin{equation}
\sigma_0(\zeta)\equiv \frac{d^5\sigma_0(\zeta)}{dm_{pp}dm_{p\eta}d\phi_{\eta}d\theta_{\eta}d\psi}
\label{przekroj0}
\end{equation} 
the differential cross section for the reaction in absence of 
the polarisation of beam and target, and since the internucleon strong interaction is spin-dependent,
let us also define 
\begin{equation}
\sigma(\zeta,{\bf\vec{P}},{\bf\vec{Q}})\equiv \frac{d^5\sigma(\zeta,{\bf\vec{P}},{\bf\vec{Q}})}{dm_{pp}dm_{p\eta}d\phi_{\eta}d\theta_{\eta}d\psi}
\label{przekroj}
\end{equation}
as the differential cross section for the reaction induced by a polarised 
beam on the polarised target, where ${\bf \vec{P}}$ and ${\bf \vec{Q}}$
are the polarisation vectors of beam and target, respectively.
{\bf The beam polarisation vector ${\bf\vec{P}}$
and the target polarisation vector ${\bf\vec{Q}}$
should be referred to in the Madison frame} defined in 
Section~\ref{refi}.
In the most general case, when both ${\bf \vec{P}}$ and ${\bf \vec{Q}}$
are the non-zero vectors, the formula for $\sigma$ reads:
\begin{equation}
\sigma(\zeta,{\bf\vec{P}},{\bf\vec{Q}})=\sigma_0(\zeta)(1+\sum_{i=1}^3 P_iA_i(\zeta)+\sum_{i=1}^3 Q_iA'_i(\zeta)+\sum_{i,j=1}^{3}c_{ij}({\bf\vec{P}},{\bf\vec{Q}})C_{ij}(\zeta)),
\label{cross_section}
\end{equation}
where A$_i$ and A$_i^{\prime}$ are the beam and target analysing powers, respectively,
and 
C$_{ij}(\zeta)$ are the spin correlation coefficients. $P_i$, $Q_i$ and $c_{ij}$ 
are the expansion coefficients, which depend on the polarisation of beam and target.
In our case, where 
the target is unpolarised, 
the $Q_i$ and $c_{ij}$ coefficients disappear and we only deal with
the $A_i$ vector.
Hence Equation~\ref{cross_section}
simplifies to:
\begin{equation}
\sigma(\zeta,{\bf\vec{P}})=\sigma_0(\zeta) (1+\sum_{i=1}^3 P_i A_i(\zeta)). 
\label{cross_section_2}
\end{equation}

For in Equation~\ref{cross_section_2} the scalar product
of ${\bf\vec{P}}$ and ${\bf\vec{A}}$ is present, in the case of a transversally 
polarised beam we can rewrite this equation
in the following manner:
\begin{equation}
\sigma(\zeta,{\bf\vec{P}})=\sigma_0(\zeta) (1 + P A_y(m_{pp},m_{p\eta},\theta_{\eta},\psi) \cos(\phi_{\eta})),
\label{ay_with_cos}
\end{equation}
where $P$ is the component of the beam polarisation normal
to the incident beam, 
such that the $y$-component of polarisation vector expressed in the Madison frame
is given by $P_y = P \cos\phi_{\eta}$. 

\begin{figure}[H]
  \unitlength 1.0cm
        \begin{center}
  \begin{picture}(14.0,8.0)
    \put(2.8,0.0){
      \psfig{figure=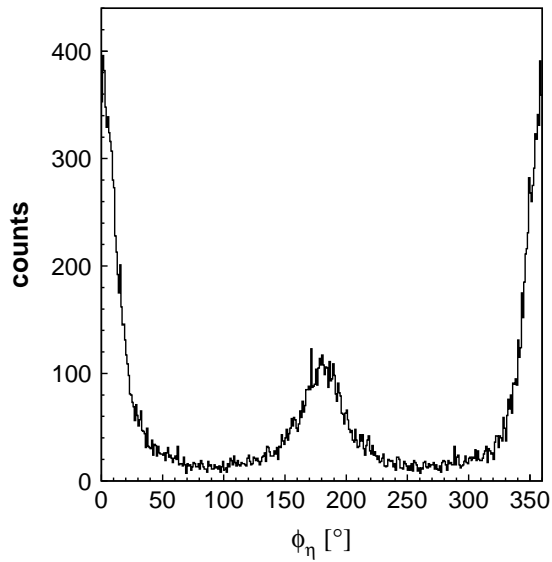,height=8.0cm,angle=0}
    }
  \end{picture}
  \caption{ {\small Distribution of the $\vec{p}p\to pp\eta$ events which can 
	be registered with the \mbox{COSY-11} setup as a function of the 
	azimuthal angle of the $\eta$ meson production in the 
	centre-of-mass system -- $\phi_{\eta}$. 
                 }
 \label{bialka}
  }
        \end{center}
\end{figure}

At this point we would like to mention that the acceptance 
of the \mbox{COSY-11} facility allows to register only events 
scattered near the horizontal plane. In the analysis the 
azimuthal angle $\phi_{\eta}$ was restricted to values 
of $\cos\phi_{\eta}$ ranging between 0.87 and 1. This is 
depicted in Figure~\ref{bialka}, where we show the shape of the acceptance
of the \mbox{COSY-11} system for the registration of the events 
with the $\eta$ meson production as a function of the azimuthal 
angle $\phi_{\eta}$.  
One can see that the acceptance for events with
\mbox{$\phi_{\eta}\in\left[30^{\circ},330^{\circ}\right]$} 
is relatively very small. Moreover, the acceptance 
is much larger for the cases where the $\eta$ meson 
is emitted to the left ($\phi_{\eta}\approx 0^{\circ}$)
than when it is emitted to the right side of the polarisation 
plane ($\phi_{\eta}\approx 180^{\circ}$).
Therefore we decided to consider in our analyses only the events
with the $\eta$ meson production to the left side with respect to the polarisation plane. 
For such a case \mbox{$\cos\phi_{\eta}\approx 1$},
and considering that the beam was vertically polarised along 
the $Oy^{acc}$ axis (polarisation vector given by Equation~\ref{polar_vector})
we can rewrite Formula~\ref{ay_with_cos}:
\begin{equation}
\sigma(m_{pp},m_{p\eta},\theta_{\eta},\psi,P)\approx \sigma_0(m_{pp},m_{p\eta},\theta_{\eta},\psi) (1 + P A_y(m_{pp},m_{p\eta},\theta_{\eta},\psi)).
\label{cross_3}
\end{equation}

It is worth mentioning that in the case 
of the $\vec{p}p\to pp\eta$ reaction $A_x(\zeta)$ and $A_y(\zeta)$
are equivalent, because the rotation of the whole system 
by $\pi$/2 around the $Oz$ axis does not lead to a change of the scattering
pattern. Therefore, in case of spin 1/2 particles, there are only two 
independent polarisation observables: $A_y(\zeta)$ and $A_z(\zeta)$ -- the 
transversal and longitudinal analysing power, the latter 
being of no significance in the experiments considered here.

From Equation~\ref{cross_3} it follows that the vector analysing power $A_y(\zeta)$
may be understood as a measure of the relative deviation between the
differential cross section for the experiments with and without polarised beam
(in the absence of the target polarisation), normalized 
to the beam polarisation:
\begin{equation}
A_y(\zeta)=\frac{1}{P}\frac{\sigma(\zeta,P)-\sigma_0(\zeta)}{\sigma_0(\zeta)}.
\label{definicjaAy}
\end{equation}

 
\section{
Practical formula for the analysing power
}
\label{form}
\vspace{3mm}
{\small 
The practical recipe for the calculation of the analysing power is derived.  
}
\vspace{5mm}

In this section we consider the meson production 
where the beam consists of 
vertically polarised protons  and the 
production takes place on an unpolarised proton 
target.  

\begin{figure}[H]
  \unitlength 1.0cm
        \begin{center}
  \begin{picture}(14.5,5.5)
    \put(1.1,0.0){
      \psfig{figure=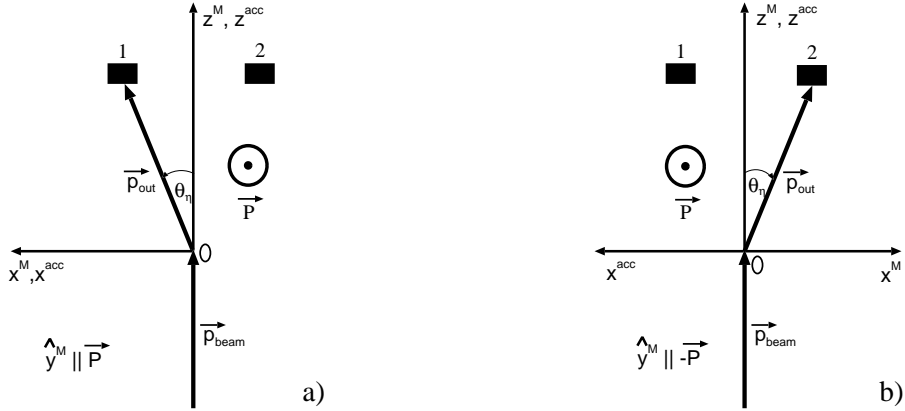,height=5.5cm,angle=0}
    }
  \end{picture}
  \caption{ \small {Schematic picture of the $\eta$ meson production
to the left (a) and right side (b) 
with respect to the polarisation 
plane. The polarisation vector $\bf{\vec{P}}$ is pointing 
up (out of the paper), 
which is denoted by 
$\bigodot$. Black squares are 
the virtual $\eta$ ``detectors".}  
 \label{sss}
  }
        \end{center}
\end{figure}

Naturally, the Madison frame defined in Section~\ref{refi}
is not fixed in space, as is schematically 
shown in Figure~\ref{sss}. In the left panel 
there is depicted a situation when the $\eta$
meson is produced to the left side with respect 
to the polarisation plane, whereas the right panel
shows the production of the  
$\eta$ meson to the right side with respect to the polarisation plane, 
defined in Section~\ref{definitions}. 
While the Madison frame changes depending on $\vec{p_{\eta}}$ the 
accelerator frame remains unalterred in space. 
{\bf One should also notice that the 
production of the $\eta$ meson to the left (right) with the polarisation vector 
of the beam pointing up in the accelerator frame is 
physically equivalent to the production of the $\eta$ meson to 
the right (left) with the beam polarisation vector pointing down.} 
In both cases the physics, meaning the 
spin-dependent nuclear interaction, is the same 
and the two abovementioned cases are equivalent, from 
which we will make use in due course.  

In Figure~\ref{sss} we presented schematically the 
production of the $\eta$ meson in the accelerator frame's plane x$^{acc}$-z$^{acc}$.
The polarisation vector of the beam of protons is pointing up, i.e. along the y$^{acc}$ axis. 
Denoting by $N(\theta_{\eta},\phi_{\eta})$\footnote{ We remind, that the
angles $\theta_{\eta}$ and $\phi_{\eta}$ are referred to the accelerator frame. } the number of the $\eta$ mesons
emitted into the solid angle around $\theta_{\eta}$ and $\phi_{\eta}$
and by N$_1$ and N$_{2}$ the numbers of the $\eta$ mesons reaching
the virtual $\eta$ ``detectors" 1 and 2 
we can write:
\begin{equation}
N_{1}^{\uparrow} = N^{\uparrow}(\theta_{\eta},0) \equiv N^{\uparrow}_{+} = \sigma_0(\theta_{\eta}) (1+P^{\uparrow}A_y(\theta_{\eta})) E(\theta_{\eta},0)  \int L^{\uparrow}dt, 
\label{aa}
\end{equation}
and
\begin{equation}
N_{2}^{\uparrow} = N^{\uparrow}(\theta_{\eta},\pi) \equiv N^{\uparrow}_{-} = \sigma_0(\theta_{\eta}) (1-P^{\uparrow}A_y(\theta_{\eta})) E(\theta_{\eta},\pi)  \int L^{\uparrow}dt, 
\label{bb}
\end{equation}
where the symbol ``$\uparrow$" stands for the spin up orientation, 
$\int L^{\uparrow}dt$ is the luminosity during the measurement with spin up, integrated 
over the whole time of the measurement, $E(\theta_{\eta},\phi_{\eta})$ is the efficiency 
of the detection setup for ``registering" the $\eta$ mesons produced into the 
solid angle around $\theta_{\eta}$ and $\phi_{\eta}$, $\sigma_0(\theta_{\eta})$ is the cross 
section for the $\eta$ meson production with absence of the beam polarisation, 
$P^{\uparrow}$ is the beam polarisation during the spin up cycles
and $A_y(\theta_{\eta})$ is the analysing power for the $\eta$ production
averaged over $m_{pp}$, $m_{p\eta}$ and $\psi$.  
The ``+" sign 
denotes the production with the polarisation vector along the 
$Oy^{M}$-axis (according to the Madison convention $P_y$ is positive in this case), and the ``$-$" sign 
is for the production with the polarisation vector 
antiparallel to the $Oy^{M}$-axis (negative $P_y$).

Now, if we flipped the polarisation vector i.e. if we made it 
pointing down (into the paper 
in Figure~\ref{sss}) we would get: 
\begin{equation}
N_{1}^{\downarrow} = N^{\downarrow}(\theta_{\eta},0) \equiv  N^{\downarrow}_{-} = \sigma_0(\theta_{\eta}) (1-P^{\downarrow}A_y(\theta_{\eta})) E(\theta_{\eta},0) \int{L^{\downarrow}dt},
\label{cc}
\end{equation}

\begin{equation}
N_{2}^{\downarrow} = N^{\downarrow}(\theta_{\eta},\pi) \equiv N^{\downarrow}_{+} = \sigma_0(\theta_{\eta}) (1+P^{\downarrow}A_y(\theta_{\eta})) E(\theta_{\eta},\pi) \int{L^{\downarrow}dt},
\label{dd}
\end{equation}
where the quantities with arrows pointing down refer to the respective
quantities from Equations~\ref{aa} and~\ref{bb}, but for the cycles with spin down mode.

From Equations~\ref{aa} and~\ref{bb} we obtain:
\begin{equation}
A_y(\theta_{\eta}) = \frac{1}{P^{\uparrow}} \frac{N^{\uparrow}_{+}/E(\theta_{\eta},0)-N^{\uparrow}_{-}/E(\theta_{\eta},\pi)}{N^{\uparrow}_{+}/E(\theta_{\eta},0)+N^{\uparrow}_{-}/E(\theta_{\eta},\pi)}.
\label{wniosek}
\end{equation} 
The same is valid for Equations~\ref{cc} and~\ref{dd}, but we have to 
change $\uparrow$ into $\downarrow$ in Equation~\ref{wniosek}.
Note, that Equation~\ref{wniosek}
is independent of the integrated luminosity.
Equation~\ref{wniosek} shows that the analysing power $A_y(\theta_{\eta})$ may be calculated
from the production with only one spin orientation (spin up or spin down), if 
both scatterings -- into left and right side in the accelerator 
plane -- can be measured. To some extend it may be 
realised by the COSY-11 detector. 
However, for the COSY-11
detector setup $E(\theta_{\eta},0) \gg E(\theta_{\eta},\pi)$ 
(see Figure~\ref{bialka}, where between \mbox{$\phi_{\eta}\in\left[330^0,30^0\right]$}
there are much more events than for \mbox{$\phi_{\eta}\in\left[150^0,210^0\right]$}), 
and therefore in practice we have to perform the measurement
with two beam spin orientations.

Assuming that the degree of polarisation during the cycles with spin down is equal to 
the degree of polarisation for cycles with spin up, which is the case 
within the $2\%$ of accuracy\footnote{See also Section~\ref{br}.} 
as has been shown in the previous measurements by means 
of the EDDA polarimeter~\cite{altmeier}, we may introduce the
average degree of polarisation $P$:
\begin{equation}
P \approx P^{\uparrow} \approx P^{\downarrow} \approx \frac{P^{\uparrow}+P^{\downarrow}}{2}.
\end{equation}   

%
%
%

Assuming that the degrees of polarisation for spin up 
and down are equal, the asymmetry between $N^{\uparrow}_{+}$ and
$N^{\downarrow}_{-}$ may be used as a measure
of the spin-dependent $\eta$ meson production.
Dividing Equation~\ref{aa} by Equation~\ref{cc}, 
and replacing $P^{\uparrow}$ and $P^{\downarrow}$ by $P$ yields to: 
\begin{equation}
\frac{N^{\uparrow}_{+}(\theta_{\eta})}{N^{\downarrow}_{-}(\theta_{\eta})} = \frac{\int{L_{\uparrow}dt}}{\int{L_{\downarrow}dt}} \hspace{0.2cm} \frac{1+P A_y(\theta_{\eta})}{1-P A_y(\theta_{\eta})}.
\label{kk}
\end{equation} 
Introducing the relative luminosity defined as: 
\begin{equation}
L_{rel}\equiv \frac{\int{L_{\uparrow}dt}}{\int{L_{\downarrow}dt}}, 
\label{lumi_rel}
\end{equation}
and solving Equation~\ref{kk} for A$_y$ we obtain:
\begin{equation}
A_y(\theta_{\eta}) = \frac{1}{P} \frac{N^{\uparrow}_{+}(\theta_{\eta})-L_{rel} N^{\downarrow}_{-}(\theta_{\eta})}{N^{\uparrow}_{+}(\theta_{\eta})+L_{rel} N^{\downarrow}_{-}(\theta_{\eta})},
\label{anal_part}
\end{equation}
a formula which is independent of the efficiency of the 
detection setup. 
In what follows, Formula~\ref{anal_part} 
will be used 
for the calculation of the analysing power.

\newpage
\clearpage
\pagestyle{fancy}
\chapter{Theory} 

\label{theory}

\section{
Theoretical models of the $\eta$ meson production in nucleon-nucleon 
collisions
} 
\label{mechanism}

\vspace{3mm}
{\small
Some phenomenological models of the $\eta$ meson production 
in the $pp\to pp\eta$ reaction are shortly reviewed. Conclusions
from the analysis of the available physical observables
along with the predictions for the analysing power 
are pointed out. 
}
\vspace{5mm}

Despite the fact that the discovery of the $\eta$ meson took place over forty years ago
\cite{pevsner:61}, its production mechanism still remains an open question~\cite{hanhard, moskal_magnus}.
After early measurements of the total cross sections for the $pp\to pp\eta$ reaction in 
bubble chamber experiments~\cite{pic,ale,col,bod,cas,alm,colt,yek}, only recently there appeared 
high-statistics, close-to-threshold data from storage rings, giving opportunity to investigate 
more accurately the 
structure, properties, production mechanism as well as the interaction 
of the $\eta$ meson with hadronic matter.  
The close-to-threshold total cross section measurements for the $pp\to pp\eta$ reaction
\cite{bergdolt:93, chiavassa:94, calen:96, calen:97, hibou:98, smyrski:00, moskal-prc, moskal:02, moskal:02-3},
investigations on the differential cross sections for this reaction
\cite{moskal-prc, abdelbary:02, calen:99, tatischeff:00, moskal:01-2}
and recently performed measurements of the analysing power for the
$\vec{p}p\to pp\eta$ reaction~\cite{balestra, winter:02-2}
made the theoretical analysis possible aiming in understanding the production process.

\subsection{Conclusions from the total cross section measurements}
\label{ggg}

\vspace{3mm}
{\small
From the total cross section measurements for the 
$pp\to pp\eta$ reaction it is inferred that amongst several 
possible scenarios the resonant current is the 
dominant one in the process of the $\eta$ meson production 
in proton-proton collisions. 
}
\vspace{5mm}

The majority of the theoretical models tries to elucidate the 
production of the $\eta$ meson within the framework 
of a one-boson-exchange formalism~\cite{nakayama, wilkin, germond, laget, santra, nakayama:01,
batinic:97, gedalin:98, vetter:91, pena:00, baru:02}, where two interacting nucleons
on the basis of a momentary energy violation  
exchange a virtual meson, which subsequently, under the 
interaction with nucleons turns into the $\eta$ meson. 
Some of the possible mechanisms which may lead to the 
$\eta$ meson creation in nucleon-nucleon collisions are
pointed out in Figure~\ref{mechanism} and will be described in the further part of this section. 
What is characteristic to this kind of approach to the meson production
is that the models do not enter into the quark-gluonic
structure of the meson and nucleons, but rather introduce 
the phenomenological parameters like for example the coupling constants for different channels, 
the scattering lengths, the effective ranges, and cut-offs. 
 
On the other hand there are also trials to explain the $\eta$ meson production mechanism 
on the basis of instanton models for QCD vacuum~\cite{kochelev:99, dillig:02}. 
In these QCD-oriented approaches the calculations are on the very elementary level,
where the effective degrees of freedom are not mesons and baryons like it was in the 
case of the one boson exchange models
but rather the constituent quarks and gluons.  
These models however are temporarily in the early stage of development, and so
we will deal here with the phenomenological meson-exchange models solely.

\begin{figure}[H]
  \unitlength 1.0cm
        \begin{center}
  \begin{picture}(14.5,5.5)
    \put(0.0,0.0){
      \psfig{figure=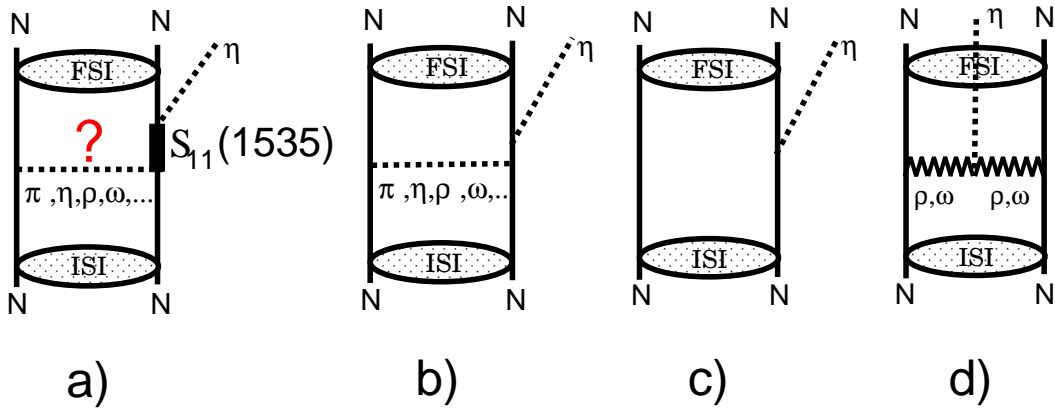,height=5.5cm,angle=0}
    }
  \end{picture}
  \caption{ \small {Possible mechanisms of the $\eta$ meson production in nucleon-nucleon collisions:
                                           (a) resonant currents, (b) nucleon currents, (c) direct production, 
                                           (d) mesonic currents.
 \label{mechanism_1}
  }}
        \end{center}
\end{figure}

\begin{figure}[H]
  \unitlength 1.0cm
  \begin{picture}(14.0,5.5)
      \put(-0.20,1.0){
         \psfig{figure=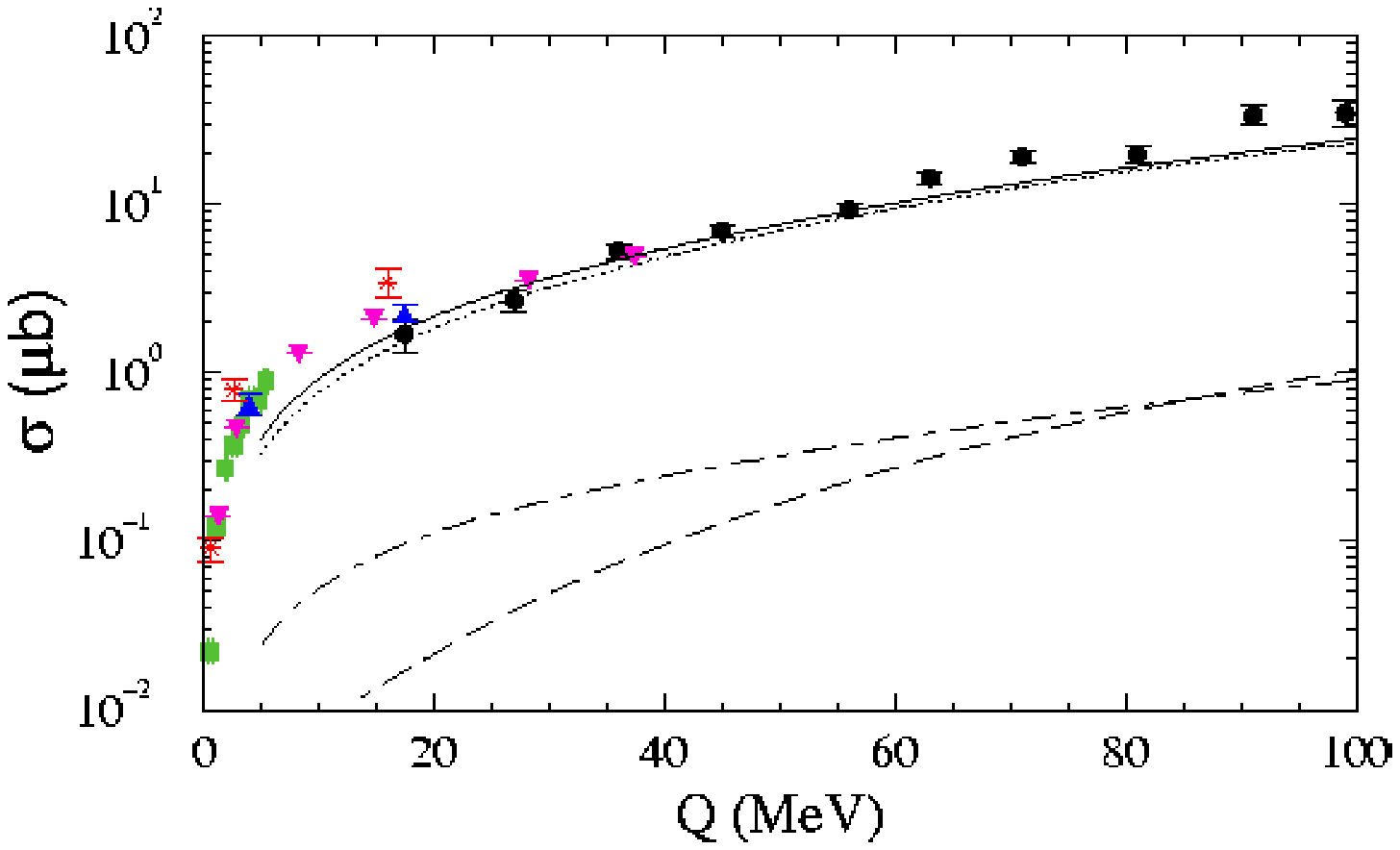,height=4.3cm,angle=0}
      }
      \put(7.30,0.7){
         \psfig{figure=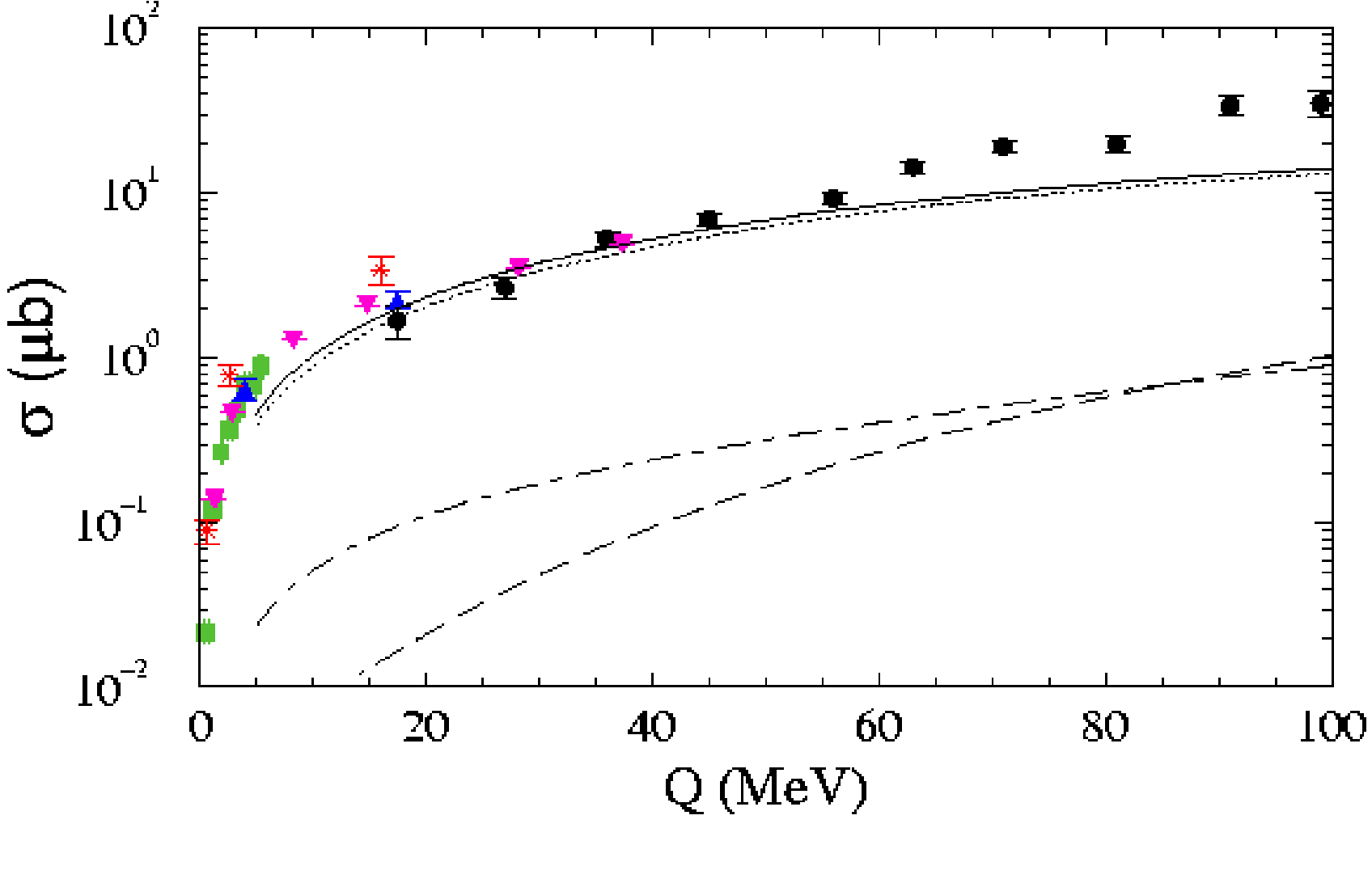,height=4.6cm,angle=0}
      }
      \put(6.3,0.2){{\normalsize {\bf a)}}}
      \put(13.8,0.2){{\normalsize {\bf b)}}}
  \end{picture}
  \caption{ \small{ Description of the close-to-threshold total cross section
		data for the $pp\to pp\eta$ reaction by means of the
		pseudoscalar meson exchange model (a) and vector meson 
		exchange model (b). Data are from references~\cite{chiavassa:94, hibou:98, calen:96, 
		smyrski:00, calen:97}. The dotted curves are the resonance current contributions, 
		the dashed lines represent the nucleonic current contribution, while 
		the dashed-dotted curves are the mesonic current contribution. The solid 
		lines show the total contribution.  
		The figure is adapted from~\cite{nakayama}.
                  }
  }
  \label{comparison}
\end{figure}

In Figure~\ref{mechanism_1} there are depicted some of the possible scenarios
for the $\eta$ meson production in the nucleon-nucleon collisions. 
Figure~\ref{mechanism_1}.a shows the resonant current, where
the $\eta$ meson is produced
in a two-step process: in the first stage the exchange of one of the pseudoscalar or vector
mesons excites the nucleon to the S$_{11}$(1535) resonance
and subsequently in the second step this resonance decays into a proton-$\eta$ pair.
Apart from the S$_{11}$(1535) resonance, also 
the contributions from the other possible resonances are considered. For example in  
the model reported in the reference~\cite{nakayama}
also P$_{11}$(1440) and D$_{13}$(1520) states,  
excited by the exchange of $\pi$, $\eta$, $\rho$, and $\omega$ mesons have been included,
however, the contribution from the S$_{11}$(1535) excitation
by the $\pi$ and $\eta$ mesons has been found to be dominant.
The S$_{11}$(1535) resonance seems to play an important role
as an intermediate state since it has a large width, covering
the threshold energy for the $pp\to pp \eta$ reaction, and couples
strongly to the proton-$\eta$ system with the branching ratio
corresponding to 30--55\%~\cite{eidelman}.
The mechanism of the excitation of the colliding proton to the  
S$_{11}$(1535) resonance remains an open issue. 
The authors of~\cite{wilkin, santra, gedalin:98} have found the 
$\rho$ meson exchange to play the dominant role in this
excitation. 
It has been stated that the $\rho$ meson exchange is 
particularly important to explain the shape of the 
angular distribution of the $pp\to pp\eta$ reaction.
On the other hand, the authors of article~\cite{nakayama}
showed that the excitation function of the total cross section for the $pp\to pp\eta$
reaction as well as the angular distribution of the emitted $\eta$ meson 
can be equally well described by the resonant mechanism, 
where the S$_{11}$(1535) is excited by the $\pi$ meson.
This is presented in Figures~\ref{comparison}.a and~\ref{comparison}.b showing the 
comparison of the theoretical description of the excitation
function for the $pp\to pp\eta$ reaction: in 
Figure~\ref{comparison}.a the pseudoscalar meson exchange
model has been used, while in Figure~\ref{comparison}.b
there are presented predictions of the vector meson dominance model.

As can be seen from these figures, both models give equally 
well descriptions of the experimental data. The underestimation
of the data by the models in the very close-to-threshold region
results mainly from neglecting the $\eta$-proton final state interaction (FSI). 
Proton-proton FSI has been accounted for in both calculations.  
Also at the higher excess energies (Q$>$60~MeV) models underestimate
the existing data, which may be due to the 
higher partial waves which were not taken into account. 
The main conclusion here is that the
resonant current (dotted curves), i.e. the mechanism shown in Figure~\ref{mechanism_1}.a,
plays the dominant role in the $\eta$ meson creation. 
However, as far as only the 
excitation function of the total cross section is being considered, we are not able to state
about the exchanged mesons in the process of the $\eta$ meson production in proton-proton 
collisions. \\ 
Contribution of the other process that is depicted in Figure~\ref{comparison}
is the nucleonic current (dashed curves)
which we deal 
with when one of the nucleons emits a meson, 
which subsequently mixes its quark-gluonic structure
with the structure of the other nucleon, and in the final state
we again finish with a proton along with the $\eta$ meson (see diagram~\ref{mechanism_1}.b).
Naturally, the final state proton may have
different properties -- momentum and/or angular momentum, etc. -- than
the initial state proton. \\
The last process taken into account in the abovementioned models
is the mesonic current (dash-dotted curves, schematically presented 
in diagram~\ref{mechanism_1}.d), when the $\eta$ meson is being produced
via the exchange and fusion of two mesons
emitted instantly from interacting nucleons.

As has been mentioned previously, in this model 
the contribution of the two latter currents is rather negligible 
with comparison to the contribution of the resonant current (see Figure~\ref{comparison}).  
%
%

Last possible scenario presented in Figure~\ref{mechanism_1}.c -- 
the direct $\eta$ production --
although not taken into account in 
the calculations of~\cite{nakayama} has been considered in 
the model of reference~\cite{baru:02}. It has been found that 
the contribution from this process is of minor
importance, being a one order of magnitude weaker than the mesonic current contribution.


\subsection{Differential cross sections}
\label{kik}

\vspace{3mm}
{\small
Results of the measurements of the differential cross section for the 
$pp\to pp\eta$ reaction are reported.  
}
\vspace{5mm}

\begin{figure}[H]
  \unitlength 1.0cm
  \begin{picture}(14.0,6.5)
      \put(-0.20,1.0){
         \psfig{figure=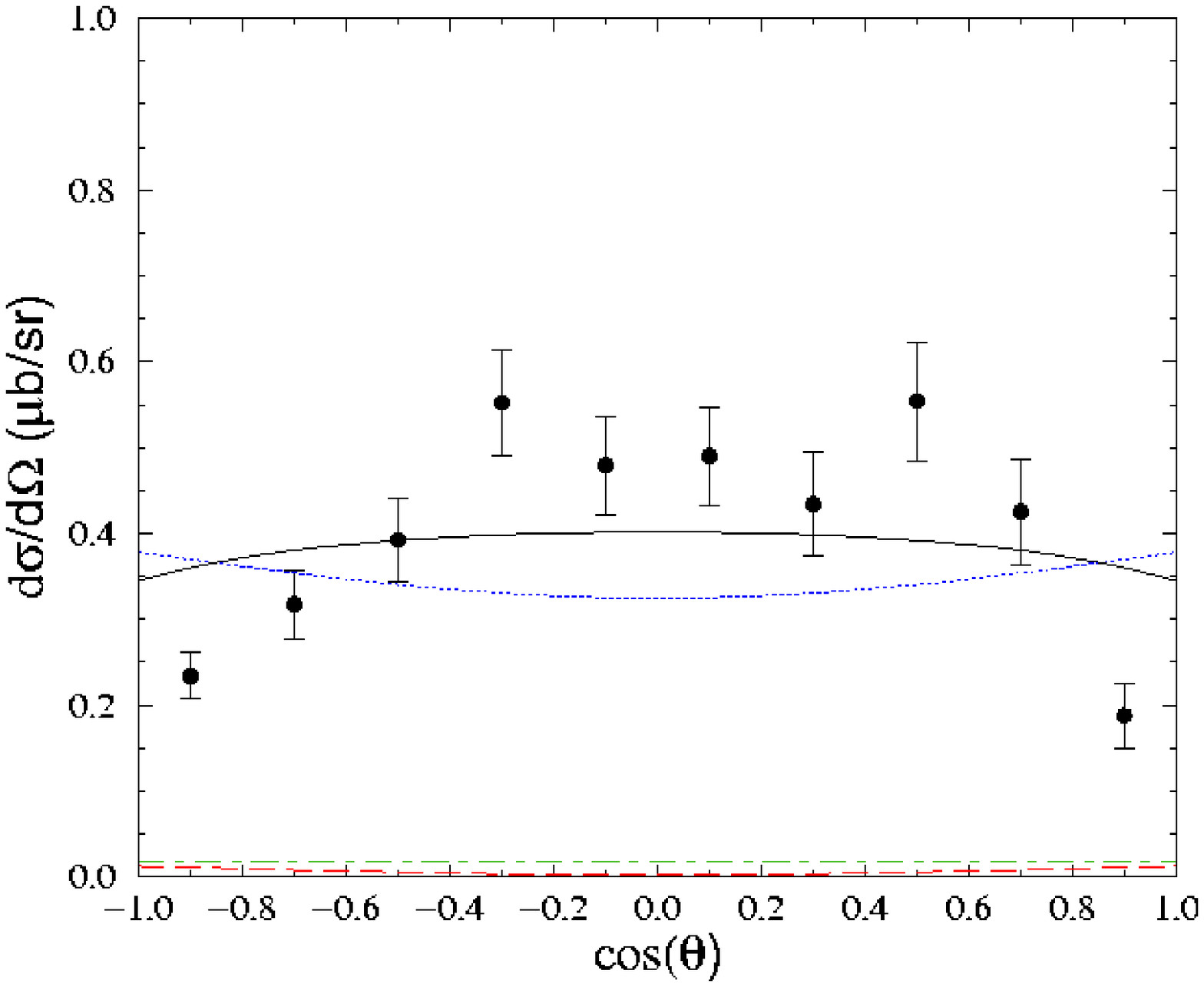,height=6.0cm,angle=0}
      }
      \put(7.60,1.0){
         \psfig{figure=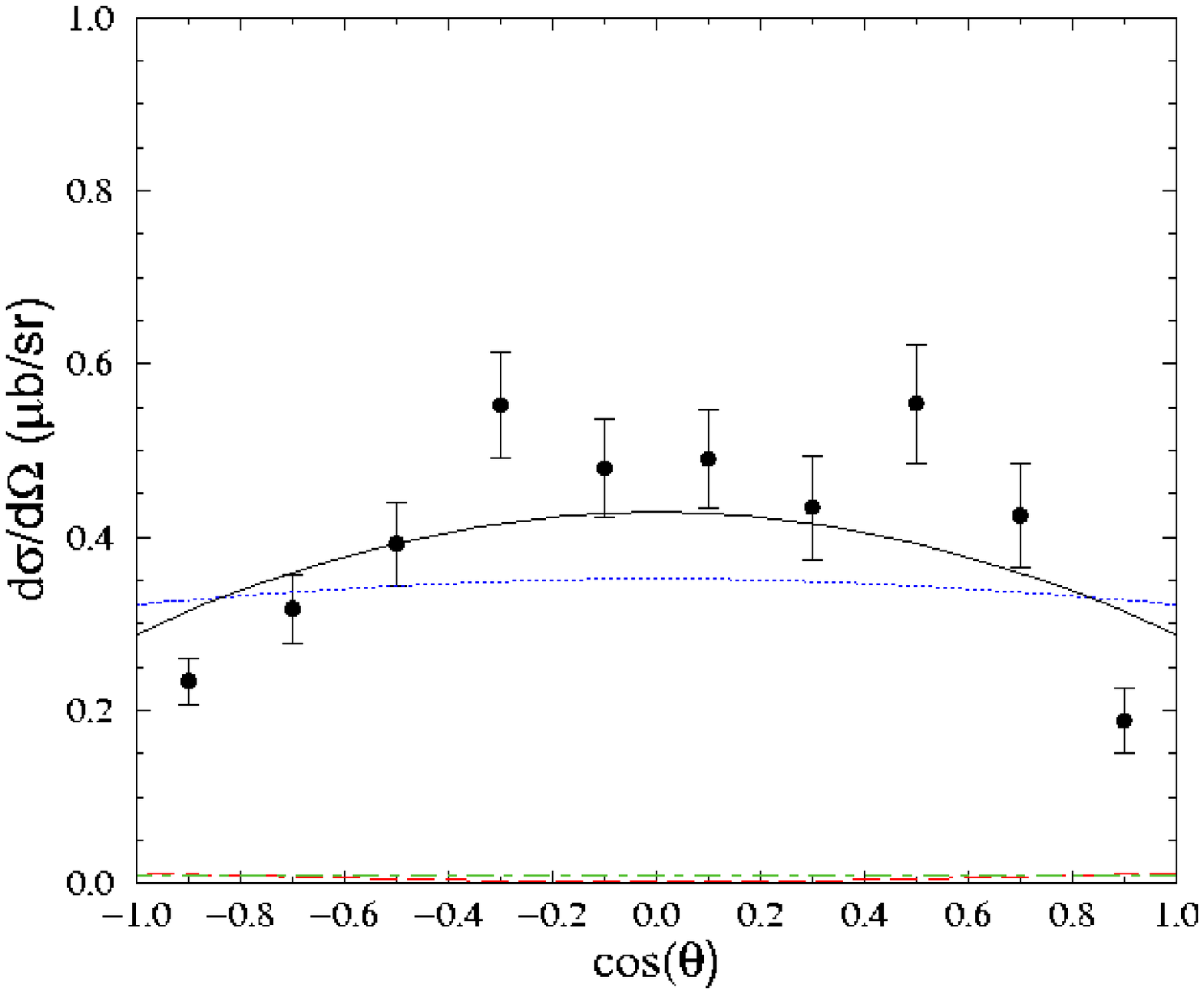,height=6.0cm,angle=0}
      }
      \put(5.8,0.9){{\normalsize {\bf a)}}}
      \put(14.0,0.9){{\normalsize {\bf b)}}}
  \end{picture}
	\vspace{-1cm}
  \caption{ \small{ Description of the angular distribution of the produced $\eta$ meson in the CM system
		in the $pp\to pp\eta$ reaction at Q~=~36~MeV within the pseudoscalar meson 
		exchange model (a) and vector meson exchange model (b).
		Data are from reference~\cite{calen:99}.
		The meaning of the curves is the same as in Figure~\ref{comparison}.   
                The figure is adapted from~\cite{nakayama}.
                  }
  }
  \label{comparison_2}
\end{figure}

Figure~\ref{comparison_2} shows one-dimensional differential spectra
of the cross sections for 
the $pp\to pp\eta$ process at Q~=~36~MeV, namely the angular distributions
of the emitted $\eta$ meson in the centre-of-mass (CM) frame. Left panel 
presents the theoretical predictions of the 
pseudoscalar meson exchange model confronted with the data. The description of the 
curves shown in this figure is the same as that of Figure~\ref{comparison}. 
The data have the tendency 
to bend down in the range of extreme values of the $\cos(\theta)$.
On the other hand, the contribution from the resonant 
current of the pseudoscalar meson exchange model 
(see Fig.~\ref{comparison_2}.a) has the opposite 
tendency, i.e. it bends upwards. However, the interference
with the remaining two currents -- nucleonic and mesonic --
bends the overall contribution downwards. 
The behavior of the experimental data, as can be seen in Fig.~\ref{comparison_2}.a, 
is not quite well reproduced by the resultant interference.

In the case when the main contribution to the excitation 
of the S$_{11}$ resonance originates from the $\rho$ meson exchange
(Figure~\ref{comparison_2}.b) the experimental data are slightly 
better reproduced, especially at the edges of the $\cos(\theta)$ range.
This might indicate that the vector meson exchange mechanism was better in 
explaining the shape of the angular distributions for the 
$pp\to pp\eta$ process.


\begin{figure}[H]
  \unitlength 1.0cm
  \begin{picture}(14.0,7.5)
      \put(0.00,1.0){
         \psfig{figure=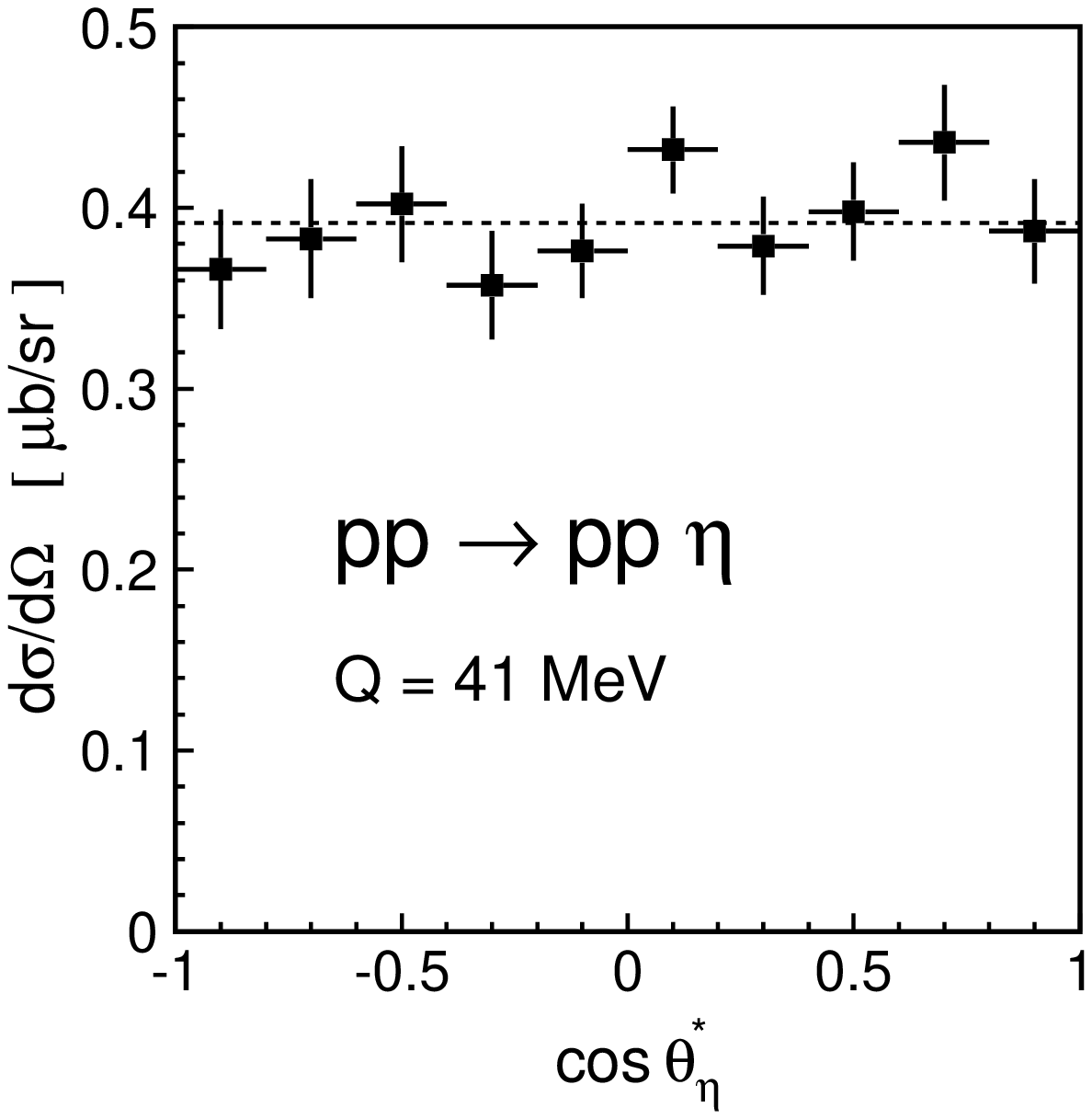,height=7.5cm,angle=0}
      }
      \put(7.50,1.0){
         \psfig{figure=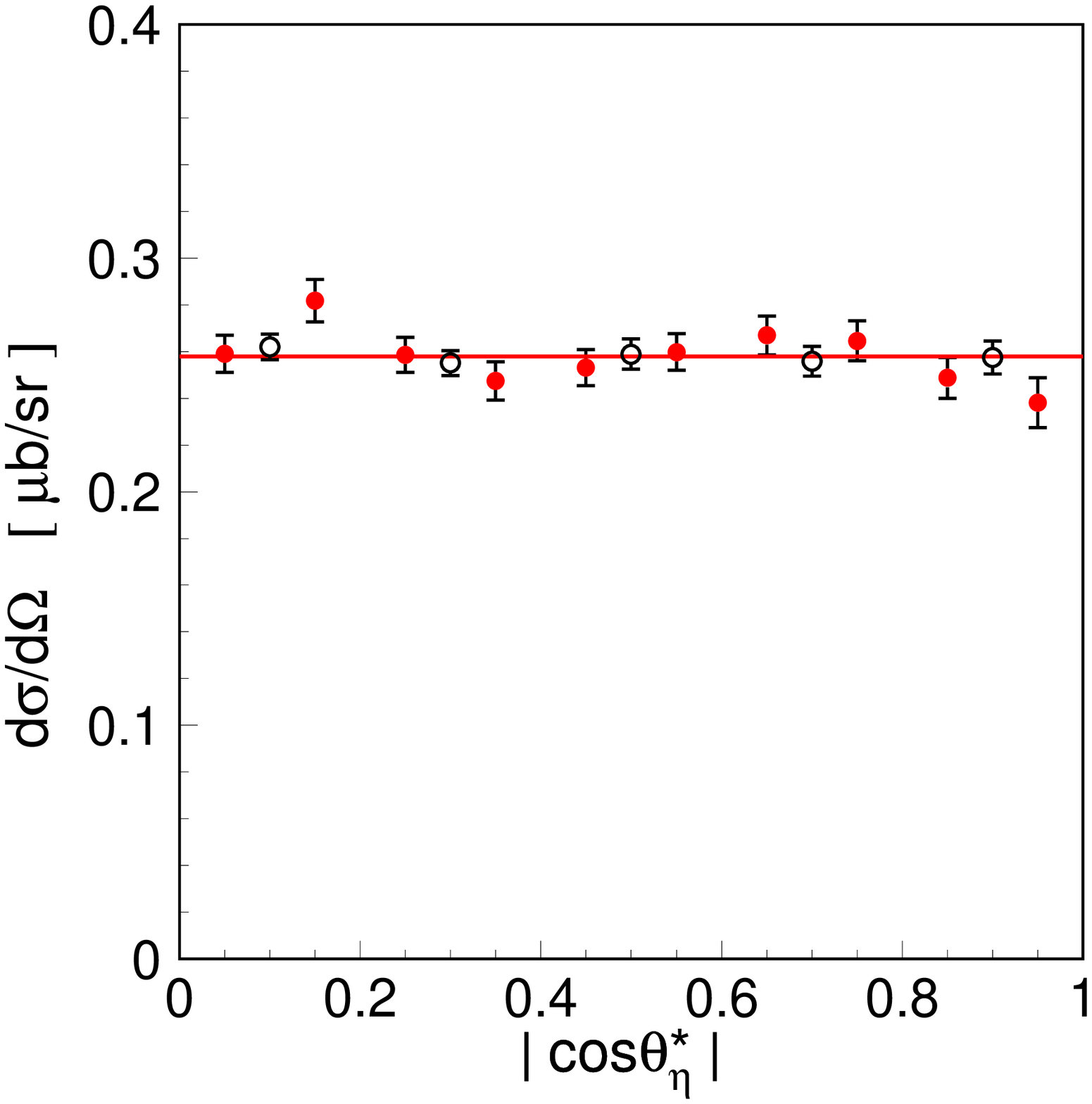,height=6.5cm,angle=0}
      }
      \put(6.5,0.8){{\normalsize {\bf a)}}}
      \put(14.0,0.8){{\normalsize {\bf b)}}}
  \end{picture}
  \caption{ \small { Angular distribution for the $\eta$ meson emission in the CM system
         for the $pp\to pp\eta$ reaction (a) 
         as measured by the COSY/TOF collaboration~\cite{abdelbary:02}
	 at Q~=~41~MeV, 
	 (b) as obtained by the COSY-11 collaboration~\cite{moskal-prc}
	 at Q~=~15.5~MeV (full circles) and by the  
          COSY/TOF collaboration~\cite{abdelbary:02} at $Q=15$~MeV
	 (open circles).
         The solid line is the fit of the Legendre polynomials.
         Figures are adapted from~\cite{moskal-prc}.
         } 
  }
  \label{pawla_rozklad}
\end{figure}

Here, a few words of explanation are required.
At the time when the work of~\cite{nakayama} has been released
the only measurements of the differential cross section
for the $pp\to pp\eta$ reaction available were
the early results of H.~Calen et al.~\cite{calen:99}.
Unfortunately, the detector used in these measurements -- the
PROMICE/WASA detector~\cite{calen:99} --
had relatively poor angular acceptance in the side-range
of $\theta$ angle (i.e. the $\cos(\theta)$ around 0 value).
Later on, the measurements have been repeated by means of the COSY/TOF
experimental setup~\cite{abdelbary:02} at the excess energy of Q~=~41~MeV
and show, within the error bars, the isotropic behavior over the whole range of $\cos\theta$ --
see Figure~\ref{pawla_rozklad}.a.

Therefore, as can be seen from comparison of Figures~\ref{comparison_2} and~\ref{pawla_rozklad}.a,
the theoretical predictions suitable for description of the flat differential cross sections
are either those of the total contribution of the pseudoscalar meson exchange model or 
those of the resonance current contribution of the vector exchange model.

The newer data on the 
angular distribution for the $pp\to pp\eta$ reaction at Q~=~15.5~MeV~\cite{moskal-prc}
and Q~=~15~MeV~\cite{abdelbary:02}, both presented in Figure~\ref{pawla_rozklad}.b, 
show the flat distribution of the $\eta$ meson emission angle
in the CM frame. The discussed angular distributions suggest that in the close-to-threshold 
region, i.e. up to the excess energy of about Q~=~50~MeV, the $\eta$ meson
is predominantly produced in the $s$ wave and the 
higher partial waves are rather suppressed.

\subsection{Isospin dependence of the hadronic $\eta$ meson production}
\label{zagorzany}

\vspace{3mm}
{\small
It is shown that the measurements of the isospin dependence of
the total cross section for the $NN\to NN\eta$ reaction 
help to diminish the number of possible mesons in the excitation 
of the S$_{11}$(1535) resonance. 
}
\vspace{5mm}

The conclusion from the analysis of the results on the total and differential cross section 
for the $pp\to pp\eta$ reaction presented in 
Sections~\ref{ggg} and~\ref{kik} is that the production of the $\eta$
meson in proton-proton collisions proceeds predominantly via the excitation 
of one of the colliding protons to the resonant state S$_{11}$(1535).
However, the relative contributions from the 
$\pi$, $\eta$, $\rho$, and $\omega$ meson exchanges still remain to be determined.

For the $NN\to NN\eta$ reaction there are two independent 
total cross sections: $\sigma_0$ and $\sigma_1$, 
corresponding to the total isospin of two initial state nucleons
I~=~0 and I~=~1, respectively. The $pp\to pp\eta$ reaction corresponds 
to the pure $\sigma_1$ cross section. The $pn\to pn\eta$ reaction
is a mixture of I~=~0 and I~=~1 initial states, as the proton's and 
neutron's isospins may sum up to the total 
isospin 0 or 1. The theoretical considerations involving the 
Clebsch-Gordan coefficients yield: 
\begin{equation}
\sigma(pn\to pn\eta) = \frac{1}{2} (\sigma_0 + \sigma_1). 
\label{sekowa}
\end{equation}

\begin{figure}[H]
  \unitlength 1.0cm
        \begin{center}
  \begin{picture}(14.0,8.0)
    \put(3.0,0.0){
      \psfig{figure=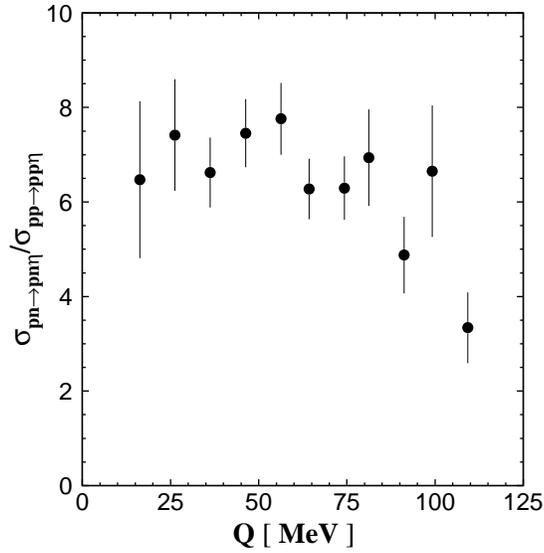,height=8.0cm,angle=0}
    }
  \end{picture}
  \caption{ \small { The ratio R$_{\eta}=\sigma(pn\to pn\eta)/\sigma(pp\to pp\eta)$
	of the total cross section for the quasi-free $pn\to pn\eta$ to the 
	total cross section for the $pp\to pp\eta$ reaction 
	as the function of the excess energy. 
	The original data are from reference~\cite{calen_pn}. 
 \label{siary}
  }}
        \end{center}
\end{figure}

The total cross section for the quasi-free $pn\to pn\eta$ reaction
has been determined in the excess energy range 
between Q~=~16 and 109~MeV by the WASA/PROMICE collaboration~\cite{calen_pn}.
The ratio R$_{\eta}=\sigma(pn\to pn\eta)/\sigma(pp\to pp\eta)$
has been found to equal about 6.5 in the quoted 
excess energy range, as depicted in Figure~\ref{siary}.

From relation~\ref{sekowa}, the fact that $\sigma(pp\to pp\eta)=\sigma_1$, 
and the definition of the ratio R$_{\eta}=\sigma(pn\to pn\eta)/\sigma(pp\to pp\eta)$
we get: 
\begin{equation}
\sigma_0 = (2R_{\eta}-1) \sigma_1.
\end{equation}
Therefore, the measured ratio $R_{\eta}=6.5$ implies that 
the production of the $\eta$ meson 
with the total isospin I~=~0 exceeds the production 
with the isospin I~=~1 by a factor of 12.

Using the short range approximation, where the 
S wave cross sections are given by~\cite{wilkin-tsl,calen_pn}
\begin{equation}
\sigma(pp\to pp\eta) = \sigma_1(pn\to pn\eta) = C |t_{\pi}+t_{\eta}+t_{\rho}+t_{\omega}|^2 |\psi_{I=1}(0)|^2,  
\label{stroze}
\end{equation}
and 
\begin{equation}
\sigma_0(pn\to pn\eta) = C |-3t_{\pi}+t_{\eta}+3t_{\rho}-t_{\omega}|^2 |\psi_{I=0}(0)|^2, 
\label{stroze2}
\end{equation}
where $t_i$ stands for the strength of the different meson exchanges. 
The relative phases of the meson exchanges were taken to be real. The factor $C$,
which includes the phase space and particles interaction,
was assumed to be the same for both isospins. The squared ratio
between I~=~0 and I~=~1 wave functions $\psi(r)$ at the zero distance between 
the particles was calculated to be approximately constant
and equals 0.8~\cite{wilkin-tsl}. 
Therefore the ratio
\begin{equation}
\frac{|-3t_{\pi}+t_{\eta}+3t_{\rho}-t_{\omega}|^2}{|t_{\pi}+t_{\eta}+t_{\rho}+t_{\omega}|^2}\approx 15, 
\label{magura}
\end{equation}
which is a relatively large value, suggesting 
the dominance of the isovector meson exchange ($\pi$ and $\rho$)
over the isoscalar meson exchange for the 
$\eta$ meson hadronic production.

\section{How do the analysing power measurements 
enable a distinction between the possible scenarios?}

\vspace{3mm}
{\small
The hitherto performed measurements of the analysing power
for the $\vec{p}p\to pp\eta$ reaction are quoted. 
The available model predictions for the A$_y$ at different 
excess energies as calculated based on different meson exchange models
are shown. 
It is pointed out how the measurements of the analysing power 
enable to infer about the production mechanism of the $\eta$ 
meson in hadronic collisions. 
}
\vspace{5mm}

What has been described previously implicates that more limitations have to be added to the
models in order to extract the information
which meson -- pseudoscalar $\pi$  meson or vector $\rho$ meson -- 
plays the most important role in the 
excitation of the intermediating S$_{11}$(1535) resonance.
 
\begin{figure}[H]
  \unitlength 1.0cm
  \begin{picture}(16.0,9.0)
      \put(-0.80,1.0){
         \psfig{figure=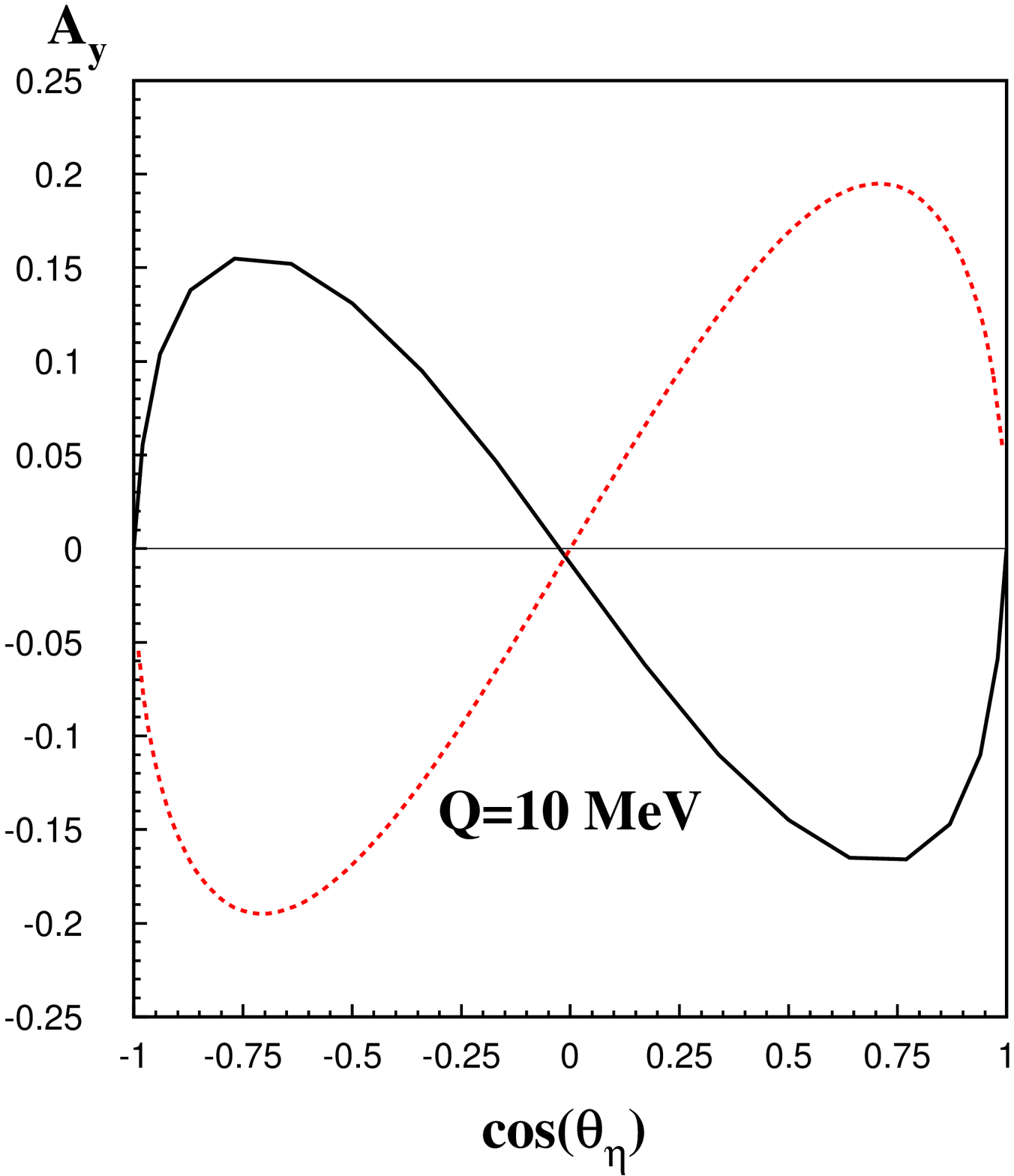,height=8.0cm,angle=0}
      }
      \put(6.70,1.0){
         \psfig{figure=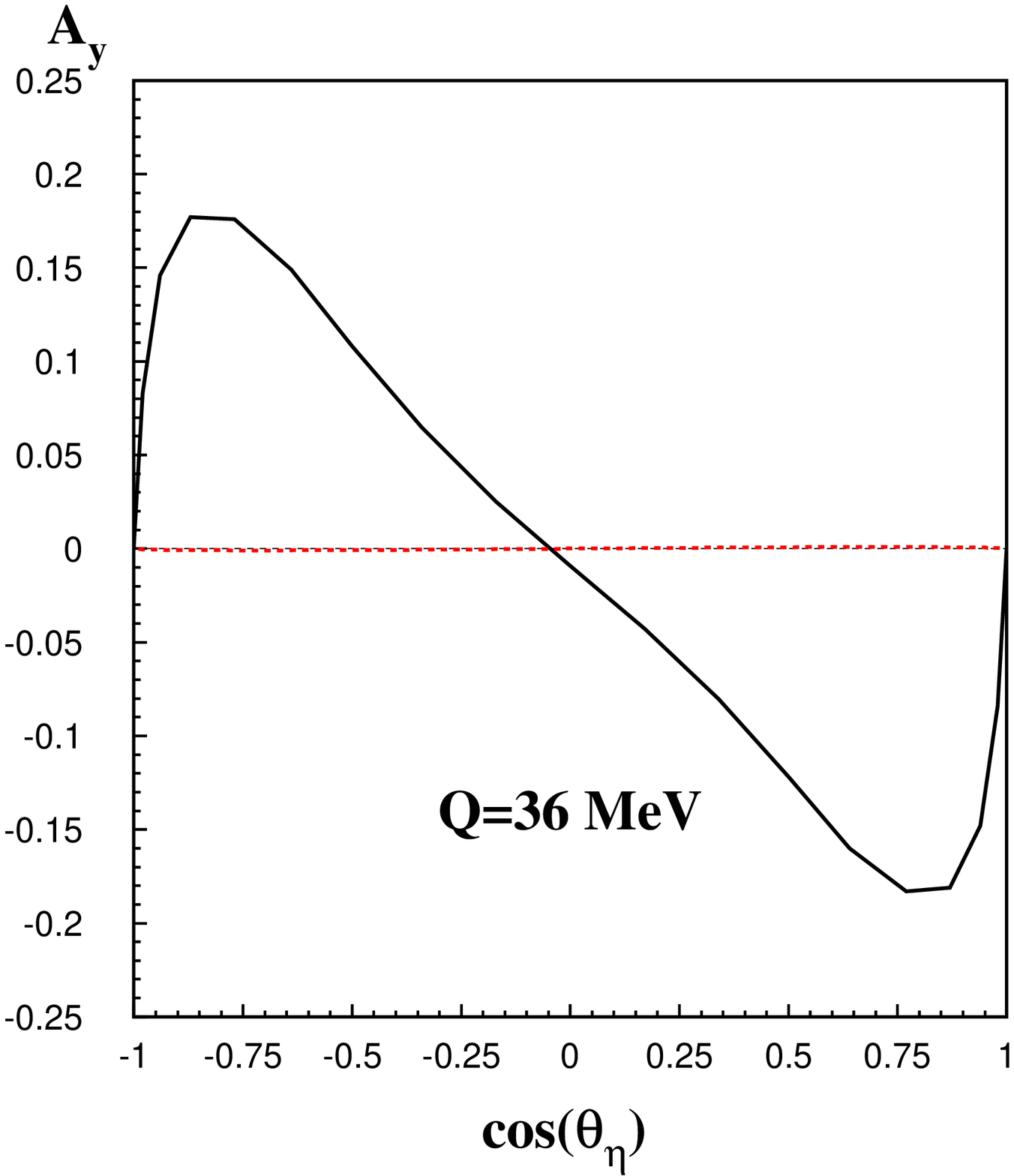,height=8.0cm,angle=0}
      }
      \put(6.5,0.8){{\normalsize {\bf a)}}}
      \put(14.0,0.8){{\normalsize {\bf b)}}}
  \end{picture}
  \caption{ \small Predictions of the analysing power for the
        $\vec{p}p\to pp\eta$ reaction as a function of
        the cosine of centre-of-mass polar angle of $\eta$ -- $\cos(\theta_{\eta})$ --
        at Q~=~10~MeV (a) and Q~=~36~MeV (b). Full lines are the
        predictions of the pseudoscalar meson exchange model~\cite{nakayama}
        whereas the dotted lines represent the results
        of the calculations based on the vector meson exchange~\cite{wilkin}.
  \label{ay10_30}
  }
\end{figure}

One solution
would be verification of the polarisation observables -- like 
the analysing power or spin correlation functions -- given by 
different models. At present there exist two models that predict
the energy dependence of the beam analysing power for the $\vec{p}p\to pp \eta$ reaction
\cite{wilkin, nakayama}.
There are significant differences between the predictions
based on these models visible in Figure~\ref{ay10_30}: e.g. the relative sign and the magnitude of the
analysing power. Measurements of this observable
might therefore serve in establishing the valid mechanism of the
$\eta$ meson production.

\begin{figure}[H]
  \unitlength 1.0cm
        \begin{center}
  \begin{picture}(14.0,8.0)
    \put(3.0,0.0){
      \psfig{figure=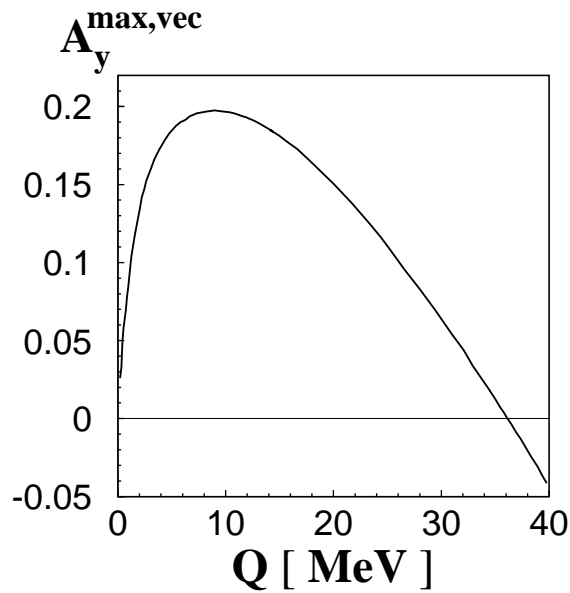,width=8.0cm,angle=0}
    }
  \end{picture}
  \caption{ \small{Predictions for the value of A$_y^{max}$ for 
	the $\vec{p}p\to pp\eta$ reaction as a function of excess energy.
	In the estimations, the vector meson exchange of reference~\cite{wilkin}
	has been applied. 
 \label{koby}}
  }
        \end{center}
\end{figure}

According to the model predictions of reference~\cite{wilkin} the analysing power 
for the $\vec{p}p\to pp\eta$ reaction in the close-to-threshold region 
should have the following form:
\begin{equation}
A_y(Q,\theta_{\eta}) = A_y^{max}(Q) \sin(2\theta_{\eta}),
\label{ay_wilkin}
\end{equation}
where $\theta_{\eta}$ is the angle of the $\eta$ meson 
emission in the centre-of-mass frame.
Equation~\ref{ay_wilkin} has been obtained under the 
assumption of only $\rho$ meson exchange in the
excitation mechanism of the S$_{11}$(1535) resonance.
The amplitude $A_y^{max}$ in reference~\cite{wilkin}
is parameterized by means of the dimensionless parameter
$\eta$, namely:
\begin{equation}
A_y^{max} = -1.5 [(-3\pm 0.5)\eta^2 + (16\pm 4)\eta^{4}], 
\label{costam}
\end{equation} 
where 
\begin{equation}
\eta = \sqrt{\frac{4M}{2mM+m^2} Q}.
\label{parametr_eta}
\end{equation}   
In Equation~\ref{parametr_eta} $m=547.75\pm 0.12$~MeV~\cite{eidelman} denotes the $\eta$ 
meson mass, 
while M~=~$938.27203\pm 0.00008$~MeV is the proton mass~\cite{eidelman}. 
The shape in the form of $\sin(2\theta_{\eta})$ was obtained from
the parameterization of the $\gamma p\to p\eta$
differential cross sections~\cite{krusche1, krusche2}
with the Legendre polynomials, and by the assumption that 
vector-meson-induced reaction amplitudes may be  
obtained from the amplitudes for photoproduction, 
using the proper Jacobians~\cite{wilkin}. 
The predicted values for $A_y^{max}$ as 
function of the excess energy Q are depicted in Figure~\ref{koby}.

The characteristic feature here is that
A$_y^{max}$ is peaked at Q~=~10~MeV, where
COSY-11 has the largest acceptance.
This feature has been one of the main motivations
to perform the measurement of the analysing power
at this particular excess energy value. At 
Q of about 36~MeV A$_y^{max}\approx 0$, and we also decided 
to investigate this region of the excess energy.  
Note that at this value of excess energy 
the A$_y^{max}$ changes its sign and 
becomes negative for the higher excess energies
in the close-to-threshold region.

The model independent considerations of the 
spin observables can be found in reference~\cite{naka-love}.
The details of the pseudoscalar model assumptions 
and its prediction concerning the analysing power
for the $\vec{p}p\to pp\eta$ reaction has been described in~\cite{nakayama, nakayama2}. 

\begin{figure}[H]
  \unitlength 1.0cm
  \begin{picture}(16.0,9.0)
      \put(-0.80,1.0){
         \psfig{figure=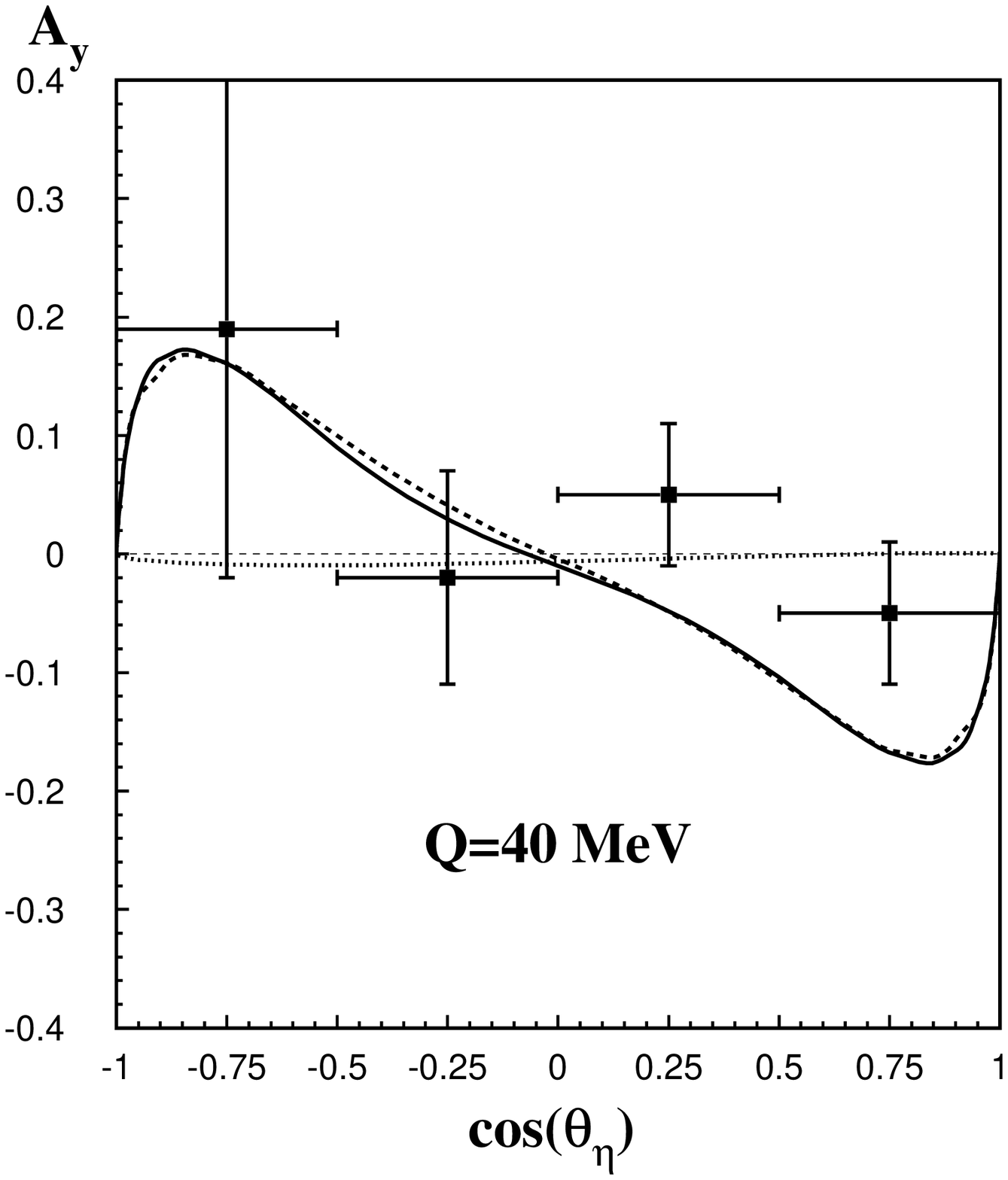,height=8.0cm,angle=0}
      }
      \put(6.70,1.0){
         \psfig{figure=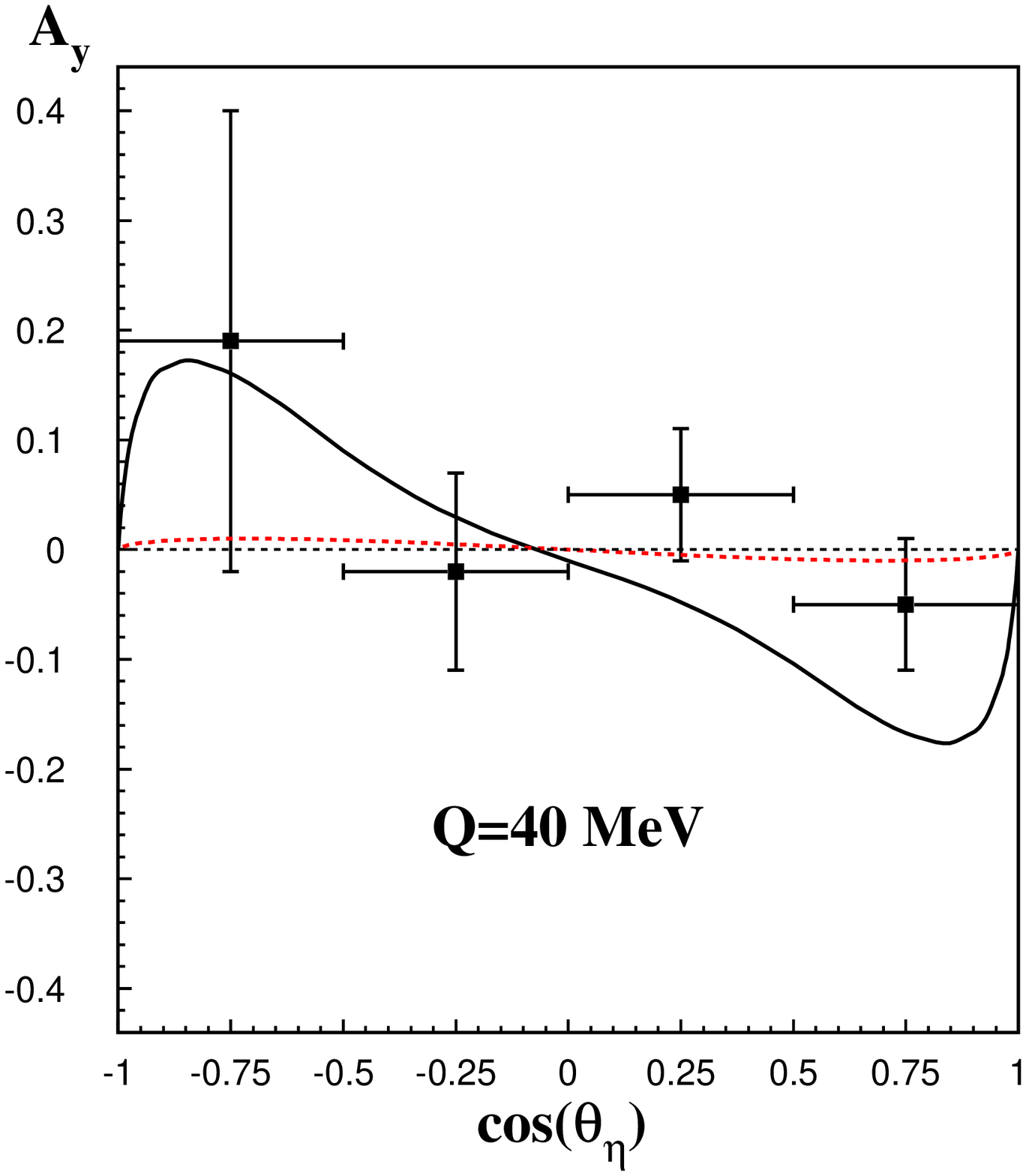,height=8.0cm,angle=0}
      }
      \put(6.5,0.8){{\normalsize {\bf a)}}}
      \put(14.0,0.8){{\normalsize {\bf b)}}}
  \end{picture}
  \caption{ \small{(a) Analysing power for the $\vec{p}p\to pp\eta$ reaction at Q~=~40 MeV.
        Meaning of the curves is explained in the text.
        (b) Comparison of experimental data of the analysing power for
        the $\vec{p}p\to pp\eta$ reaction at Q~=~40~MeV with the predictions of pseudoscalar
        (solid line) and vector meson exchange models (dashed line). } 
  }
  \label{czyzyk}
\end{figure}

The first test measurement of the analysing power 
for the $\vec{p}p\to pp\eta$ reaction at the excess energy 
of Q~=~40~MeV has been performed by the COSY-11
collaboration in the year 2001. The method of the analysis and the results have been 
reported in~\cite{winter:02-2, winter:02-1-en} and are depicted in Figure~\ref{czyzyk}.

Figure~\ref{czyzyk}.a shows the comparison of the experimental 
results with the theoretical predictions~\cite{nakayama3} of the individual 
partial waves contributions in the $\eta$ meson production.
Production of the $\eta$ meson in the $s$ wave solely would 
force analysing power to vanish~\cite{nakayama3}, and therefore
any non-zero result must involve the production of the $\eta$ meson
in higher partial wave.  
The dotted curve corresponds to the case, where the $\eta$ meson 
is created in the $s+p$ waves, while the dashed curve
represents the $\eta$ meson production in the $s+p+d$ wave.
The solid line is the full result of the model, where also 
the higher partial waves of the $\eta$ meson production are
taken into account. Unfortunately the accuracy of the data does
not allow to state firmly whether the $\eta$ 
meson is produced in the $s$ or $s+p$ wave and therefore to 
cope with this problem further investigations were required.
One should also mention that the data of~\cite{winter:02-2}
need to be reanalyzed in order to get rid of the occasional 
error, however as has been checked in the analysis described
in this dissertation, the corrected results should not differ from the 
results presented in Figure~\ref{czyzyk} of more than 
one standard deviation. 


Figure~\ref{czyzyk}.b shows the comparison of the data
with theoretical predictions of the pseudoscalar meson 
exchange model (solid line) and vector meson exchange model (dashed curve). 
Both model predictions lie within around 2$\sigma$ distance from experimental data, and therefore
at the level of accuracy
obtained during the first measurement of the analysing power 
we were not able to distinguish
between two different hypotheses of the $\eta$ production.
Further investigations were necessary and constitute the subject of this dissertation.


\newpage
\clearpage
\pagestyle{fancy}
\chapter{Experiment} 
\label{experiment}


The laboratory aparatus which enabled the experiments presented in this 
dissertation comprises the storage ring COSY,
which provides the accelerated polarised proton beams, and the 
detection setup COSY-11 used to register and identify the products 
of reaction. Both facilities will be described in the following sections.

The COSY-11 detection setup has been described in details in many previous publications, 
therefore here we will present it only very briefly emphasizing 
the detectors relevant for measurements described in this thesis
and aspects connected with the beam polarisation.  

\section{Cooler Synchrotron COSY}
\label{cosy}

\subsection{General properties}

\vspace{3mm}
{\small
The rough description of the cooler synchrotron COSY is given. 
}
\vspace{5mm}

\begin{figure}[p]
  \unitlength 1.0cm
  \begin{picture}(14.0,21.5)
      \put(0.00,1.0){
         \psfig{figure=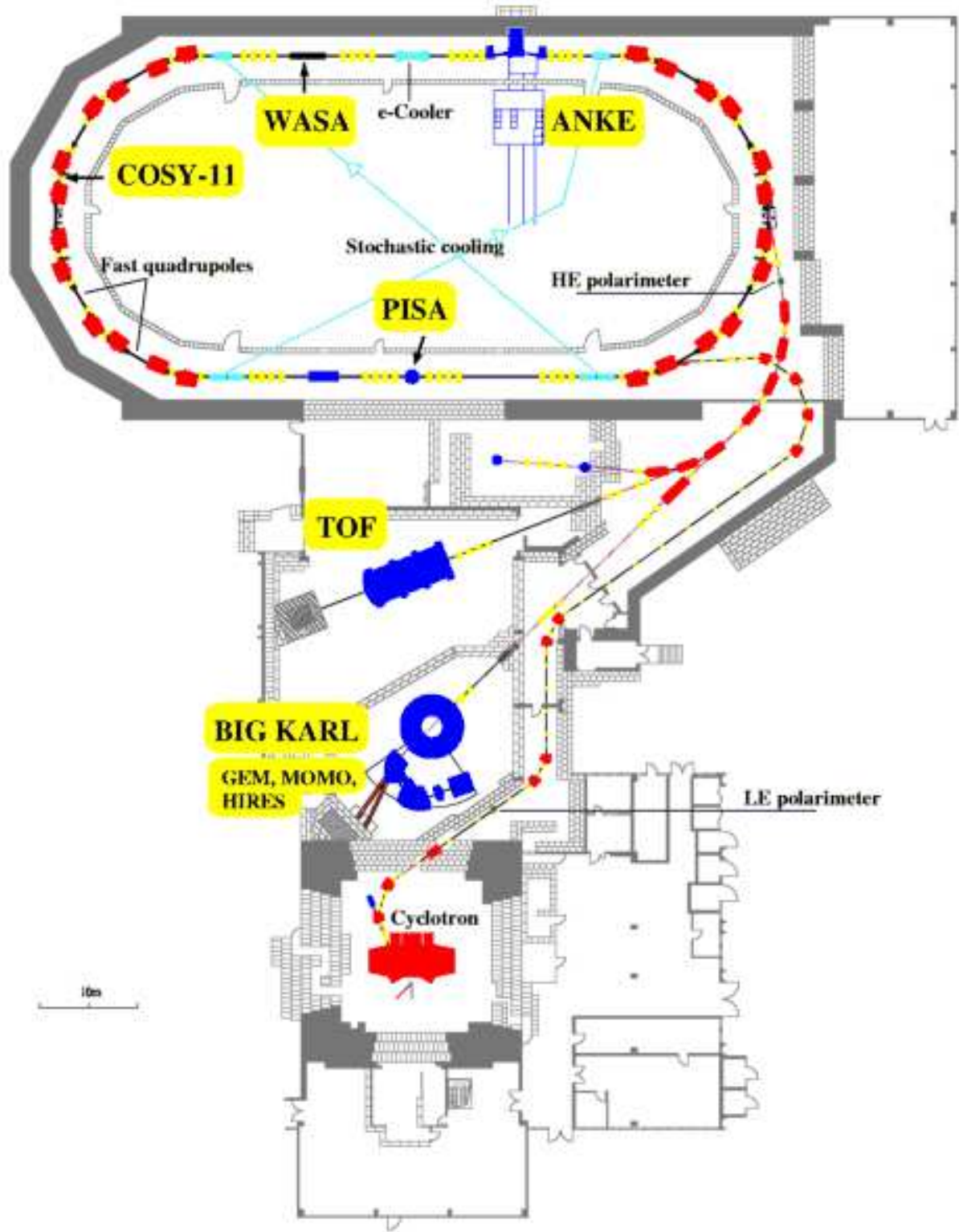,height=20.5cm,angle=0}
      }
  \end{picture}
  \caption{ \small{ COoler SYnchrotron COSY - the floorplan. 
		Several internal (ANKE~\cite{anke}, COSY-11~\cite{cosy11},
		PISA~\cite{pisa})
as well as external experiments (GEM~\cite{gem}, HIRES~\cite{hires},
MOMO~\cite{momo}, TOF~\cite{cosy-tof,tof}) are making use of the COSY beams. 
Recently 
the 4$\pi$ detection setup 
WASA~\cite{wasa} has been installed.
                  }
  \label{akcelerator}
  }
\end{figure}

The cooler synchrotron COSY~\cite{Maier,webcosy} -- a storage ring designed 
to accelerate the polarised and unpolarised beams of protons and deuterons --
is operated by the Institute
of Nuclear Physics (IKP) in the Research Center J\"ulich in Germany. This device, 
sketched in Figure~\ref{akcelerator},  
consists of the ion source that provides the polarised or unpolarised
ions of H$^{-}$ and D$^{-}$ to be further preaccelerated in the low energy cyclotron 
JULIC, which accelerates these beams up to the energies of 45~MeV for H$^{-}$ ions 
and 75~MeV for D$^{-}$~\cite{dit}. 
After the preacceleration, the beam is extracted 
and guided through the 100~m long beamline to be injected into the
storage ring COSY, that further accelerates the beams  
to the demanded momenta in the range from 0.3~GeV/c up to 3.7~GeV/c.   

The synchrotron consist of two arcs, containing in total 24 dipole magnets, 
and two straight sections, each of about 40~m length.
The total length of the COSY ring is 183.4~m. 

The synchrotron COSY is equipped with two systems of beam cooling:
the electron and stochastic cooling~\cite{cool}.
In order to increase the intensity of the beams of polarised protons
the electron cooler is being used. 
Using the electron cooler and the stacking 
technique it was possible to store
10$^{10}$ polarised protons during the runs 
concerned in this thesis. 
The stochastic cooling for COSY~\cite{stocha,stock-cool} is designed to reduce the 
emmitance of the proton beams in the momentum range 
between 1.5 and 3.7~GeV/c. The system consists of one pickup tank 
of 4~m length and a kicker of 2~m length for each the vertical and horizontal planes. 
The stochastic cooling is generally used for internal target experiments 
in order to achieve the equilibrium conditions between target heating 
and stochastic cooling. 
For more details on the working principle 
of the COSY electron and stochastic cooler the reader is referred to~\cite{cool,stock-cool}.

\subsection{Production of the polarised proton beam}

\vspace{3mm}
{\small
The source of the polarised proton beam is described along with the
method of polarised beam production at the cooler synchrotron 
COSY.  
}
\vspace{5mm}

The source of the polarised proton beam~\cite{weidmann} is of 
a Colliding-Beam Source type, that provides the polarised $\vec{H}^-$ ions 
in a direct charge-exchange process of colliding beams:
\begin{equation}
\vec{H}^0 + Cs^0 \to \vec{H}^- + Cs^+.
\label{proc}
\end{equation} 
Here, the neutral nuclear polarised beam of hydrogen produced 
in the atomic-beam source
meets a fast neutral Cs$^0$ beam. The cross section for the process~\ref{proc}
is large because the binding energy of the electron in 
Cs (3.9 eV) is close to the binding energy 
of the electron in hydrogen's 2S state (3.4 eV). 
Additionally the $\vec{H}^0$ atoms are rather slow, which 
increases the probability of the charge-exchange reaction. 
For injection the nuclear polarised anions are used rather than H$^+$, as it has been 
experimentally proven that stripping off the electrons from H$^-$ during injection
into the COSY is about an order of magnitude more efficient than 
stacking injection of protons using a bunched beam~\cite{weidmann}.  

\begin{figure}[h]
  \unitlength 1.0cm
  \begin{picture}(14.0,8.5)
      \put(0.25,0.0){
         \psfig{figure=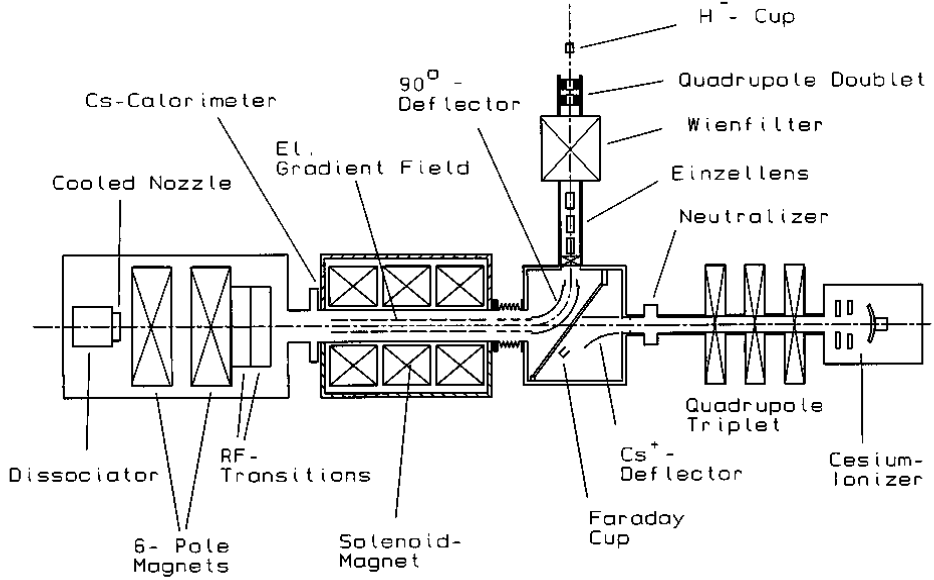,height=8.5cm,angle=0}
      }
  \end{picture}
  \caption{ \small{ The polarised ion source for COSY. Figure is adapted from~\cite{weidmann}. 
                  }
  \label{ion}
  }
\end{figure}

The polarised ion source, presented in Figure~\ref{ion}, consists of dissociator, two groups of sextupole 
magnets, the radio frequency transitions, the solenoid magnet, cesium 
ionizer, colliding zone, the 90$^0$ deflector, and a Wienfilter. 
In the dissociator the H$_2$ molecules are dissociated into atoms, which 
subsequently through the cooled nozzle reach the sextupole magnets region. 
The first sextupole magnet 
produces the electron state polarisation of the atoms, and 
subsequently atoms with the electron spin state m$_J=-1/2$ are 
defocused and only atoms with m$_J$=+1/2 survive. The second
sextupole magnet acts as the lense to focus the hydrogen atoms into 
the radio frequency transition units, which produce the nuclear polarisation. 
Afterwards the polarised $\vec{H}^0$ atoms collide with 
a cesium beam moving in the opposite direction. In the charge-exchange 
region where the reactions~\ref{proc} occur the nuclear polarisation of hydrogen is preserved by the 
strong longitudinal magnetic field at the order of 1.7~kG. 
For this value of magnetic field the polarisation of protons at the level of circa 95\% 
can be expected~\cite{weidmann}. The polarised $\vec{H}^-$ anions 
are subsequently deflected by 90$^{\circ}$ in the magnetic deflector and guided 
into the extraction system, passing on the way 
the Wienfilter, which may be rotated around the 
beam axis, and therefore can select the required spin orientation of 
the $\vec{H}^-$ ions, and also separate the anions from the electrons 
and other background particles. 

Subsequently the $\vec{H}^-$ beam is 
injected into the cyclotron JULIC at the energy of 4.5~keV. The cyclotron preaccelerates them  
up to energies of 45~MeV. After that, the beam of $\vec{H}^-$ anions is guided through the 
beam lines and reaches the stripping injector~\cite{budker}, where the $\vec{H}^-$
anions are stripped off two electrons and furthermore injected into the COSY-Ring
as polarised protons with the selected polarisation orientation.

With this method the intensity of circa 10$^{10}$ stored polarised protons 
with a degree of polarisation over 65\% have been obtained~\cite{stockhorst} 
during the experiments reported in this work\footnote{Nowadays it is 
possible to achieve over 90\% polarisation for the internal experiments
and circa 80\% for the external ones~\cite{gebel}.}. 

\subsection{Acceleration of the polarised proton beam}
\label{ii}

\vspace{3mm}
{\small
Acceleration of the polarised proton beams is  
described. Superimposed are the methods of overcoming the 
intrinsic and imperfection resonances. The table of all possible 
depolarising resonances at the COSY ring is presented. 
}
\vspace{5mm}

The acceleration of the polarised beams is challenging due to the number of  
depolarising resonances that occur during this process, depending 
on the demanded momentum of the beam.

The polarisation vector of the beam of protons stored in the ring precesses 
around the magnetic fields encountered along the particles' orbit\footnote{In the case when
the precession repeats on each revolution the particles are on the so called
{\it spin closed orbit}.}. 
In the ideal ring, where the 
dipole magnetic fields are vertical, the number of spin precessions during one turn 
(so called spin tune) is given by~\cite{bargman}: 
\begin{equation}
\nu = \gamma G, 
\label{spin_tune}
\end{equation} 
where $\gamma$ is the Lorentz factor, and G denotes
the gyromagnetic anomaly -- for protons $G=1.79285$. 
Depolarising spin resonances arise when the frequency of the precession of the
polarisation vector is such that upon each revolution the difference between
the phases of the polarisation vector and the depolarising magnetic field
vector is the same at the point where these two meet.
At the COSY ring there are two main types of depolarising resonances~\cite{stockhorst}:
\begin{itemize}
\item
{\bf imperfection resonances}, caused by magnetic field errors, arising if the magnets are  
slightly misaligned or if there are vertical orbit distortions. These resonances
occur at the beam momenta for which $\gamma G$ has an integer value;
\item
{\bf intrinsic resonances}, caused by the radial fields due to vertical focusing of the 
beam, occurring when $\gamma G \pm (Q_y-2)$ equals an integer, where $Q_y$ is the betatron 
vertical tune, i.e. the number of vertical oscillations of the beam per one 
revolution. 
\end{itemize}
The resonances connected with the horizontal betatron tune $Q_x$, occurring when 
$\gamma G \pm (Q_x-2)$ equals an integer are not important at COSY\footnote{As long
as the coupling between the horizontal and vertical betatron motion is
not present.} for the  
vertically polarised beams, because the horizontal betatron motion is 
driven by the vertical magnetic fields of the quadrupoles.
Also we will not 
consider here the induced resonances, arising due to the longitudinal 
oscillating magnetic fields.

Table~\ref{resonances}~\cite{lehrach} shows the chart of 
all depolarising resonances and the corresponding proton beam 
momenta p$_{beam}$ in the momentum range of COSY.    
Resonance strengths $\epsilon_r$ and ratios of preserved polarisation
$P_f/P_i$ are shown. In the simulations~\cite{lehrach} an energy 
gain per turn equal to 0.7~keV has been assumed.  
Number of intrinsic resonances depends on the superperiodicity S, the parameter 
which describes the machine setting of the quadrupole magnets. 
If all magnets are operated with the same quadrupole settings, the 
superperiodicity S~=~6 and only one intrinsic resonance 
occurs. However, this setting does not allow to accelerate the beam 
up to the maximum beam momentum. For this purpose a special setting 
of the magnets is required~\cite{lehrach} with S~=~2, thus four additional 
intrinsic resonances appear, all listed in Table~\ref{resonances}.   
The occurance of the imperfection resonances does not depend upon 
the superperiodicity.

\begin{table}[H]
 \begin{center}
   \begin{tabular}{|c|c|c|c|c|c|}
    \hline
      Resonance & $\gamma G$ &  S   &  p$_{beam}$  &  $\epsilon_r$ & $P_f/P_i$ \\
        type    &            &      &   (MeV/c)    &  ($10^{-3}$)  &	       \\
    \hline
	IMP	&    2       & ---  &   463.8      &   0.95        &  $-1.00$    \\
	INT     &   6$-$Q$_y$  &  2   &   826.9      &   0.26        &   0.20    \\
	IMP	&    3       & ---  &   1258.7     &   0.61        &  $-0.88$    \\
	INT     &   0+Q$_y$  &  2   &   1639.3     &   0.21        &   0.43    \\ 
	IMP	&    4       & ---  &   1871.2     &   0.96        &  $-1.00$    \\
	INT     &   8$-$Q$_y$  & 2;6  &   2096.5     &   1.62        &  $-1.00$    \\
	IMP	&    5       & ---  &   2442.6     &   0.90        &  $-1.00$    \\
	INT     &   2+Q$_y$  &  2   &   2781.2     &   0.53        &  $-0.74$    \\ 
	IMP	&    6       & ---  &   2996.4     &   0.46        &  $-0.58$    \\
	INT     &  10$-$Q$_y$  &  2   &   3208.9     &   0.25        &   0.25    \\
    \hline
   \end{tabular}
     \caption{ {\small Imperfection (IMP) and intrinsic resonances (INT) at the synchrotron COSY.
     \label{resonances}
         }
         }
 \end{center}
\end{table}

The ratio of the preserved polarisation $P_f/P_i$ depends on the 
strength of a resonance (see Table~\ref{resonances}). 
Each time the beam is crossing the imperfection resonances, the polarisation
may be completely lost if the strength of the resonance is not sufficient to flip
the spin of all beam particles. At some $\epsilon_r$ value, for a certain 
imperfection resonance and certain momentum spread of the beam $\Delta p/p$, 
the ratio $P_f/P_i$ is equal to $-1$, which means that crossing the resonance 
with a proper strength one deals with the spin flip, with no polarisation loss. 
Simulations reported in reference~\cite{lehrach} showed, and the experiments confirmed, that the excitation of the 
vertical orbit by 1 mrad, using the horizontal correcting dipoles, is sufficient to 
adiabatically flip the spin at all imperfection resonances.

Another method to overcome the imperfection resonances involves the 
use of spin rotators -- so called siberian snakes. 
At the synchrotron COSY the solenoids of the electron cooler, acting as the partial 
siberian snake, are able to rotate the spin around the longitudinal axis by 
circa 8$^{\circ}$ at the maximum momentum of COSY~\cite{lehrach}. It has been experimentally 
proven that a rotation angle of less than 1$^0$ leads to a total spin flip 
with no polarisation loss at all five imperfection resonances.

In order to avoid the intrinsic resonances 
the technique of tune jumping is applied. A sudden tune jump 
increases significantly the crossing speed 
of the resonance and therefore the beam ``jumps over" the intrinsic resonance
with less than 5\% polarisation loss at the strongest intrinsic resonances
and no more than 1\% loss at all other intrinsic resonances~\cite{lehrach}.

The experiments discussed in this dissertation have been performed 
at the beam momenta of p$_{beam}=2.010$ and 2.085~GeV/c. 
Thus, according to Table~\ref{resonances}, three imperfection 
and two intrinsic resonances had to be crossed. Finally, 
as we shall see in Section~\ref{polarisation}, the obtained polarisation degree
was equal to about 68\% and 66\% at the lower and higher beam momentum, respectively.

\section{COSY-11 facility}
\label{cosy-11}

\vspace{3mm}
{\small
The detectors needed for 
identification of the $pp\to pp\eta$ and $pp\to pp$ reactions are described. 
The trigger conditions for both reactions are given. 
}
\vspace{5mm}

\begin{figure}[H]
  \unitlength 1.0cm
  \begin{picture}(14.0,14.0)
    \put(0.0,0.0){
       \psfig{figure=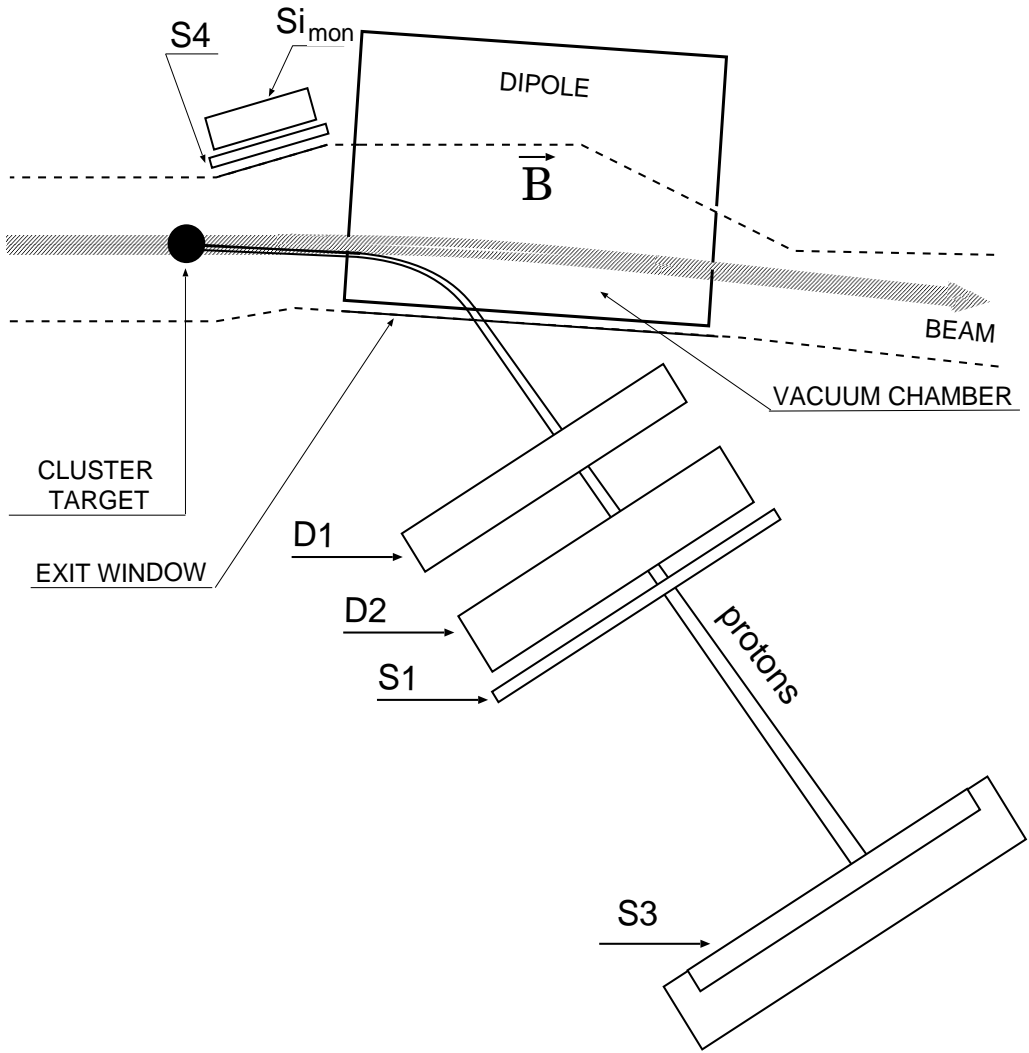,width=14.0cm,height=14.0cm,angle=0}
    }
  \end{picture}
  \caption{ \small {Schematic top view of the COSY-11 detection setup~\cite{Brauksiepe}. Only detectors
            needed for measurements of the $\vec{p}p\to pp\eta$ and $\vec{p}p\to pp$ reactions
            are shown. S1, S3, and S4 denote the scintillator detectors; D1 and D2 stand for the drift chambers; 
            Si$_{mon}$ are the silicon detectors for measurements of the elastic
            scattering. Figure is adapted from~\cite{moskal4}}.  
 \label{koza}
  }
\end{figure}

The experiments described in this dissertation  
have been performed by means of the COSY-11 detector setup~\cite{Brauksiepe}
presented in Figure~\ref{koza}. COSY-11 is an internal experiment
at the cooler synchrotron and storage ring COSY, designed to study 
the production processes, the structure, and interactions of the 
mesons in the 1~GeV mass range. The facility makes use of a regular
COSY dipole magnet, acting as a magnetic momentum spectrometer for charged
reaction products.

A vertically polarised proton beam has been scattered on the H$_2$ molecules
from an internal cluster target~\cite{dombrowski,khoukaz} installed in front of
the COSY magnet. Reaction
products possess lower momenta than the beam protons, therefore these are bent more
in the magnetic field of the dipole. Positively
charged ejectiles leave the scattering chamber~\cite{Brauksiepe}
through the thin exit window\footnote{This exit window (1870$\times$76 mm$^2$) is made of
a 30$\mu$m layer of aluminum foil, and two crossed unidirectional sheets
of carbon fibers soaked in epoxy resin of 150$\mu$m width, acting as an outward carrier material.
The choice of materials with low nuclear charge reduces straggling in the
exit window to values below the resolution of the detection system~\cite{Brauksiepe}.}
reaching the detection system operating under the
atmospheric pressure.

Trajectories of the positively charged protons, which are bent to the
inside part of the ring, are measured by means of two planar drift chambers~\cite{Brauksiepe,gugulski}
D1 and D2, presented schematically in Figure~\ref{koza}. These drift chambers
are spaced by 70~cm and contain together $6+8=14$ detection planes.
The active area of the chambers is 1680~$\times$~433~mm$^2$. The D1 chamber
contains 6 detection planes; two planes with vertical wires, two
with wires inclined by +31$^{\circ}$, and last two with wires
inclined by $-31^{\circ}$. The D2 drift chamber has additional two planes
in the back part with vertical wire orientation. The wires of the consecutive planes
of each pair are shifted by half of the cell width (20~mm) in order to
resolve the left-right ambiguity of passing through particles.
Drift chambers operate with a gas mixture
of 50\% argon and 50\% ethane at slightly more than atmospheric pressure.
The particle trajectories are reconstructed using the computer code
MEDUZA~\cite{sokol}. The code allows for reconstruction of multiple
track events.

The particle trajectories are traced through the magnetic field
of the dipole back to the interaction point. 
In this way the momenta of the particles can be determined
with precision of 4~MeV/c (standard deviation)~\cite{moskal-prc}. 

Leaving the drift chambers, the positively charged reaction products
pass the scintillator hodoscope S1~\cite{Brauksiepe}, consisting of
16 scintillation modules with dimensions of 450$~\times$~100~$\times$~4~mm$^3$
made of Bicron BC 404~\cite{bicron}. The scintillators are arranged 
vertically and read out at both sides by photomultipliers. The time resolution 
of this hodoscope in the range between 160 and 220 ps has been measured~\cite{magnus_diploma}, 
depending on the hit position. This detector delivers the start signal 
for the time-of-flight measurement.  

The stop signal of the time-of-flight measurement is generated in 
the S3 scintillator wall~\cite{Brauksiepe} (made of a 220~$\times$~100~$\times$~5~cm$^3$
scintillator Bicron BC 404), situated at a distance
of circa 9~m from S1.
The scintillator wall 
generates a light signal upon a charged particle crossing its volume. This signal 
is read out by a matrix of 217 photomultipliers. The scintillator
wall and the photomultipliers are separated with 4~cm air gap. The centre of 
gravity of the pulse height from individual photomultipliers 
is calculated in order to resolve the hit position 
of a particle. The restriction of few photomultipliers responding 
allows to separate two or more hit positions in the scintillator wall.
 
Time-of-flight measurement together with a reconstructed momentum of a particle
yields a possibility of indentification of the particle by calculating its mass.    
Hence, the four momenta of the outgoing particles can be determined. 
This information, in connection with the known four momenta of 
initial state particles allows to calculate the missing mass of the undetected 
particle or system of particles. The missing mass method will be described 
in Section~\ref{masa_brakujaca}. 

The trigger condition for registering the $\vec{p}p\to pp\eta$ reaction is:
\begin{equation}
T_{\vec{p}p\to pp\eta} = (S1_{\mu\geq 2} \vee S1_{\mu=1,high}) \wedge S3_{\mu>2},  
\label{trigger1}
\end{equation}
where $\mu$ denotes the multiplicity of segments in the S1 detector, 
and also the multiplicity of the photomultipliers that have fired 
in the S3 detector. 
Subscript {\it high} stands for the high energy deposition in the S1 scintillator module, 
corresponding to the case when two or more particles cross a single module.  

Similarly, for registering the $\vec{p}p\to pp$ reaction 
a trigger condition of the type
\begin{equation}
T_{\vec{p}p\to pp} = S1_{\mu=1} \wedge S4
\label{trigger2}
\end{equation}
has been set, where S4 is a scintillator detector used for registering
the recoil protons from the elastic scattering. Protons passing through this 
detector reach the granulated silicon-pad detector Si$_{mon}$ consisting of
144 silicon pads with dimensions of 22$\times$4.5$\times$0.28~mm$^3$
arranged in three layers one above the other. 
Two of the layers are located in the front part of the detector 
and one in the back.  

For completeness we would like to mention that there are other detectors building up the 
COSY-11 facility like the silicon pad detector~\cite{czyzyk-dip,bilger} for the spectator protons
from the quasi-free $pd\to ppnX$ reactions~\cite{janusz,moskal-jpg,moskal14}, the neutron 
detector~\cite{moskal-ikp,czyzyk-dip,przerwa-dip,rozek}, the Cherenkov detector~\cite{siemaszko}, 
the C-shaped hexagonal chamber~\cite{smyrski-kom} and the auxiliary scintillator
detectors~\cite{Brauksiepe,moskal4}.
However, as these detectors were not used during the experiments reported in this
dissertation we omitted their description herein. The interested reader is 
referred to the publications quoted above. 
\newpage
\clearpage
\pagestyle{fancy}
\chapter{Data Analysis} 

\label{dat_anal}

\section{
Calibration of the detection system
} 
\label{calibration}

\subsection{Time-space calibration of the drift chambers}
\label{dccalib}

\vspace{3mm}
{\small
The method of time-space function derivation
for the drift chambers is described. 
}
\vspace{5mm}

The drift chambers described in Section~\ref{cosy-11}
operate with a gas mixture of 50\% argon and 50\% ethane
at slightly more than atmospheric pressure. Upon crossing of a charged particle
through the gas-filled volume the electron clusters are  
generated and move towards the anode wires. The drift time 
of the electron clusters to the anode wire is a measure of the 
minimum distance between the sense wire and the trajectory of the particle.   
This relation is called a time-space calibration of the drift chamber 
and has to be determined from the experimental data. 

The drift velocity varies with the atmospheric conditions~\cite{raport-thomas}. 
Therefore, in order to perform the time-space calibration 
the experimental 
data have been divided into 32 groups for each energy, 
each group corresponding to about 6 hours of measurement. 
The time-space calibration~\cite{moskal4} has been performed iteratively
starting with an approximate time-space function $d(t)$. 
From this function and from measured drift times the points where the particle 
crossed the middle of the cells 
have been found for each detection plane and the straight line corresponding to the
particle's trajectory has been fitted. For such obtained 
trajectory of the particle for an $i$-th event 
the distance $d'_i$ to the sense wire has 
been calculated and the average of the differences 
\begin{equation}
\Delta d(t) = \frac{\sum_{i=1}^{n}{(d_i(t) - d'_i(t))}}{n}
\label{db}
\end{equation}
obtained from the mentioned sample of the experimental data  
have been used in order to correct the time-space function. 
In the Equation~\ref{db} $d_i$ denotes the distance between 
the particle trajectory and  the sense wire for an $i$-th event, 
reconstructed using the $d(t)$ calibration function.  
Subsequently, new time-space functions $d(t)+\Delta d(t)$ have been used for further
iteration. 
The procedure has to be repeated until the $\Delta d$ corrections 
becomes negligible. 


\begin{figure}[H]
  \unitlength 1.0cm
        \begin{center}
  \begin{picture}(14.0,8.0)
    \put(3.0,0.0){
      \psfig{figure=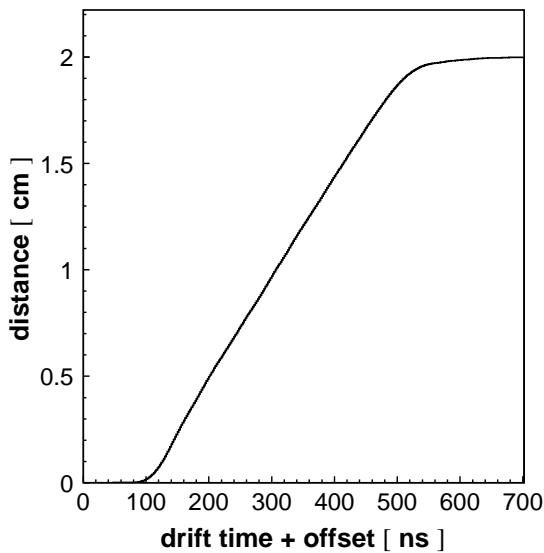,height=8.0cm,angle=0}
    }
  \end{picture}
  \caption{ {\small Distance from the sense wire as a function of the
        drift time for an arbitrarily chosen cell of DC1 -- so
        called time-space calibration of the drift chambers. 
                 }
 \label{dc_calibb}
  }
        \end{center}
\end{figure}

Figure~\ref{dc_calibb}.a presents the time-space calibration of an
arbitrarily chosen sense wire of the DC1. The linearity of the
distance from the sense wire as a function of the drift
time may be seen in the range between circa 100 and 500 ns.
The value of 100~ns corresponds to the time offset
introduced by the electronics.

\subsection{Time-of-flight calibration}
\label{s1s3}

\vspace{3mm}
{\small
Time calibration of the scintillator detectors used for 
time-of-flight measurement is presented. 
}
\vspace{5mm}

As it was mentioned in Section~\ref{cosy-11} the S1 and S3 
counters serve as a start and stop detectors, respectively, for the 
time-of-flight measurement. As the S1 scintillator consists of
16 modules, in which signals are read from both sides 
by photomultipliers and the signals from S3 scintillator
are read by a matrix of 217 photomultipliers, in order to 
obtain the credible time-of-flight information, the relative time 
offsets of the electronics for each photomultiplier 
have to be determined. 

For the S1 time calibration, we used the events 
where particles were crossing adjacent modules. 
In this case we can assume that in both modules
the signal is generated at the same time. 
The TDC value of the single photomultiplier of the S1 detector can be expressed as follows~\cite{moskal4}:
\begin{equation}
TDC_{s1} = t_{s1} + t_y + t_{s1}^{walk}(PM) + t_{s1}^{offset}(PM) - t_{trigger}, 
\label{s1cal}
\end{equation}
with t$_{s1}$ denoting the real time of the signal generation
in the S1 detector and $t_y$ standing for the time that the signal needs
for passing the distance between the hit position in the scintillator module 
and the edge of the scintillator. The $t_{s1}^{walk}$(PM) 
is the time walk effect correction\footnote{By the {\it time walk 
effect} we understand the variation of the registered TDC time 
with the amplitude of the signal typical for leading edge discrimination~\cite{tani}. This effect can be corrected 
by adding an offset $t_{s1}^{walk}$(PM)$=\frac{1}{\sqrt{ADC}}$, with 
{\it ADC} denoting the signal charge value~\cite{moskal4}.}, $t_{s1}^{offset}$(PM) denotes the 
time offset of the electronics for a given photomultiplier, and 
$t_{trigger}$ is the time when the trigger pulse started the readout of TDC modules. 

\begin{figure}[H]
  \unitlength 1.0cm
  \begin{picture}(14.0,7.5)
      \put(0.00,1.0){
         \psfig{figure=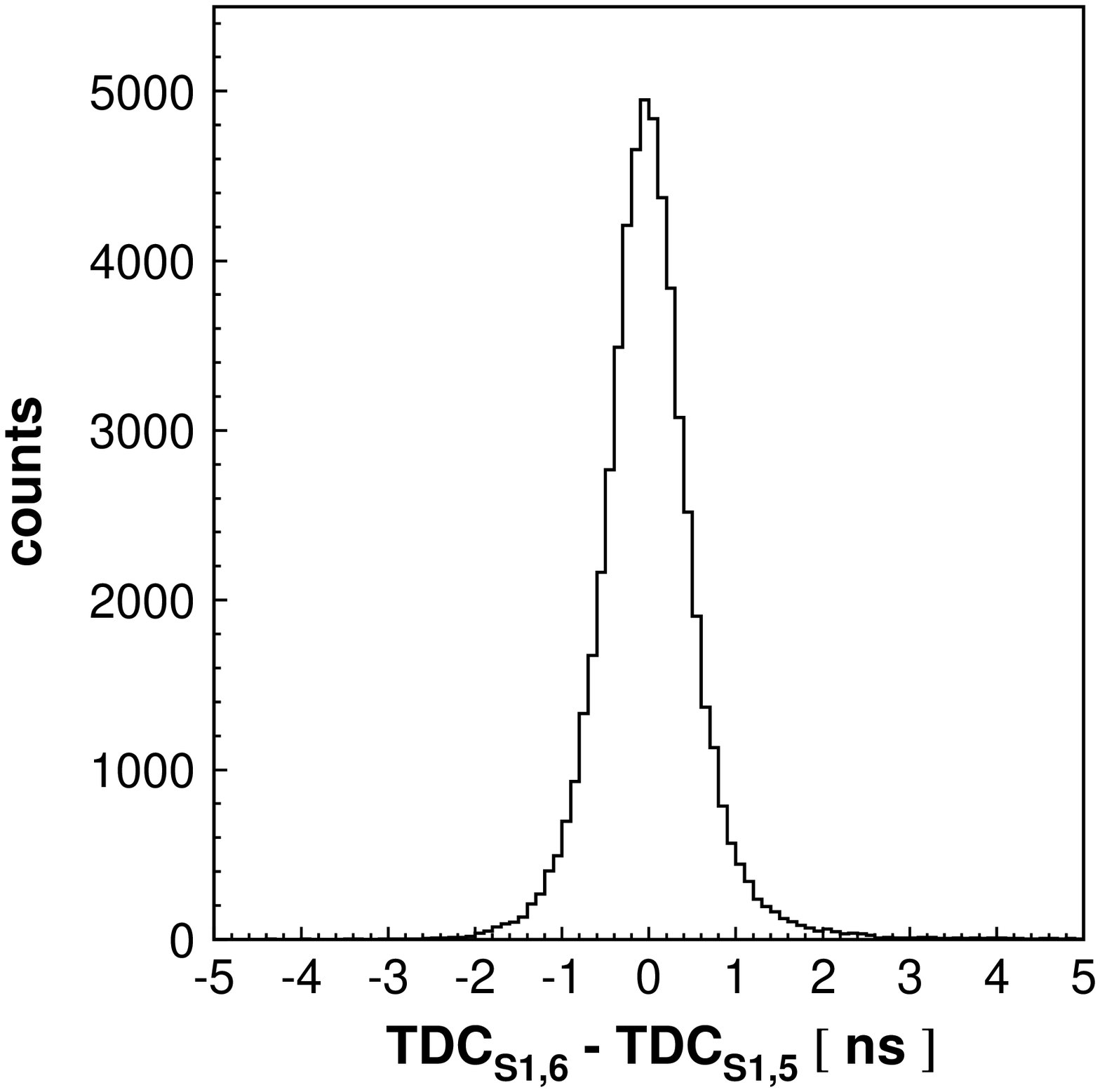,height=6.5cm,angle=0}
      }
      \put(7.50,1.0){
         \psfig{figure=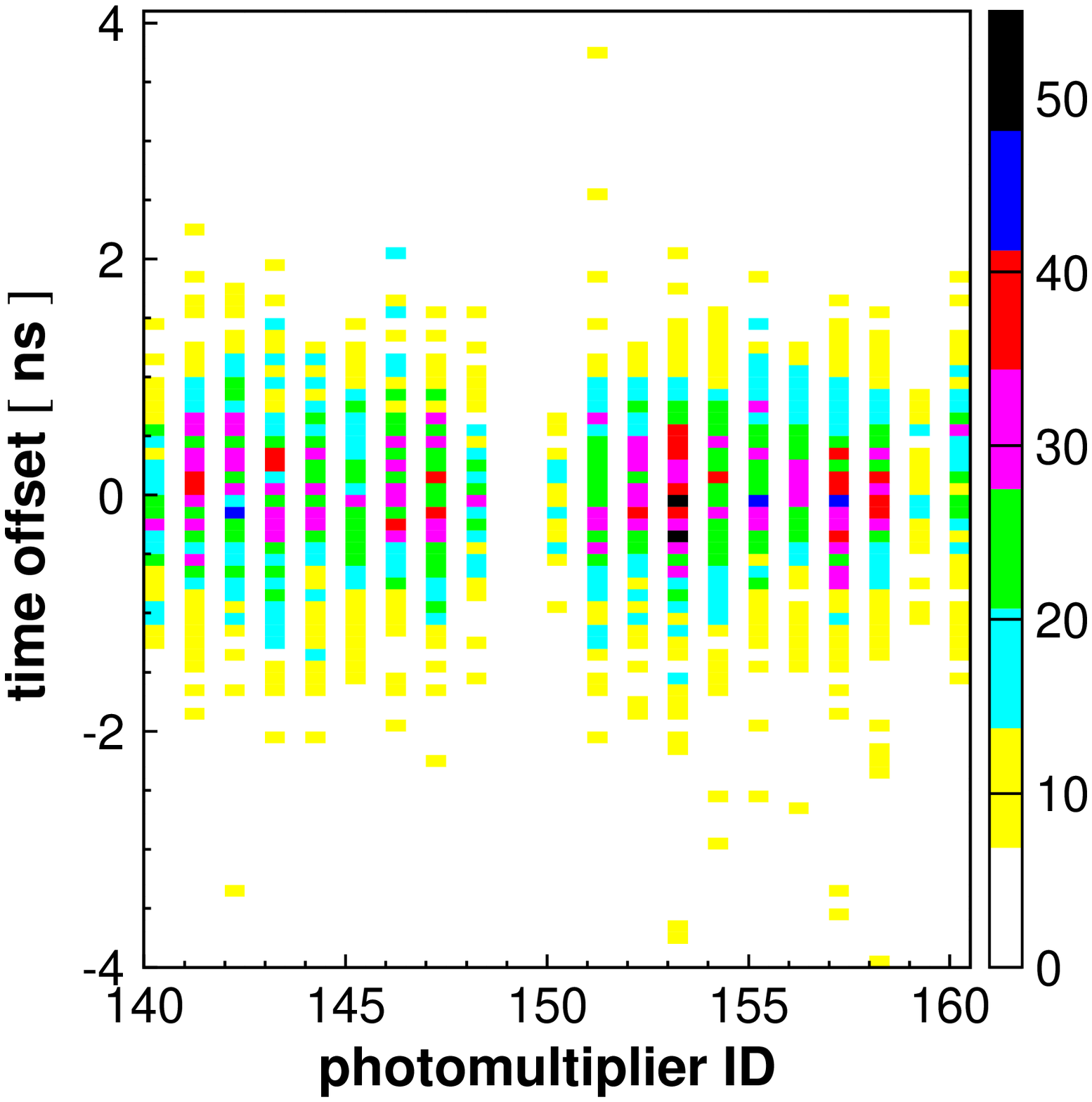,height=6.5cm,angle=0}
      }
      \put(6.5,0.8){{\normalsize {\bf a)}}}
      \put(14.0,0.8){{\normalsize {\bf b)}}}
  \end{picture}
  \caption{ {\small (a) Determined TDC time difference between 
	arbitrarily chosen 6$^{\textrm{th}}$ and 5$^{\textrm{th}}$ module of the S1 scintillator
	detector after the calibration. (b) Difference between time-of-flight measured 
	on the S1-S3 path from the signals registered in S1 and S3 detectors, and the 
	time-of-flight between S1 and S3 calculated from the reconstructed 
	momentum of the particle.   
  \label{calib_s1s3}
  }}
\end{figure}

After the drift chamber calibration,
we have chosen the events with only one track reconstructed
and a signal in two neighboring modules of the S1 detector.
The time offset for the first module in S1 (the one
standing most to the inner part of the ring) has been set
arbitrarily. Then, the time difference: $t_{s1,2}^{offset}=$ TDC$_{S1,2}$ - TDC$_{S1,1}$
has been taken as a correction for the time offset between the second and the first
module of the S1 detector. The same procedure has been repeated
for all modules. Figure~\ref{calib_s1s3}.a depicts the TDC time difference between
arbitrarily chosen
6$^{\textrm{th}}$ and 5$^{\textrm{th}}$ module of the S1 counter upon single particle crossing
through the overlap region, after the correction for the time offset of
the electronics and photomultipliers. One can see that this time difference is peaked
around the zero value, which confirms the correctness of the
time calibration of the S1 scintillator.

Having adjusted the time offsets of the individual photomultipliers 
in the S1 detector and hence having established the common start
for the time-of-flight measurement we need to adjust the time 
offsets for each of the 217 photomultipliers in the S3 detector
in order to set up the common stop.   
Similarly as it was in the case of the S1 detector, the  
TDC value for the individual electronic channel, corresponding to the 
single photomultiplier in the S3 detector reads:
\begin{equation}
TDC_{s3} = t_{s3} + t_{pos} + t_{s3}^{walk}(PM) + t_{s3}^{offset}(PM) - t_{trigger}. 
\label{s3cal}
\end{equation}
Here, the $t_{pos}$ denotes the time that light signals 
need to pass from the scintillation origin down to the
photomultiplier photocathodes. 

In Equations~\ref{s1cal} and~\ref{s3cal} the $t_{trigger}$
values are the same, and so the time-of-flight value 
\begin{equation}
TOF(PM) = t_{s3}(PM) - t_{s1}
\label{tof}
\end{equation}
does not depend on the triggering time.     
Hence, the only unknown quantities in the TOF calculation 
are the time offsets for the individual photomultipliers in 
the S3 detector -- $t_{s3}^{offset}(PM)$. These values determined 
for each photomultiplier can be 
extracted from comparing the TOF values with the time-of-flight
calculated from the reconstructed momenta of the particles.  
The differences after the calibration for the photomultipliers no. 140-160 
are presented in Figure~\ref{calib_s1s3}.b.
In order to extract the $t_{s3}^{offset}(PM)$ values the one-track 
events with an identified proton has been chosen. 

Having established all the time offsets of the photomultipliers 
in the S1 and S3 scintillators, the time-of-flight
were calculated as TOF~=~$t_{s3}-t_{s1}$. One should mention that 
the t$_{s1}$ have been taken as the  
average of times from the 
upper and lower photomultipliers of the hit module, 
and $t_{s3}$ were the weighted mean times for the 
proper cluster of photomultipliers in the S3 detector that 
registered the light pulses. For more details on the 
S1 and S3 calibration the reader is referred to~\cite{moskal4}.

\subsection{Position of the drift chambers}
\label{pozycja_komor}

\vspace{3mm}
{\small
A method for the search of the optimal position 
of the drift chambers is presented. 
}
\vspace{5mm}

\begin{figure}[H]
  \unitlength 1.0cm
  \begin{picture}(14.0,5.0)
      \put(0.00,1.0){
         \psfig{figure=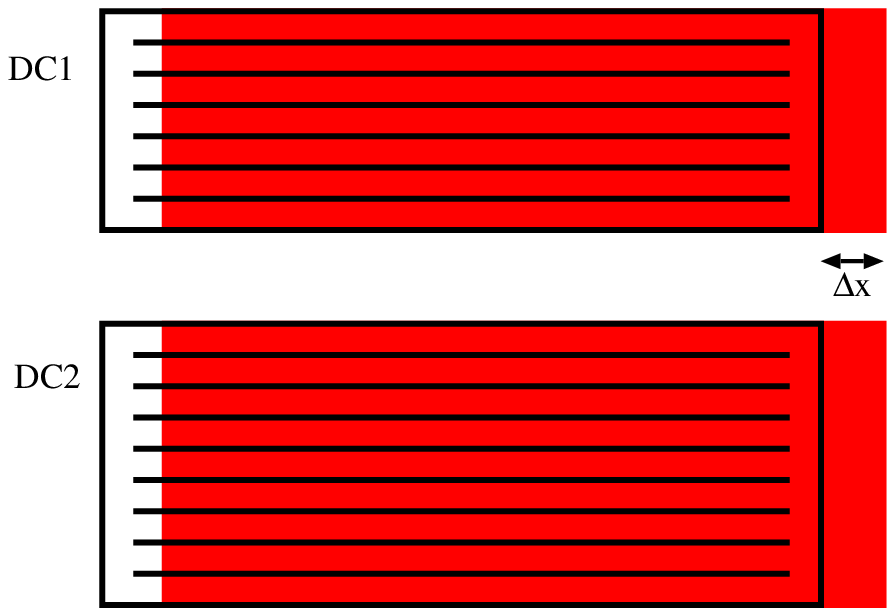,height=4.0cm,angle=0}
      }
      \put(7.70,0.8){
         \psfig{figure=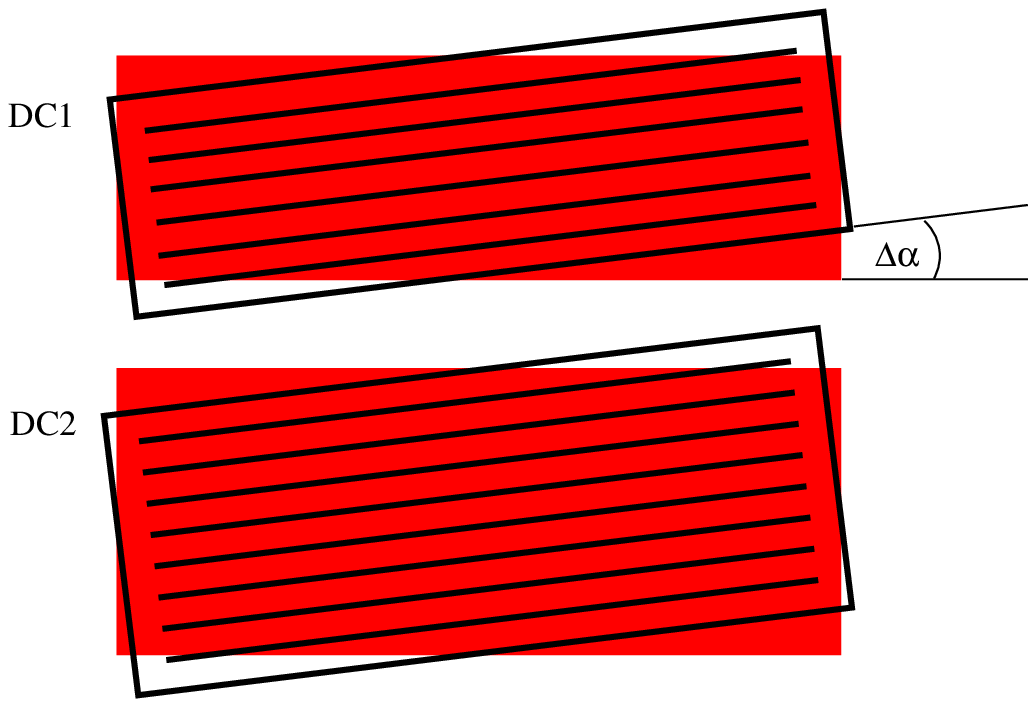,height=4.3cm,angle=0}
      }
      \put(6.5,0.2){{\normalsize {\bf a)}}}
      \put(14.0,0.2){{\normalsize {\bf b)}}}
  \end{picture}
  \caption{ \small{ Schematic illustration of the parameters $\Delta x$ (a)
                and $\Delta \alpha$ (b) mentioned in the text.
                  }
  \label{komory}
  }
\end{figure}

The time-space calibration of two drift chambers
D1 and D2 has been described in 
Section~\ref{dccalib}. It is very important to 
know the exact position of these drift chambers 
in order to achieve the 
required resolution in the momentum 
reconstruction of outgoing protons. 
This position can be parameterized for example 
introducing the parallel shift of drift chambers, which in the following we will denote
by $\Delta x$, and their inclination $\Delta \alpha$. 
Both parameters are illustrated in Figure~\ref{komory}. 

The parameters
$\Delta x$ and $\Delta \alpha$ 
influence the resolution of reconstructed four momenta of outgoing
particles, and consequently influence also indirectly
the missing mass resolution $\sigma\left(mm^2\right)$.     
The studies of the missing mass resolution as a 
function of $\Delta x$ and $\Delta \alpha$ have been performed 
on the data set for the excess energy of Q~=~10~MeV in order to optimize the position of drift chambers. 
We have varied these parameters in the range of $\left[-1cm;2cm\right]$
for $\Delta x$, and $\left[-0.015^{\circ};0.055^{\circ}\right]$ for $\Delta \alpha$, 
investigating the width of the missing mass spectra as a function of these parameters.  
The minimum value of the width of the $\eta$ peak in the missing mass squared histogram
(which we will present in Section~\ref{masa_brakujaca})
has been found for $\Delta x=0.35$~cm and $\Delta \alpha=0.025^{\circ}$ and  
equals $\sigma\left(mm^2\right)=0.00164$~GeV$^2$/c$^4$.
This corresponds to the value of the width of $\eta$ missing mass peak 
which is $\sigma\left(mm\right)=1.5$~MeV/c$^2$.

%
%
%
%

For the analysis at the excess energy of Q~=~36~MeV the same optimal parameters 
have been chosen.

It is important to note that the determined values of 
$\Delta x$ and $\Delta \alpha$ are in a very good agreement with the result 
obtained in the preceeding COSY-11 analyses~\cite{winter:02-1-en,winter-phd}.

\subsection{Beam-target relative settings}
\label{beamshift}

\vspace{3mm}
{\small
It is shown that the relative position of the beam 
and target influence the reconstructed momenta of the 
particles. A search for the optimal setting is 
presented. 
}
\vspace{5mm}

\begin{figure}[H]
  \unitlength 1.0cm
  \begin{picture}(14.0,5.0)
      \put(0.00,1.0){
         \psfig{figure=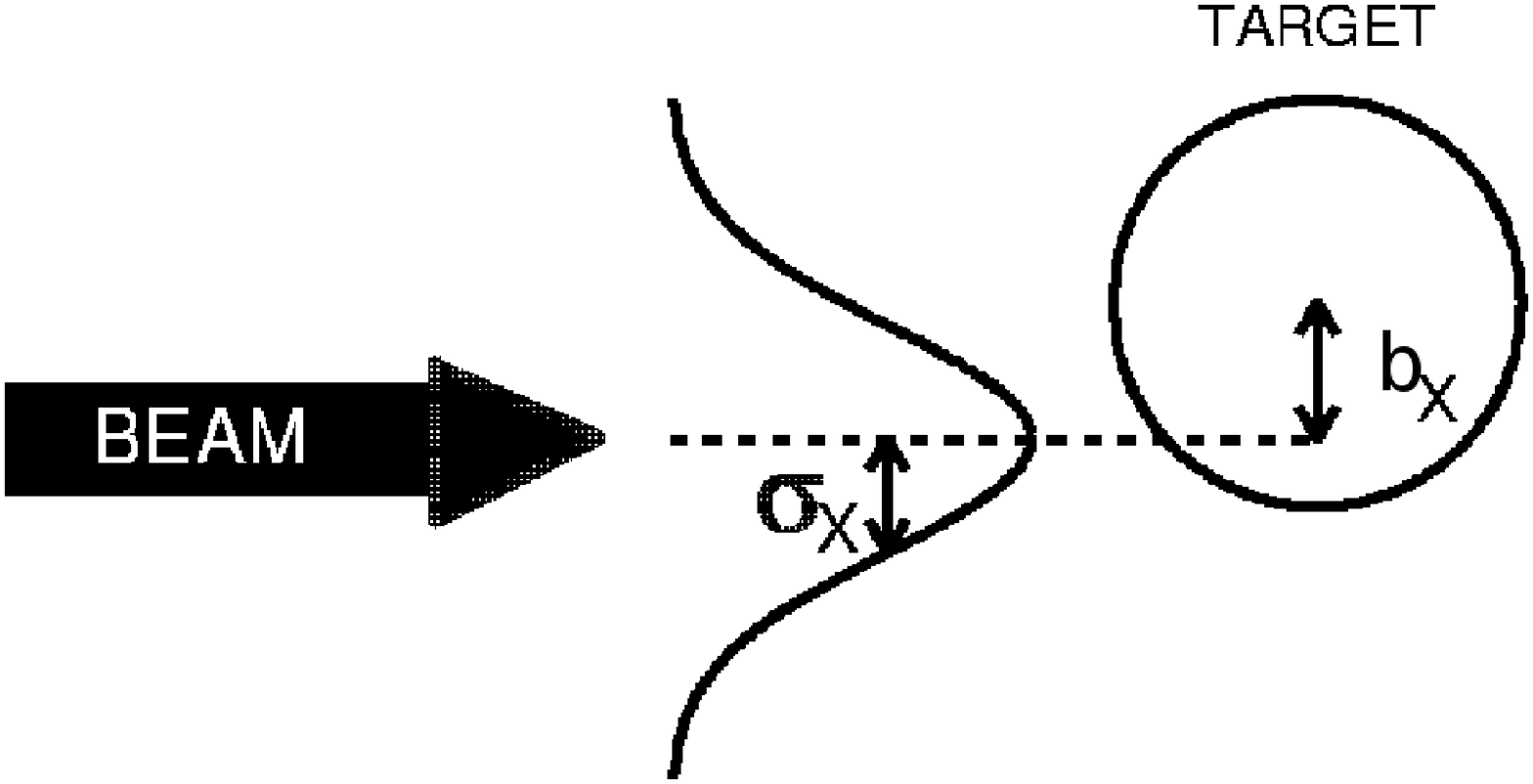,height=4.0cm,angle=0}
      }
      \put(8.50,0.8){
         \psfig{figure=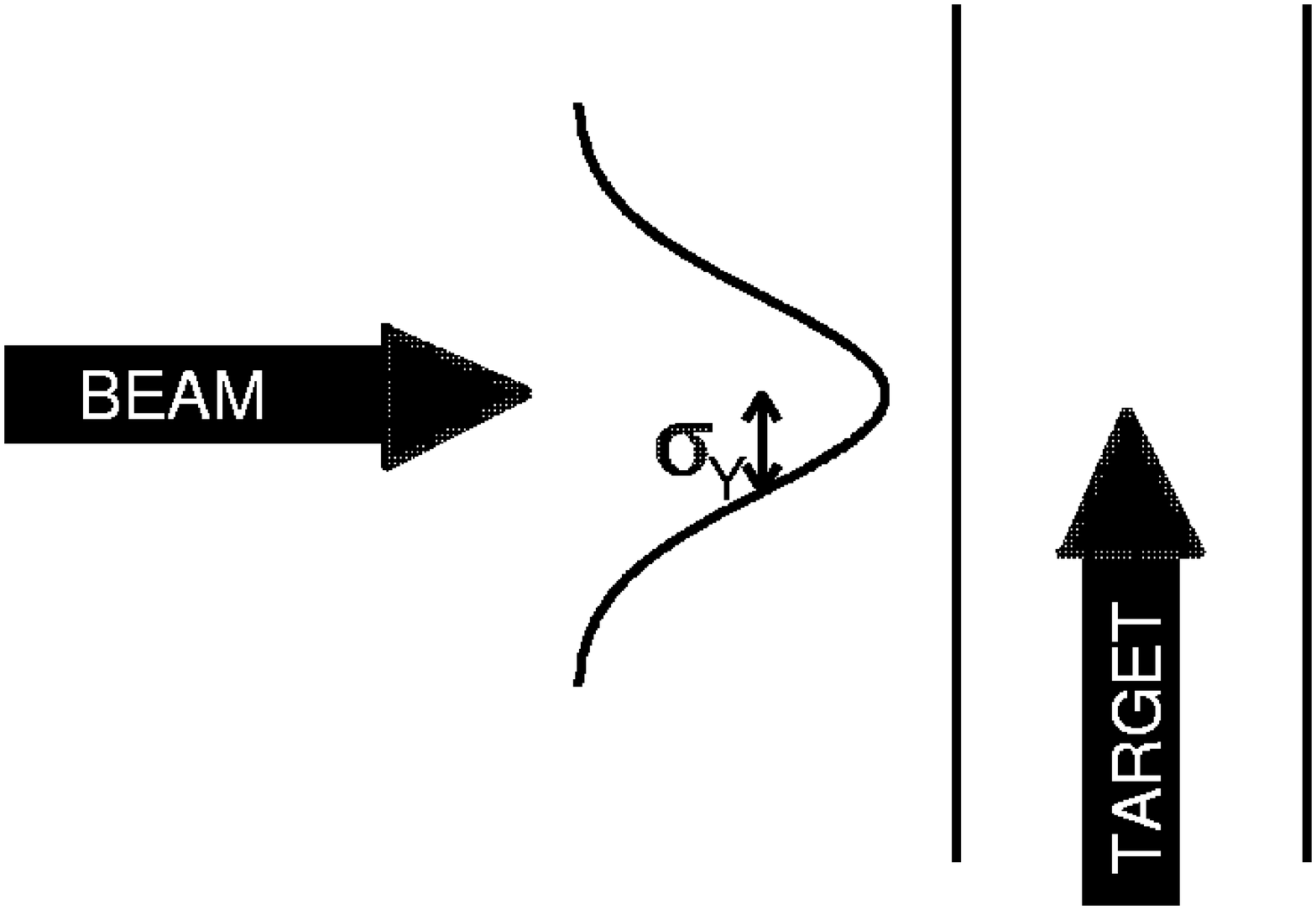,height=4.0cm,angle=0}
      }
      \put(6.5,0.2){{\normalsize {\bf a)}}}
      \put(14.0,0.2){{\normalsize {\bf b)}}}
  \end{picture}
  \caption{ \small{ Schematic illustration of the beam and
        target dimensions and their relative position:
        (a) view from above, (b) side view. Figure is adapted from~\cite{moskal-hab}. 
                  }
  \label{gaussy_pawla}
  }
\end{figure}

Another parameter that influences the resolution of the reconstruction of 
four momenta of protons are the dimensions of beam and target 
and also the relative position between them. 
In the procedure of momentum reconstruction it is assumed
that the interaction vertex is a point where the infinitesimally thin 
target intersects with the absolutely thin beam of protons. However, in reality 
both beam and target are of finite dimensions. Therefore, the reactions 
may be initiated in some finite volume, size of which depends on the 
abovementioned parameters, introducing a spreading to the reconstructed four
momenta of protons. In the analysis, following the studies reported in~\cite{moskal_nim,moskal-hab}, we used the approach that  
the target may be described by a cylinder of radius $r=4.5$~mm with uniformly distributed protons~\cite{dombrowski,khoukaz}, 
and the density of the protons in the beam
is given by the two dimensional gaussian distribution, as 
schematically illustrated in Figure~\ref{gaussy_pawla}. The $\sigma_X$ and $\sigma_Y$ denote spreadings
of the beam distribution in horizontal and vertical directions, respectively.   
Basing on the previous measurements~\cite{moskal_nim}, in our Monte Carlo studies we 
assumed these parameters to equal $\sigma_X=0.2$~cm and $\sigma_Y=0.5$~cm. 
Distance between the centre of the target and the centre of proton beam distribution 
is denoted in Figure~\ref{gaussy_pawla}.a by $b_X$. 

\begin{figure}[H]
  \unitlength 1.0cm
  \begin{picture}(14.0,7.5)
      \put(2.00,0.0){
         \psfig{figure=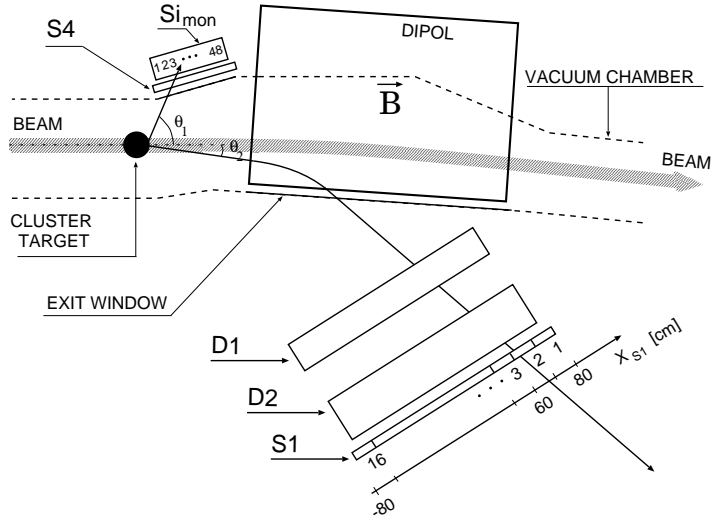,height=7.0cm,angle=0}
      }
  \end{picture}
  \caption{ \small{ Schematic view of the COSY-11 detection setup.
        Only detectors used for the measurement of the proton-proton elastic
        scattering are shown. Figure is adapted from~\cite{moskal4}. 
                  }
  \label{cosy11_polar}
  }
\end{figure}

Let us denote by $\Delta_X$ the deviation between 
the center of the reaction region and the nominal 
position of the target. 
In order to determine the position of the center of the reaction region 
we used the elastically scattered protons, registered 
by means of the detectors shown in Figure~\ref{cosy11_polar}.
Forward scattered protons from the elastic process are detected by the
scintillator detector S1 and the drift chambers D1 and D2. 
The recoil protons passing through the scintillator S4
reach the granulated position-sensitive silicon detector Si$_{mon}$.

\begin{figure}[p]
  \unitlength 1.0cm
  \begin{picture}(14.0,19.5)
      \put(0.00,13.5){
         \psfig{figure=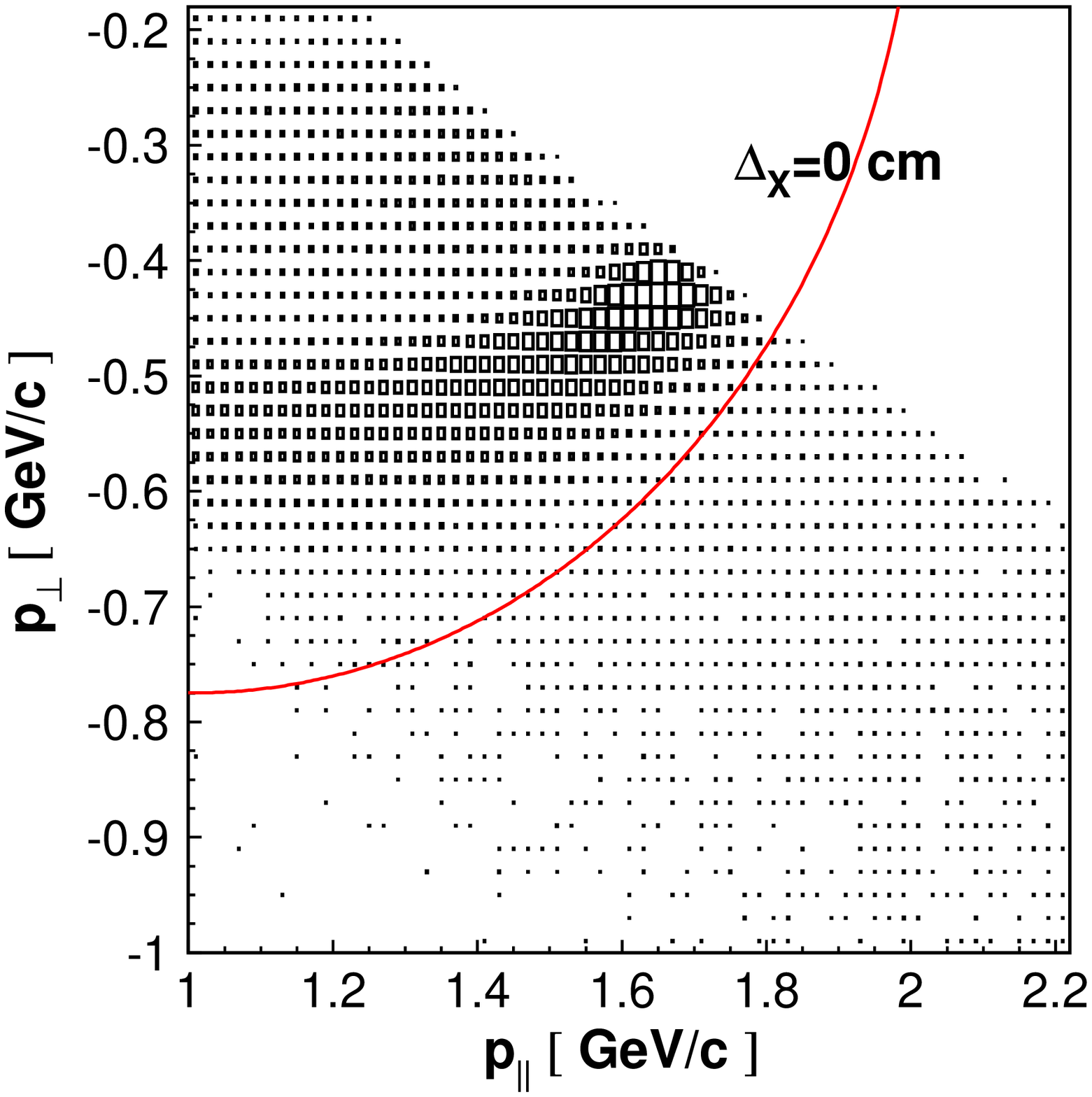,height=6.0cm,angle=0}
      }
      \put(7.70,13.5){
         \psfig{figure=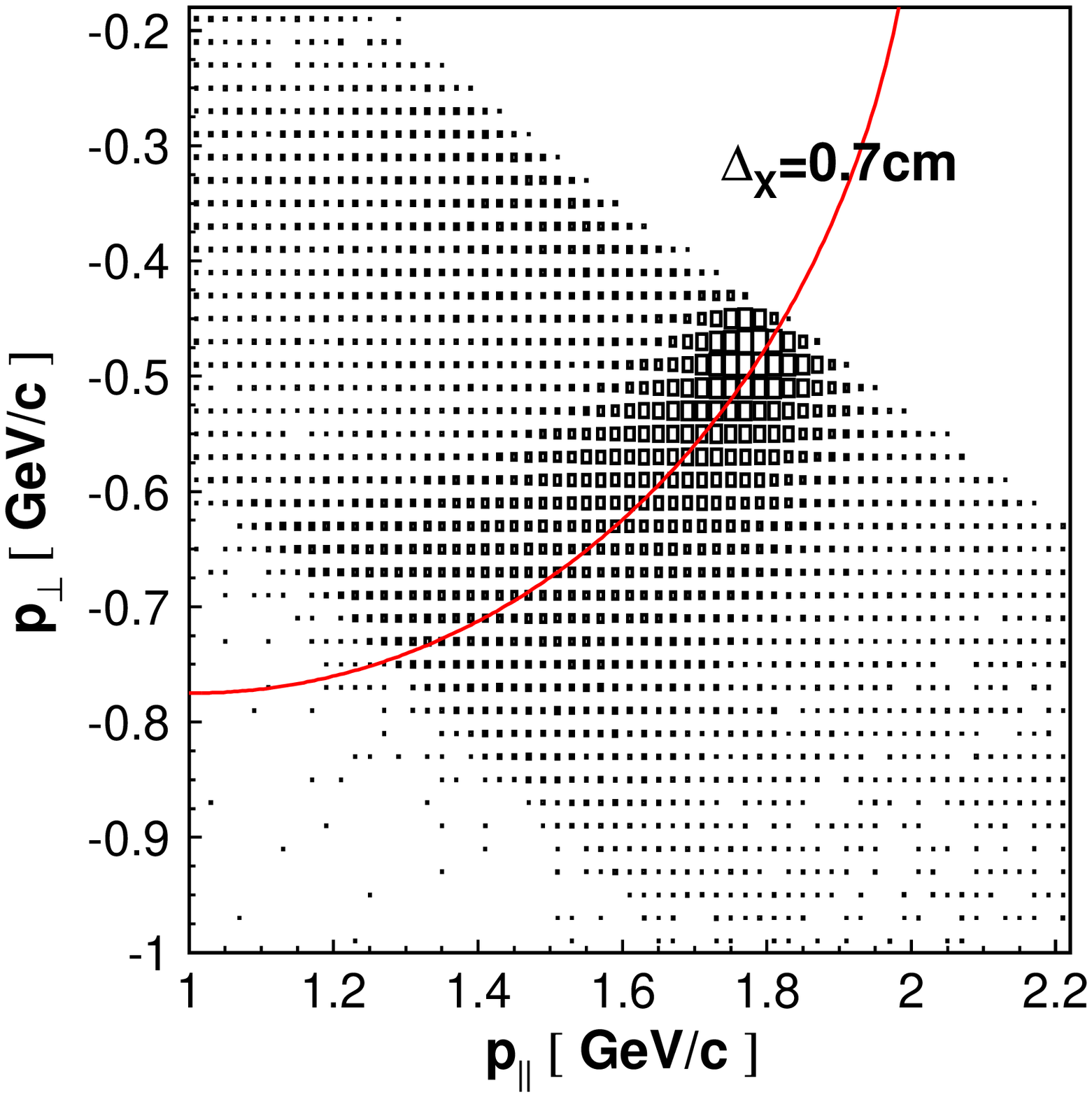,height=6.0cm,angle=0}
      }
      \put(0.00,7.5){
         \psfig{figure=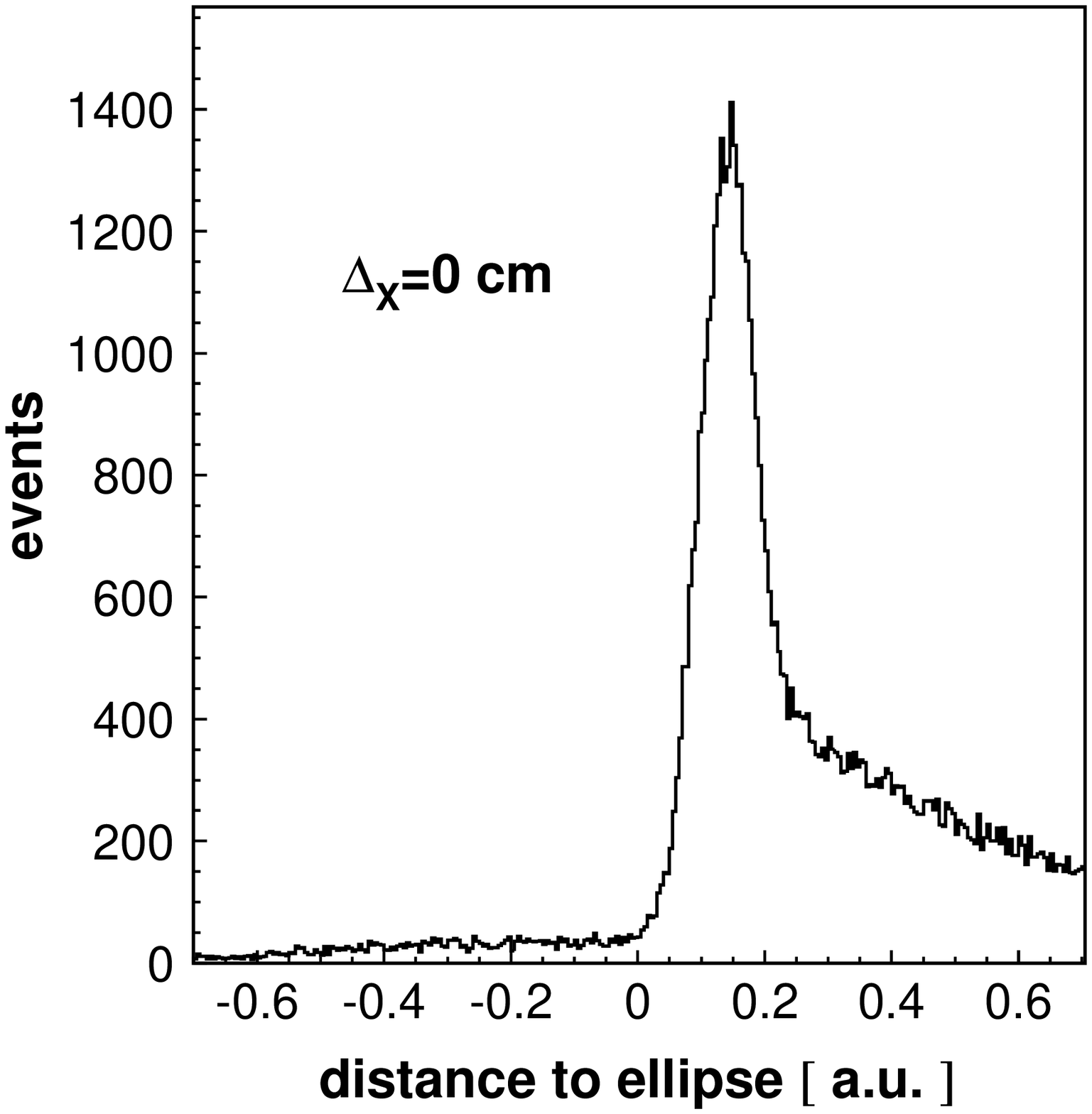,height=6.0cm,angle=0}
      }
      \put(7.70,7.5){
         \psfig{figure=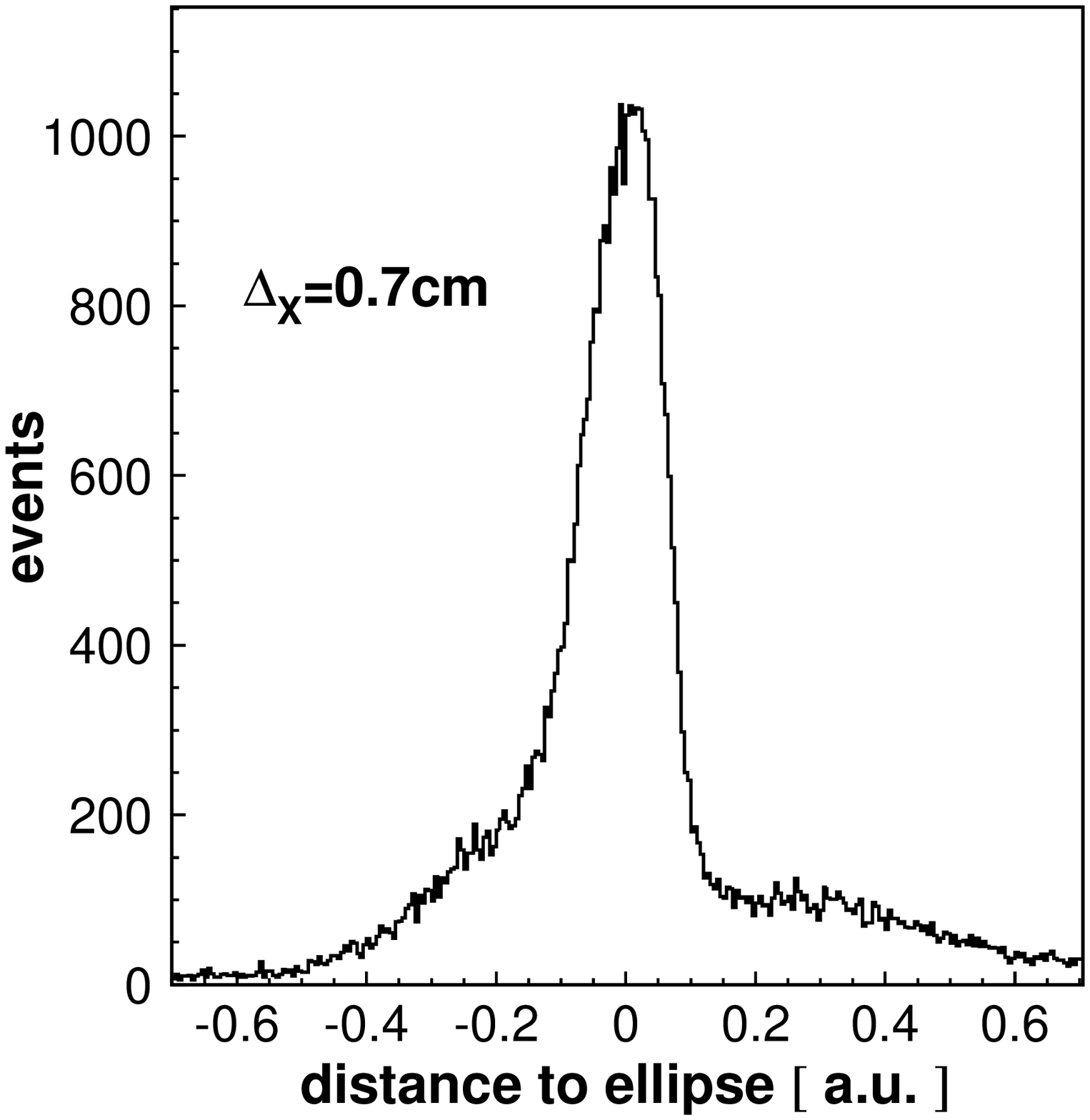,height=6.0cm,angle=0}
      }
      \put(3.85,1.3){
         \psfig{figure=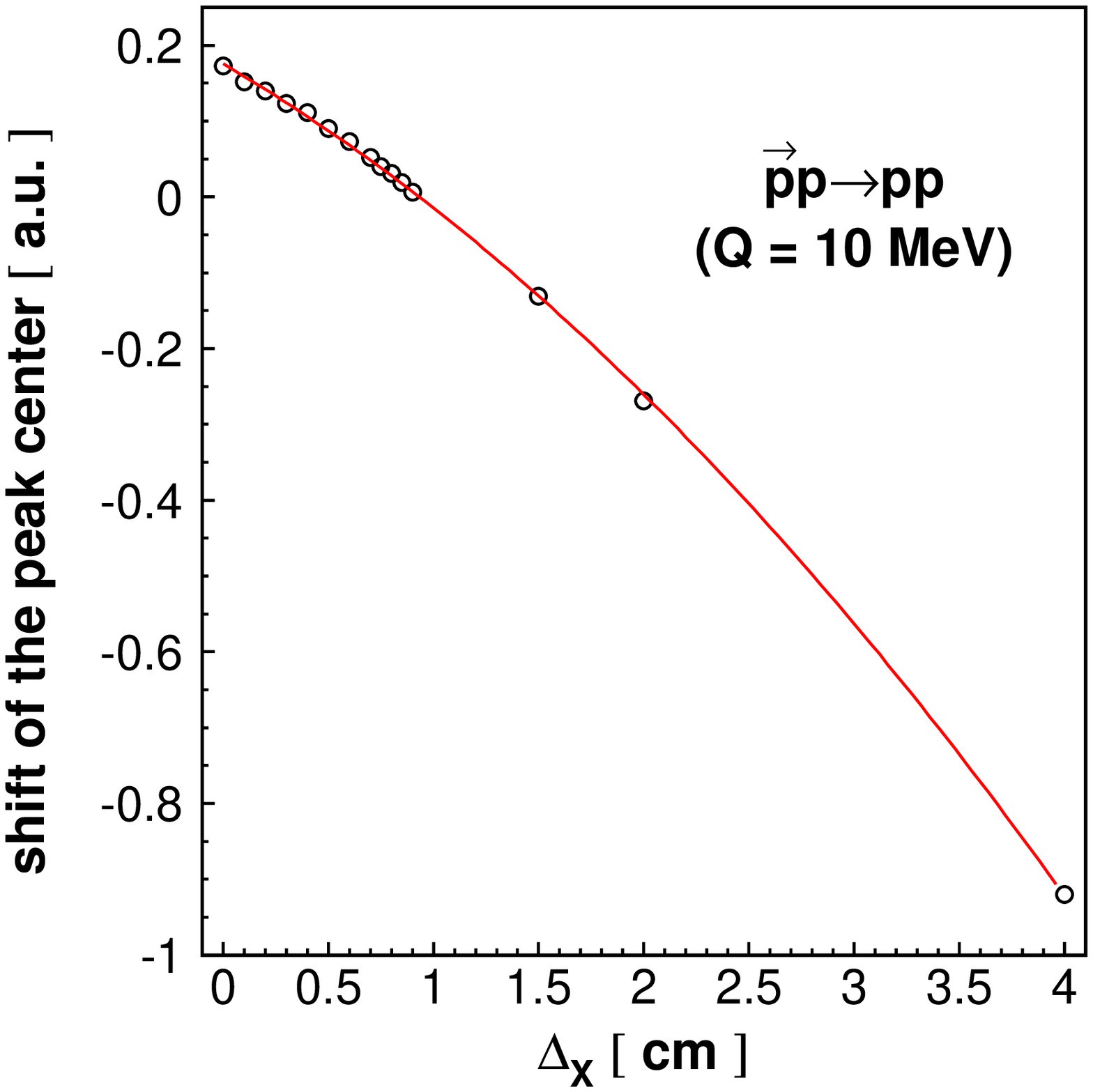,height=6.0cm,angle=0}
      }
      \put(5.8,13.3){{\normalsize {\bf a)}}}
      \put(13.3,13.3){{\normalsize {\bf b)}}}
      \put(5.8,7.3){{\normalsize {\bf c)}}}
      \put(13.3,7.3){{\normalsize {\bf d)}}}
      \put(9.35,1.1){{\normalsize {\bf e)}}}
  \end{picture}
        \hspace{-1cm}
  \caption{ \small{ Perpendicular versus parallel momentum component of the 
	elastically scattered protons measured 
	at the beam momentum of p$_{beam}=2.010$~GeV/c for 
	 $\Delta_X$~=~0 (a) and $\Delta_X$~=~0.7~cm (b). The solid line represents 
	the theoretical kinematical ellipse for this particular value of the beam momentum. 
	Enhancement of the event distribution shows the elastically scattered 
	protons over the constant background originating from the production reactions.
	Figures (c) and (d) show the projections
	of the event distribution along the theoretical ellipse for data from figures (a) and (b) respectively.
	(e) Distance 
	between the theoretical ellipse and the centre of the points 
	distribution from the (p$_{\perp}$, p$_{\parallel}$) plot versus the 
	parameter $\Delta_X$. 
                  }
  \label{beamshits}
  }
\end{figure}

As the proton scattered forward under the $\theta_2$ angle -- passes through 
the stack of drift chambers D1 and D2, its trajectory may be reconstructed and subsequently traced back 
to the reaction vertex, through the known map of the magnetic field
inside the COSY dipole. This procedure yields the momenta of the particles 
passing through the drift chambers. 
Further, the reconstructed momentum vector may be decomposed at the reaction point into two components: 
the transversal (p$_{\perp}$) and the longitudinal one (p$_{\parallel}$),  
with respect to the momentum of the beam protons.  
The two body kinematics puts some constraints on these 
two components, namely the set of points $\left(p_{\parallel},p_{\perp}\right)$ 
should form an ellipse~\cite{byckling} in the momentum plane. 
The size of this ellipse depends on the $\sqrt{s}$ -- the total energy 
in the centre-of-mass system. A part of one arm of this theoretical ellipse 
for the $\sqrt{s}=2.43$~GeV, corresponding to the proton beam momentum of p$_{beam}=2.010$~GeV/c, is 
depicted in Figures~\ref{beamshits}.a and~\ref{beamshits}.b
as the solid line.

The value of the reconstructed particle's momentum depends on the correctness
of the assumption of the reaction vertex position. 
Shifting the reaction vertex upwards the beam line presented 
in Figure~\ref{cosy11_polar} 
would yield in higher reconstructed momenta
of the particles, as we decrease the curvature of the trajectory inside 
the dipole magnet.
On the other hand, a parameter $\Delta_X$ with the negative 
values (shift of the reaction vertex downwards in Figure~\ref{cosy11_polar})
would induce a higher curvature of the tracks inside the dipole magnet, 
hence lower momenta of the reaction products.  
The values of the components of the reconstructed momentum vector
correspond to a point in the momentum plane p$_{\perp}$ -- p$_{\parallel}$.
This point is lying inside or outside the expected
theoretical ellipse, depending whether the reconstructed particle's momentum 
is lower or greater than the actual one.

Following the studies of~\cite{moskal_nim} we have been searching
for the optimal position of the reaction vertex, by changing the 
$\Delta_X$ parameter in the range from 0 to 4~cm.
In Figure~\ref{beamshits}.a a distribution of the reconstructed
momenta of the elastically scattered protons is presented, for the case where $\Delta_X$ was set
to $\Delta_X=0$. The enhancement of the event distribution is assigned to the
elastically scattered protons. The background is mainly due to many body
reactions. As one can see, the reconstructed
momenta of the elastically scattered protons are lying inside the theoretical kinematical
ellipse, which means that the reconstruction yields too small values of the particle's momentum.
This is also seen in Figure~\ref{beamshits}.c, where we presented the projection of the event distribution
along the kinematical ellipse.
One can notice the shift of the centre of the distribution
outside the zero value. 

For the same data sample and with 
$\Delta_X=0.7$~cm the components of the reconstructed particle's momenta 
are lying on the kinematical ellipse, as can be seen in Figure~\ref{beamshits}.b.
Figure~\ref{beamshits}.d shows the projection of the data set onto the kinematical 
ellipse for this case. Indeed, the event distribution is centered around 
the zero value. This means that for the data sample we considered here, the 
optimal value of the $\Delta_X$ parameter equals $\Delta_X=0.7$~cm.

The same procedure has been differentially performed for the whole sample of data. 
The experimental data have been divided into 32 groups, corresponding to 
about 6 hours of measurement, and for each out of these 32 groups 
the value of the $\Delta_X$ parameter has been determined
using a $\chi^2$ test, where $\chi^2$ was calculated 
between the experimental event distribution and the theoretical 
kinematical ellipse.

In Figure~\ref{beamshits}.e 
we have presented the relation between the parameter $\Delta_X$ 
and the shift of the centre of the peak, 
originating from the projection of the experimental points 
along the kinematical ellipse.
This dependence have been 
studied in the $\Delta_X$ range from 0 to 4~cm, and it was found that it 
may be described fairly well by a polynomial function of 
second order, the one presented in the Figure~\ref{beamshits}.e. 
One can see in this figure that the zeroth shift of the peak centre
corresponds to the value of $\Delta_X\approx$~0.7~cm.

\subsection{Missing mass technique}
\label{masa_brakujaca}

\vspace{3mm}
{\small
The idea of the missing mass technique is sketched. 
The software presort of the events with two protons
in the final state is given.   
We also present the missing mass spectra for both excess energies. 
}
\vspace{5mm}

As the $\eta$ meson is a non-charged short living particle, 
its direct registration in any of the nowadays known detectors is impossible
due to the very low distance that is passed by this particle before it 
decays\footnote{The Particle Data Group~\cite{yao} gives the value of the 
total width of the $\eta$ meson $\Gamma_{\eta}=1.30~\pm~0.07$~keV, which 
yields the mean life-time of the $\eta$ meson equal to $\tau_{\eta}=\left(5.10\pm0.29\right)\cdot 10^{-19}$~s, 
corresponding to $c\tau_{\eta}=1.53\cdot 10^{-10}$~m.}. 
To our best knowledge, nowadays three methods
are applied in the worldwide experiments in order
to identify the $\eta$ meson.  
First of them is the missing mass technique, 
which will be described in the following part of this section.  
Second method is the reconstruction of the four momentum of the $\eta$ meson
by means of the identification of its decay products, e.g. $2\gamma$ rays. 
Third method is the combination of the two latter techniques, namely, 
the measurement of the $2\gamma$ decay in coincidence with the directions (and/or energies)
measurements of two final particles. This method, for example, will be used 
at the recently installed WASA-at-COSY experiment~\cite{wasa}.

The idea of the missing mass technique involves the 
theorem of the four momentum conservation. 
Let us consider the $pp\to ppX$ reaction with the 
accelerated proton beam and the proton target with zero momentum value. 
Denoting by $P_b=\left(E_b,\vec{p}_b\right)$, $P_t=\left(E_t,0\right)$,
$P_1=\left(E_1,\vec{p}_1\right)$, and $P_2=\left(E_2,\vec{p}_2\right)$
the four momenta of the beam, target, and two outgoing protons, respectively, 
one can calculate the missing mass ($m_X$) of an undetected particle or system of 
particles in the exit channel
according to the formula:
\begin{eqnarray}
      m_{X}^{2} =  E_{X}^{2} - \vec{p}_{X}^{~2}  =  \left(P_{b} + P_{t} - P_{1} - P_{2}\right)^{2} =
\nonumber \\
 = \left(E_{b} + E_{t} - E_{1} - E_{2}\right)^{2} - \left(\vec{p}_{b} + \vec{p}_{t} - \vec{p}_{1} - \vec{p}_{2}\right)^{2}.
\label{miss_form}
\end{eqnarray}

After the adjustment of the drift chambers' parameters 
$\Delta x$ and $\Delta \alpha$ and the parameter $\Delta_X$,
and having performed the time calibration of the S1 and S3 scintillators, 
as well as the calibration of the drift chambers (described 
in Sections~\ref{dccalib}--~\ref{beamshift}) one can  
reconstruct the momenta of the outgoing particles 
and also their velocities. 
Having the momenta ($\vec{p}$) and velocities ($\beta$) 
of the particles one can calculate their squared invariant masses applying the 
relativistic formula:
\begin{equation}
m^2 = \frac{(\vec{p})^2\left(1-\beta^2\right)}{\beta^2}. 
\label{mass}
\end{equation}

\begin{figure}[H]
  \unitlength 1.0cm
  \begin{picture}(14.0,7.0)
      \put(3.00,0.0){
         \psfig{figure=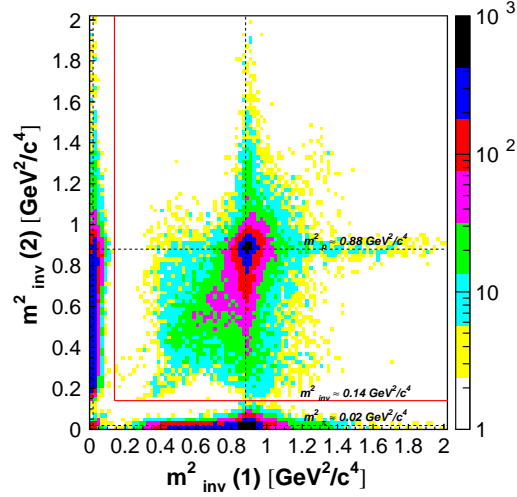,height=7.0cm,angle=0}
      }
  \end{picture}
  \caption{ \small{ Squared masses of two simultaneously measured particles 
		in the exit channel. Events from the measurement
		of the $\vec{p}p\to pp\eta$ reaction at the excess energy of Q~=~10~MeV
		are presented.  
                  }
  \label{inv_mass_plot}
  }
\end{figure}

Figure~\ref{inv_mass_plot} shows 
the distribution of the invariant masses of two particles 
measured in coincidence at the excess energy of Q~=~10~MeV. Over the background 
arise four ``isles" of the events with $\pi^{+}$-$\pi^{+}$, $\pi^{+}$-proton, proton-$\pi^{+}$, 
and proton-proton identified in the final state.
For the further analysis only events with two protons 
were taken into account. The software cut on the squares of the missing masses
of particles is denoted with the solid lines. The dotted lines show the 
values of the $\pi^{+}$ meson and proton squared masses.   

\begin{figure}[H]
  \unitlength 1.0cm
  \begin{picture}(14.0,7.0)
      \put(0.00,0.8){
         \psfig{figure=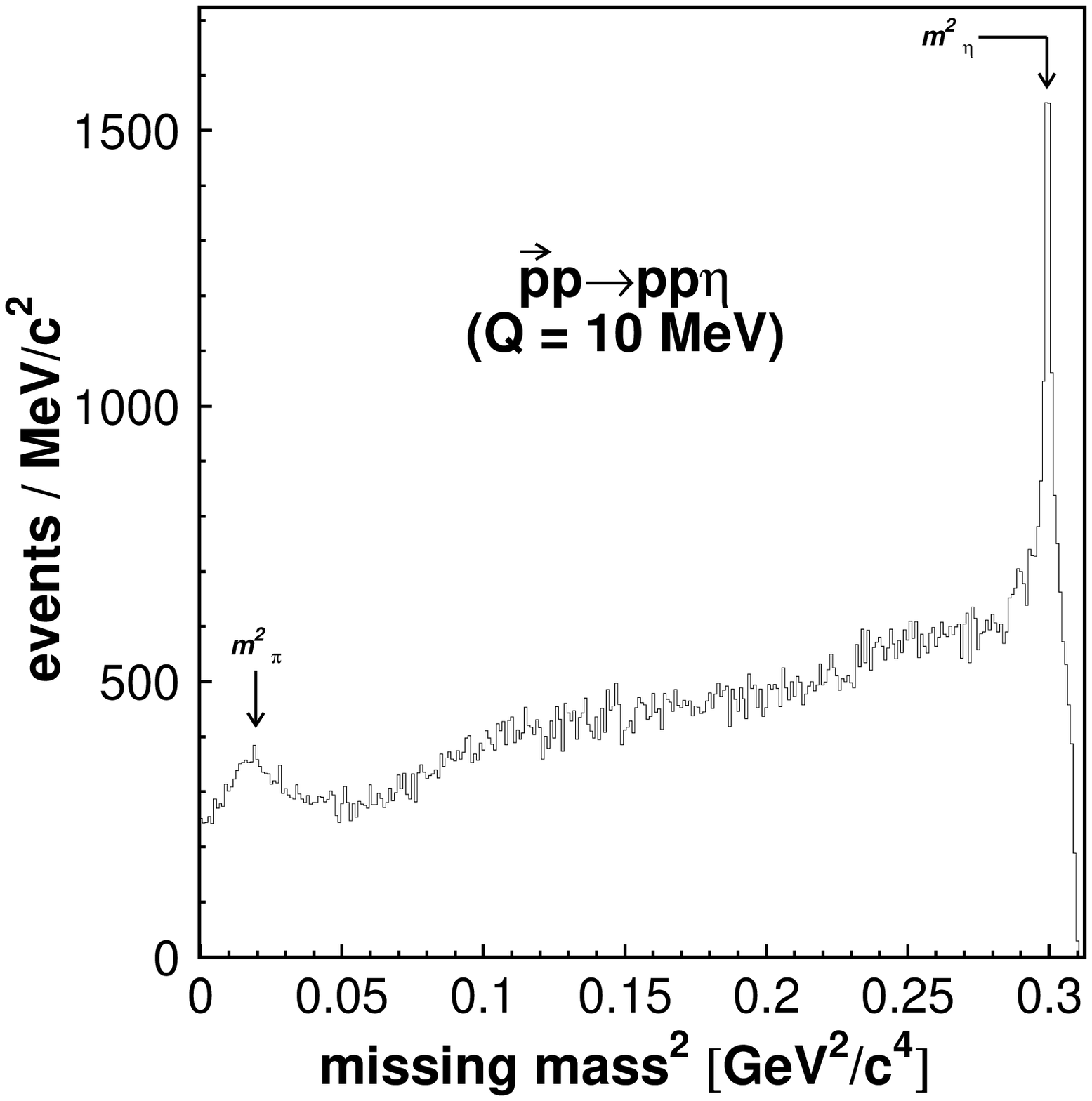,height=7.0cm,angle=0}
      }
      \put(7.50,0.8){
         \psfig{figure=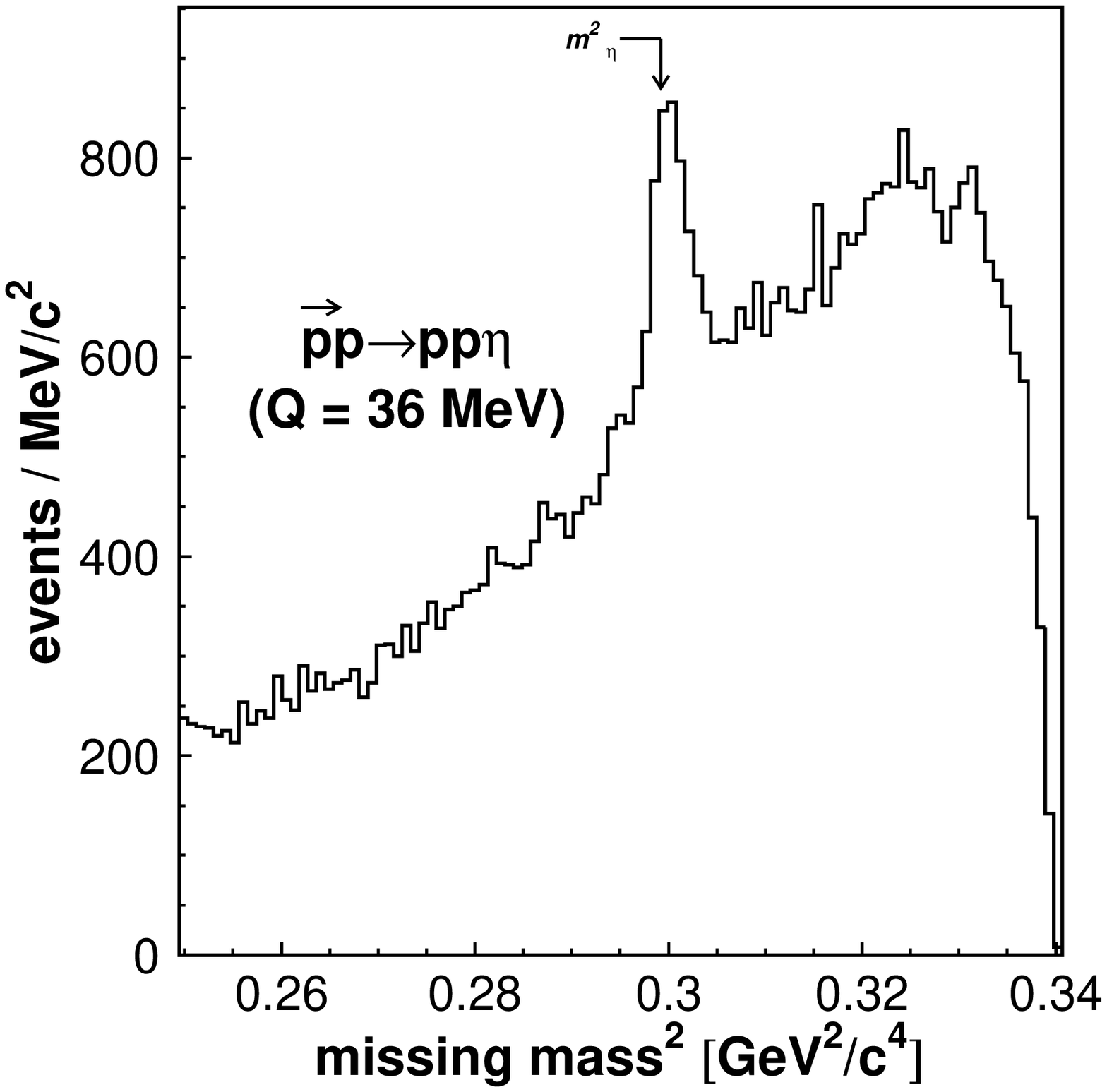,height=7.0cm,angle=0}
      }
      \put(6.5,0.6){{\normalsize {\bf a)}}}
      \put(14.0,0.6){{\normalsize {\bf b)}}}
  \end{picture}
	\vspace{-0.5cm}
  \caption{ \small{ (a) Spin-averaged spectrum of the square value of the missing mass 
	for the $\vec{p}p\to pp\eta$ reaction at the excess energy of Q~=~10~MeV, 
	as measured by means of the COSY-11 detector setup. 
	(b) The same, but for the excess energy of Q~=~36~MeV.  
                  }
  \label{miss_spectra}
  }
\end{figure}

\vspace{-2mm}
Knowing all the four momenta -- $P_b$, $P_t$, $P_1$, and $P_2$ --   
and applying Equation~\ref{miss_form} one can calculate the missing mass 
of X for each event separately. 
The spin-averaged missing mass spectra for the $\vec{p}p\to pp\eta$ reaction
as measured using the COSY-11 detection setup are presented
in Figure~\ref{miss_spectra}.
Over the wide background, originating mainly
from the multipionic production, clear $\eta$ peaks
are visible for both measurements. There are around
3000 and 1500 $\eta$ events at Q~=~10 and 36~MeV, respectively,
integrated over both spin orientations.

The differences in the shape of the background are due to the
different trigger setting between two measurements, and also 
due to the fact that in both analyses different software cuts
have been applied. 
It is also visible that the signal 
from the $\eta$ meson production at Q~=~10~MeV is closer to the kinematical limit than
the signal at Q~=~36~MeV. This is due to the fact that the  
kinematical limit m$_{max}$ 
\begin{equation}
m_{max}=m_{\eta}+Q, 
\label{kin_limit}
\end{equation}
equals\footnote{In the calculations
the $\eta$ meson mass $m_{\eta}=547.75$~MeV/$c^{2}$~\cite{eidelman} has been used.} 
\mbox{m$_{max}^{2}$= 0.311~MeV$^{2}$/c$^{4}$} at Q~=~10~MeV and it equals 
\mbox{m$_{max}^{2}$= 0.340~MeV$^{2}$/c$^{4}$} at Q~=~36~MeV.

\vspace{0.5cm}
\section{
Identification of the spin up and down modes
}
\label{spin_up_down}

\vspace{3mm}
{\small
It is shown how we resolve the problem of the identification 
of the spin state of the polarised proton beam using the elastically scattered events. 
}
\vspace{5mm}

A few words of explanation are required concerning the method
of spin up and down identification. 
In principle the polarisation state could have been
determined from the known spin orientation in the polarised ion source
and the number of imperfection resonances, mentioned in Section~\ref{ii}, to
be crossed in order to accelerate the beam up to the required momentum.
However, even without a precise knowledge of the beam optics during 
the acceleration process it is possible to determine the spin state of 
accelerated protons.

To enable an offline assigment of the polarisation mode
two scaler channels have been used, working in the sequence mode.
Whenever there was a beam in the spin state A one of the scaler channels 
has been working while the other one remained idle, and vice-versa.

In order to identify the spin state A with spin up or down mode
the detector system of Figure~\ref{cosy11_polar} for registration of the elastic scattering
in the accelerator plane has been used.

For the elastically scattered
protons at the beam momentum of $p_{beam}=2.010$~MeV/c, 
which reach the drift chambers D1 and D2 (see Figure~\ref{cosy11_polar})
the scattering angle $\theta_{2}$ may vary from 35$^{\circ}$ to circa 80$^{\circ}$~\footnote{Which
is restricted by the acceptance of COSY-11 setup for the $pp\to pp$ reaction -- 
see also Figure~\ref{theta_s1}.} in the centre-of-mass system.
Here we deal only with scattering to the right with respect to the polarisation plane.
For spin up mode, where the polarisation vector of the proton beam is
pointing along the $Oy^{acc}$ axis defined in Section~\ref{refi}, the
formula~\ref{cross_3} -- integrated over $m_{pp}$, $m_{p\eta}$, and $\psi$ angle -- reads:
\begin{equation}
\sigma\left(\theta_2,P\right)=\sigma_0\left(\theta_2\right) \left(1 - P A_y\left(\theta_2\right)\right)\qquad \textrm{for} \cos\theta_2\in\left(35^{\circ},80^{\circ}\right), \phi\approx 180^{\circ} ;
\label{cross__1}
\end{equation}
while for spin down it yields:
\begin{equation}
\sigma\left(\theta_2,P\right)=\sigma_0\left(\theta_2\right) \left(1 + P A_y\left(\theta_2\right)\right)\qquad \textrm{for} \cos\theta_2\in\left(35^{\circ},80^{\circ}\right), \phi\approx 180^{\circ} ;
\label{cross__2}
\end{equation}
where the $\pm$ sign in front of $P A_y(\zeta)$ term
is subject to the Madison convention quoted in Section~\ref{refi}.
As the number of registered events is proportional
to the $\sigma(\zeta,P)$, from Equations~\ref{cross__1} and~\ref{cross__2} we expect to register more
elastic scatterings during the spin down mode, since
A$_y\left(\theta_2\right)$ is positive for the quoted $\theta_2$ ranges~\cite{altmeier}.
Starting with the unidentified spin orientations A and B, we divided
the available range of $\theta_2$ into 10 bins, each of 4$^{\circ}$ width.
For each bin and both spin orientations the numbers of elastic scatterings N$_{A}\left(\theta_2\right)$ and N$_{B}\left(\theta_2\right)$
have been determined and exemplary results for the experiment
performed at Q~=~10~MeV are presented in Table~\ref{tabelka_elastyczne}.

\begin{table}[H]
 \begin{center}
   \begin{tabular}{|c|c|c|}
    \hline
      $\theta_2 [^{\circ}]$ & N$_{A}\left(\theta_2\right)$ & N$_{B}\left(\theta_2\right)$  \\
    \hline
        40 & 200054 $\pm$ 836 & 353972 $\pm$ 1106 \\
        44 & 159864 $\pm$ 748 & 275061 $\pm$ 983 \\
        48 & 128808 $\pm$ 679 & 214802 $\pm$ 874 \\
        52 & 103183 $\pm$ 612 & 162223 $\pm$ 761 \\
        56 &  71068 $\pm$ 513 & 105549 $\pm$ 616 \\
        60 &  65849 $\pm$ 493 &  90496 $\pm$ 575 \\
        64 &  61430 $\pm$ 469 &  76328 $\pm$ 520 \\
        68 &  53476 $\pm$ 437 &  61810 $\pm$ 467 \\
        72 &  37984 $\pm$ 364 &  42024 $\pm$ 384 \\
        76 &  14784 $\pm$ 232 &  15694 $\pm$ 236 \\
    \hline
   \end{tabular}
     \caption{ {\small Number of elastic scatterings N$_{A}\left(\theta_2\right)$ and N$_{B}\left(\theta_2\right)$
                for the unidentified spin orientations A and B as a function
                of the CM scattering angle $\theta_2$ as measured during
                the experiment with $p_{beam}$~=~2010~MeV/c (Q~=~10~MeV).
          }
     \label{tabelka_elastyczne}
         }
 \end{center}
\end{table}

Having a look at the numbers in Table~\ref{tabelka_elastyczne} one can notice
that for each bin of the scattering angle $\theta_2$
we have:
\begin{equation}
N_{B}\left(\theta_2\right) > N_{A}\left(\theta_2\right),
\end{equation}
therefore one can identify mode A with the spin up orientation and
mode B with spin down. The difference between N$_{A}\left(\theta_2\right)$
and N$_{B}\left(\theta_2\right)$ is undoubtfully statistically significant. 
The same method was used for identification of the spin up and
down modes for the experiment at Q~=~36~MeV.

\clearpage
\section{
Calculation of the relative luminosity
} 
\label{luminosity}
\label{mmm}

\vspace{3mm}
{\small
A method of the determination of the relative 
luminosity is presented. A detection subsystem used 
for this purpose is described. Systematical errors of this method 
are evaluated. 
}
\vspace{5mm}

In order to determine the relative luminosity $L_{rel}$
of Equation~\ref{lumi_rel},
the scattering in the
polarisation plane (see page~\pageref{kosne}) had been used, as  
in this plane the differential cross section for any nuclear reaction 
induced by the strong interaction
does not depend on the magnitude of beam polarisation. 
This is a consequence of the fact, that the parity in strong interactions 
is conserved. For detailed explanation the reader is referred to Appendix~\ref{parity}. 
 
Therefore, whenever in the experiment we are restricted to the scattering 
in the polarisation plane the result of the measurement should only
depend on the polar angle $\theta$, and is independent of the 
beam polarisation. Thus, the number of reactions -- $n(t)$ -- 
registered in the polarisation plane within a time interval $t$ 
may be used as the measure of the integrated luminosity over this time interval: 
\begin{equation}
\int_{0}^{t} L(t') dt' \sim n(t).
\label{n_od_t}
\end{equation}

\begin{figure}[h]
  \unitlength 1.0cm
        \begin{center}
  \begin{picture}(14.5,8.5)
    \put(0.0,0.0){
      \psfig{figure=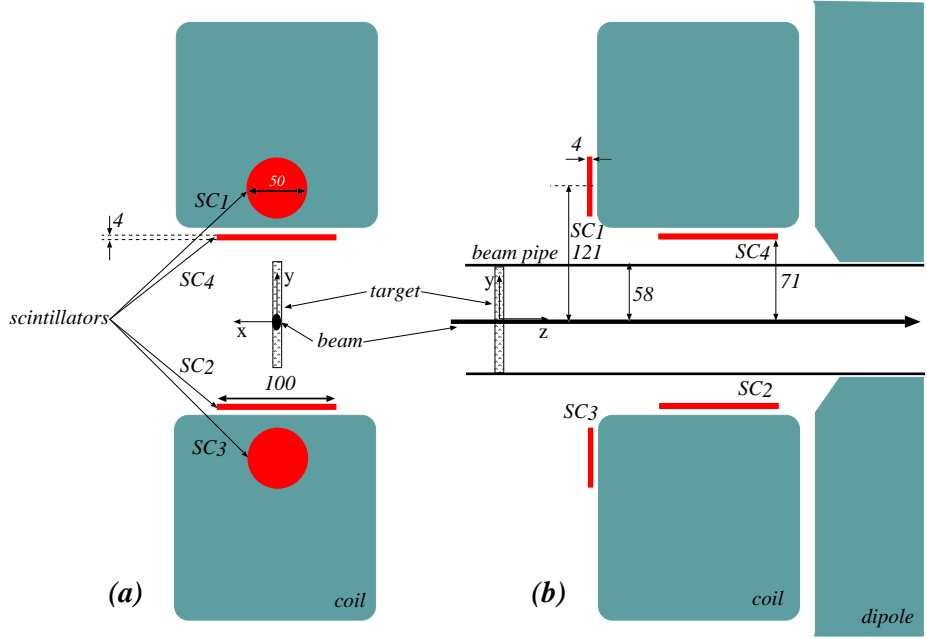,height=8.5cm,angle=0}
    }
  \end{picture}
  \caption{ {\small Schematic view of the detection system dedicated 
	for the relative luminosity determination, 
	as mounted at the COSY-11 section of the COSY ring.  
	(a) front view, (b) side view. Beam is circulating 
	along the z-axis and target is along the y-axis.
                 }
 \label{uklad_do_swietlnosci}
  }
        \end{center}
\end{figure}

In order to determine the relative luminosity, 
the detector system schematically presented 
in Figure~\ref{uklad_do_swietlnosci} had been used.  
It consists of four scintillator detectors: two round-shaped, placed 
vertically in front of the COSY-11 magnet, and two 
square-shaped located horizontally between the magnet coils. 
The location of the scintillators has been chosen in order 
to conform the kinematic conditions 
of the elastically scattered protons
in the polarisation plane. 
The coincidence rate
originates mainly from the elastic proton-proton scattering, which 
constitutes about 75\%~\cite{czyzyk_proc} of all reactions. 
The rest of about 25\% comes from multibody reactions.
The number of coincidences of the
following type:
\begin{equation}
(SC_1 \wedge SC_2) \vee (SC_3 \wedge SC_4).
\label{coincidences}
\end{equation}
was measured.

Denoting by $n_{\uparrow}$ and $n_{\downarrow}$ numbers of coincidences
integrated over the time of the measurement 
during the cycles with spin up and down, respectively, we get from Equation~\ref{lumi_rel}
and relation~\ref{n_od_t}:
\begin{equation}
L_{rel} = \frac{\int{L_{\uparrow}dt}}{\int{L_{\downarrow}dt}} = \frac{n_{\uparrow}}{n_{\downarrow}}. 
\label{blaaa}
\end{equation}

The numbers of coincidences n$_{\uparrow}$ and n$_{\downarrow}$
and also the relative luminosities $L_{rel}$ for runs at Q~=~10~MeV and Q~=~36~MeV
have been determined using this method and are presented in Table~\ref{swietlnosci}. 

\begin{table}[H]
 \begin{center}
   \begin{tabular}{|c|c|c|c|c|}
    \hline
      Q[MeV] & n$_{\uparrow}$ & n$_{\downarrow}$ &  $L_{rel}$ &  $L_{rel}^{corrected}$\\
    \hline
	10 & 5732570 $\pm$ 2394 & 5942438 $\pm$ 2438 & 0.96468 $\pm$ 0.00056 & 0.98468  \\
	36 & 5657318 $\pm$ 2378 & 5874621 $\pm$ 2424 & 0.96301 $\pm$ 0.00057 & 0.98301  \\
    \hline
   \end{tabular}
     \caption{ {\small Relative luminosities, and numbers needed for their calculation for runs at excess 
	energies Q~=~10 and Q~=~36~MeV. $L_{rel}^{corrected}$ are the values of the relative luminosities, corrected 
	for the shift of the target, as explained in the text below. 
	The errors indicated in the table are the statistical uncertainties only. 
	The systematic errors are evaluated in the text.  }
     \label{swietlnosci}
         }
 \end{center}
\end{table}

The detectors were installed centrally around the nominal target 
position ($\Delta_X=0$), however one should 
keep in mind that the $\Delta_X$ found in Section~\ref{beamshift}
equals $0.7$~cm. This introduces false asymmetries 
into the calculations of $L_{rel}$, and so, the values of $L_{rel}$ have to be 
corrected for these false asymmetries. In order to do so, Monte-Carlo 
simulations have been performed, using the GEANT3-based code~\cite{geant}, containing the geometry of
the luminosity system from Figure~\ref{uklad_do_swietlnosci}, and taking into account
the phase-space distribution of the events modified by the differential cross sections~\cite{albers} and analysing powers 
for the $\vec{p}p\to pp$ reaction~\cite{altmeier}. In the simulations 
the real values of the polarisations (quoted in the following section) were taken 
into account. Subsequently, the same number of events have been generated 
for spin up and down modes, and indeed, it has been found that the 7~mm shift  
introduces false asymmetry that changes the value of the $L_{rel}$
from $L_{rel}$~=~1 (for the symmetric system) to $L_{rel}$~=~0.98. 
Therefore the value of $\Delta L=1-0.98=0.02$ has to be added
to the $L_{rel}$ in order to obtain the real values of the luminosities, 
which in Table~\ref{swietlnosci} we called $L_{rel}^{corrected}$.

It appears from results given in Table~\ref{swietlnosci} that the corrected 
relative luminosities differ from 1 by circa 2\%, although
the spin was flipped from cycle to cycle.
The overall integrated luminosity during spin down cycles is greater
than the luminosity for spin up.
This apparent inconsistency may be explained
by the fact that the measurements were always started with the
spin down orientation, however the breaks in the data taking
(due to the target regeneration, beam optimization
and other events aiming in improvement of the quality of measurement)
happened accidentally with the same probability for both
spin adjustments\footnote{Indeed, during the experiment there were circa
$n~=~16$ breaks in the operation
of the COSY accelerator. Therefore, in average, there were around 8 cycles with spin
down more than the ones with spin up. The whole time of measurement for experiment
at Q~=~10~MeV lasted \mbox{T~=~829201~s}. The time period of spin up or spin down
cycle for the run at Q~=~10~MeV was equal to \mbox{t~=~1200~s}.
The rough estimation gives the total number of cycles (spin up and down
altogether) equal k~=~T/t~$\approx$~690 -- out of which, we assume, there were 349 cycles
with spin down and 341 with spin up.
Therefore the relative difference in the
time of measurement with spin down and spin up mode equals to circa 2\%.
The same inference is valid for the experiment at the excess energy Q~=~36~MeV.
}.

Statistical uncertainties were calculated according to the 
rule of the error propagation applied to Equation~\ref{blaaa}: 
\begin{equation}
\sigma(L_{rel}) = \sqrt{\left(\frac{\partial L_{rel}}{\partial n_{\uparrow}}\right)^{2} \sigma(n_{\uparrow})^{2} + \left(\frac{\partial L_{rel}}{\partial n_{\downarrow}}\right)^{2} \sigma(n_{\downarrow})^{2}}, 
\label{blaaaaa}
\end{equation}
where 
\begin{equation}
\sigma(n_{\uparrow(\downarrow)}) = \sqrt{n_{\uparrow(\downarrow)}}
\label{ccccc}
\end{equation}
are the statistical uncertainties of $n_{\uparrow(\downarrow)}$.
After inserting the partial derivative and relations~\ref{ccccc} into
Equation~\ref{blaaaaa}, the formula simplifies to:
\begin{equation}
\sigma(L_{rel}) = L_{rel} \sqrt{\frac{1}{n_{\uparrow}} + \frac{1}{n_{\downarrow}}}.
\label{blad_L}
\end{equation}

After these remarks on the statistical uncertainties we will 
discuss the systematic errors. 
The main source in the systematic error of the relative luminosity determination 
originates from the uncertainty of the position of the scintillator detectors from Figure~\ref{uklad_do_swietlnosci}.
A possible non-symmetrical adjustment of the detectors
with respect to the polarisation plane would lead to 
the false asymmetries and hence to the error of the $L_{rel}$
determination.
In order to estimate the systematic error the conservative assumption has been made 
that the centres of the 
scintillators from Figure~\ref{uklad_do_swietlnosci} were shifted from the nominal position by $\Delta$r~=~4~mm. 
In order to estimate the systematic uncertainty of $L_{rel}$, again we have performed
Monte-Carlo simulations, taking into account all the factors quoted when discussing the
influence of the $\Delta_X$ on the $L_{rel}$.  
Numerical evaluation brings the value
of 1\% for the systematic uncertainty of $L_{rel}$.

\section{
Beam polarisation
} 
\label{polarisation}

\vspace{3mm}
{\small
The methods of the determination of beam polarisation 
along with the used detection systems are described. 
All the methods are basically based on the asymmetry measurements
for the $\vec{p}p\to pp\eta$ reaction.  
}
\vspace{5mm}

During two measurements of the analysing power
at different excess energies, three independent methods
have been used in order to determine the beam polarisation.
In the run at the excess energy of Q~=~10~MeV the COSY-11 polarimeter has been used as 
the main equipment to extract information about
the value of the polarisation degree. During this run in parallel the polarimeter of the COSY team~\cite{bauer} 
has also been used as the auxiliary detector to monitor the polarisation 
and to verify the information from the COSY-11 polarimeter.  

During the measurements performed at the excess energy of Q~=~36~MeV we managed to get access to the 
EDDA polarimeter~\cite{bauer}, commonly used by many experimental groups 
to perform the exact determination of the degree of polarisation at the COSY accelerator. 

All three methods of the polarisation monitoring will be 
described in this section.

\subsection{ Measurement at the excess energy of Q~=~10~MeV
}
\label{bbb}

\vspace{3mm}
{\small
The methods for the determination of the degree of polarisation 
by means of the internal COSY polarimeter as well as the COSY-11 
polarimeter will be described. 
}
\vspace{5mm}

\subsubsection{The internal COSY beam polarimeter}
\begin{figure}[H]
  \unitlength 1.0cm
        \begin{center}
  \begin{picture}(14.5,6.5)
    \put(2.6,0.0){
      \psfig{figure=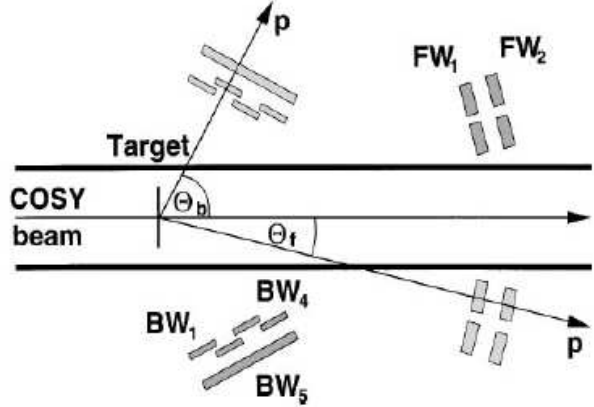,height=6.5cm,angle=0}
    }
  \end{picture}
  \caption{ {\small  Schematic view of the COSY polarimeter. 
	For the short-cuts' explanation see text. Figure is adapted from~\cite{bauer}.
 \label{bauer_uk}
                 }
  }
        \end{center}
\end{figure}

One of the polarimeters used for the polarisation monitoring in this experiment was a 
polarimeter of the COSY team~\cite{bauer}. Subsequently we will refer to it as to the 
COSY polarimeter.  
A schematic view of this setup is presented in Fig.~\ref{bauer_uk}.

Protons from the polarised beam are scattered on the protons from an internal CH$_{2}$
fiber target and subsequently they reach the detectors presented in Fig.~\ref{bauer_uk}.
There is a kinematical correlation between the angles of the
forward ($\Theta_f$) and backward ($\Theta_b$) scattered protons, namely:
\begin{equation}
\tan\Theta_{f,lab} \tan\Theta_{b,lab} = \frac{2 m_p c^2}{T_p + 2 m_p c^2},
\label{tangensy}
\end{equation}
where m$_p$ denotes the proton mass, while T$_p$ is
proton's kinetic energy. The geometry of this polarimeter
was designed such that
due to this kinematical restriction the 
polarimeter can operate within the beam energy range from
300~MeV up to several GeV.

The detector setup may be rotated via remote control
around the beam axis by $\Delta\phi=290^{0}$ in order
to eliminate the false asymmetries. For the
true asymmetry measurements, the $\Delta\phi$ ranging from 23$^{0}$ (for $\Theta_{f}=11^{0}$)
to 13$^{0}$ (for $\Theta_{f}=19^{0}$) has been used.
For the dimensions and
other details of the detector geometry the reader is
referred to~\cite{bauer}.

The principle of the determination of the 
degree of polarisation by means of this setup is based on the 
asymmetry ($\epsilon$) measurements of the quasi-free $\vec{p}p$ elastic scattering
realized by using a CH$_{2}$ fiber target.
All the detectors presented in Figure~\ref{bauer_uk} are
scintillator detectors.
The beam line is the symmetry axis of the detector setup.
FW$_{1}$, FW$_{2}$, and their symmetric mirrors
constitute the forward detector to register the fast proton, while
the recoil protons are detected in the conjugate backward arm, consisting
of four detectors BW$_{1}$, ..., BW$_{4}$, and a larger common
detector BW$_{5}$ which helps to get rid of the accidental
coincidences and give the trigger signal along with the
FW$_1$ and FW$_2$.
Thus, the trigger conditions to be fulfilled for calculation
of the asymmetry are:
\begin{equation}
FW_1 \wedge FW_2 \wedge BW_5,
\end{equation}
for left and right arm of the polarimeter.

The asymmetry $\epsilon(\Theta)$:
\begin{equation}
\epsilon(\Theta) = \frac{N_+(\Theta) - N_-(\Theta)}{N_+(\Theta) + N_-(\Theta)}
\label{epsilon}
\end{equation} 
is calculated using the geometrical averages N$_{\pm}$ 
defined as follows: 
\begin{equation}
N_{+} \equiv \sqrt{N^{\uparrow}_{+} N^{\downarrow}_{+}} = C (1+P A_y(\theta_{\eta})),
\label{gg}
\end{equation}
and
\begin{equation}
N_{-} \equiv \sqrt{N^{\uparrow}_{-} N^{\downarrow}_{-}} = C (1-P A_y(\theta_{\eta})),
\label{hh}
\end{equation}
where $C$ is the constant depending on the detector efficiencies and the 
relative luminosity during spin up and down modes. The arrows 
denote the spin mode of the beam particles.  

 Thus, the degree of polarisation may be calculated as:
\begin{equation}
P=\frac{1}{A(\Theta)} \epsilon(\Theta),
\label{polaryzacja_bauer} 
\end{equation}
where $A(\Theta)$ is the analysing power for  
the proton-proton elastic scattering. 
Please note, that the degree of polarisation calculated in this way is independent of the
efficiencies of the detectors.

In order to correct the above formula for false asymmetries, originating
from the beam misalignment, the following formula has been applied~\cite{ohlsen}: 
\begin{equation}
P=\frac{1}{A(\Theta)} \frac{\epsilon(\Theta)-\epsilon'(\Theta)}{1-\epsilon(\Theta) \epsilon'(\Theta)},
\label{polaryzacja_bauer_2}
\end{equation}  
where $\epsilon'(\Theta)$ stands for the false asymmetries, 
calculated from equation~\ref{epsilon}, 
for a measurement with an unpolarised proton beam.
The results of polarisation measurement 
with the COSY polarimeter~\cite{lorentz} are presented in Figure~\ref{polaryzacja}
in the following section. Here, we would only like to mention that 
the main source of the systematic error in the 
calculations of $P$ arises from the beam misalignment and this 
has been estimated to be less than 4\%~\cite{lorentz}.

It's worth mentioning that the construction of this polarimeter 
allows to calculate the polarisations for spin up and down separately.
It occurs that during the experiment the discrepancy between 
both polarisations was less than $\pm 3\%$~\cite{lorentz}. 
In order to compare the polarisation values obtained with this 
equipment with the results of the COSY-11 polarimeter
for time periods when the COSY polarimeter has been in operation 
the average values of the polarisations during spin up and down cycles 
have been calculated.

\subsubsection{COSY-11 polarimeter}
\label{kkk}

The second method devoted to the evaluation of the degree of polarisation 
made use of the COSY-11 detector setup. This method is based 
on the asymmetry measurement for the elastic $\vec{p}p\to pp$ process.  
The COSY-11 detectors that served for this purpose are presented in 
Figure~\ref{cosy11_polar} of Section~\ref{beamshift}.
%
%
%
%

\begin{figure}[H]
  \unitlength 1.0cm
  \begin{picture}(14.0,7.5)
      \put(-0.30,1.0){
         \psfig{figure=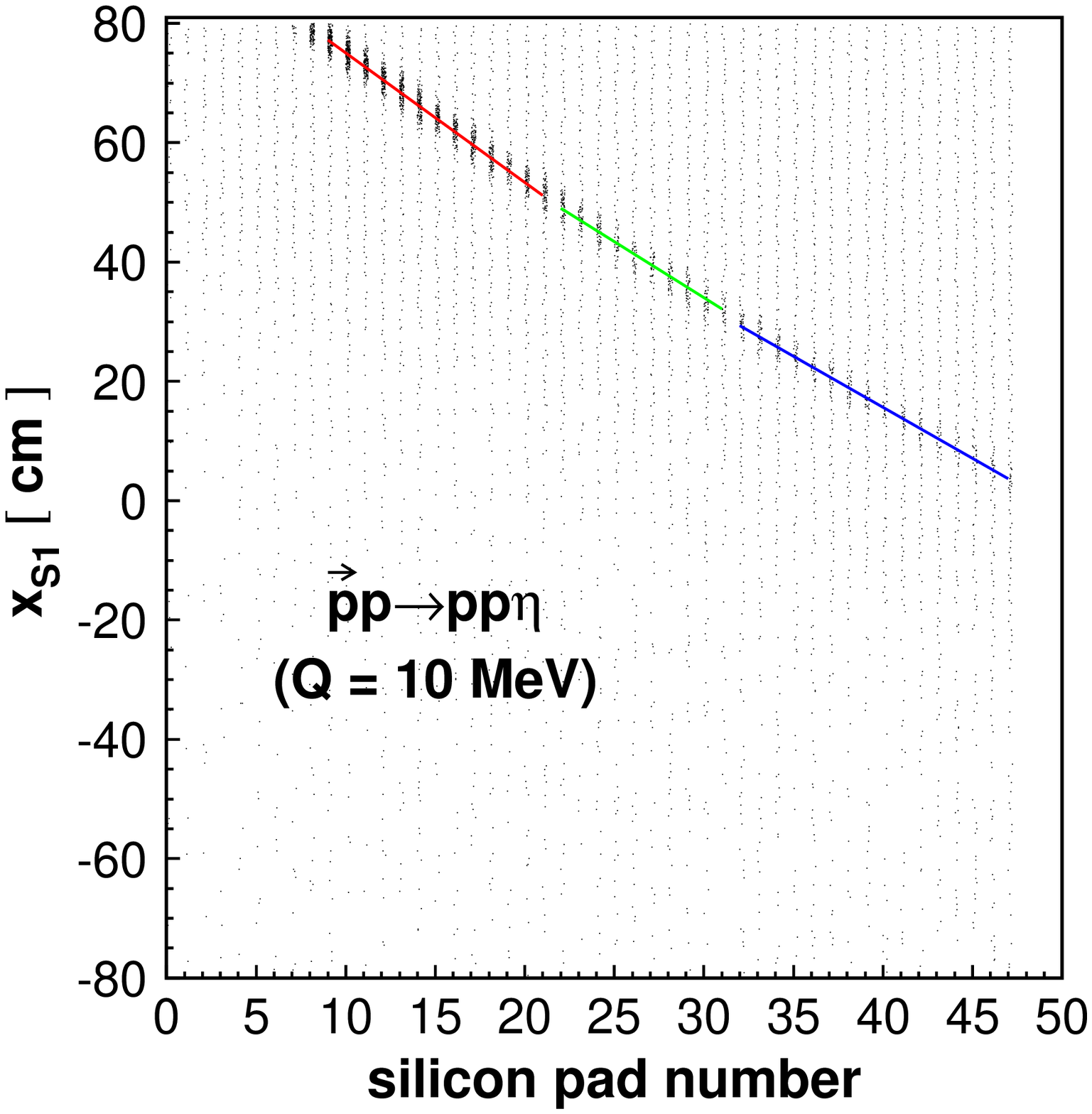,height=7.0cm,angle=0}
      }
      \put(7.50,1.0){
         \psfig{figure=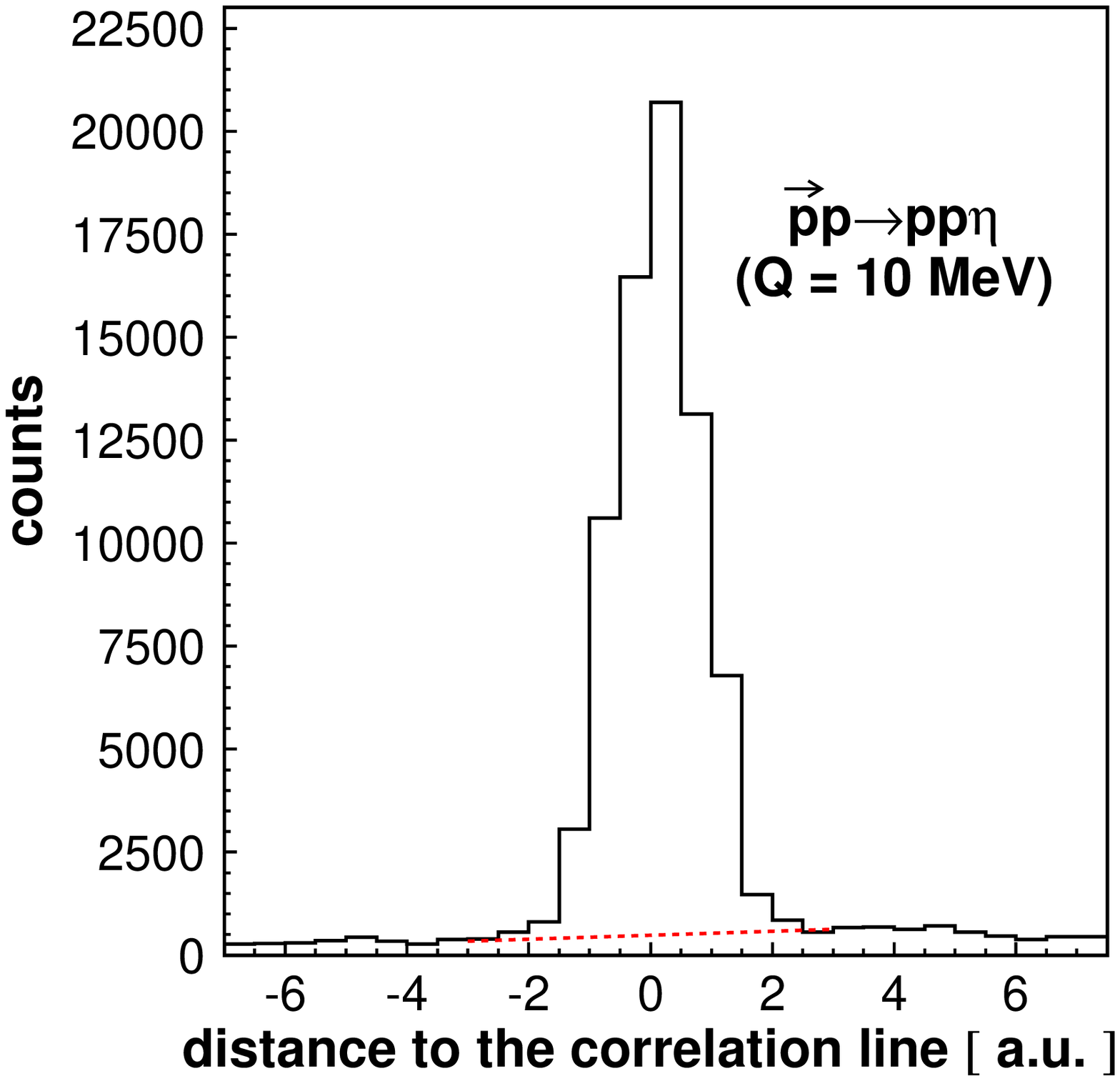,height=7.0cm,angle=0}
      }
	\put(6.5,0.2){{\normalsize {\bf a)}}}
        \put(14.0,0.2){{\normalsize {\bf b)}}}
  \end{picture}
  \caption{ \small{ (a) Correlation plot for the proton-proton 
	elastic scattering at the beam momentum p$_{beam}$~=~2010~MeV/c.
        For the explanation see text. 
        (b) Exemplary spectrum of the distance 
	of the events to the correlation curve of Figure (a) for the arbitrarily chosen 
	angle $\Theta_{CM}=47^{0}$. The dotted line denotes the approximated background limit.  
                  }
  \label{korelacja}
  }
\end{figure}

The elastically scattered events have been identified based 
on the constraints given by the two body kinematics. 
The relation~\ref{tangensy} for $\Theta_1$ and $\Theta_2$ 
puts constraints on the event distribution, 
and according to that relation there is a correlation 
between the hit position in S1 -- x$_{S1}$ -- and 
the index of the silicon pad that gave a signal. This correlation
may be seen in Figure~\ref{korelacja}.a as an  
enhancement in the density of the event distribution.
The events lying outside the correlation curve originate from inelastic
reactions, which occasionally fulfill the trigger conditions 
for the elastic scattering:
\begin{equation}
T_{elas} = S1 \wedge S4. 
\label{trigger_2}
\end{equation} 
The background subtraction is necessary for the 
evaluation of $N_{+}(\Theta)$ and $N_{-}(\Theta)$ -- the numbers of elastically 
scattered events during cycles with spin up and down, respectively.

To facilitate an easy background subtraction, the correlation curve has been divided
into three ranges as shown in Figure~\ref{korelacja}.a and in each range it 
was approximated by a straight line. Event distribution visible in this figure 
has been projected along the correlation line, and an exemplary spectrum,  
for $\Theta_{CM}=47^{0}$,
is depicted in figure~\ref{korelacja}.b. Next, the linear interpolation of 
the background has been performed (see dashed line in Figure~\ref{korelacja}.b), 
for each range of $\Theta_{CM}$ angle. 
As it is seen in Figure~\ref{korelacja}.b, the background constitutes only
about 5\% of the signal, and since it is smooth and flat on both sides
of the signal peak we assumed, conservatively, that the systematic 
error due to the assumption of the linearity of the background
is less than 10\%. Hence, the overall systematic error due to the 
background subtraction is not greater than 0.5\%. Another source of the systematic error
of the number of events in the individual $\Theta_{CM}$ ranges originates from 
the possible beam position misalignment, which may have been 
different for spin up and down modes. In our estimations we have assumed 
conservatively a 2~mm shift of the beam between spin up and spin down modes, a choice which 
was dictated by the geometrical dimensions of the reaction region~\cite{moskal-hab}.
The Monte-Carlo simulations performed with this 2~mm shift of the beam
revealed the value of 0.5\% of the systematic error.  
Therefore, the overall systematic error
of the number of events in the individual $\Theta_{CM}$ ranges we estimate to be less than 1\%.

After the background subtraction events lying within 2.5 or less distance from the correlation line 
are regarded to be elastics. 
Number of elastic events for scattering during spin up and down cycles
allows to determine the asymmetry defined in Equation~\ref{epsilon}.

\begin{figure}[H]
  \unitlength 1.0cm
  \begin{picture}(14.0,7.5)
      \put(0.00,1.0){
         \psfig{figure=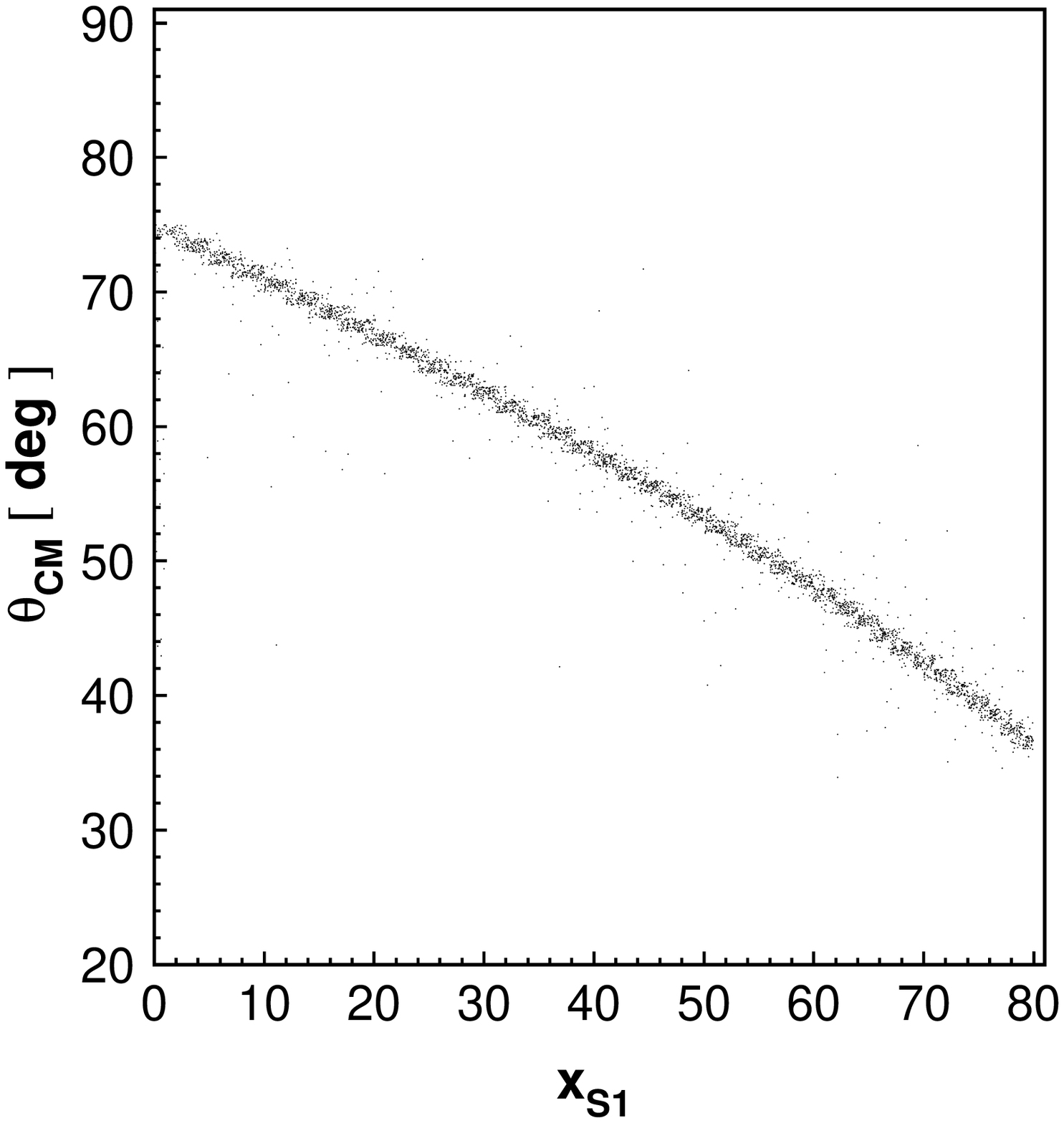,height=7.0cm,angle=0}
      }
      \put(7.50,1.0){
         \psfig{figure=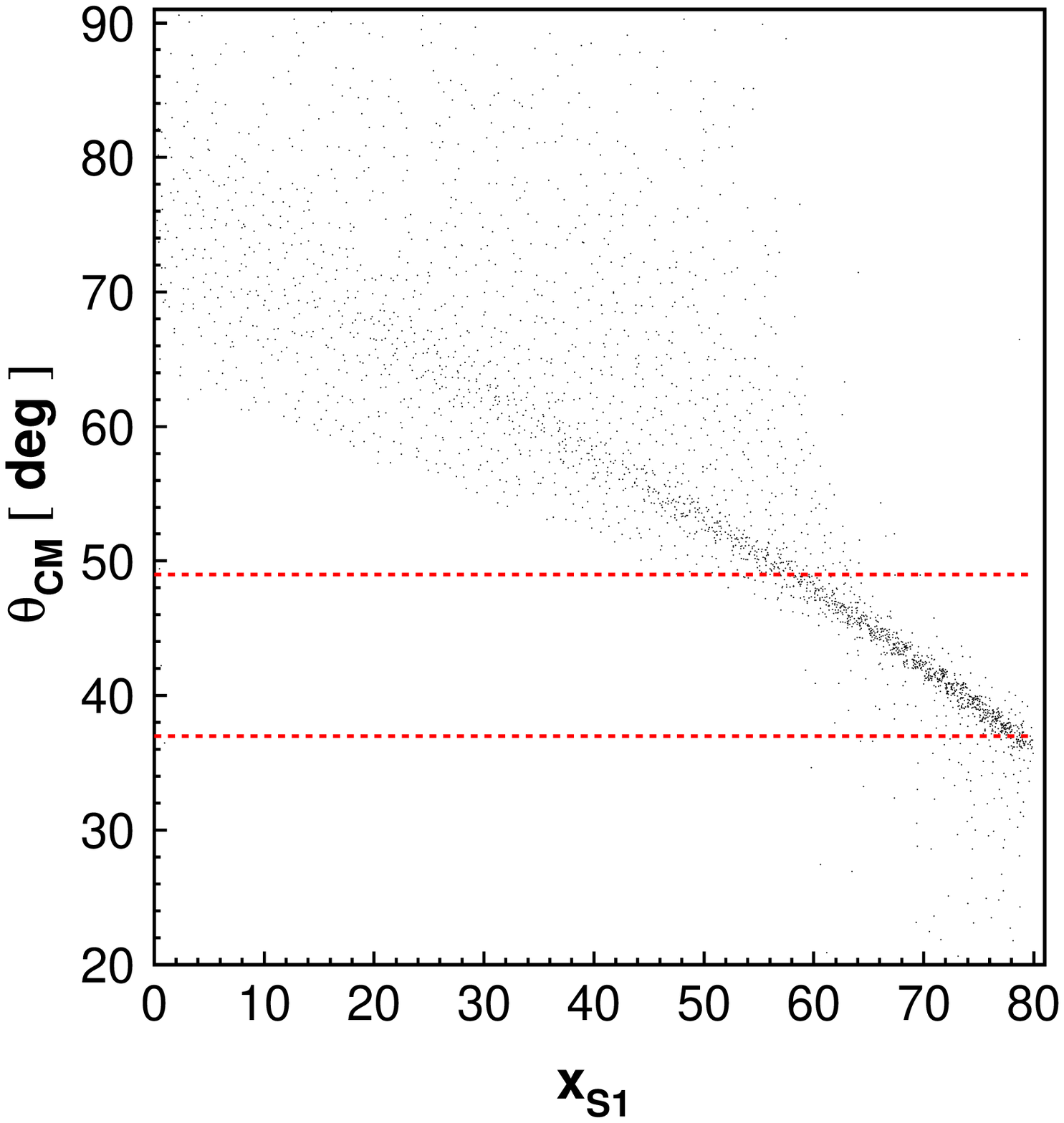,height=7.0cm,angle=0}
      }
      \put(6.5,0.2){{\normalsize {\bf a)}}}
      \put(14.0,0.2){{\normalsize {\bf b)}}}
  \end{picture}
  \caption{ \small{ Relation between the centre-of-mass scattering angle and the position
        in the S1 detector for the $\vec{p}p\to pp$ elastic scattering at $p_{beam}~=~2010$~MeV/c: 
	(a) Monte-Carlo simulations, (b) experiment. Dotted lines are the boundaries 
	of the practical range where the polarisation is calculated. The reason for
	a poor resolution outside this region is explained in the text.  
                  }
  \label{theta_s1}
  }
\end{figure}

Figure~\ref{theta_s1} depicts  the relation between the centre-of-mass scattering angles
and the position in the S1 detector
for the $\vec{p}p\to pp$ reaction at the beam momentum of $p_{beam}=2010$~MeV/c.
The scattering angles for which both protons may still be measured in coincidence by S1 and $Si_{mon}$
ranges from around 35 to circa 75 degrees, as depicted in Figure~\ref{theta_s1}.a.
However, the practical partition of the CM scattering angles that may be used
for the polarisation evaluation is squeezed to the range 
of (37;49) degrees (see Figure~\ref{theta_s1}.b). This is
due to the strong dependence of the cross section on the scattering CM angle and also
due to the deterioration of the momentum determination for protons passing the dipole only on its edge.
The experimental distribution of the $\Theta_{CM}$ angle as the function of the hit position in 
the S1 scintillator is shown in Figure~\ref{theta_s1}.b.

\begin{figure}[h]
  \unitlength 1.0cm
        \begin{center}
  \begin{picture}(14.0,9.0)
    \put(2.5,0.0){
      \psfig{figure=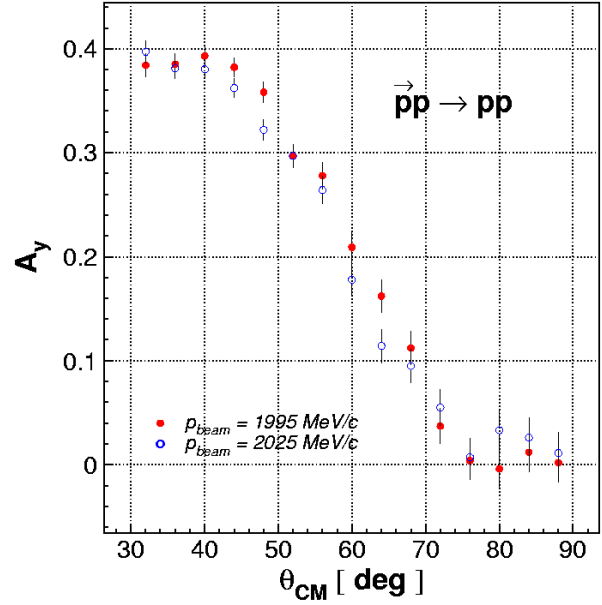,height=9.0cm,angle=0}
    }
  \end{picture}
  \caption{ {\small  
	Analysing power for the $pp\to pp$ elastic scattering
        as measured by the EDDA collaboration at the beam momenta
of
$p_{beam}=1995$~MeV/c and
        $p_{beam}=2025$~MeV/c. In the COSY-11 experiment the beam momentum was
        set to $p_{beam}=2010$~MeV/c, which is the middle of the range of beam momenta
        presented in this figure.
                 }
 \label{analysing}
  }
        \end{center}
\end{figure}

 This range was divided into
three
sections of 4 degrees width: (37;41), (41;45) and
(45;49) degrees, choice which was dictated to facilitate an easy  
comparison with the data of the EDDA collaboration.  
As there were no EDDA measurements of the analysing power 
at the  beam momentum $p_{beam}=2010$~MeV/c, we used the 
average of the $\vec{p}p\to pp$ analysing powers measured at $p_{beam}=1995$~MeV/c
and $p_{beam}=2025$~MeV/c~\cite{altmeier}
(see Figure~\ref{analysing}).
The obtained average values of the analysing power 
for the $\vec{p}p\to pp$ process at the beam 
momentum of $p_{beam}=2010$~MeV/c are given in Table~\ref{tab1}.
It is worth noting that values of $A_y$ for $p_{beam}=2010$~MeV/c and 
$p_{beam}=2025$~MeV/c are about the same within the statistical uncertainties. 

\begin{table}[H]
\begin{center}
\begin{tabular}{|c|c|c|}
\hline
$i$ & $\Theta_{i}$ [$^{\circ}$] & A$_y$ \\
\hline
1 & (37;41) & 0.385~$\pm$ 0.013 \\
2 & (41;45) & 0.376~$\pm$ 0.013 \\
3 & (45;49) & 0.348~$\pm$ 0.013 \\
\hline
\end{tabular}
\caption{ {\small Analysing power values at a beam momentum of $p_{beam}=2010$~MeV/c 
for the corresponding ranges of the CM scattering angle.
\label{tab1}}}
\end{center}
\end{table}

For the purpose of the analysis it is enough to know only the value of polarisation
averaged over the time of measurement. However, in order to check the 
performance of the used polarimeters we have also performed the differential measurements
of the degree of polarisation.  
The number of scatterings during spin up $N_{+}(\Theta)$ (scattering to the right side with 
respect to the polarisation plane) and spin down $N_{-}(\Theta)$ (scattering to the left side) 
has been determined for periods of 20 cycles, corresponding to
circa 1.5~h of measurement. 
The values of $N_{+}$ and $N_{-}$ have been normalized to the  
corresponding luminosities. 
We used the following formula to
calculate the degree of polarisation:
\begin{equation}
P=\frac{\sum_{i=1}^{3}{P(\Theta_i)}}{3},
\label{srednia}
\end{equation}
where
\begin{equation}
P(\Theta_{i})=\frac{1}{A_{y}(\Theta_{i})} \frac{N_{+}(\Theta_{i})-N_{-}(\Theta_{i})L_{rel}}{N_{+}(\Theta_{i})+N_{-}(\Theta_{i})L_{rel}}.
\label{polar_ok}
\end{equation}
The ranges of $\Theta_i$ were chosen as indicated in 
Table~\ref{tab1}. $L_{rel}$ are the relative luminosities defined
in equation~\ref{blaaaaa}, calculated separately for each time interval of polarisation averaging. 
The method of  $L_{rel}$ calculation has been described in Chapter~\ref{luminosity}.

\begin{figure}[H]
  \unitlength 1.0cm
        \begin{center}
  \begin{picture}(14.5,9.5)
    \put(1.7,0.0){
      \psfig{figure=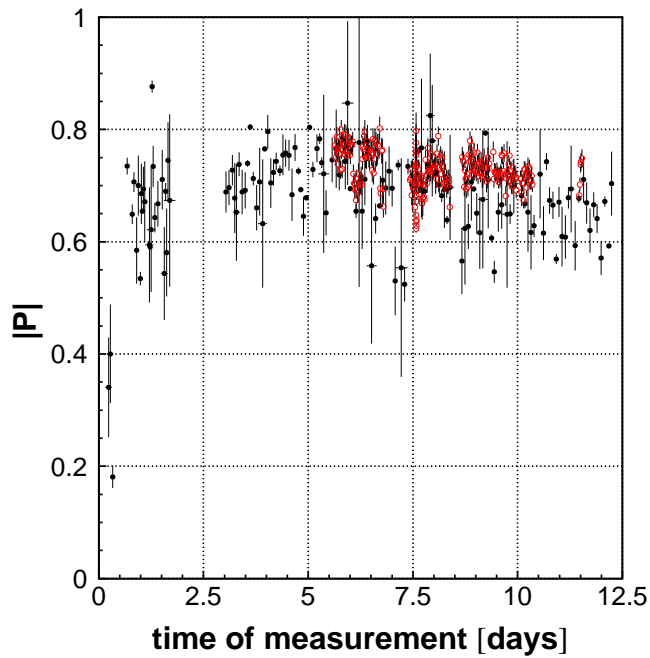,height=9.5cm,angle=0}
    }
  \end{picture}
  \caption{ {\small The degree of polarisation versus the time of the measurement
        as obtained during the experiment  
        at the excess energy of Q~=~10~MeV.
        Open circles denote results obtained by means of the
        COSY polarimeter~\cite{lorentz}, whereas the full points were determined
        using the COSY-11 setup.
                 }
 \label{polaryzacja}
  }
        \end{center}
\end{figure}

The confrontation of the polarisation values determined with the COSY and COSY-11 polarimeters 
is shown in Figure~\ref{polaryzacja}. 
The open circles are the averaged values (averaged over spin up and down periods)
as obtained by means of the COSY polarimeter, whereas the full circles 
show the results of the 
COSY-11 polarimeter.
It is worth mentioning that the presentation
is only for the sake of the comparison between the results obtained by means of both 
abovementioned methods. 
Due to the technical problems the COSY polarimeter
has been operating only for about 1/3 of the period 
of measurement.
Over the most of this time the results
from both polarimeters are in a good agreement.
Low polarisation values in the first day of the run  
was due to the ongoing polarisation 
development by the accelerator team. Errors in the 
Figure~\ref{polaryzacja} are the total errors, containing both statistical 
and systematic uncertainties.

Finally, based on the COSY-11 polarimeter the polarisation 
value for the full time of measurement has been computed 
and equals:
\begin{equation}
P = 0.680 \pm 0.007, 
\label{polar_q10}
\end{equation}
where the statistical error of the polarisation has been given. 
The systematic error of polarisation determination depends on 
the uncertainty of the analysing power $A_y(\Theta_i)$, 
the error of the relative luminosity $L_{rel}$, and 
the uncertainty of the scattering yields $N_{\pm}(\Theta_i)$. 
The systematic uncertainty of the $A_y(\Theta_i)$ 
equals 1.2\%~\cite{altmeier}, the  
$L_{rel}$ systematic error has been estimated in Section~\ref{mmm} to be about 1\%, 
and the systematic uncertainties of the 
determination of $N_{\pm}(\Theta_i)$ is not greater than 1\%.
Applying the error propagation method for the systematic uncertainties to Equation~\ref{polar_ok}
yields the following formula for the systematic uncertainty of the polarisation:
\begin{equation}
\Delta P = |\frac{\partial P}{\partial A_y}| \Delta A_y + |\frac{\partial P}{\partial L_{rel}}| \Delta L_{rel} + |\frac{\partial P}{\partial N_{+}}| \Delta N_{+} + |\frac{\partial P}{\partial N_{-}}| \Delta N_{-},
\label{anal_error_2}
\end{equation}
with
\begin{equation}
\frac{\partial P}{\partial A_y} = -\frac{1}{A_y^2} \frac{N_{+}-L_{rel} N_{-}}{\left(N_{+}+L_{rel} N_{-}\right)^2},
\nonumber
\end{equation}
\begin{equation}
\frac{\partial P}{\partial L_{rel}} = -\frac{1}{A_y} \frac{2N_{+}N_{-}}{\left(N_{+}+L_{rel} N_{-}\right)^2},
\nonumber
\end{equation}
\begin{equation}
\frac{\partial P}{\partial N_{+}} = \frac{1}{A_y} \frac{2 L_{rel} N_{-}}{\left(N_{+}+L_{rel} N_{-}\right)^2},
\nonumber
\end{equation}
\begin{equation}
\frac{\partial P}{\partial N_{-}} = -\frac{1}{A_y} \frac{2 L_{rel} N_{+}}{\left(N_{+}+L_{rel} N_{-}\right)^2}.
\label{parcjalne2_error}
\end{equation}
Inserting the partial derivatives~\ref{parcjalne2_error} into 
Equation~\ref{anal_error_2} brings the overall systematic error of polarisation to $\Delta P=8\%$.

\subsection{ Measurement at the excess energy of Q~=~36~MeV 
}
\label{br}

\vspace{3mm}
{\small
The detector setup EDDA and the method of the measurement of the degree of polarisation 
by means of this aparatus will be described. 
}
\vspace{5mm}

For the polarisation monitoring during this run we made use of 
the EDDA detector setup~\cite{altmeier}.
This setup has commonly been used at the COSY synchrotron as
an internal high-energy polarimeter since November 1997\footnote{
It is worth mentioning that apart from the monitoring of the polarisation
at the COSY storage ring, EDDA is a well operating detector that performed many
measurements of the $pp\to pp$ reactions. The EDDA collaboration
measured the excitation functions $d\sigma/d\Omega(p_p,\Theta_{CM})$ for the unpolarised
proton-proton elastic scattering~\cite{albers}. Data has been gathered for 
108 different proton kinetic energies, ranging from 240~MeV up to 2577~MeV,
and are available at~\cite{EDDA_strona}.
The EDDA group has also performed the measurements of the excitation function
for the analysing power $A_N(p_p,\Theta_{CM})$~\cite{altmeier}
for 77 different proton beam kinetic energies, ranging from 436~MeV
up to 2492~MeV~\cite{EDDA_strona_2}. The centre-of-mass scattering angles
of protons -- $\Theta_{CM}$ -- from 30$^{0}$ up to 90$^{0}$
have been covered. Measurements of the analysing power made use of the
polarised CH$_2$ fiber target~\cite{eversheim}.
Recently, 
the spin correlation coefficients have been measured~\cite{bauer2, EDDA_strona_3} 
at 34 values of proton's kinetic energy ranging from 504~MeV up to 2493~MeV.
}.

\begin{figure}[H]
  \unitlength 1.0cm
        \begin{center}
  \begin{picture}(14.5,4.5)
    \put(2.2,0.0){
      \psfig{figure=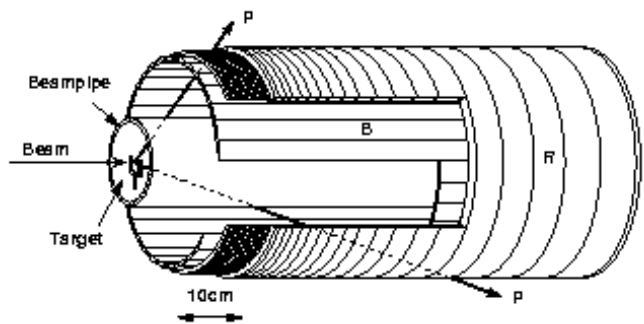,height=4.5cm,angle=0}
    }
  \end{picture}
  \caption{ {\small EDDA detector. 
        Figure is adapted from~\cite{altmeier}.
}
 \label{det_EDDA}
} 
        \end{center}
\end{figure}

The schematic view of the EDDA detector is presented in 
Figure~\ref{det_EDDA}. Here we will only sketch the principle of the 
polarisation measurement. For more information about the detector components and for the explanation  
of detector's operation the reader is referred to~\cite{altmeier,schwarz}.
The EDDA detector consists of two cylindrical double layers that 
covers $\Theta_{CM}$ range from 30$^{0}$ to 90$^{0}$ for the $\vec{p}p\to pp$
reaction~\cite{altmeier}. The construction of the detector is such that 
about 82\% of the full solid angle are covered. The  
inner part is made of helically wound 
scintillator fibers of 2.5 mm diameter, while the outer part
consists of 32 scintillator bars (B) surrounded by semirings (R).
Target used in the experiment was an atomic beam 
target~\cite{eversheim}, which can operate 
in the polarised and unpolarised mode. In the measurements of the polarisation 
for COSY-11 run it has been operating in the latter mode.

Signals from the outgoing protons are registered in the detectors
and subsequently reconstruction of the reaction vertex is performed
with the resolution better than 2 mm in all three space directions~\cite{altmeier}.
Tracks of particles are reconstructed and the calculation of 
kinematical variables is possible. Next, the asymmetries $\epsilon$ for different 
$\Theta_{CM}$ angles can be calculated, 
using the Formula~\ref{epsilon} with corrections for false asymmetries.   
Polarisation is then given by:
\begin{equation}
P = \frac{\epsilon}{A_y \cos(\phi)}, 
\label{pol_edda}
\end{equation}
where A$_y$ is the analysing power for the $\vec{p}p\to pp$ reaction, determined 
in previous experiments with the EDDA detector setup,
and $\phi$ is the azimuthal position of the hit in the detector.


As the acceptance of the EDDA detector for registering 
events scattered to the left and to the right side  
is large,
it was possible to extract the information of the
polarisations for spin up and spin down ($P^{\uparrow}$ and $P^{\downarrow}$, respectively)
separately. It was found that~\cite{ulbrich}
\begin{eqnarray}
P^{\uparrow} = 0.642\pm 0.004\pm 0.008, \\
\nonumber 
P^{\downarrow} = 0.684\pm 0.004\pm 0.008,
\label{polaryzacje}
\end{eqnarray}
where the statistical and systematic errors are 
given, respectively. 
Thus, the average polarisation degree equals:
\begin{equation}
P = 0.663\pm 0.003\pm 0.008.
\label{mean_polar}
\end{equation}
In our analysis we used this averaged value due to the reasons 
explained in Section~\ref{form}. Determination of the polarisation 
degree for both spin orientations by the EDDA collaboration allowed
us to estimate the error we made by this averaging. Similarly as 
during the measurement at Q~=~10~MeV~\cite{lorentz} this amounts
to about $\pm$~3\% only. 
\section{
Determination of the background free production rates
for the $\vec{p}p\to pp\eta$ reactions 
} 
\label{production_rates}

\vspace{3mm}
{\small
We show the way of the determination of the production yields
used for reckoning the analysing power. Different methods 
at both excess energies have been used. Differences are pointed 
out in the text. 
}
\vspace{5mm}

In Section~\ref{calibration} we presented the missing mass spectra
for the $\vec{p}p\to pp\eta$ reaction at two excess energies:
Q~=~10 and 36~MeV. The histograms in Fig~\ref{miss_spectra}  
include all events measured for both beam spin orientations.
In order to determine $A_y$ as a function of 
the centre-of-mass polar angle of the $\eta$ meson emission -- $\theta_{\eta}$ --
we have to divide the events according to the $\theta_{\eta}$ range and 
consider events measured during spin up and down modes separately. 
Optimizing the statistics and the expected shape of the analysing power function,
the range of the centre-of-mass polar angle of the $\eta$ meson emission
for both excess energies has been divided into four bins, and the results were integrated over
the remaining four kinematical variables introduced in Section~\ref{biecz}. 

Here we will describe the methods of background 
subtraction from the missing mass spectra determined 
for both excess energies. We shall start with 
the description of the procedure for the excess energy Q~=~36~MeV, 
where the shape of the background could be parameterized by 
a polynomial function of the second order. Then we will describe the more sophisticated methodology of 
the background subtraction at Q~=~10~MeV.   

\subsection{Q~=~36~MeV}
\label{back_q36}

In order to extract the production rates 
the missing mass spectra have been determined as a function 
of the cosine of the centre-of-mass emission angle of the $\eta$ meson, 
for spin up and down modes separately.
The values of $A_y$ will be derived in accordance with the Madison convention 
quoted in Section~\ref{refi}. Subsequently we will 
apply the notation defined in Section~\ref{form}
with $N_{+}^{\uparrow}$ and $N_{-}^{\downarrow}$
denoting the production yields to the left side with respect to the 
polarisation plane during spin up and down modes, 
respectively.

\begin{figure}[H]
  \unitlength 1.0cm
        \begin{center}
  \begin{picture}(14.0,8.0)
    \put(3.0,0.0){
      \psfig{figure=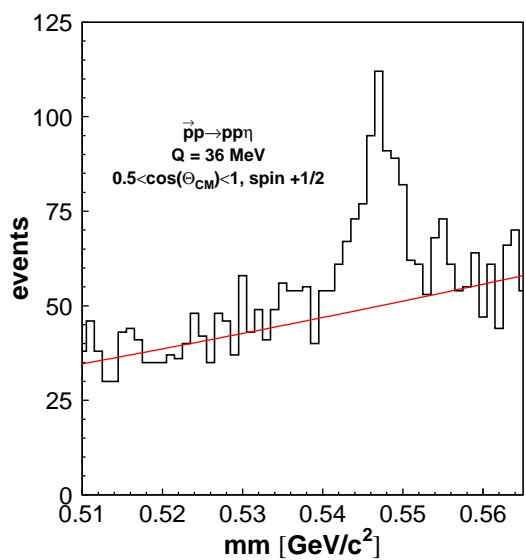,height=8.0cm,angle=0}
    }
  \end{picture}
  \caption{ \small { A histogram of missing mass for 
	the $\vec{p}p\to pp\eta$ reaction at Q~=~36~MeV. Spectrum 
	for spin up mode and $\cos\theta_{\eta}\in\left[0.5,1\right]$
	is shown.  
}
 \label{miss_up_05do1}
  }
        \end{center}
\end{figure}

An exemplary spectrum of the missing mass
for spin up mode and \mbox{$\cos\theta_{\eta}\in\left[0.5,1\right]$} 
is presented in Figure~\ref{miss_up_05do1}.
Next, we made the assumption that the background\footnote{Originating
from the multipionic reactions: $pp\to pp 2\pi^{0}$, 
$pp\to pp \pi^{+}\pi^{-}$, $pp\to pp 3\pi^{0}$, $pp\to pp\pi^{0}\pi^{+}\pi^{-}$, $pp\to pp 4\pi^{0}$, 
$pp\to pp 2\pi^{0}\pi^{+}\pi^{-}$, and $pp\to pp 2\pi^{+} 2\pi^{-}$.} shape  
may be described by a polynomial of second order, and 
the function
\begin{equation}
F_{backgr}(mm) = a + b~mm + c~mm^2,  
\label{back_36}
\end{equation} 
where $mm$ stands for the missing mass, has been fitted to the 
experimental histograms in the ranges outside the $\eta$ 
peak with $a$, $b$, and $c$ treated as free parameters. 
%
%
%
%
%

\begin{figure}[p]
  \unitlength 1.0cm
  \begin{picture}(14.0,23.0)
      \put(0.00,17.9){
         \psfig{figure=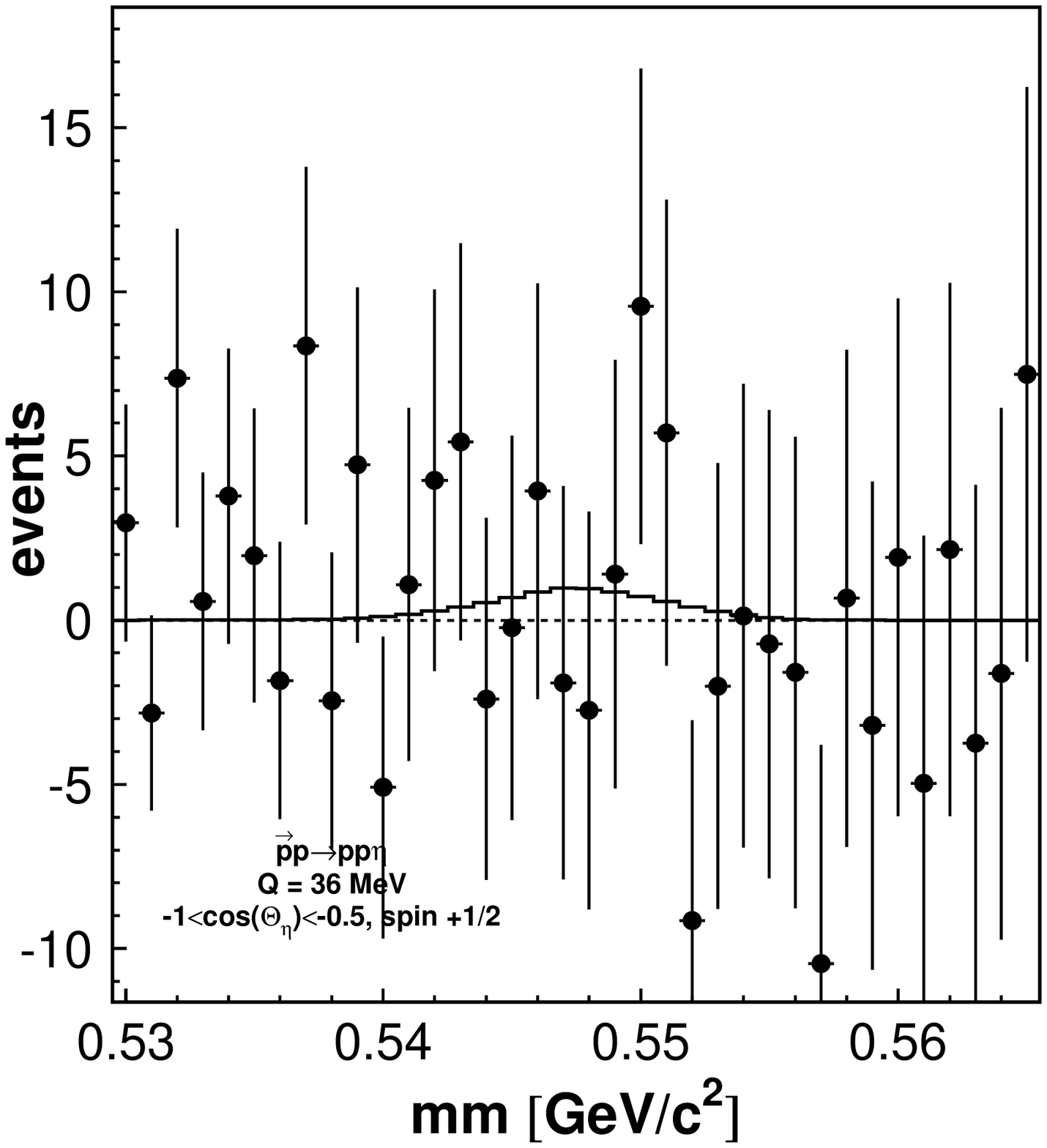,height=6.0cm,angle=0}
      }
      \put(7.50,17.9){
         \psfig{figure=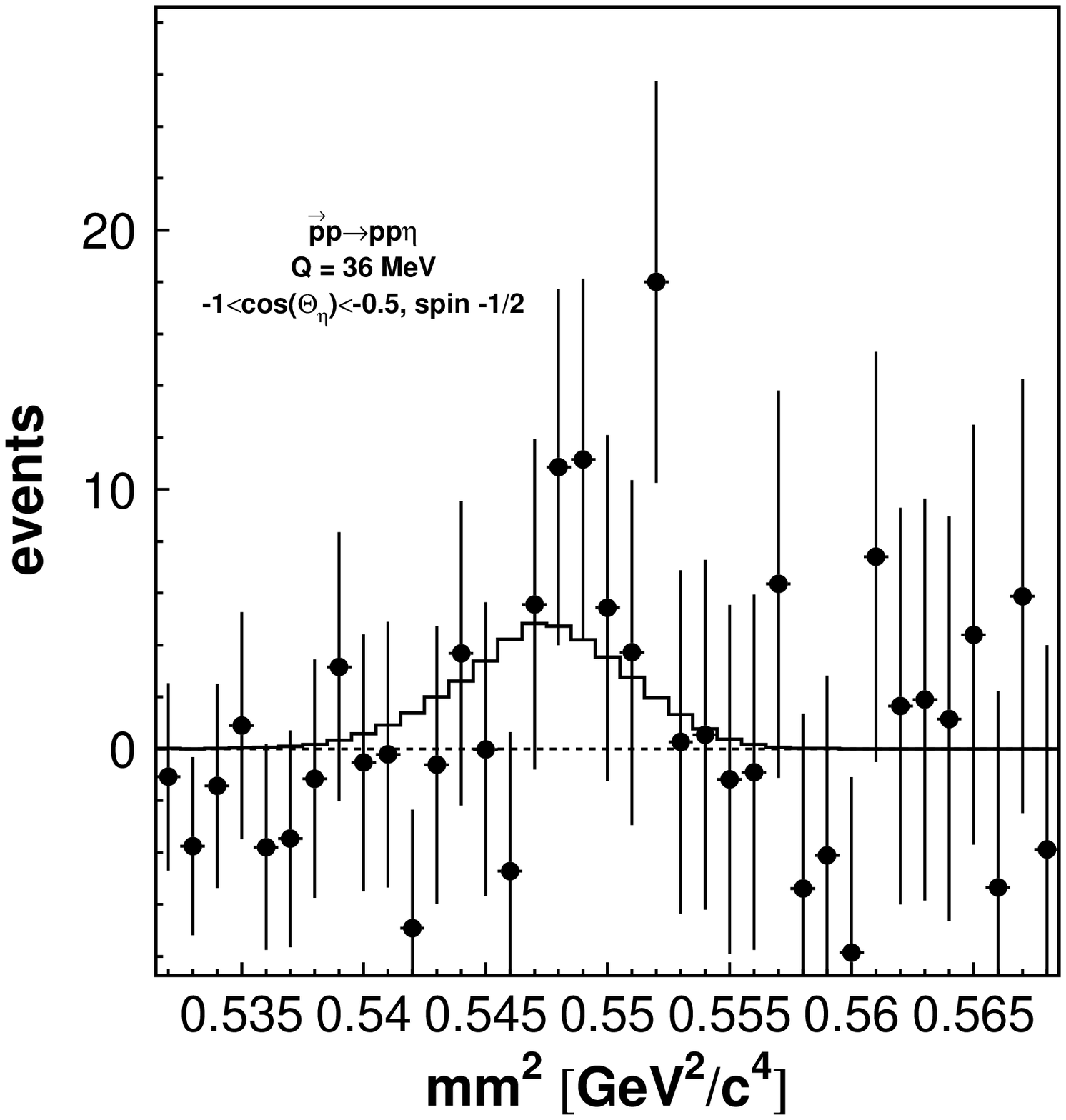,height=6.0cm,angle=0}
      }
      \put(0.00,11.9){
         \psfig{figure=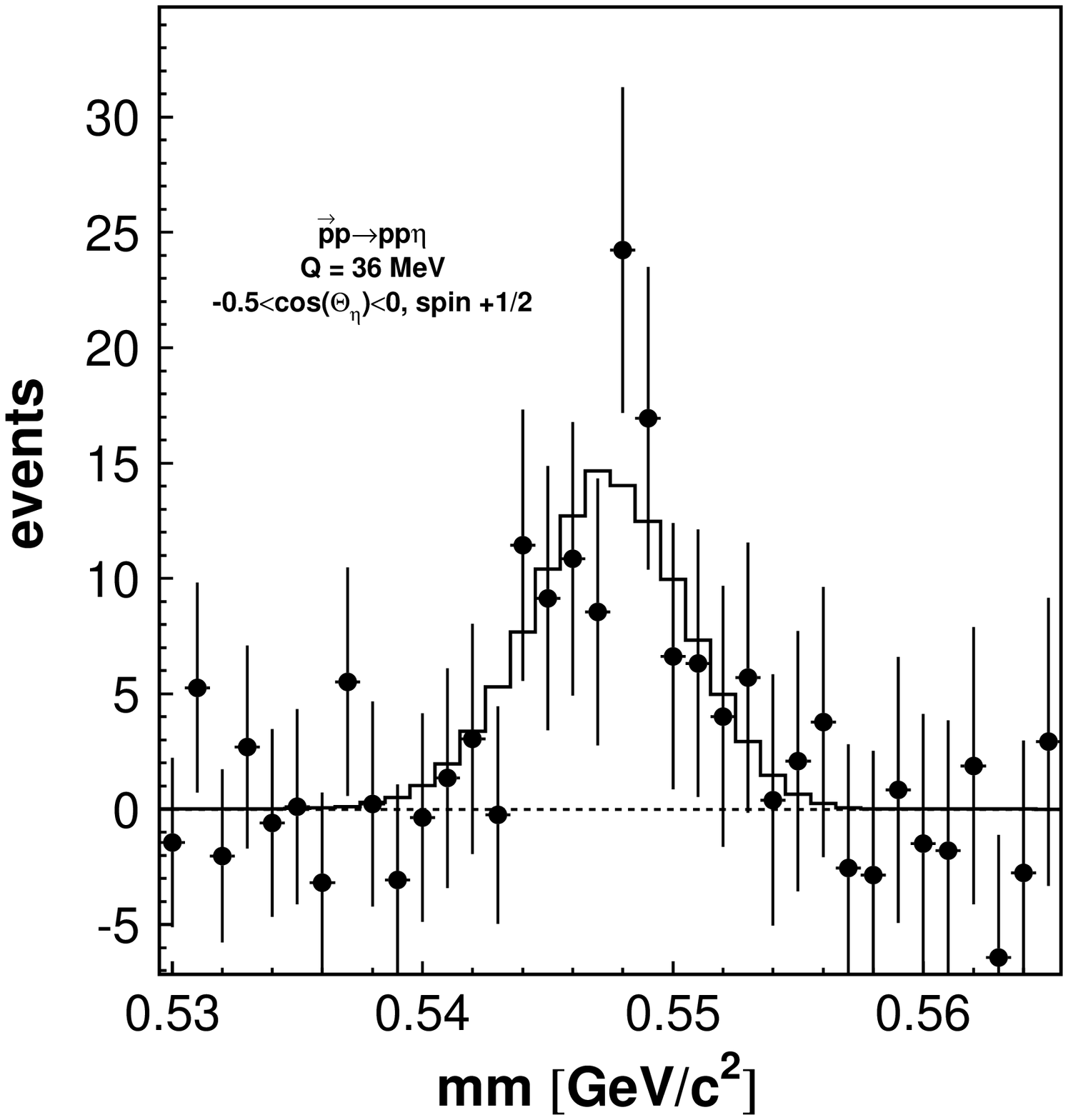,height=6.0cm,angle=0}
      }
      \put(7.50,11.9){
         \psfig{figure=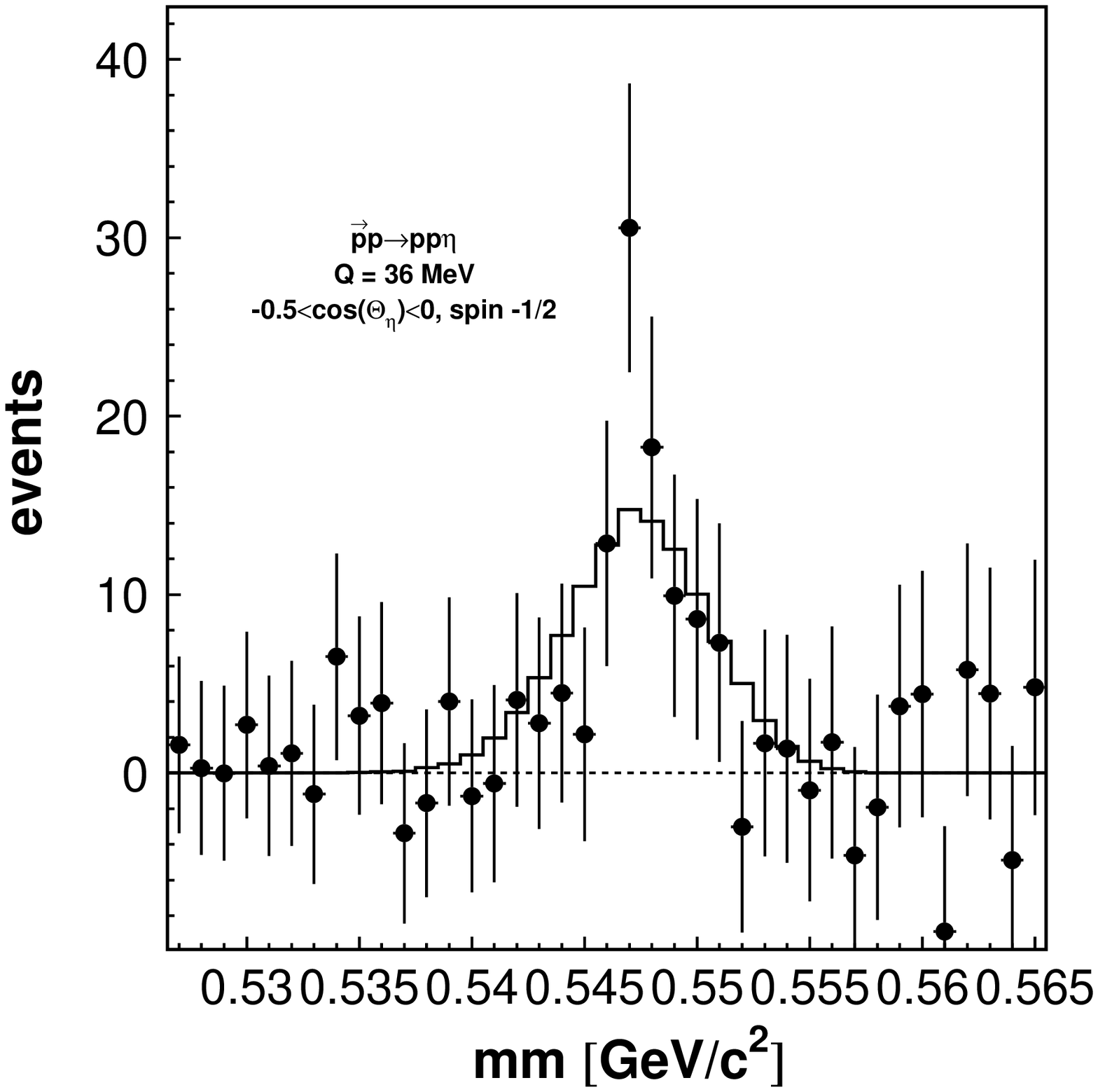,height=6.0cm,angle=0}
      }
      \put(0.00,5.9){
         \psfig{figure=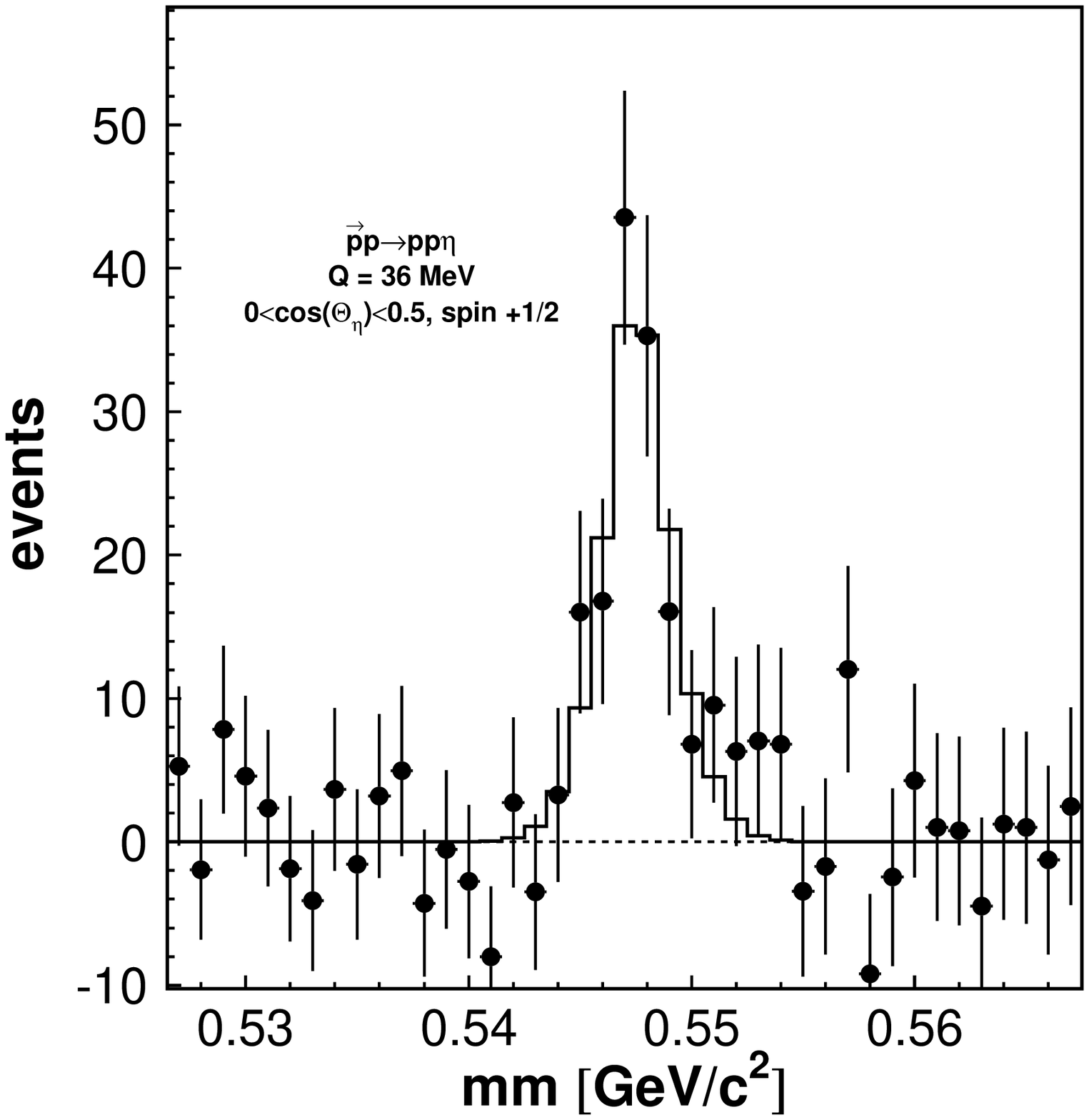,height=6.0cm,angle=0}
      }
      \put(7.50,5.9){
         \psfig{figure=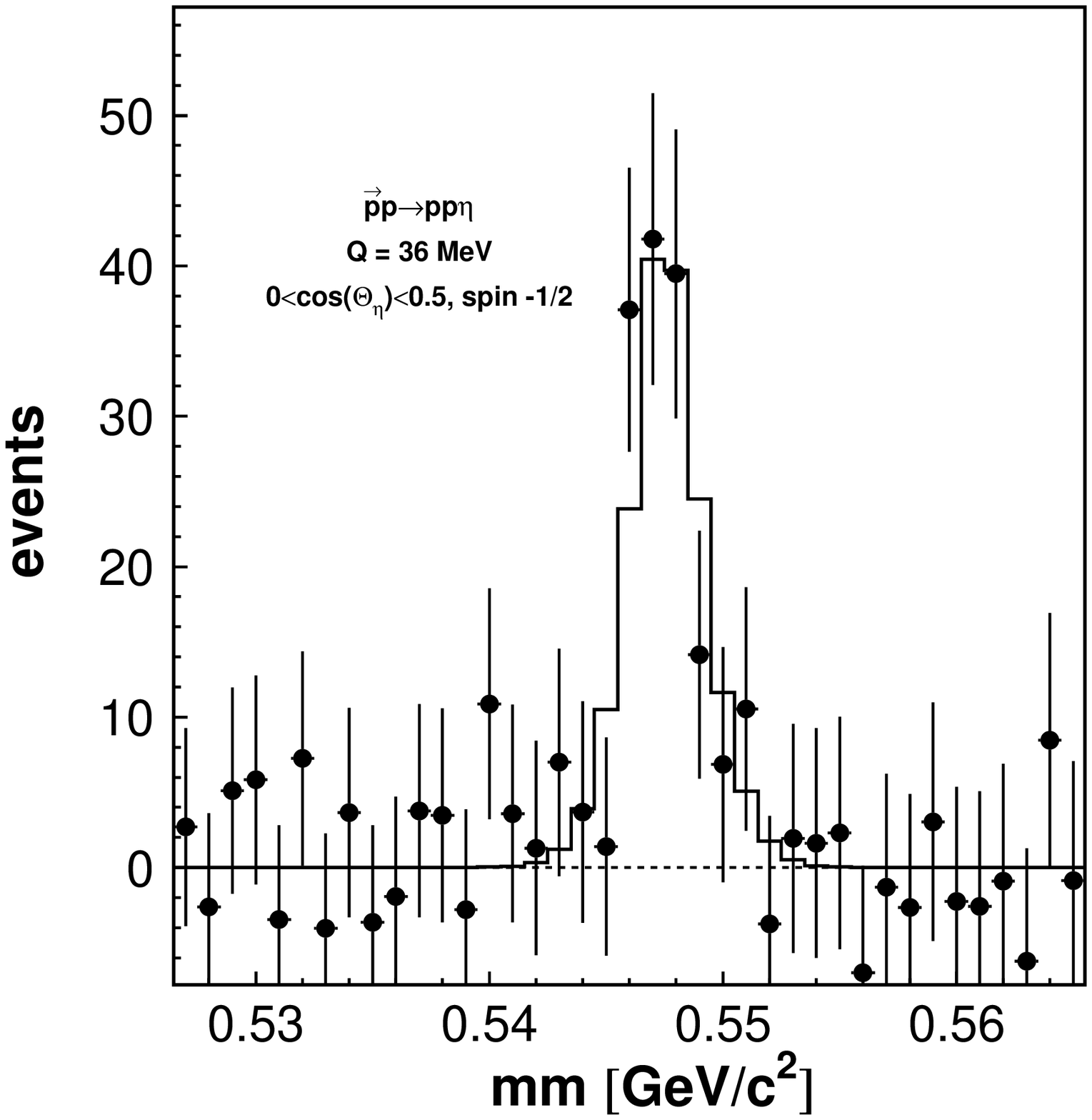,height=6.0cm,angle=0}
      }
      \put(0.00,-0.1){
         \psfig{figure=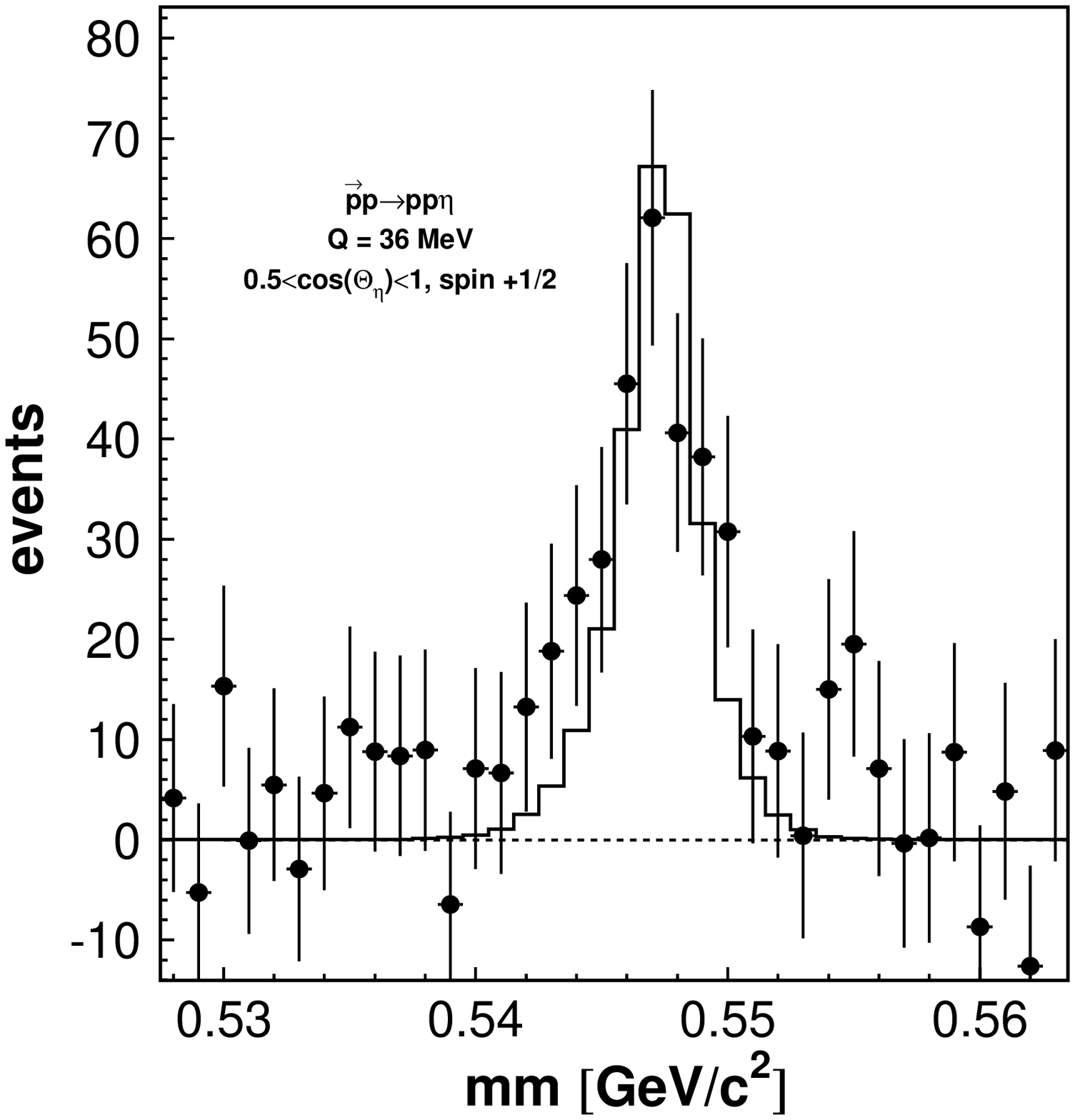,height=6.0cm,angle=0}
      }
      \put(7.50,-0.1){
         \psfig{figure=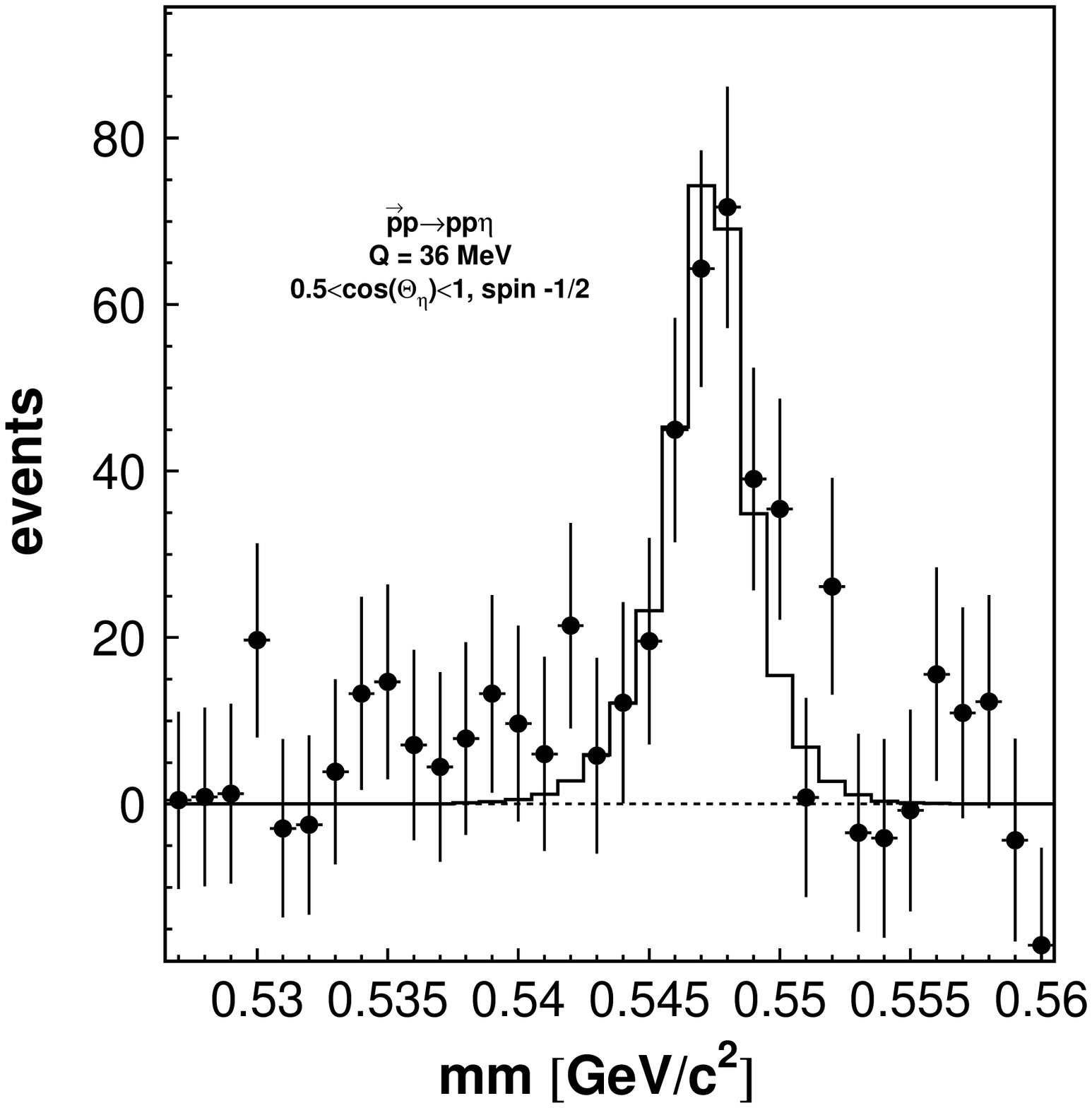,height=6.0cm,angle=0}
      }
      \put(5.8,17.8){{\normalsize {\bf a)}}}
      \put(13.3,17.8){{\normalsize {\bf e)}}}
      \put(5.8,11.8){{\normalsize {\bf b)}}}
      \put(13.3,11.8){{\normalsize {\bf f)}}}
      \put(5.8,5.8){{\normalsize {\bf c)}}}
      \put(13.3,5.8){{\normalsize {\bf g)}}}
      \put(5.8,-0.2){{\normalsize {\bf d)}}}
      \put(13.3,-0.2){{\normalsize {\bf h)}}}
  \end{picture}
        \hspace{-1cm}
  \caption{ \small{  Missing mass spectra for the $\vec{p}p\to pp\eta$ reaction at Q~=~36~MeV
for different ranges of $\cos\theta_{\eta}$
as measured with spin up (a--d) and spin down modes (e--h).
Dots represent the experimental results along with their statistical errors.
The solid lines show the best fit of the signal function to the experimental data.
                  }
  \label{up_and_down_36}
  }
\end{figure}

Afterwards, the background evaluated in such a way 
has been subtracted from the 
experimental spectra for each histogram separately.   The resulting background-free histograms of
missing mass are presented in Figures~\ref{up_and_down_36}.a-h as 
full circles. 

Due to low acceptance of the COSY-11 system for the production of the $\eta$ meson 
in the backward directions in the centre-of-mass system at the 
excess energy of Q~=~36~MeV (missing mass bin 
for $\cos\theta_{\eta}\in\left[-1,-0.5\right]$), and as a consequence
of the problems with the background subtraction for this bin, resulting
in the huge systematical errors, we decided to omit this bin from 
further analysis and determine the analysing powers at this excess energy for 
$\cos\theta_{\eta}$ larger than -0.5.  


Subsequently, the simulations of the shape of the missing mass spectrum for different ranges 
of $\cos\theta_{\eta}$ have been done 
using a program based on the GEANT3 code~\cite{geant}.
This program contains the exact geometry of the COSY-11 detector system
as well as the precise map of the magnetic field of the dipole magnet. Also the momentum and spatial beam 
spreads, multiple scattering, and other instrumentation effects have been taken into account. 
Generated events which fulfilled conditions equivalent to the experimental trigger have been
analysed in the same way as the experimental data and corresponding histograms 
for the $pp\to pp\eta$ reaction have been determined resulting in 
three missing mass functions $g_{pp\to pp\eta}(mm,\cos\theta_{\eta})$, 
where $mm$ denotes the missing mass.
Further, to the experimental histograms presented in Figure~\ref{up_and_down_36}
the functions 
\begin{equation}
G^{\uparrow(\downarrow)}(mm,\cos\theta_{\eta}) = \gamma^{\uparrow(\downarrow)}(\cos\theta_{\eta}) g_{pp\to pp\eta}(mm,\cos\theta_{\eta})
\label{funkcje_g}
\end{equation} 
have been fitted, for spin up and down separately. 
Note that the functions $g_{pp\to pp\eta}(mm,\cos\theta_{\eta})$
describing the missing mass shape
are the same for spin up and down modes, and vary only
with the $\cos\theta_{\eta}$. On the other hand, 
the free parameters of the fit -- $\gamma^{\uparrow(\downarrow)}(\cos\theta_{\eta})$ --
might be different, depending on the spin orientation. 
The number of identified $\eta$ mesons for each $\cos\theta_{\eta}$ bin 
and for spin up and down modes have been calculated from   
the determined $\gamma^{\uparrow(\downarrow)}(\cos\theta_{\eta})$ parameters, namely:
\begin{equation}
N_{+(-)}^{\uparrow(\downarrow)}(\cos\theta_{\eta}) = \gamma^{\uparrow(\downarrow)}(\cos\theta_{\eta}) \int_{0}^{mm_{max}}{g_{pp\to pp\eta}(mm,\cos\theta_{\eta}) d(mm)}.
\label{number_of_eta}
\end{equation}
The statistical errors $\sigma(N_{+(-)}^{\uparrow(\downarrow)})(\cos\theta_{\eta})$ have been estimated 
based on the formula:
\begin{equation}
\sigma(N_{+(-)}^{\uparrow(\downarrow)}(\cos\theta_{\eta})) = \sigma(\gamma^{\uparrow(\downarrow)}(\cos\theta_{\eta})) \int_{0}^{mm_{max}}{g_{pp\to pp\eta}(mm,\cos\theta_{\eta}) d(mm)},
\label{number_of_etai_errorr}
\end{equation}
where $\sigma(\gamma^{\uparrow(\downarrow)}(\cos\theta_{\eta}))$ are the estimates of the 
$\gamma^{\uparrow(\downarrow)}(\cos\theta_{\eta})$ parameters uncertainties (standard deviations)
determined by means of the MINUIT minimization package~\cite{minuit}. 
Results are presented in Table~\ref{liczba_eta_36}. 

\begin{table}[H]
 \begin{center}
   \begin{tabular}{|c|c|c|}
    \hline
       $\cos\theta_{\eta}$ & $N_+^{\uparrow}(\cos\theta_{\eta})$ & $N_-^{\downarrow}(\cos\theta_{\eta})$ \\ 
    \hline
         $\left[-0.5;0\right)$  &  {\bf 103} $\pm$ {\bf 16} &  {\bf 100} $\pm$ {\bf 18} \\
         $\left[0;0.5\right)$   &  {\bf 144} $\pm$ {\bf 16} &  {\bf 153} $\pm$ {\bf 18} \\
         $\left[0.5;1\right]$   &  {\bf 259} $\pm$ {\bf 24} &  {\bf 296} $\pm$ {\bf 28} \\
    \hline
   \end{tabular}
     \caption{ {\small Number of identified $\eta$ mesons in the $\vec{p}p\to pp\eta$ reaction 
	at the excess energy of Q~=~36~MeV 
        as the function of the spin orientation and the cosine 
	of the centre-of-mass polar angle of the $\eta$ meson emission -- $\theta_{\eta}$.
         }
     \label{liczba_eta_36}
         }
 \end{center}
\end{table}

\subsection{Q~=~10~MeV}
\label{back_q10}

\begin{figure}[H]
  \unitlength 1.0cm
        \begin{center}
  \begin{picture}(14.0,8.0)
    \put(3.0,0.0){
      \psfig{figure=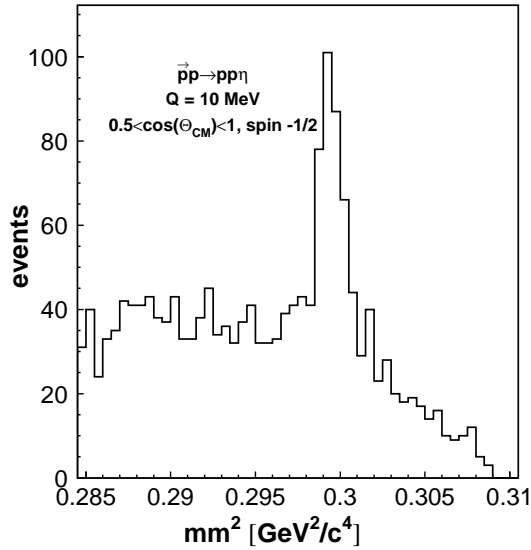,height=8.0cm,angle=0}
    }
  \end{picture}
  \caption{ \small { A histogram of missing mass squared for 
        the $\vec{p}p\to pp\eta$ reaction at Q~=~10~MeV for  
        selected $\cos\theta_{\eta}$ range as measured during spin down mode. 
}
 \label{miss_test}
  }
        \end{center}
\end{figure}

At this excess energy the upper-energetic tale of the $\eta$ peak 
disperses to the edge of the kinematical limit, as 
presented in Figure~\ref{miss_test}. 
This makes the identification of the upper part of the multipionic 
background more inacurate, and hence the method of the background subtraction
similar to the one presented in~\ref{back_q36} would be biased by the larger
systematical uncertainties.  

Therefore, to separate the actual production rates from the
background, the reactions with multipionic production 
as well as the events with the $\eta$ meson production have been simulated in 
order to reconstruct the shape of the missing mass spectra.
For simulations, the same program as the one described in Section~\ref{back_q36}
has been used, but this time apart from the $pp\to pp\eta$ reaction
also the background reactions like $pp\to pp~2\pi, 3\pi, 4\pi$
have been generated.

In order to perform the credible comparison between experiment and the simulations, in the Monte-Carlo calculations  
the geometry of drift chambers, as well as the position
and geometrical parameters of the target have been fixed to be the same as for the experimental data analysis.  

Let the $f_{pp\to pp2\pi}(mm^{2},\sigma_p,\Delta p,\cos\theta_{\eta}$), 
$f_{pp\to pp3\pi}(mm^{2},\sigma_p,\Delta p,\cos\theta_{\eta}$), 
$f_{pp\to pp4\pi}(mm^{2},\sigma_p,\Delta p,\cos\theta_{\eta}$), 
and $f_{pp\to pp\eta}(mm^{2},\sigma_p,\Delta p,\cos\theta_{\eta}$)
denote the generated background functions for double, triple, fourfold pion production and 
for the $\eta$ production, respectively, where $mm^{2}$ stands for the
missing mass squared, whereas $\sigma_p$ and $\Delta p$ are the momentum spread and the deviation of the beam momentum 
from its nominal value.
The functions of the type:
\begin{eqnarray}
f^{\uparrow(\downarrow)}(mm^2,\sigma_p,\Delta p,\cos\theta_{\eta}) = 
\alpha^{\uparrow(\downarrow)}(\cos\theta_{\eta})~f_{pp\to pp2\pi}(mm^{2},\sigma_p,\Delta p,\cos\theta_{\eta}) + \\
\nonumber
 \beta^{\uparrow(\downarrow)}(\cos\theta_{\eta})~f_{pp\to pp3\pi}(mm^{2},\sigma_p,\Delta p,\cos\theta_{\eta}) + \\ 
\nonumber
\gamma^{\uparrow(\downarrow)}(\cos\theta_{\eta})~f_{pp\to pp4\pi}(mm^{2},\sigma_p,\Delta p,\cos\theta_{\eta}) + \\
\nonumber
\delta^{\uparrow(\downarrow)}(\cos\theta_{\eta})~f_{pp\to pp\eta}(mm^{2},\sigma_p,\Delta p,\cos\theta_{\eta}),  
\label{fit_function}
\end{eqnarray}
with $\alpha(\cos\theta_{\eta}), \ldots \delta(\cos\theta_{\eta})$ as free parameters have been
fitted to the experimental spectra of the missing mass squared
using the MINUIT~\cite{minuit} minimization package. 
For Q~=~10~MeV 
also the bin for $\cos\theta_{\eta}\in\left[-1,-0.5\right]$ has been considered
in the analysis, as the statistic was sufficient to extract 
the numbers of events also in this particular bin. The fit has been 
performed simultaneously to 8 histograms of missing mass squared, 
each for different spin-$\cos\theta_{\eta}$ range configuration. 
It is important to point out that for a given $\cos\theta_{\eta}$ bin the shapes of the generated
missing mass histograms, i.e. the functions $f_{pp\to pp2\pi}(mm^{2},\sigma_p,\Delta p,\cos\theta_{\eta})$, $f_{pp\to pp3\pi}(mm^{2},\sigma_p,\Delta p,\cos\theta_{\eta}$),
$f_{pp\to pp4\pi}(mm^{2},\sigma_p,\Delta p,\cos\theta_{\eta})$ and $f_{pp\to pp\eta}(mm^{2},\sigma_p,\Delta p,\cos\theta_{\eta})$ 
depend only on the width of the momentum distribution $\sigma_p$ and 
the shift from the optimal beam momentum $\Delta p$. 
The $\chi^2$ of the fit has been minimized as a
function of 34 parameters: amplitudes 
$\alpha(\theta_i)$,...,$\delta(\theta_i)$ of the generated background and signal reactions
(4 for each histogram) and 2 parameters responsible for the spread and the absolute
value of the beam momentum. 
However, it is worth noting that altogether the histograms contained 480 points.

\begin{figure}[p]
  \unitlength 1.0cm
  \begin{picture}(14.0,23.0)
      \put(0.00,17.9){
         \psfig{figure=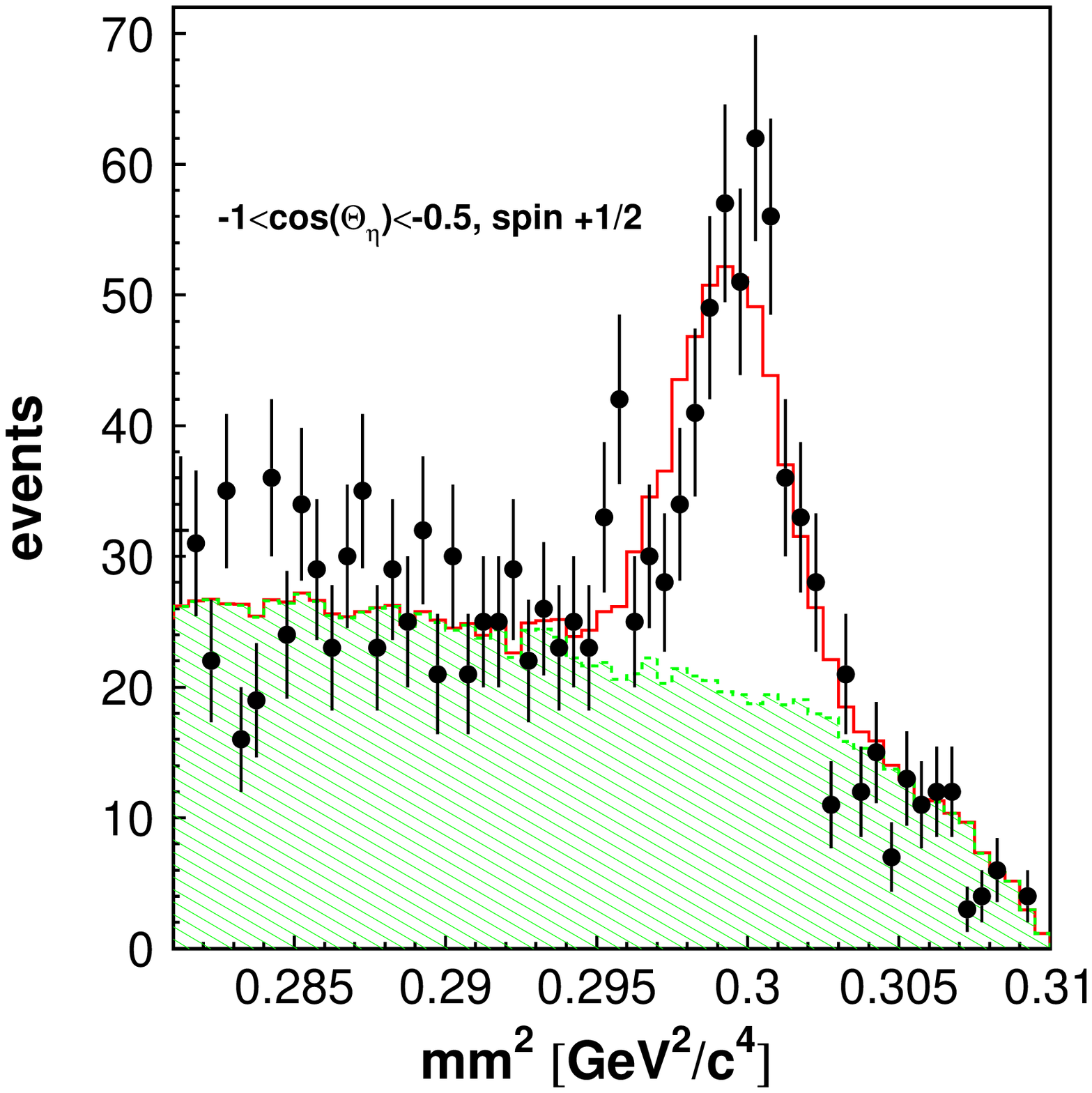,height=6.0cm,angle=0}
      }
      \put(7.50,17.9){
         \psfig{figure=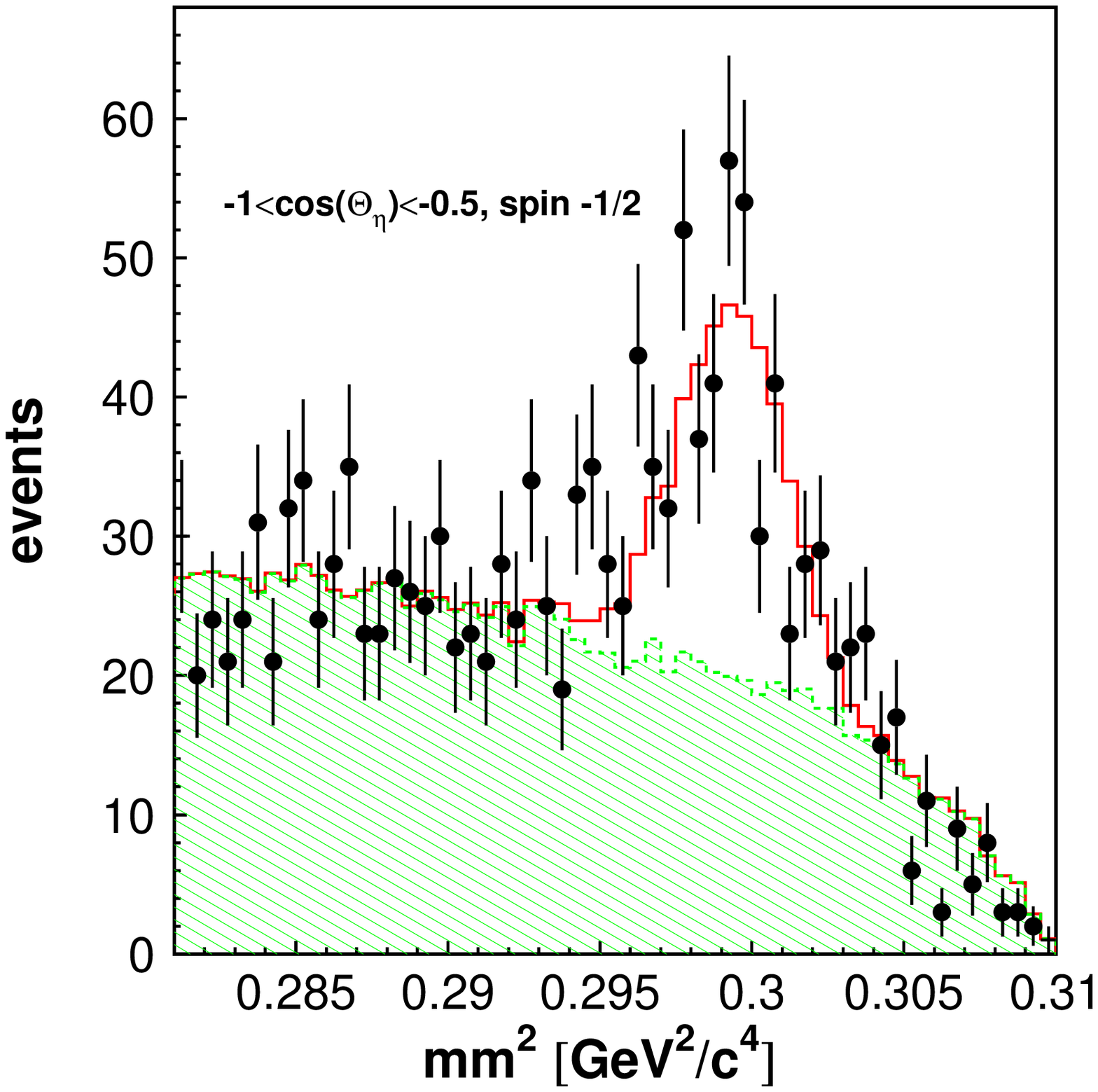,height=6.0cm,angle=0}
      }
      \put(0.00,11.9){
         \psfig{figure=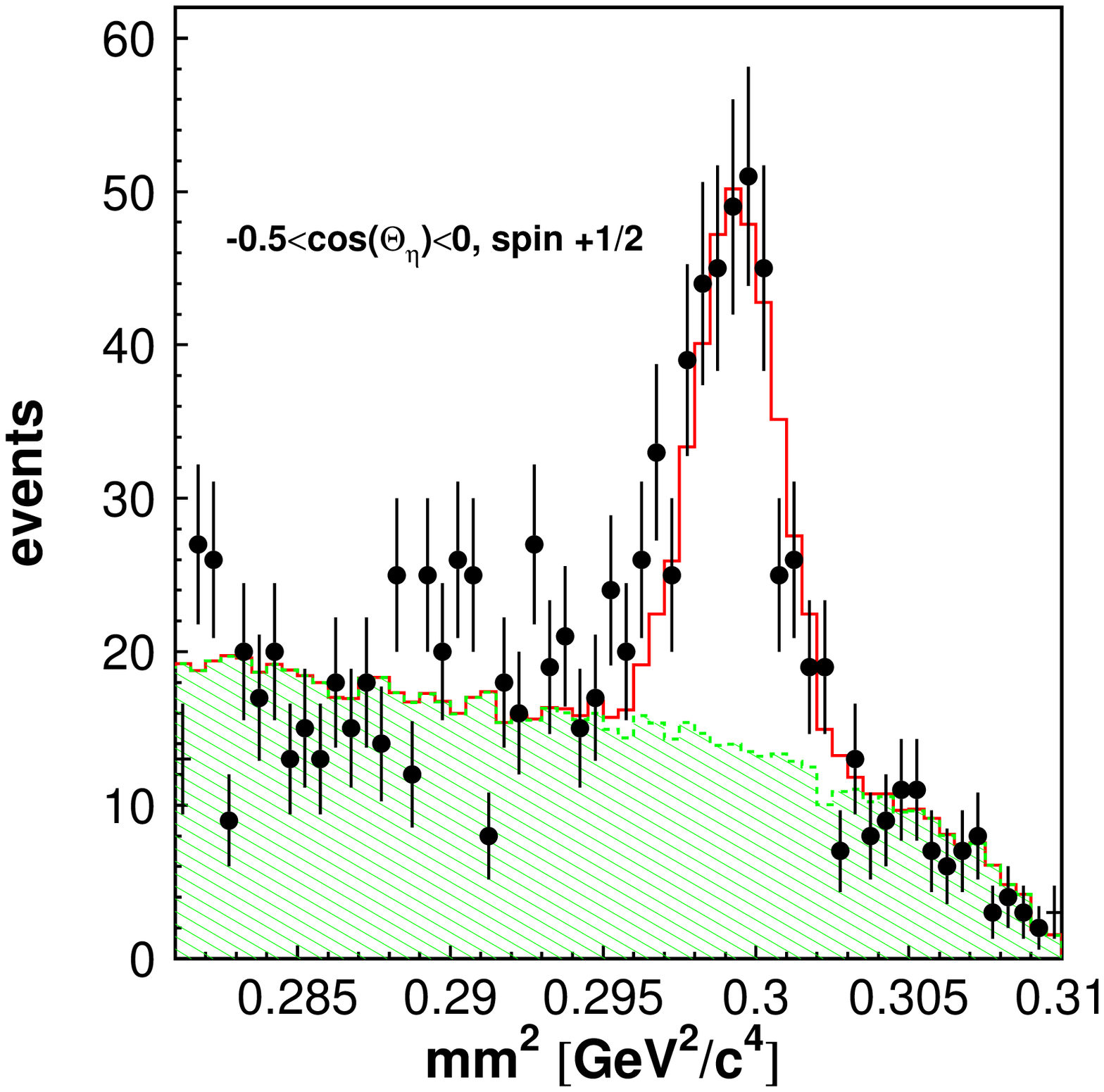,height=6.0cm,angle=0}
      }
      \put(7.50,11.9){
         \psfig{figure=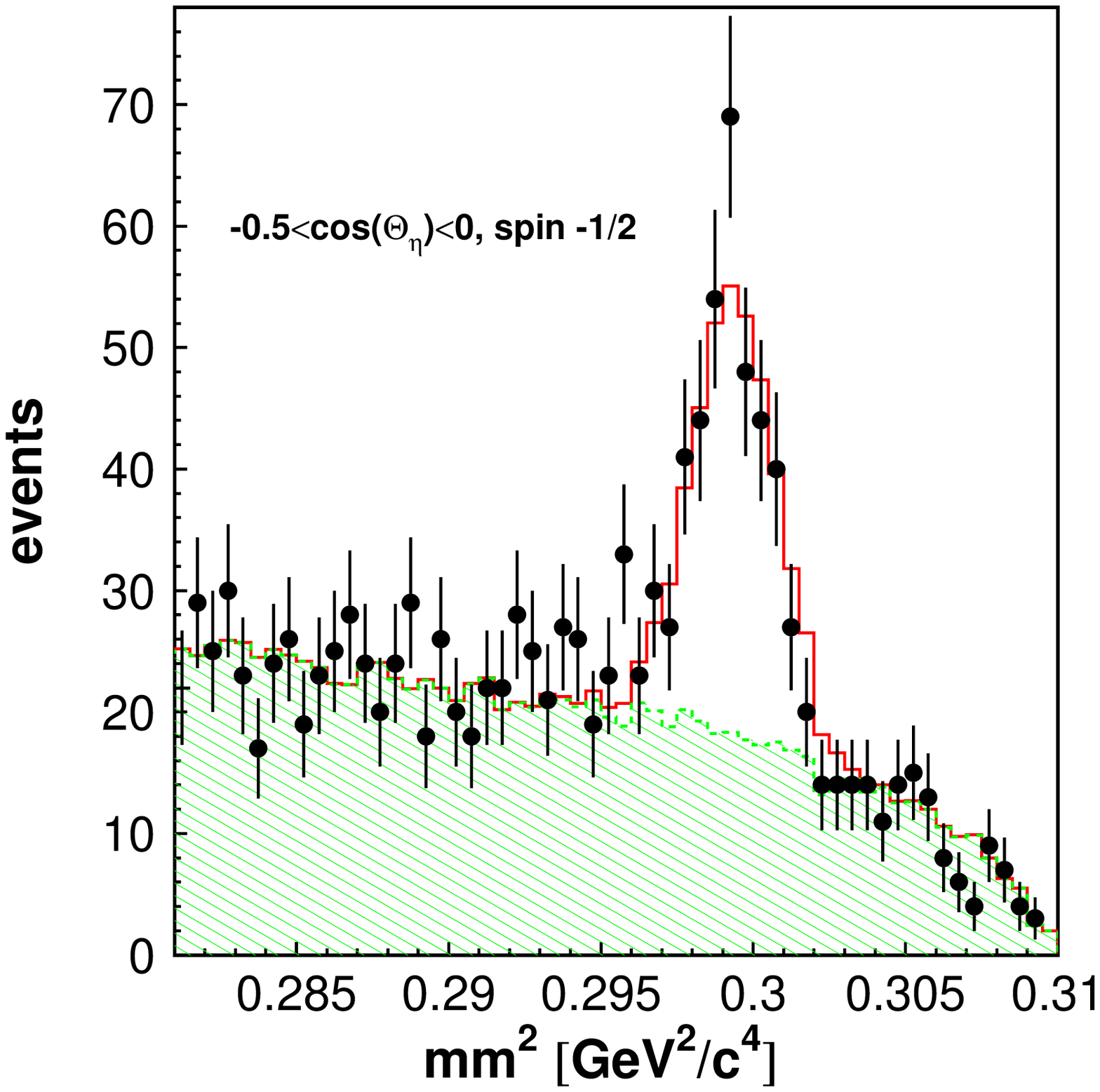,height=6.0cm,angle=0}
      }
      \put(0.00,5.9){
         \psfig{figure=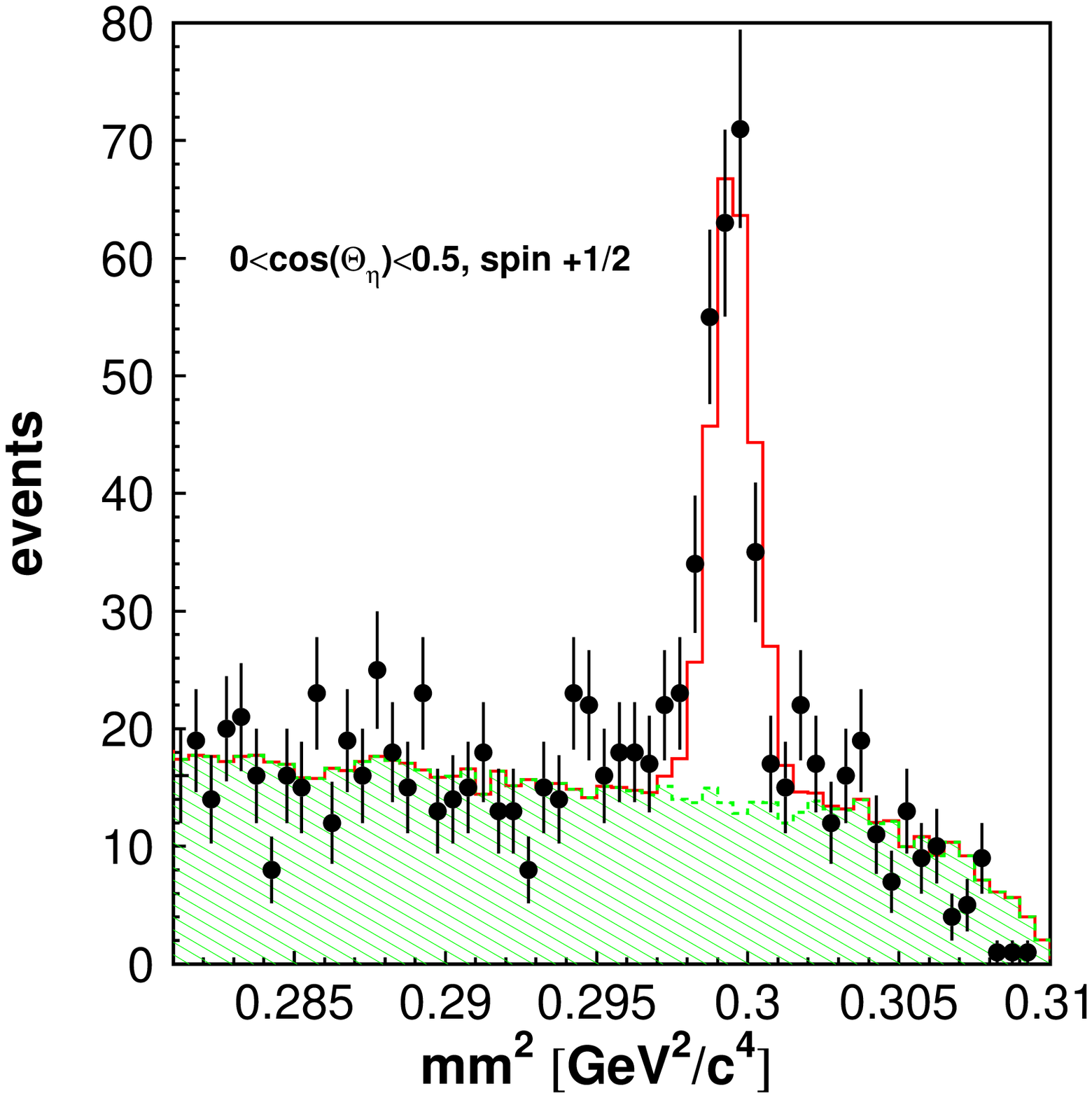,height=6.0cm,angle=0}
      }
      \put(7.50,5.9){
         \psfig{figure=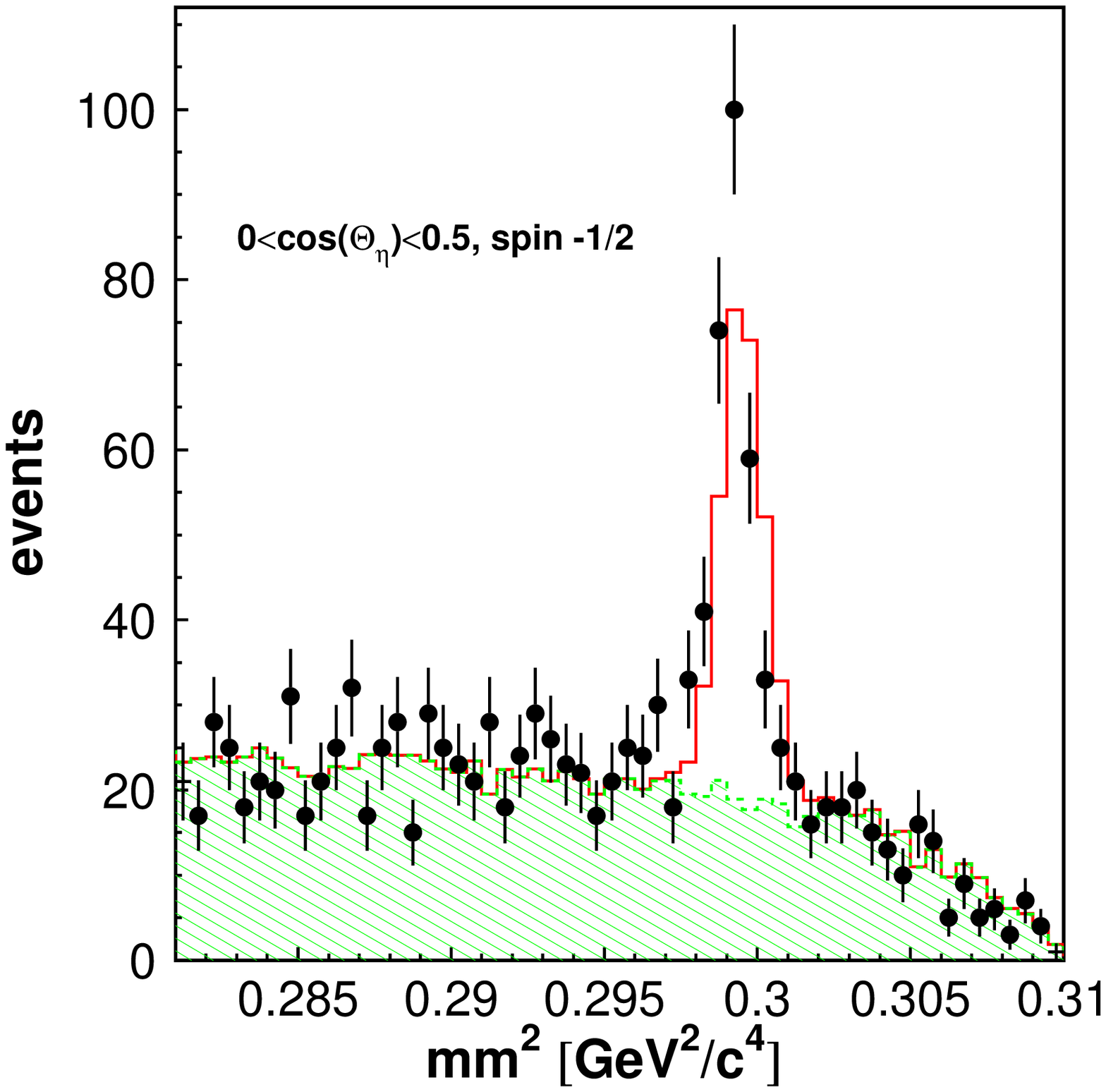,height=6.0cm,angle=0}
      }
      \put(0.00,-0.1){
         \psfig{figure=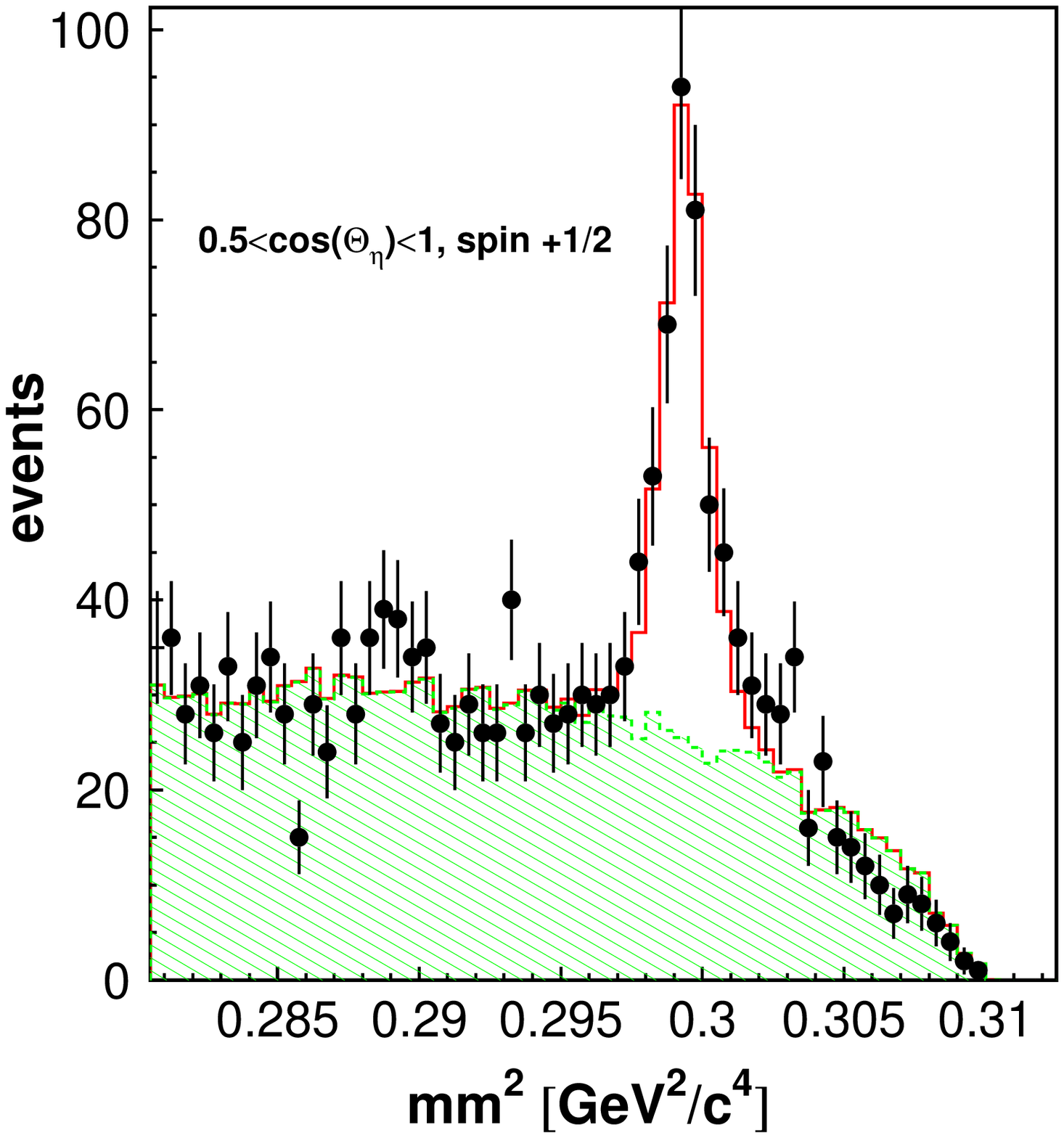,height=6.0cm,angle=0}
      }
      \put(7.50,-0.1){
         \psfig{figure=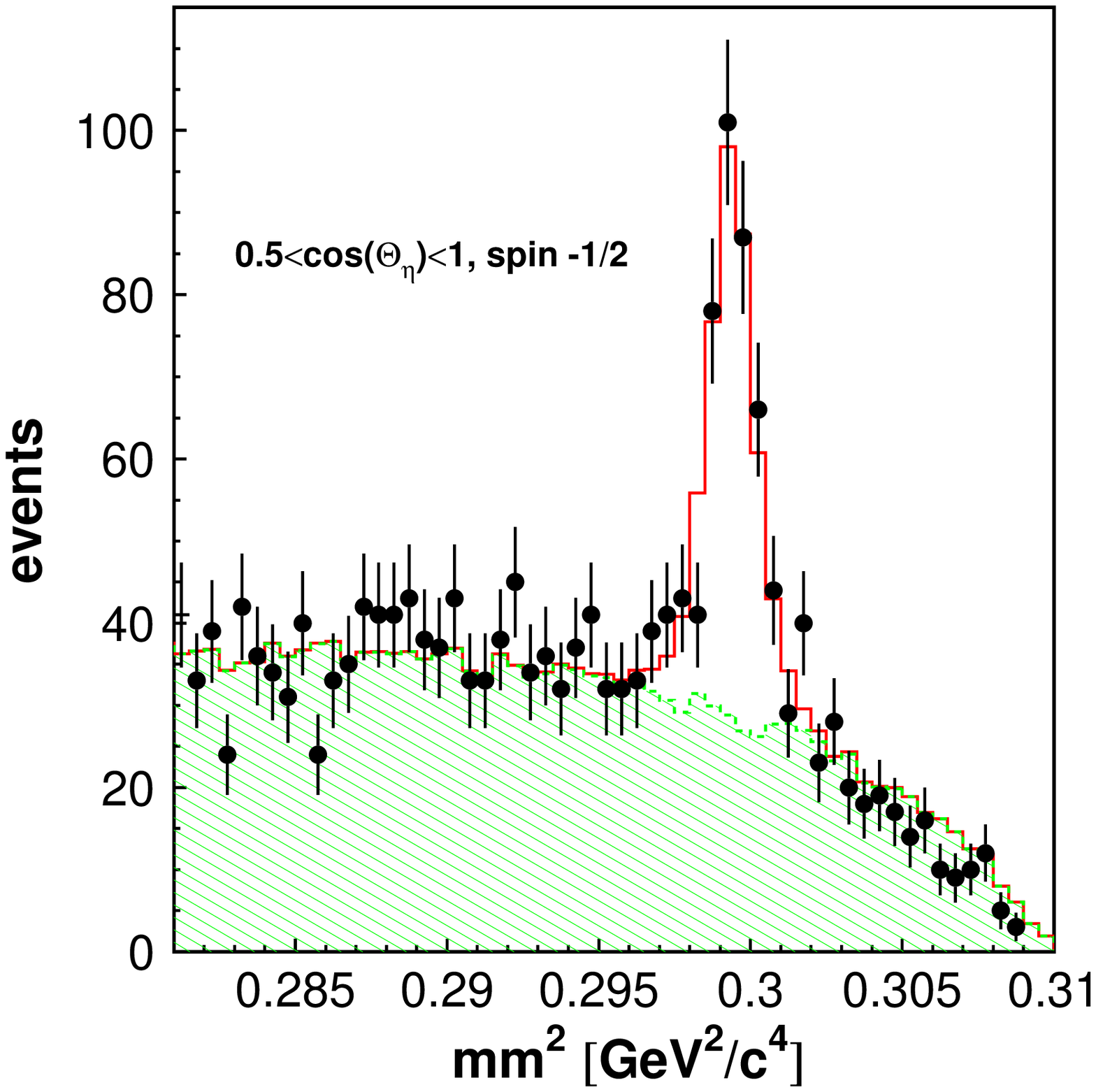,height=6.0cm,angle=0}
      }
      \put(5.8,17.8){{\normalsize {\bf a)}}}
      \put(13.3,17.8){{\normalsize {\bf e)}}}
      \put(5.8,11.8){{\normalsize {\bf b)}}}
      \put(13.3,11.8){{\normalsize {\bf f)}}}
      \put(5.8,5.8){{\normalsize {\bf c)}}}
      \put(13.3,5.8){{\normalsize {\bf g)}}}
      \put(5.8,-0.2){{\normalsize {\bf d)}}}
      \put(13.3,-0.2){{\normalsize {\bf h)}}}
  \end{picture}
	\hspace{-1cm}
  \caption{ \small{  Missing mass spectra for different ranges of $\cos\theta_{\eta}$
for spin up (a--d) and spin down (e--h).
Dots represent the experimental data along with their statistical errors. Shaded part shows the generated mutli-pionic
background. The solid line is the best fit of the sum of the signal and background to the experimental data. 
                  }
  \label{up_and_down}
  }
\end{figure}

In Fig.~\ref{up_and_down} the missing masses for the individual
$\cos\theta_{\eta}$ subranges are shown.  
Full dots denote the experimental data,
the shaded parts of the histograms depict the multipionic background, whereas the solid line
represents the best fit of the Monte-Carlo data to the experimental spectra.
The minimum value of $\chi^2$ for this fit divided by the number 
of degrees of freedom was determined to be 1.6.

Numbers of $\eta$ mesons
$N_{+(-)}^{\uparrow(\downarrow)}(\cos\theta_{\eta})$ for the individual ranges of $\cos\theta_{\eta}$
for spin up and down cycles have been calculated as:
\begin{equation}
N_{+(-)}^{\uparrow(\downarrow)}(\cos\theta_{\eta}) = \delta^{\uparrow(\downarrow)}(\cos\theta_{\eta}) \int_{0}^{mm_{max}^{2}}{f_{pp\to pp\eta}(mm^{2},\sigma_p,\Delta p,\cos\theta_{\eta}) d(mm^{2})}.
\label{number_of_eta}
\end{equation}
The statistical errors $\sigma(N_{+(-)}^{\uparrow(\downarrow)}(\cos\theta_{\eta}))$
have been estimated based on the formula analogous to Equation~\ref{number_of_etai_errorr}:
\begin{equation}
\sigma(N_{+(-)}^{\uparrow(\downarrow)}(\cos\theta_{\eta})) = \sigma(\delta^{\uparrow(\downarrow)}(\cos\theta_{\eta})) \int_{0}^{mm_{max}^{2}}{f_{pp\to pp\eta}(mm^{2},\sigma_p,\Delta p,\cos\theta_{\eta}) d(mm^{2})},
\label{error_of_eta}
\end{equation}
where $\sigma(\delta^{\uparrow(\downarrow)}(\cos\theta_{\eta}))$ 
are again the estimates of the standard deviations 
of the $\delta^{\uparrow(\downarrow)}(\cos\theta_{\eta})$ parameters. 
The obtained number of $\eta$ mesons per $\cos\theta_{\eta}$ range are
quoted in Table~\ref{liczba_eta}. 

\begin{table}[H]
 \begin{center}
   \begin{tabular}{|c|c|c|}
    \hline
       $\cos\theta_{\eta}$ & $N_+^{\uparrow}(\cos\theta_{\eta})$ & $N_-^{\downarrow}(\cos\theta_{\eta})$ \\ 
    \hline
         $\left[-1;-0.5\right)$ &  {\bf 306} $\pm$ {\bf 27} &  {\bf 250} $\pm$ {\bf 26} \\
         $\left[-0.5;0\right)$  &  {\bf 267} $\pm$ {\bf 22} &  {\bf 260} $\pm$ {\bf 24} \\
         $\left[0;0.5\right)$   &  {\bf 198} $\pm$ {\bf 18} &  {\bf 208} $\pm$ {\bf 19} \\
         $\left[0.5;1\right]$   &  {\bf 279} $\pm$ {\bf 23} &  {\bf 286} $\pm$ {\bf 25} \\
    \hline
   \end{tabular}
     \caption{ {\small Number of registered $\eta$ mesons 
        as the function of the cosine of the centre-of-mass 
	$\eta$ emission angle ($\theta_{\eta}$) and spin 
	orientation of the beam.
         }
     \label{liczba_eta}
         }
 \end{center}
\end{table}

In the table only the statistical uncertainties are presented. The systematic 
errors for the extraction of number of events 
have been estimated by comparison of the values in Table~\ref{liczba_eta} with the number of events determined using the other method, 
namely assuming that the background is linear in the range of missing mass peak. 
In such a way for each $\cos\theta_{\eta}$ bin we obtained
values which differed from the numbers quoted in Table~\ref{liczba_eta}
by no more than 1.5\%. This value we assigned to the systematic error 
of N$^{\uparrow(\downarrow)}(\cos\theta_{\eta})$ at both excess energies.  
\section{
Results -- the analysing power
}
\label{analysing_powers}

\vspace{3mm}
{\small
The results of the analysing power determination are 
summarized in the table and in the figures. 
}
\vspace{5mm}

Having presented the methodology of $L_{rel}$, $P$ and
$N_{+(-)}^{\uparrow(\downarrow)}(\cos\theta_i)$ determination we can now
equate the analysing power for both excess energies.
For this purpose we shall exploit Formula~\ref{anal_part},
which reads:
\begin{equation}
A_y(\cos\theta_{\eta}) = \frac{1}{P} \frac{N^{\uparrow}_{+}(\cos\theta_{\eta})-L_{rel} N^{\downarrow}_{-}(\cos\theta_{\eta})}{N^{\uparrow}_{+}(\cos\theta_{\eta})+L_{rel} N^{\downarrow}_{-}(\cos\theta_{\eta})}.
\label{anal_part_b}
\end{equation}

Relative luminosities can be found in Table~\ref{swietlnosci},
average polarisations have been determined in Section~\ref{polarisation}, and finally
Tables~\ref{liczba_eta_36} and~\ref{liczba_eta} provide the values of background free productions rates
for Q~=~36 and 10~MeV, respectively.
The analysing powers for the $\vec{p}p\to pp\eta$ reaction which
were calculated basing on these numbers are shown in 3$^{\textrm{rd}}$ column of Table~\ref{tabelka_zd}.
Values presented in the 4$^{\textrm{th}}$ and 5$^{\textrm{th}}$ column of this table are the mean 
values over a $\cos\theta_{\eta}$ bin of the theoretical
predictions for $A_y$ according to the models with the dominance 
of the pseudoscalar and vector meson exchanges, respectively.
In the derivation we took advantage of the fact that 
for an isotropic distribution of the differential cross section
                 $\frac{d\sigma}{d\theta}\left(\theta\right)$~=~const, the average
                 analysing power over an angular range $\theta$
                 is an arithmetical average of the analysing powers for the individual
                 $\theta_n$ subranges of $\theta$ -- for 
proof see Appendix~\ref{prove}. 

\begin{table}[H]
 \begin{center}
   \begin{tabular}{|c|c|c|c|c|}
    \hline
    \hline
      Q[MeV] & $\cos\theta_{\eta}$ & A$_y(\cos\theta_{\eta})$ & A$_y^{pseud}(\cos\theta_{\eta})$ & A$_y^{vec}(\cos\theta_{\eta})$ \\
    \hline
    \hline
           & $\left[-1;-0.5\right)$ & {\bf 0.163} $\pm$ {\bf 0.099} $\pm$ {\bf 0.022} & 0.133  & $-0.170$ \\
  {\bf 10} & $\left[-0.5;0\right)$  & {\bf 0.035} $\pm$ {\bf 0.091} $\pm$ {\bf 0.012} & 0.067  & $-0.092$ \\
           & $\left[0;0.5\right)$   & ${\bf -0.021}$ $\pm$ {\bf 0.095} $\pm$ {\bf 0.011} & $-0.082$ &  0.092 \\
           & $\left[0.5;1\right]$   & ${\bf -0.003}$ $\pm$ {\bf 0.088} $\pm$ {\bf 0.009} & $-0.144$ &  0.170 \\
    \hline
           & $\left[-1;-0.5\right)$ &  ---  & 0.147  & -0.001 \\
  {\bf 36} & $\left[-0.5;0\right)$  & {\bf 0.039} $\pm$ {\bf 0.179} $\pm$ {\bf 0.012} & 0.046  &  0.000 \\
           & $\left[0;0.5\right)$   & ${\bf -0.029}$ $\pm$ {\bf 0.122} $\pm$ {\bf 0.010} & $-0.062$ &  0.000 \\
           & $\left[0.5;1\right]$   & ${\bf -0.084}$ $\pm$ {\bf 0.100} $\pm$ {\bf 0.011} & $-0.154$ &  0.001 \\
    \hline
    \hline
   \end{tabular}
     \caption{ {\small Analysing power for the $\vec{p}p\to pp\eta$ reaction
                determined at the excess energies Q~=~10 and 36~MeV.  A$_y^{pseud}$ and A$_y^{vec}$
                are the theoretical predictions according to the models with the 
                pseudoscalar and vector meson exchange dominance, described in Section~\ref{mechanism}.
		The first quoted error is statistical, whereas the second one is systematic.
		The estimation of the systematic errors will be given in the next section.  
       }
     \label{tabelka_zd}
         }
 \end{center}
\end{table}

Statistical uncertainties of A$_y$ were calculated according to the rule of error propagation,
and the following formula has been applied for their determination:
\begin{footnotesize}
\begin{equation}
\sigma(A_y)=\sqrt{\left(\frac{\partial A_y}{\partial L_{rel}}\right)^2 \sigma^2\left(L_{rel}\right) + \left(\frac{\partial A_y}{\partial P}\right)^2 \sigma^2\left(P\right) + \left(\frac{\partial A_y}{\partial N^{\uparrow}_{+}}\right)^2 \sigma^2\left(N^{\uparrow}_{+}\right) + \left(\frac{\partial A_y}{\partial N^{\downarrow}_{-}}\right)^2 \sigma^2\left(N^{\downarrow}_{-}\right) }
\label{anal_error}
\end{equation}
\end{footnotesize}
with 
\begin{equation}
\frac{\partial A_y}{\partial L_{rel}} = -\frac{1}{P} \frac{2N^{\uparrow}_{+}N^{\downarrow}_{-}}{\left(N^{\uparrow}_{+}+L_{rel} N^{\downarrow}_{-}\right)^2},
\nonumber
\end{equation}
\begin{equation}
\frac{\partial A_y}{\partial P} = -\frac{1}{P^2} \frac{N^{\uparrow}_{+}-L_{rel} N^{\downarrow}_{-}}{\left(N^{\uparrow}_{+}+L_{rel} N^{\downarrow}_{-}\right)^2}, 
\nonumber
\end{equation}
\begin{equation}
\frac{\partial A_y}{\partial N^{\uparrow}_{+}} = \frac{1}{P} \frac{2 L_{rel} N^{\downarrow}_{-}}{\left(N^{\uparrow}_{+}+L_{rel} N^{\downarrow}_{-}\right)^2},
\nonumber
\end{equation}
\begin{equation}
\frac{\partial A_y}{\partial N^{\downarrow}_{-}} = -\frac{1}{P} \frac{2 L_{rel} N^{\uparrow}_{+}}{\left(N^{\uparrow}_{+}+L_{rel} N^{\downarrow}_{-}\right)^2}.
\label{parcjalne_error}
\end{equation}
It is worth mentioning, that the partial contributions to the overall statistical error from  
the $L_{rel}$ and $P$ uncertainties were about two orders of magnitude smaller than the 
contributions from $N^{\uparrow}_{+}$ and $N^{\downarrow}_{-}$ uncertainties.

\begin{figure}[H]
  \unitlength 1.0cm
  \begin{picture}(14.0,7.5)
      \put(0.00,1.0){
         \psfig{figure=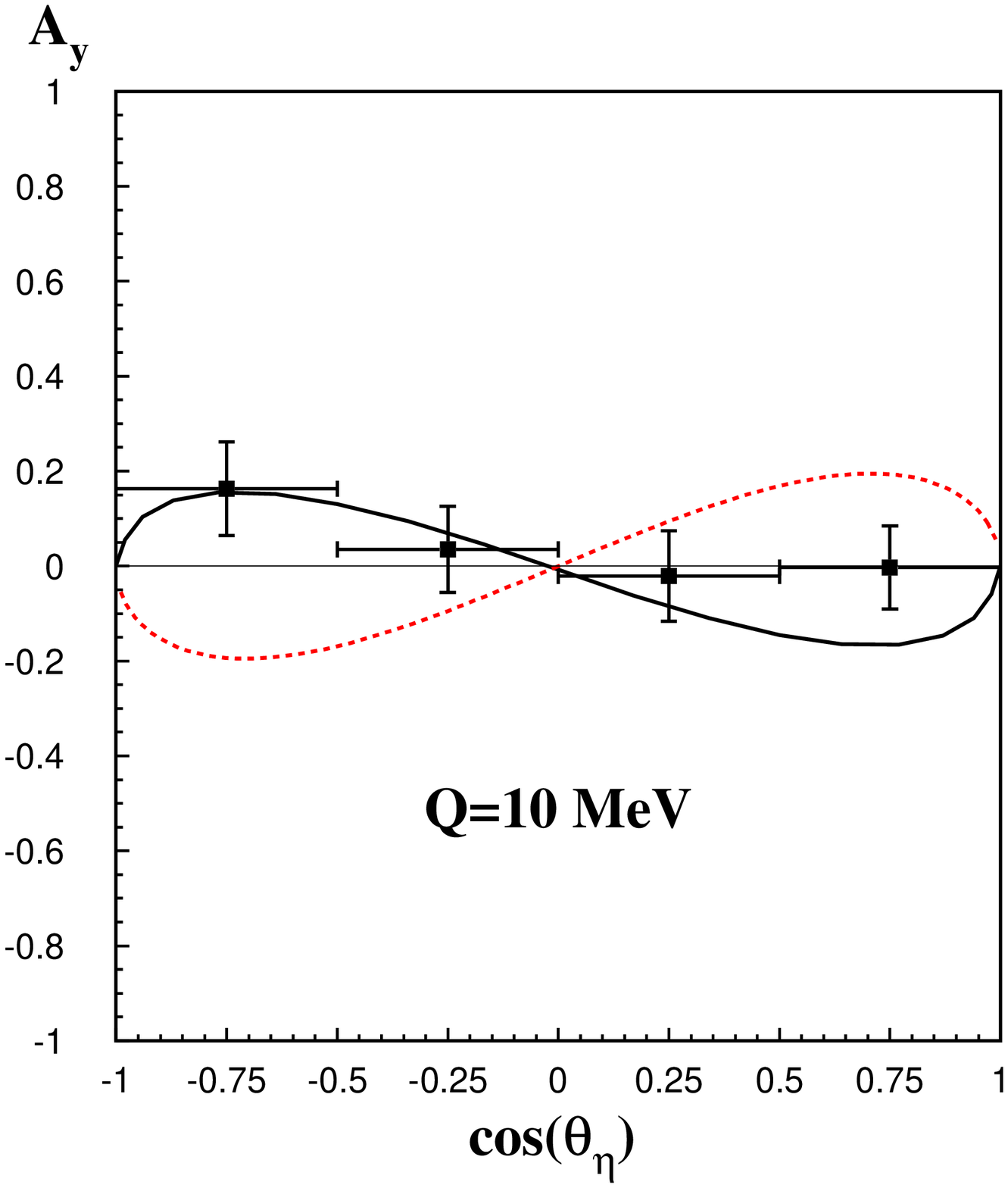,height=6.5cm,angle=0}
      }
      \put(7.50,1.0){
         \psfig{figure=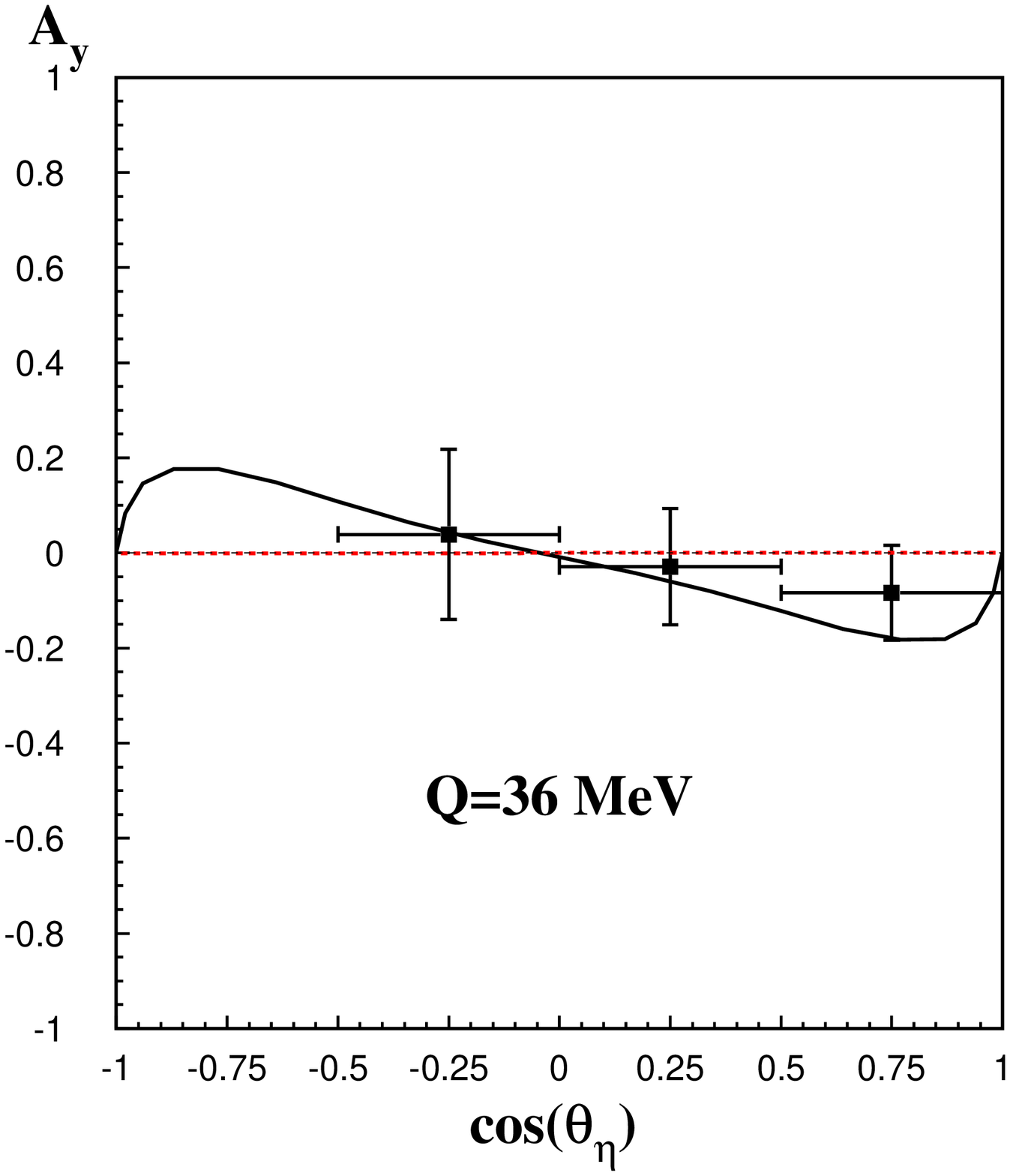,height=6.5cm,angle=0}
      }
      \put(6.5,0.8){{\normalsize {\bf a)}}}
      \put(14.0,0.8){{\normalsize {\bf b)}}}
  \end{picture}
  \caption{ {\small Analysing power for the  $\vec{p}p\to pp\eta$ reaction as function of  
        the cosine of polar emission angle of the $\eta$ meson in the centre-of-mass system for 
        Q~=~10~MeV (a) and Q~=~36~MeV (b). Full lines are the
        predictions based on the pseudoscalar meson exchange model~\cite{nakayama}
        whereas the dotted lines represent the results
        of the calculations based on the vector meson exchange~\cite{wilkin}.
	In the right panel the predictions of the vector meson exchange dominance
	model are consistent with zero. 
	Error bars in both panels of the figure show the statistical uncertainties only. 
	Systematic errors are estimated in next section.  
  \label{analysing_rysunek}
  }}
\end{figure}

The analysing power from Table~\ref{tabelka_zd} along with statistical errors and
the models predictions for Q~=~10 and 36~MeV are depicted
in Figure~\ref{analysing_rysunek}.a and~\ref{analysing_rysunek}.b, respectively.

\section{
Systematic uncertainties of the analysing power
}
\label{systematic}

\vspace{3mm}
{\small
We estimate the systematic uncertainties for the analysing power
at both excess energies. 
}
\vspace{5mm}

The physical quantities needed to calculate the analysing power $A_y$,
according to the Equation~\ref{anal_part_b},
are the average beam polarisation $P$, the relative
luminosities between spin up and down modes $L_{rel}$,
and the production rates $N^{\uparrow(\downarrow)}$ for spin 
up and down, respectively.

The systematic uncertainty of polarisation 
for Q~=~10~MeV has been estimated in Section~\ref{kkk} and equals 
$\Delta$P(Q~=~10~MeV)~=~0.055. The same quantity for the excess energy of
Q~=~36~MeV was found to be $\Delta$P(Q~=~36~MeV)~=~0.008~\cite{ulbrich}. Differences 
in the systematic uncertainties between the measurements are due to the fact 
that different detector setups were used in the determination of the degree of polarisation. 
The systematic uncertainty of the relative luminosity has been found 
in Section~\ref{mmm} and equals $\Delta$L$_{rel}=1\%$ for both 
excess energies. Finally, the systematic error of the 
determination of the production rates has been estimated in 
Section~\ref{back_q10} and amounts to $\Delta$N$_{+(-)}^{\uparrow(\downarrow)}=1.5\%$,
the same value for both excess energies.

The values quoted above have been used to determine the overall systematic
uncertainty of A$_y$. 
The contributions from different error sources were added analogously 
to Formula~\ref{anal_error_2}, exchanging P$\longleftrightarrow$ A$_y$,
N$_{+}\longleftrightarrow$ N$_+^{\uparrow}$ and N$_{-}\longleftrightarrow$ N$_-^{\downarrow}$.
This procedure yielded the values of the systematic 
uncertainties of A$_y$ quoted in Table~\ref{tabelka_zd}.

At this point it is worth mentioning that the 
beam misaligment between spin up and down modes has no significant influence upon the 
determination of $N_{+(-)}^{\uparrow(\downarrow)}$ values. Monte-Carlo simulations
analogous to the ones presented in Sections~\ref{mmm} and~\ref{kkk} have been performed and the 
systematic error of $N_{+(-)}^{\uparrow(\downarrow)}$ originating from
the beam misaligment was found to be negligible in comparison with $\Delta$N$_{+(-)}^{\uparrow(\downarrow)}$.   

It has also been confirmed that the systematic uncertainty of 
$N_{+(-)}^{\uparrow(\downarrow)}$ originating from different binnings of histograms 
from Figures~\ref{up_and_down_36} and~\ref{up_and_down} is negligible with comparison to the quoted 1.5\% 
uncertainty originating from the determination of 
$N_{+(-)}^{\uparrow(\downarrow)}$ using different background models.

\newpage
\clearpage
\pagestyle{fancy}
\chapter{Interpretation of the experimental results} 
\label{interpretation}

\section{Test of the predictions for A$_y$}

\vspace{3mm}
{\small
Using the statistical inference it is shown that the
derived values of the analysing power disagree with 
the predictions of the vector meson exchange models 
at a significance level of $\alpha=0.006$.  
The predictions of the pseudoscalar meson exchange dominance
model are in line with the data at a significance level of  
of 0.81. The values of the amplitude of the analysing power 
 are extracted at both excess energies. 
}
\vspace{5mm}

In order to determine the statistical significance of our 
results we have compared the predictions of the models with the 
experimentally determined values of the analysing power for both excess energies.  
For both theoretical hypotheses a value of the 
$\chi^{2}$ have been calculated according to the formula:

\begin{equation}
\chi^2 = \sum_{i=1}^7 \left(\frac{A_{y,i}^{model} - A_{y,i}^{exp}}{\sigma\left(A_{y,i}^{exp}\right)}\right)^{2},
\label{test_chi2}
\end{equation} 
with $\sigma\left(A_{y,i}^{exp}\right)$ denoting the experimental
uncertainty of the $A_{y,i}^{exp}$ value
and $i$ enumerating the points for the excess energies of Q~=~10 and 36~MeV.
Altogether there are 7 experimental points determined in both experiments. 
The $A_{y,i}^{model}$ are the values of the analysing power 
calculated according to the pseudoscalar and vector meson exchange models, 
averaged over the same angular divisions as was performed for the experimental data. 
The value of the analysing power for an angular range 
$\theta$ has been taken as the mean value of the individual analysing powers predicted
for this range (see Appendix~\ref{prove}). 
These values along with the experimental values 
of $A_{y,i}^{exp}$ are quoted in Table~\ref{tabelka_zd}.

We have obtained the values of $\chi^{2}$ equal to 3.78 and 
19.32 for the pseudoscalar and vector meson exchange models, respectively. 
Taking into account that there are 7 degrees of freedom,  
the reduced value of the $\chi^2$ for the
pseudoscalar meson exchange model was found to be $\bar{\chi}^{2}_{psd}=0.54$,
which corresponds to
a significance level $\alpha_{psd}=0.81$, whereas
for the vector meson exchange model $\bar{\chi}^{2}_{vec}=2.76$ resulting in
a significance level $\alpha_{vec}=0.006$~\footnote{The significance levels
as the functions of the $\bar{\chi}^2$ values have been taken from~\cite{taylor}}. 

\begin{figure}[H]
  \unitlength 1.0cm
  \begin{picture}(14.0,7.5)
      \put(0.00,1.0){
         \psfig{figure=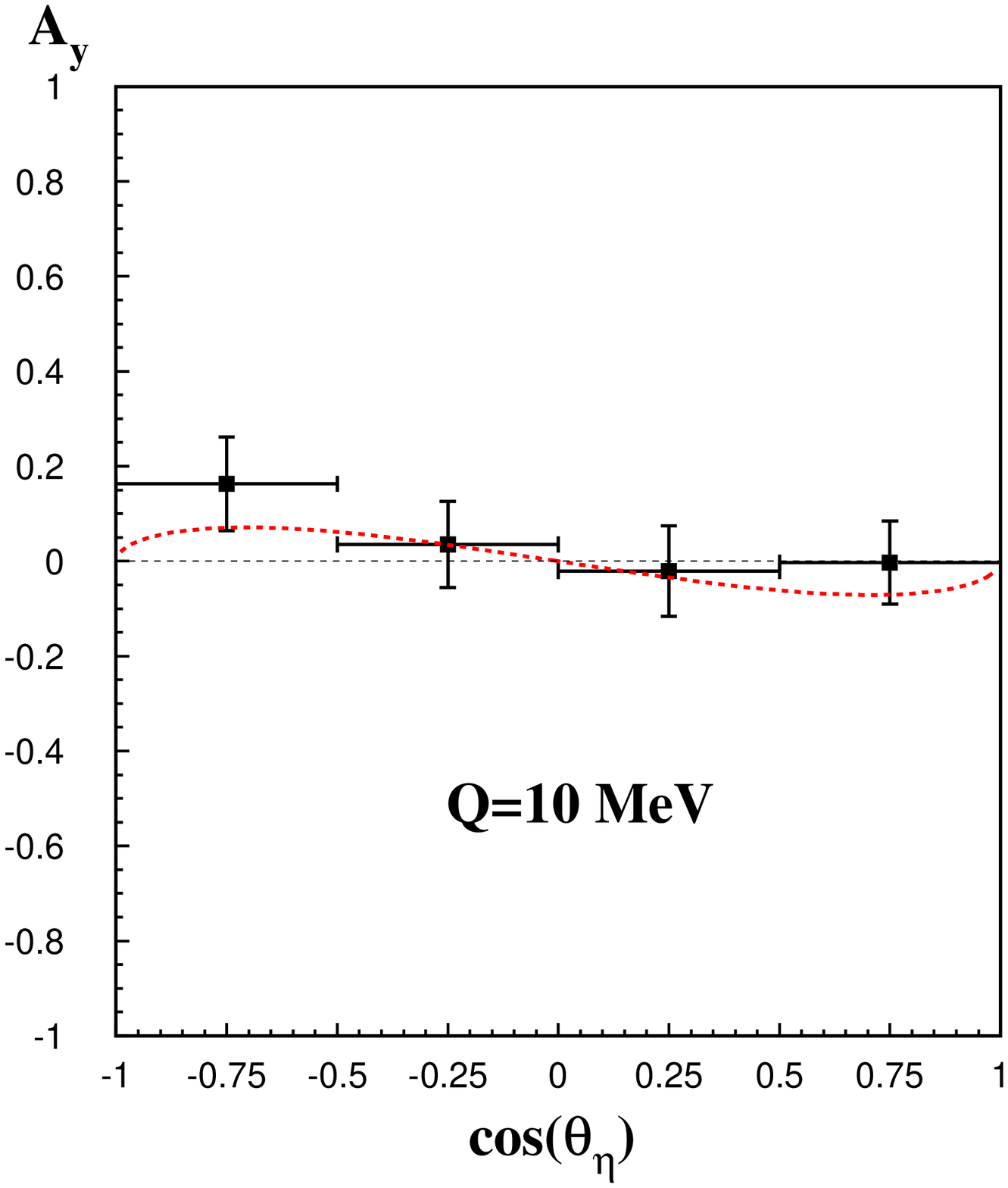,height=6.5cm,angle=0}
      }
      \put(7.50,1.0){
         \psfig{figure=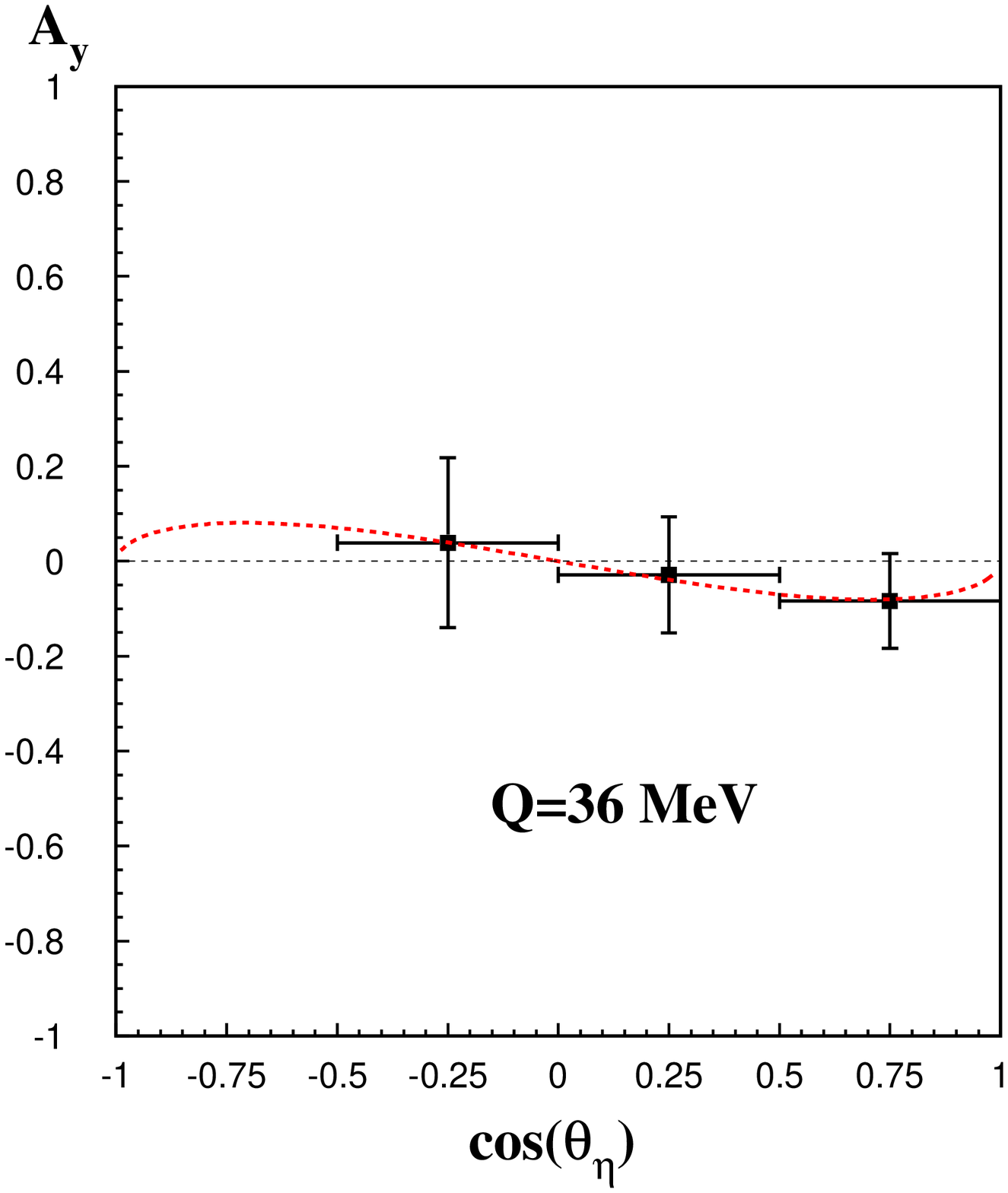,height=6.5cm,angle=0}
      }
      \put(6.5,0.8){{\normalsize {\bf a)}}}
      \put(14.0,0.8){{\normalsize {\bf b)}}}
  \end{picture}
  \caption{ {\small Fit of the function $A_y(Q,\theta_{\eta})=A_y^{max,vec}(Q) \sin2\theta_{\eta}$ to the experimental 
	results of analysing power for 
        Q~=~10~MeV (a) and Q~=~36~MeV (b). The purpose of the fit was to extract the 
	amplitude $A_y^{max,vec}$ for both excess energies.  
  \label{fit_ay_max_wilk}
  }}
\end{figure}

Assuming that the predictions of the vector meson exchange model regarding
the shape of the angular dependence of the analysing power are correct, we have
also performed the $\chi^2$ test in order to determine the 
amplitude of the angular dependence of the analysing power.
As mentioned in Section~\ref{mechanism}, the angular distribution of the
analysing power may be parametrized with the following equation:
\begin{equation}
A_y\left(Q,\theta_{\eta}\right)=A_y^{max,vec}(Q) \sin{2\theta_{\eta}},
\label{eq2}
\end{equation}
where the amplitude $A_y^{max,vec}(Q)$ is a function of the excess energy Q,
and its energy dependence is shown in Figure~\ref{koby} as the dotted line. 
A fit of the abovementioned function to the experimental results has been performed
at both excess energies, with 
A$_y^{max,vec}$ as the only free parameter. We have found 
A$_y^{max,vec}~=~-0.071~\pm$~0.058 for 
Q~=~10~MeV, and A$_y^{max,vec}~=~-0.081~\pm$~0.091 for Q~=~36~MeV. 
The best fit functions of Equation~\ref{eq2} to the experimental 
data are presented in Figure~\ref{fit_ay_max_wilk}. 

\begin{figure}[H]
  \unitlength 1.0cm
  \begin{picture}(14.0,5.5)
      \put(-.3,1.0){
         \psfig{figure=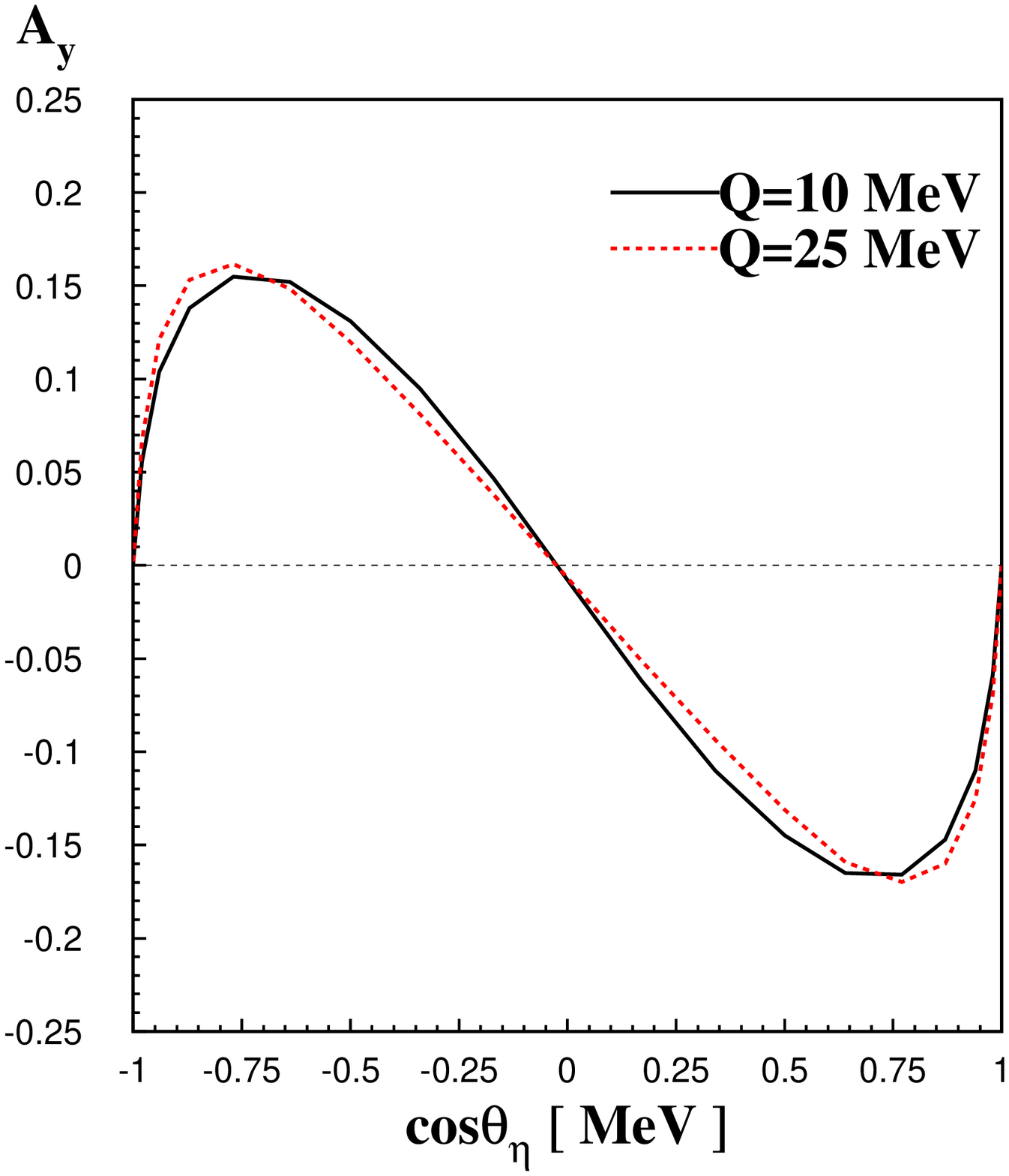,height=5.0cm,angle=0}
      }
      \put(4.7,1.0){
         \psfig{figure=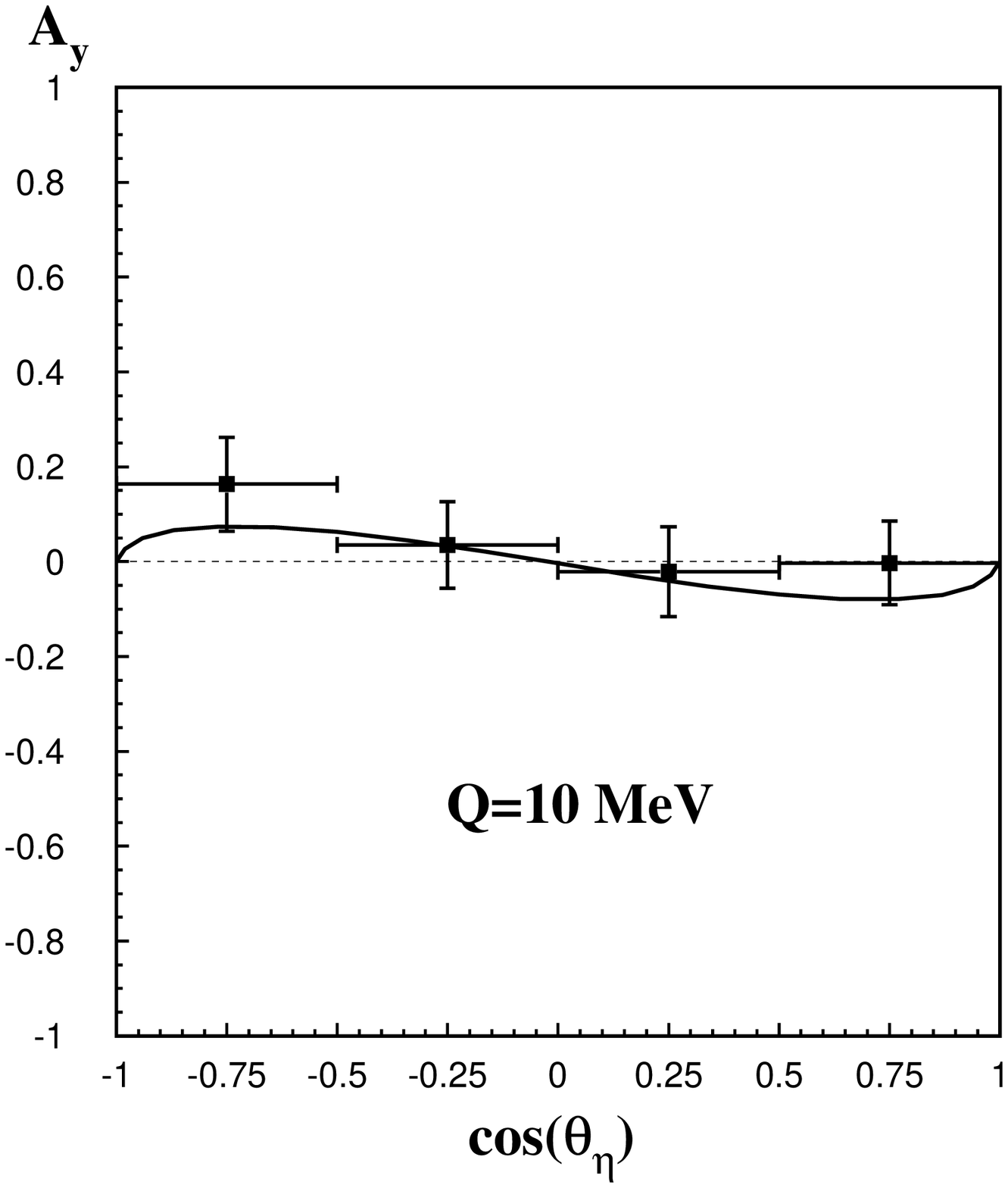,height=5.0cm,angle=0}
      }
      \put(9.6,1.0){
         \psfig{figure=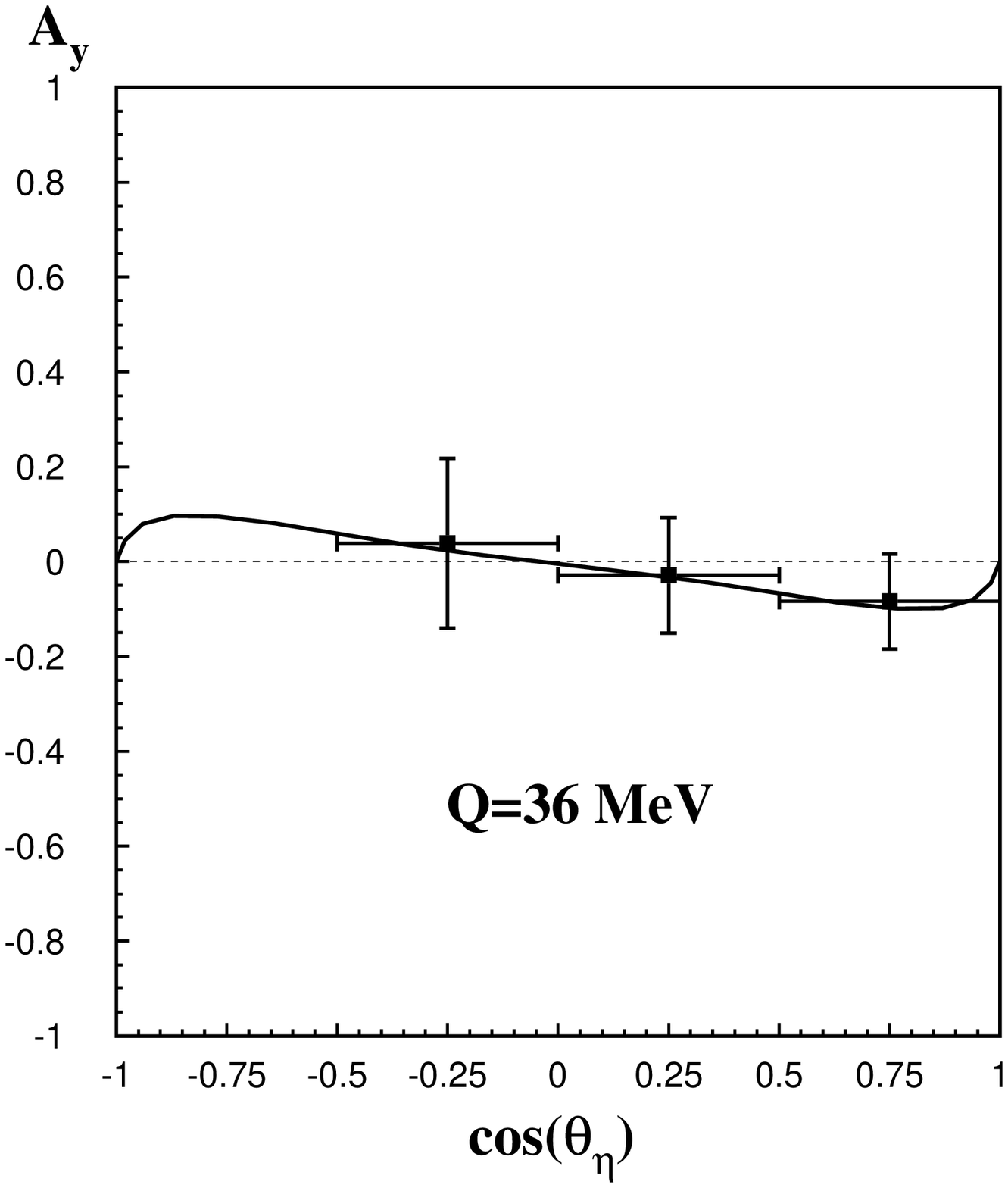,height=5.0cm,angle=0}
      }
      \put(4.2,0.8){{\normalsize {\bf a)}}}
      \put(9.2,0.8){{\normalsize {\bf b)}}}
      \put(14.2,0.8){{\normalsize {\bf c)}}}
  \end{picture}
\vspace{-0.6cm}
  \caption{ {\small (a) Comparison of the shape of the angular distribution of A$_y$ 
	based on the pseudoscalar meson exchange dominance model 
	for arbitrarily chosen Q~=~10~MeV and 25~MeV. 
	Predictions at Q~=~25~MeV were normalized to those of Q~=~10~MeV. 
	The numerical values of $A_y(Q,\theta_{\eta})$ have been made 
	available by the authors of the pseudoscalar meson exchange dominance model~\cite{nakayama,kanzo_priv}.  
	(b) Fit of the A$_y$ function for Q~=~10~MeV to the experimental data.
	(c) Similar fit, but for excess energy of Q~=~36~MeV.  
  \label{fit_ay_max_nak}
  }}
\end{figure}

Similar studies have been performed for the pseudoscalar meson exchange model. 
Although authors of this model do not give an explicit analytical prescription 
of the analysing power's angular dependence, comparison of the predictions for different 
excess energies leads to the conclusion that they do not differ much in shape.
This may be seen  
in Figure~\ref{fit_ay_max_nak}.a, where the theoretical line for 
Q~=~10~MeV is compared  with the theoretical line for Q~=~25~MeV. The latter curve 
has been normalized such that the  difference between the predictions 
for these two excess energies is at its smallest.  
One can see in Figure~\ref{fit_ay_max_nak}.a, that the differences 
are indeed much smaller than 
the experimental accuracy of the determined analysing power. 
Therefore, we may assume that the pseudoscalar meson exchange model's predictions
with respect to the analysing power have the same 
shape for all excess energies within the close-to-threshold region, and only the amplitude of this function varies 
for different excess energies. 
Thus, by analogy to Equation~\ref{eq2} we may write that
\begin{equation}
A_y(Q,\theta_{\eta}) = A_y^{max,psd}(Q) f(\theta_{\eta}),
\label{tychy}
\end{equation}
where $f(\theta_{\eta})$ is the function presented in Figure~\ref{fit_ay_max_nak}.a.

Based on this assumption we have performed a fit of
the predicted theoretical functions for Q~=~10 and 36~MeV
to the experimental data at corresponding excess energies, 
minimising the $\chi^2$ value with $A_y^{max,psd}$
varied as a free parameter. 
The amplitudes
\mbox{A$_y^{max,psd}~=~-0.074~\pm$~0.062} for Q~=~10~MeV 
and \mbox{A$_y^{max,psd}~=~-0.096~\pm$~0.108} for Q~=~36~MeV have been derived. 

\vspace{4mm}

\begin{figure}[H]
  \unitlength 1.0cm
  \begin{picture}(14.0,7.5)
      \put(0.00,1.0){
         \psfig{figure=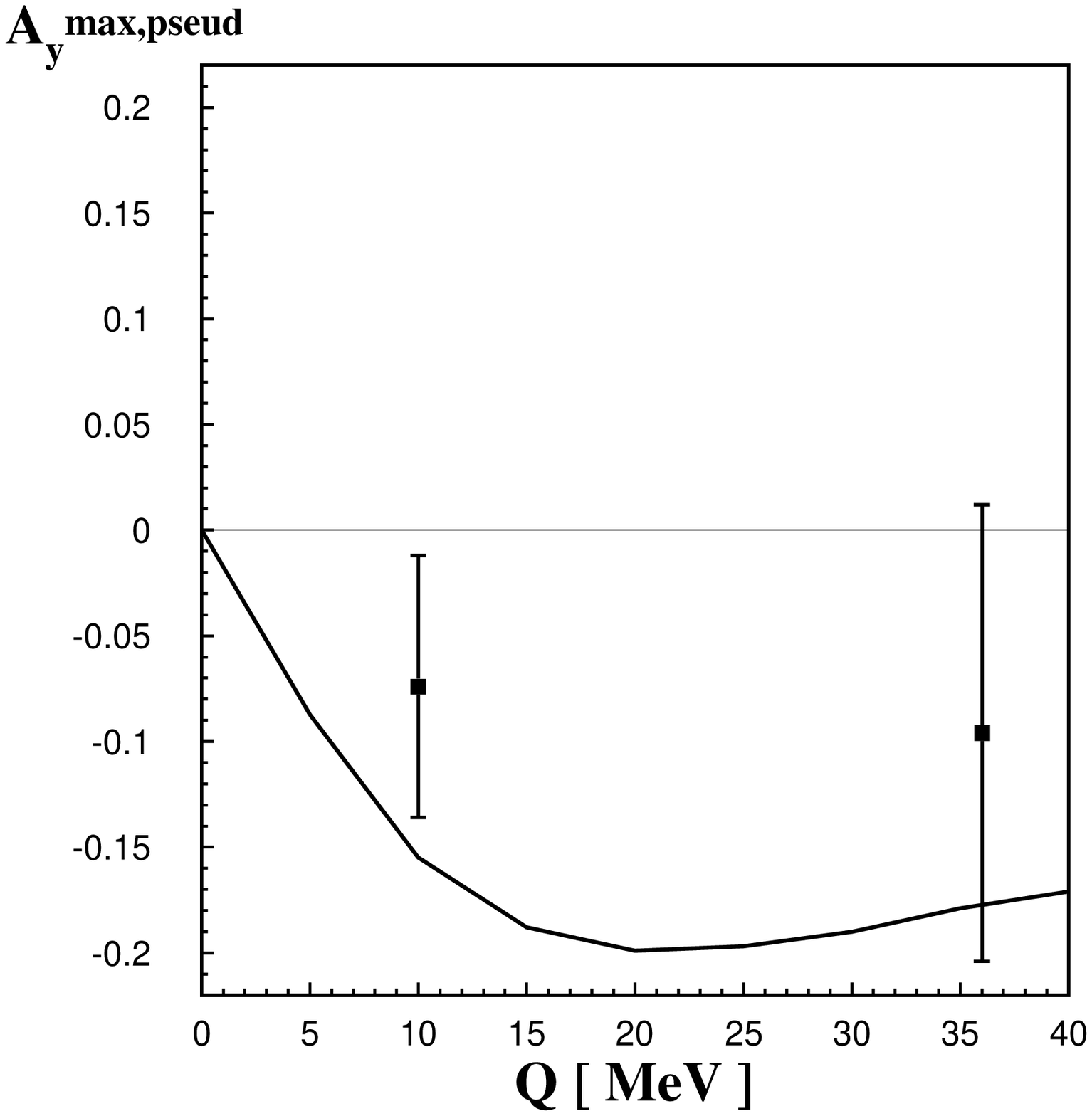,height=6.5cm,angle=0}
      }
      \put(7.50,1.0){
         \psfig{figure=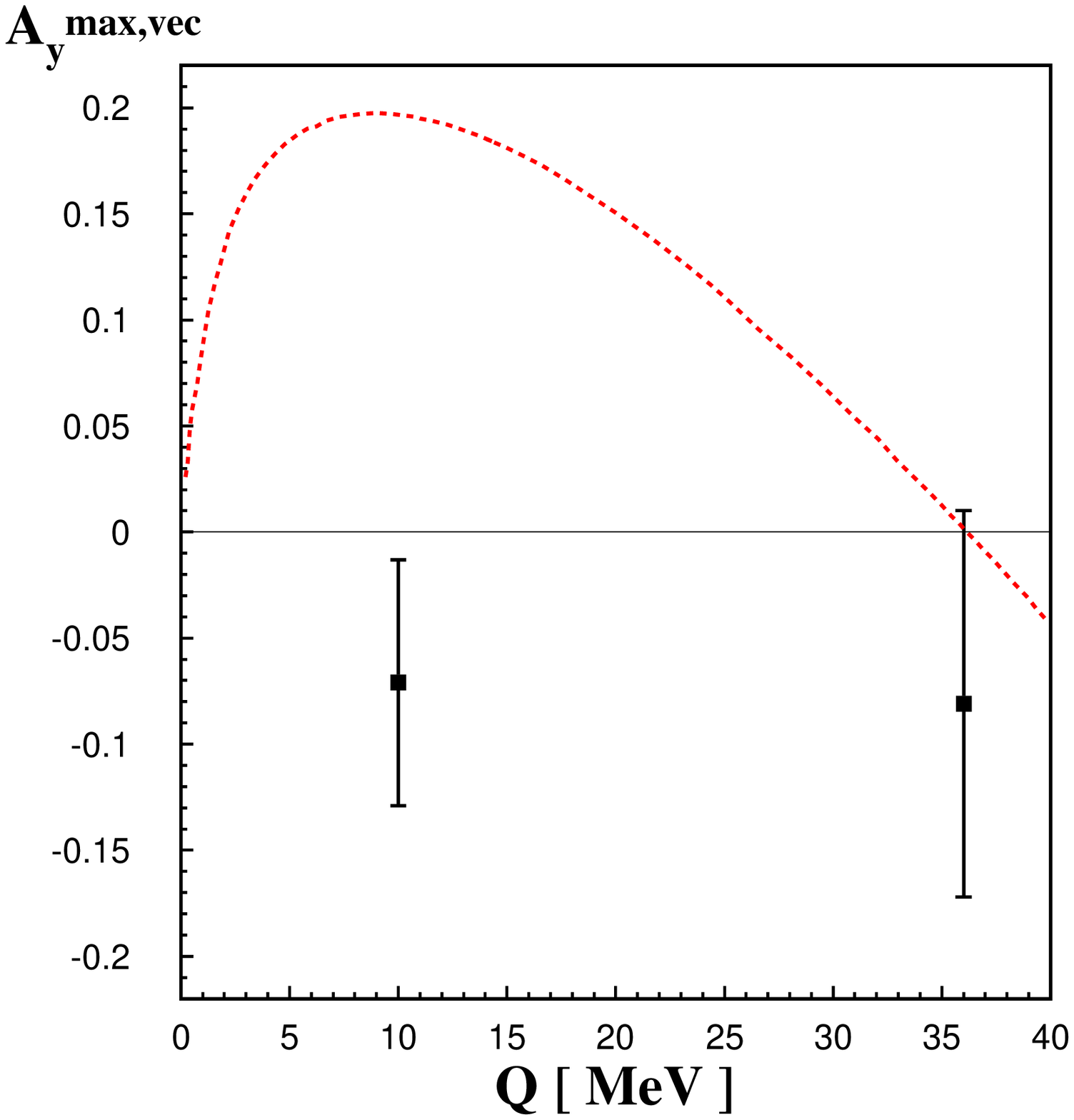,height=6.5cm,angle=0}
      }
      \put(6.5,0.8){{\large {\bf a)}}}
      \put(14.0,0.8){{\large {\bf b)}}}
  \end{picture}
  \caption{ {\small Theoretical predictions for the amplitudes of the analysing power
	angular dependence in the close-to-threshold region confronted with the 
	amplitudes determined in  the experiments at Q~=~10 and Q~=~36~MeV.
	The solid line in Figure (a) shows the prediction based on the pseudoscalar meson 
	exchange dominance model~\cite{nakayama}, while the dotted line in 
	Figure (b) corresponds to the prediction of the vector meson dominance model of reference~\cite{wilkin}. 
  \label{ay_maxxx}
  }}
\end{figure}

The values of the analysing power amplitudes together with the 
theoretical predictions are depicted in Figure~\ref{ay_maxxx}.a
for the pseudoscalar meson exchange model and in Figure~\ref{ay_maxxx}.b
for the vector meson dominance model. 

One can see in this figure that although the amplitude of the vector meson 
exchange model for the excess energy of Q~=~36~MeV lies within about one standard deviation
from the experimental result, the predicted $A_y^{max}$ for 
Q~=~10~MeV is 4.3~$\sigma$ away from the data point.
In the pseudoscalar meson
exchange dominance model predictions of $A_y^{max}$ lie within one standard deviation from the experimental data.

Another observation that can be made from the results shown in Figure~\ref{ay_maxxx} is that the 
extracted $A_y^{max}$ for different models 
are of about the same value. This is due to the fact that the 
pseudoscalar and vector meson exchange models predictions
do not differ much in the shapes of the angular dependencies of the analysing power.

\section{Mechanism of the $\eta$ meson production in nucleon-nucleon collisions}

\vspace{3mm}
{\small
The most probable mechanism of the $\eta$ meson production in 
proton-proton collisions is presented. 
}
\vspace{5mm}

As pointed out in Section~\ref{ggg},
it has been established by the theoretical analysis of the data
from the total cross section measurements for the $pp\to pp\eta$ reaction,
that the resonant current constitutes the main contribution
to the production mechanism of the $\eta$ meson in proton-proton
collisions in the close-to-threshold energy region.
It appears that amongst the available resonances the S$_{11}$(1535) resonance plays the most important
role as an intermediate state. Excitation of the proton to this resonance may proceed by either one of the
light pseudoscalar or vector meson exchanges. However, as far as only the data on the
total cross section are concerned an univocal statement cannot be made 
on which out of the spectrum of possible mesons gives rise to the
S$_{11}$(1535) resonance. Here, the possible particles to be exchanged
are the $\pi, \eta, \rho$, and $\omega$ mesons.

Some limitations to the models have been made by the measurements of
the total cross sections for the quasi-free $pn\to pn\eta$
reaction~\cite{calen_pn}. It has been found that the
$\sigma(pn\to pn\eta)/\sigma(pp\to pp\eta)$ ratio is rather constant
in the wide excess energy range from 16 to 109~MeV, and equals about 6.5. From 
such strong isospin dependence of the production amplitudes 
it has been deduced
that the excitation of the S$_{11}$(1535) resonance proceeds
via the exchange of the isovector mesons, hence
this discovery discarded the $\eta$ and $\omega$ mesons as
possible intermediate particles.
Therefore, there remained two possible particles relevant to excite
the S$_{11}$ resonance: the $\pi$ meson, belonging to the light pseudoscalars meson's
nonet and the $\rho$ meson -- a member of the light vector meson's nonet.

\begin{figure}[H]
  \unitlength 1.0cm
        \begin{center}
  \begin{picture}(14.5,6.5)
    \put(5.3,0.0){
      \psfig{figure=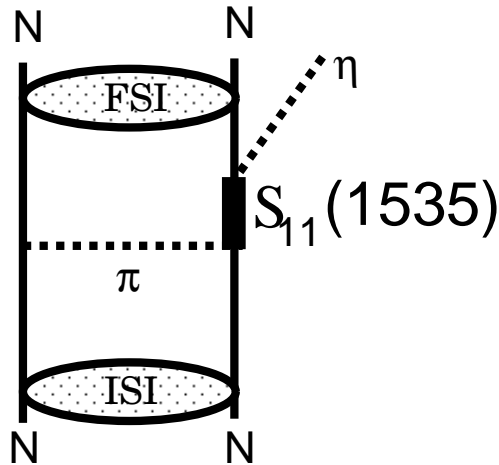,height=6.5cm,angle=0}
    }
  \end{picture}
  \caption{ \small { The most probable mechanism of $\eta$ meson
        production in the nucleon-nucleon collisions: excitation
        of the nucleon to the S$_{11}$(1535) resonance via the
        exchange of a $\pi$ meson, and its further decay into a nucleon-$\eta$ system.
 \label{pion_figure}
}
  }
        \end{center}
\end{figure}

We claim to solve this ambiguity by the measurement of the angular dependence
of the analysing power presented in this dissertation.
The results are presented in Table~\ref{tabelka_zd} of Chapter~\ref{analysing_powers},
as well as in Figure~\ref{analysing_rysunek}.
The $\chi^{2}$ analysis
of the results allows us to reject the predictions of the vector meson dominance
model at a significance level of 0.006. Predictions of the pseudoscalar
meson exchange model are in line with data at the significance level of 0.81. 
This fact, together with the abovementioned
presented inference~\cite{moskal_latest}
indicate that {\bf in nucleon-nucleon collisions
the $\eta$ meson is produced predominantly by the exchange of a $\pi$ meson}.
The most probable process of $\eta$ meson production is presented in Figure~\ref{pion_figure}.


\section{$A_y$ and the final state of the pp$\eta$ system}
\vspace{3mm}
{\small
Based on the experimental data we infer on the 
final state of the $\eta$ meson in the $pp\eta$ system. 
}
\vspace{5mm}

The analysing power determined in this thesis is within the statistical 
accuracy consistent with zero, at both excess energies.
This result implies that {\bf the $\eta$ meson is predominantly produced in the s-wave
in the close-to-threshold region}~\cite{nakayama3}. 
This observation is in agreement with the results
of the analysing power measurements performed by the DISTO collaboration~\cite{balestra}
where, interestingly, in the far-from-threshold energy region A$_y$ was also found
to be consistent with zero within
one standard deviation.
\newpage
\clearpage
\pagestyle{fancy}
\chapter{Perspectives} 
\label{perspectives}

\vspace{3mm}
{\small
Some ideas on extending the experiments presented in this 
thesis are pointed out. An experiment for resolving the 
final state partial waves of the $pp\eta$ system is proposed. 
It is mentioned that the measurements
with the new WASA-at-COSY facility would significantly increase 
the statistics.  
}
\vspace{5mm}

The improvement of the statistics concerning the data on the analysing power 
for the $\vec{p}p\to pp\eta$ reaction would be possible by means of the 
measurement of this observable with the recently brought to operation 4$\pi$
detector WASA-at-COSY~\cite{wasa}. 
Due to the installation of a 
pellet target, high luminosities for the experiments with 
polarised proton beams are expected to be achieved, and are estimated to be  
at the order of 6$\cdot$10$^{29}$cm$^{-2}$s$^{-1}$.  
This would yield around 20000 $\eta$ events per day measured 
at an excess energy of 10~MeV and about 70000 events at an 
excess energy of 36~MeV. Therefore, a one-week measurement of the analysing power  
using the WASA-at-COSY detection setup would yield the result
with circa 20 times better statistical significance\footnote{Evaluation valid for the 
experiment performed at the excess energy of Q~=~36~MeV} than measurements reported in this dissertation.   
The letter of intent for such an experiment
has already been published by the COSY-11 collaboration~\cite{czyzyk_winter}
and is awaiting realisation.

\begin{figure}[H]
  \unitlength 1.0cm
        \begin{center}
  \begin{picture}(14.5,6.5)
    \put(2.2,0.0){
      \psfig{figure=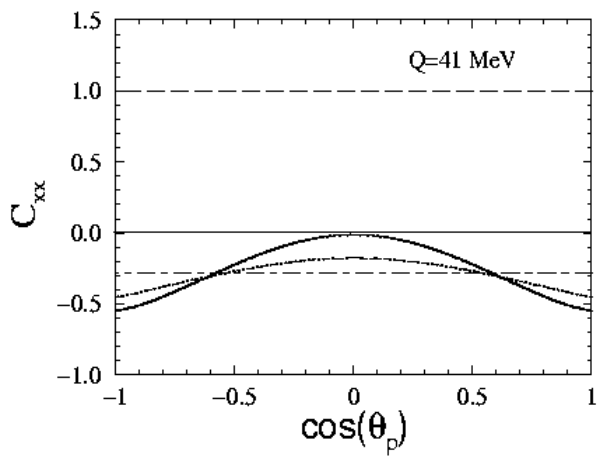,height=6.5cm,angle=0}
    }
  \end{picture}
  \caption{ \small { Predictions of the spin correlation coefficient $C_{xx}$
	for the $\vec{p}\vec{p}\to pp\eta$ reaction as a function of the 
	final proton angle in the overall centre-of-mass system at the excess energy of Q~=~41~MeV.
	The meaning of the curves is explained in the text.  
}
 \label{cxx_figure}
  }
        \end{center}
\end{figure}

Another interesting experiment would be the measurement of the spin 
correlation coefficients C$_{xx}$ for the $\vec{p}\vec{p}\to pp\eta$
reaction. As proposed  
by Nakayama et al.~\cite{nakayama3}, measurements of this 
observable would be helpful to extract the contributions of the 
individual partial waves for the $pp\to pp\eta$ reaction in the
close-to-threshold energy region. Namely, restricting to the
$\eta$ meson production in the $s$-wave\footnote{Which is confirmed by the
results presented in this thesis.} and final proton-proton state in 
the $S$ and $P$ waves there are only three partial waves\footnote{Here the following
notation of the partial waves has been used: $^{2S+1}L_J\to^{2S'+1}L'_{J'}l$, with
S, L, and J denoting the total spin, orbital angular momentum, and total angular momentum
of two protons in the initial state. The prime values are corresponding quantities in the final state.
$l$ stands for the orbital momentum of the $\eta$ meson with respect to
the pair of protons.
The values of the angular momenta are expressed in the spectroscopic notation: L~=~S,~P,~D,$\ldots$
and l~=~s,~p,~d,$\ldots$} that can possibly 
contribute to the $pp\to pp\eta$ reaction~\cite{moskal-hab}: $^3P_0\to^1\!\!S_0s$, 
$^1S_0\to^3\!\!P_0s$, and $^1D_2\to^3\!\!P_2s$.
Denoting the amplitudes of these
transitions by $\alpha, \beta$, and $\gamma$, respectively, yields~\cite{nakayama3}:
\begin{equation}
\frac{d\sigma}{d\Omega}=|\alpha|^{2}  + k^{2} \left(|\beta+\gamma|^2 +3x^2\left(|\gamma|^2-2Re\left(\beta\gamma^*\right)\right)\right),
\nonumber
\end{equation} 
\begin{equation}
\frac{d\sigma}{d\Omega}A_y = 0, 
\nonumber
\end{equation} 
\begin{equation}
\frac{d\sigma}{d\Omega}C_{xx}=|\alpha|^2 - k^{2}\left(|\beta+\gamma|^2+3x^2\left(|\gamma|^2-2Re\left(\beta\gamma^*\right)\right)\right),
\label{partial}
\end{equation}
with $k$ and $p$ standing for the relative momentum in the final and 
initial proton-proton system, respectively, and $x=\hat{k}\cdot\hat{p}$.
From the above equation it is clear that the quantity 
$\frac{d\sigma}{d\Omega}\left(C_{xx}+1\right)$ depends only 
on the amplitude $\alpha$.  
Therefore measuring the spin correlation functions in the 
region where the differential cross sections for the $pp\to pp\eta$ reaction
are known would provide the model-independent information 
about the magnitude of the $^3P_0\to^1\!\!S_0s$ transition.
In a similar way, the combination $\frac{d\sigma}{d\Omega}\left(C_{xx}-1\right)$ different than zero would 
indicate the presence of higher partial waves in the proton-proton final state. 
Authors of~\cite{nakayama3} also provide the predictions 
for the $C_{xx}$ angular dependence, which are presented in Figure~\ref{cxx_figure}. 

The dashed line in Figure~\ref{cxx_figure} assumes $^3P_0\to^1\!\!S_0s$ transition only, 
leading to the constant value of C$_{xx}$ equal to 1. The dash-dotted line 
apart from $^3P_0\to^1\!\!S_0s$ transition takes into account also the 
transition $^1S_0\to^3\!\!\!P_0s$. In this case the authors also obtained a constant value
of C$_{xx}\approx-0.3$.   
Inclusion of the $^1D_2\to^3\!\!P_2s$ contribution results in the dotted line. 
Taking into account the higher partial waves would end up with the angular dependence 
of C$_{xx}$ being given by the solid curve in Figure~\ref{cxx_figure}.

The abovementioned experiment has been proposed by the COSY-11 collaboration~\cite{czyzyk_winter}
to be performed on the new WASA-at-COSY detector setup. However there are many 
technical challenges to be fulfilled before the determination of this measurement will become
possible. 

\newpage
\clearpage
\pagestyle{fancy}
\chapter{Summary} 
\label{summary}

We have presented the theoretical background, the method of measurement 
and the results of experiments aiming in determination of the 
analysing power for the $\vec{p}p\to pp\eta$ reaction in the close-to-threshold 
energy regime. Measurements have been performed utilizing the 
polarised proton beam of the COSY accelerator, the cluster jet target
delivering jets of H$_2$ molecules, and the COSY-11 
experimental facility used to register and identify the reaction products.
For the $\eta$ meson identification the missing mass method 
has been applied. The monitoring
of relative luminosity has been realized with a dedicated 
detection subsystem, presented in Section~\ref{mmm}, measuring the 
differences in the numbers of reactions taking place in the polarisation plane  
during the spin up and down modes. The degree of polarisation has been determined 
by means of a series of measurements of asymmetries for the $\vec{p}p\to pp$ process, 
utilising three independent polarimeters. 
Experiments have been performed 
at beam momenta of p$_{beam}$~=~2.010 and 2.085~GeV/c, which 
for the $\vec{p}p\to pp\eta$ reaction correspond to the excess energies of Q~=~10 and 36~MeV, 
respectively. Results of the data analysis are summarized in Section~\ref{analysing_powers}.

For the first time ever it was possible to experimentally pin down  
the dominating production mechanism of the $\eta$ meson in nucleon-nucleon 
collisions. Our results indicate, at a significance level of 0.81, 
that the $\pi$ meson is an intermediate boson exchanged 
between colliding nucleons in order to excite one of them to the 
resonant state S$_{11}$(1535). In the latter part of the process, this baryonic resonance   
deexcites with emission of a nucleon and the $\eta$ meson, as 
presented in Figure~\ref{pion_figure}. It is important to 
note, that not only the measurements presented in this dissertation
contributed to this finding, but also many hitherto performed investigations 
by various experimental groups have been important in the understanding 
of this process~\cite{bergdolt:93, chiavassa:94, calen:96, calen:97, 
hibou:98, smyrski:00, moskal-prc, moskal:02, moskal:02-3, abdelbary:02, calen_pn}.

We have also shown that the predictions of the analysing power angular dependence 
according to the vector meson exchange model~\cite{wilkin}, where the $\rho$
meson plays the most important role as an intermediate particle 
exciting the nucleon to the S$_{11}$ state, disagree with the 
experimental data at a significance level of $\alpha_{vec}=0.006$.
This result is intuitively clear and understood, as the branching ratio
of the S$_{11}$(1535) decay into N$\pi$ channel is about 45\%, which is an order
of magnitude larger than the branching ratio for a decay 
into a $N\rho$ pair, measured to be at circa 4\%~\cite{yao}. Moreover, the coupling constant 
for the $NN\pi$ vertex is larger than the corresponding 
coupling constants for the $NN\omega$, $NN\rho$, $NN\eta$ vertices. 

One should, however, keep in mind that the interferences in 
the exchange of different types of mesons are not excluded 
and should be studied theoretically and experimentally by the 
measurement of further spin observables.

The analysing power of the $\vec{p}p\to pp\eta$ reaction 
for both excess energies studied in this work was found to be 
consistent with zero within one standard deviation. 
This may suggest that the $\eta$ meson is predominantly produced in the 
s-wave, an observation which is in agreement with the results of the 
analysing power measurements performed by the DISTO collaboration~\cite{balestra}
where, interestingly, in the far-from-threshold energy region the A$_y$
was found to be also consistent with zero within one standard deviation.

The results of this dissertation might be helpful 
in revisiting the theoretical models, in the sense 
that they may provide a new input
with respect to the coupling constants used for modeling 
the $\eta$ meson production in hadronic interaction, and  
the range of the reaction.

\newpage
\clearpage
\pagestyle{plain}
\chapter*{Acknowledgments}

\label{kolaboracja}


Here I would like to thank all the people 
who helped me during the years of Ph.~D.~studies. 

First of all I would like to express my enormous 
gratitude to \mbox{Dr hab. Pawe{\l} Moskal},
for devoting his time, the number of ideas that helped to bring 
this dissertation to life, for his patience 
and offering me help whenever I needed it.  

I am extremely grateful to Prof.~Walter~Oelert 
for giving me an opportunity to work in the COSY-11 collaboration,
for the invitation to the Research Centre J\"ulich,
and for his help.  

I would like to acknowledge Prof.~Jim~Ritman and Prof.~Kurt~Kilian for
allowing me to work in the Research Centre J\"ulich.  

I am grateful to Prof.~Bogus{\l}aw~Kamys for allowing me to prepare
this dissertation in the Faculty of Physics, Astronomy and 
Applied Computer Science of the 
Jagellonian University.

A big ''thank you" to all my colleagues from the 
COSY-11 collaboration: Dr~H.-H.~Adam, Prof.~Andrzej~Budzanowski, 
Eryk~Czerwi\'nski, Damian~Gil, Dr~Dieter~Grzonka,
Ma{\l}gorzata~Hodana, Micha{\l}~Janusz, Prof.~Lucjan~Jarczyk, 
Prof.~Bogus{\l}aw~Kamys, Dr~hab.~Alfons~Khoukaz, Pawe{\l}~Klaja, Dr~Piotr~Kowina,
Dr~hab.~Pawe{\l}~Moskal, Prof.~Walter~Oelert, Cezary~Piskor-Ignatowicz, Timo~Mersmann,  
Joanna~Przerwa, Barbara~Rejdych, Dr~Tomasz~Ro{\.z}ek, 
Dr~Thomas~Sefzick, Prof.~Marek~Siemaszko, Dr~hab.~Jerzy~Smyrski, 
Alexander~T\"aschner, Dr~Peter~Winter, Dr~Magnus~Wolke, Dr~Peter~W\"ustner 
and Prof.~Wiktor~Zipper.     
We had a lot of fun during the beamtimes! 

A lot of thanks to the COSY team for providing a good quality 
beam during the experiments.

Special acknowlegdments must be made to Dr~Bernd~Lorentz and Dr~Kay~Ulbrich 
for providing the data on the beam polarisation. 

For providing the predictions for A$_y$ special thanks are due to
Prof.~Kanzo~Nakayama, K.~S.~Man, Prof.~G\"oran~F\"aldt and Prof.~Colin~Wilkin.

I'd like to express my gratitude to the others who helped me with their ideas, 
especially to: Dr~Dieter~Grzonka, Prof.~Lucjan~Jarczyk, Dr~Piotr~Kowina,  
Prof.~Walter~Oelert, Cezary~Piskor-Ignatowicz, Dr~Thomas~Sefzick, 
Dr~hab.~Jerzy~Smyrski, Dr~Peter~Winter, Dr~Magnus~Wolke, 
and \mbox{Dr~Aleksandra~Wro\'nska}.

For careful reading and correcting this dissertation my gratitude is due to 
\mbox{Dr~Dieter~Grzonka}, Dr~hab.~Pawe{\l}~Moskal, Dr~Thomas~Sefzick and Dr~Andrew~Smith. 

I am also very grateful to Dr~Bernd~Lorentz, Dr~Hans~Stockhorst and Dr~Ralf~Gebels
for correcting some parts of this thesis.  

Thanks to all my colleagues from ``the room 03A" 
and from IKP 
for the pleasant atmosphere of the daily work. 

I am very grateful to my parents Zofia and Emil, and also 
to Julia and Anita 
for their support. 

Finally, I am indebted to Paulina~Piwowarska 
for her patience and understanding$\ldots$

\appendix

\newpage
\clearpage
\pagestyle{fancy}
\chapter{Pseudoscalar and vector mesons} 
\label{mezony}

\vspace{3mm}
{\small
Definitions of the pseudoscalar and vector mesons as well as 
differences between isoscalar and isovector mesons are given. 
The structures of the mesons building up the pseudoscalar 
and vector mesons nonet are given. 
}
\vspace{5mm}

According to QCD, mesons are bound states of quark q and antiquark $\bar{q'}$.
The quarks $q$ and $\bar{q'}$ may be the same or different. 
As quarks and antiquarks are the spin 1/2 particles, 
they may form triplet states ($\uparrow \uparrow$)
with the total intrinsic spin J~=~1, and singlet states ($\uparrow \downarrow$)
with J~=~0.

By convention, each quark is assigned positive parity and each antiquark 
has negative parity. If $L$ denotes the orbital angular momentum of the $q\bar{q'}$ pair,
then the parity of the meson built out of this pair equals P~=~$(-1)^{L+1}$.

From the three lightest quarks -- u, d, and s -- whose properties are 
quoted in Table~\ref{taba1}, nine possible
$q\bar{q'}$ combinations can be built. This set of SU(3) mesons
includes the octet and a singlet states, which can be schematically written as:
\begin{equation}
        \bf{3} \otimes \bf{\bar{3}} = \bf{8} \oplus \bf{1}.
\label{SU3}
\end{equation}

\begin{table}
 \begin{center}
   \begin{tabular}{|c|c|c|c|c|}
    \hline
     Quark  & B - baryon & Q - electric & I$_{3}$ - isospin & S - strangeness \\   
	    & number     & charge       & third component   & 		      \\
    \hline
	u   &  1/3	 &  +2/3	&  +1/2		    &	0 	      \\
	d   &  1/3	 &  $-1/3$	&  $-1/2$		    &	0 	      \\
	s   &  1/3	 &  $-1/3$	&  0		    &	$-1$ 	      \\
    \hline
   \end{tabular}
     \caption{ \small Additive quantum numbers of the SU(3) quarks. By convention, 
		the flavours of the quarks (I$_{3}$ and S) have the same 
		signs as their charges Q.  
         }
     \label{taba1}
 \end{center}
\end{table}

\begin{figure}[H]
  \unitlength 1.0cm
  \begin{picture}(14.0,7.5)
      \put(0.00,1.0){
         \psfig{figure=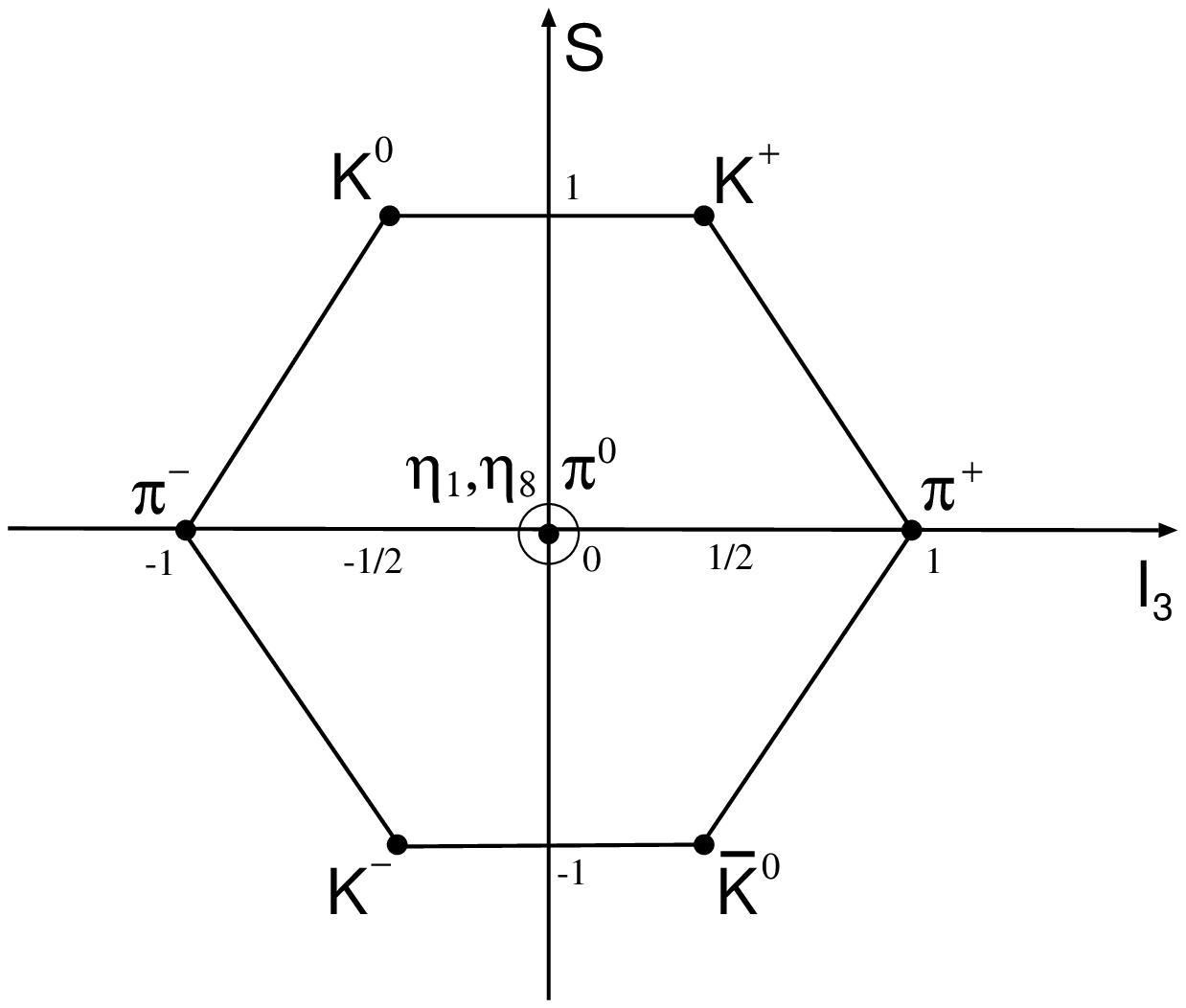,height=6.0cm,angle=0}
      }
      \put(7.50,1.0){
         \psfig{figure=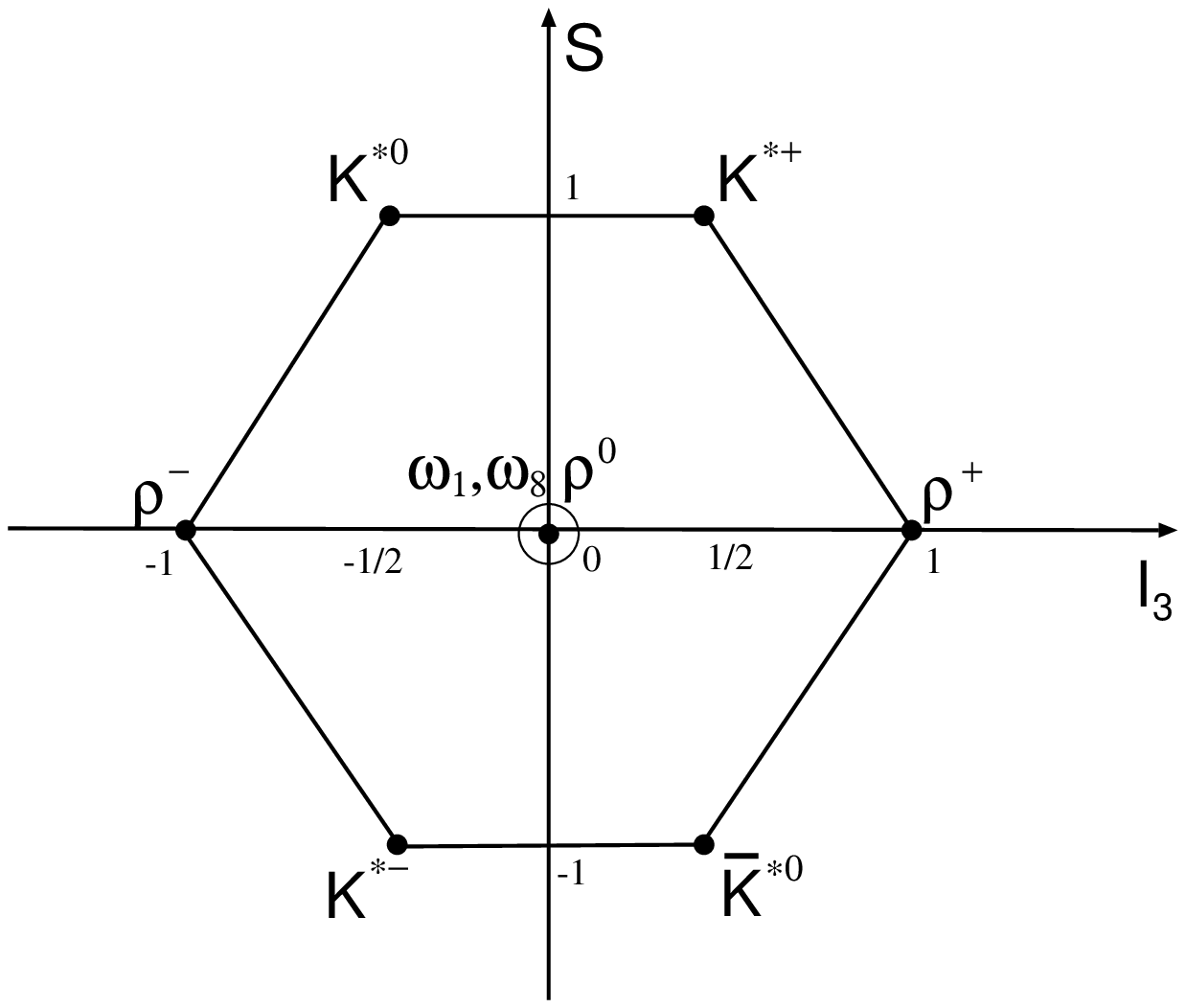,height=6.0cm,angle=0}
      }
      \put(6.5,0.8){{\normalsize {\bf a)}}}
      \put(14.0,0.8){{\normalsize {\bf b)}}}
  \end{picture}
  \caption{ {\small The nonet of the ground state (a) pseudoscalar (J$^{P}=0^{-}$) and  
		(b) vector (J$^{P}=1^{-}$) mesons. The strangeness $S$ of the meson
		is plotted versus the third component of its isospin I$_{3}$.   
		The neutral mesons at the centre of the S-I$_{3}$ plane 
		are the pure mixtures of $u\bar{u}$, $d\bar{d}$, and $s\bar{s}$ states. 
  \label{nonet}
  }}
\end{figure}

 Let us assume that $q\bar{q'}$ are the ground state combinations of quark-antiquark pairs
with the relative angular momentum L~=~0. This implies the parity $P$ of a 
such constructed meson equals $-1$.
With this condition, the mesons with internal spin J~=~0 are called 
{\it pseudoscalar mesons} and the ones with J~=~1 are called {\it vector mesons}.   
The ground state pseudoscalar and vector meson nonets are 
depicted in Figure~\ref{nonet}.

States with the same additive quantum numbers, and also the same isospin, total 
internal spin, and parity can mix. 
This is a consequence of the SU(3) symmetry breaking.
The singlet SU(3) state of the pseudoscalar meson nonet 
-- $\eta_{1}$ -- corresponds to the following combination of quarks:
\begin{equation}
        \eta_{1} = \frac{1}{\sqrt{3}} (u\bar{u} + d\bar{d} + s\bar{s}),
\label{eta1}
\end{equation}
which mixes with the $\eta_{8}$ state, belonging to the SU(3) octet:
\begin{equation}
        \eta_{8} = \frac{1}{\sqrt{6}} (u\bar{u} + d\bar{d} - 2 s\bar{s}).
\label{eta8}
\end{equation}
These two states are not the real physical objects. Real mesons 
which can be observed in the experiments --
$\eta$ and $\eta^{\prime}$ -- are the mixtures of these pure SU(3) states:
\begin{displaymath}
        \eta = cos(\theta_{psc}) \eta_{8} - sin(\theta_{psc}) \eta_{1},
\end{displaymath}
\begin{equation}
        \eta^{\prime} = sin(\theta_{psc}) \eta_{8} + cos(\theta_{psc}) \eta_{1},
\label{ety1}
\end{equation}
where a pseudoscalar mixing angle $\theta_{psc} = -15.5^{\circ}$ 
has been introduced~\cite{Bramon}.

Similarly, within the vector meson nonet the theoretical SU(3) 
states $\omega_1$ and $\omega_8$ mix with each other. This mesons have the quark content 
corresponding to the $\eta_1$ and $\eta_8$ states, respectively.
The physical states $\omega$ and $\phi$, analogous to the abovementioned
$\eta$ and $\eta^{\prime}$ pseudoscalar combinations, are the result of the 
$\omega_1$ and $\omega_8$ mixing with a mixing angle
$\theta_{vec}=37^{\circ}$~\cite{nambu}.

The quark structure of the ground state pseudoscalar and vector mesons 
together with their masses are presented in Table~\ref{taba2}.

\begin{table}
 \begin{center}
   \begin{tabular}{|c|c|c|}
    \hline
	Pseudoscalar mesons &  Quark combination & Mass [MeV] \\
    \hline
	$\pi^+$ & $u\bar{d}$ & 139.57 \\
	$\pi^-$ & $d\bar{u}$ & 139.57 \\
	$\pi^0$ & $\frac{1}{\sqrt{2}}\left(u\bar{u}-d\bar{d}\right)$ & 134.98 \\
	$K^+$ & $u\bar{s}$ & 493.68 \\
	$K^0$ & $d\bar{s}$ & 497.67 \\
	$K^-$ & $\bar{u}s$ & 493.68 \\
	$\bar{K}^0$ & $\bar{d}s$ & 497.67 \\
	$\eta$ & $A_1\left(d\bar{d}+u\bar{u}\right)+B_1\left(s\bar{s}\right)$ & 547.30 \\
	$\eta^{\prime}$ & $A_2\left(d\bar{d}+u\bar{u}\right)+B_2\left(s\bar{s}\right)$ & 957.78 \\
    \hline
    \hline
	Vector mesons &  Quark combination & Mass [MeV] \\
    \hline
	$\rho^+$ & $u\bar{d}$ & 769.3 \\
	$\rho^-$ & $d\bar{u}$ & 769.3 \\
	$\rho^0$ & $\frac{1}{\sqrt{2}}\left(u\bar{u}-d\bar{d}\right)$ & 769.3 \\
	$K^{*+}$ & $u\bar{s}$ &  891.66 \\
	$K^{*0}$ & $d\bar{s}$ &  896.10 \\
	$K^{*-}$ & $\bar{u}s$ &  891.66 \\
	$\bar{K}^{*0}$ & $\bar{d}s$ &  896.10 \\
	$\omega$& $C_1\left(d\bar{d}+u\bar{u}\right)+D_1\left(s\bar{s}\right)$ & 782.57 \\
	$\phi$ & $D_2\left(d\bar{d}+u\bar{u}\right)+D_2\left(s\bar{s}\right)$ & 1019.42 \\
    \hline
   \end{tabular}
     \caption{ \small Pseudoscalar and vector mesons as the quark-antiquark combinations. 
 Masses of the mesons are taken from~\cite{groom}.
         }
     \label{taba2}
 \end{center}
\end{table}

\vspace{2cm}

The mesons with the total isospin I~=~0, like 
the $\eta$ and $\eta^{\prime}$ mesons within the pseudoscalar 
meson nonet and the $\phi$ and $\omega$ mesons in the 
case of the vector meson nonet are called the
{\it isoscalar mesons}.

By the {\it isovector mesons} we refer to
either pseudoscalar or vector mesons with 
a total isospin I~=~1. These are the $\pi$
and $\rho$ mesons, for the pseudoscalar and 
vector meson nonet, respectively. 
\newpage
\clearpage
\pagestyle{fancy}
\chapter{Property of the average analysing power} 
\label{prove}
\newcommand{\ud}{\mathrm{d}}

\vspace{3mm}
{\small
A proof of the following theorem is presented. 
}
\vspace{5mm}

\par
\newtheorem{thm}{Theorem}
\begin{thm}
		 For an isotropic distribution of the differential cross section 
		 $\frac{d\sigma}{d\theta}\left(\theta\right)$~=~const, the average 
		 analysing power over an angular range $\Delta\Theta$ 
		 is an arithmetical average of the analysing powers for the individual 
		 $\Delta\theta_i$ subranges of $\Delta\Theta$: 

\vspace{1cm}

\begin{equation}
\bar{A_y}\left(\Delta\theta\right) = \frac{\sum_{i=1}^{n}{A_y\left(\Delta\theta_i\right)}}{n}.
\label{11}
\end{equation}
\end{thm}

\vspace{2cm}

\begin{proof}

Without loosing generality, let us consider the production
of the $\eta$ mesons during spin up mode in the plane perpendicular to the polarisation plane.
Let the $N_L\left(\Delta\Theta\right)$
and $N_R\left(\Delta\Theta\right)$
denote the number of the $\eta$ mesons
produced into the $\Delta\Theta$ range, 
symmetrically with respect to the polarisation plane
to the left and right side, respectively.
If $P$ denotes the beam polarisation then, according to the Formula~\ref{wniosek},
the averaged beam analysing power for the $\eta$ meson production into
the $\Delta\Theta$ angular range reads: 
\begin{equation}
\bar{A_y}\left(\Delta\Theta\right) = \frac{1}{P} \frac{N_L\left(\Delta\Theta\right)-N_R\left(\Delta\Theta\right)}{N_L\left(\Delta\Theta\right)+N_R\left(\Delta\Theta\right)}.
\label{232}
\end{equation}

\clearpage

Applying the Madison convention defined in Chapter~\ref{refi}
and taking into account that the production yields $N_L$ and $N_R$ are 
proportional to the cross section of Equation~\ref{cross_3} and 
under assumption that $\frac{d\sigma}{d\theta}\left(\Delta\theta\right)~=~const$ we can write:
\vspace{0.7cm}
\begin{equation}
N_L\left(\Delta\Theta\right)\sim \Delta\Theta \left(1+PA_y\left(\Delta\Theta\right)\right), 
\label{wi}
\end{equation}

\begin{equation}
N_R\left(\Delta\Theta\right)\sim \Delta\Theta \left(1-PA_y\left(\Delta\Theta\right)\right).
\label{m77}
\end{equation}

\vspace{0.7cm}
Dividing the $\Delta\Theta$ range into $n$ identical subranges $\Delta\theta_i$
we can rewrite Equations~\ref{wi} and~\ref{m77}:
\vspace{0.7cm}
\begin{equation}
N_L\left(\Delta\Theta\right) \sim \sum_{i=1}^{n}{\Delta\theta_{i} \left(1+PA_y\left(\Delta\theta_i\right)\right)},
\label{df}
\end{equation}
\vspace{0.7cm}
\begin{equation}
N_R\left(\Delta\Theta\right) \sim \sum_{i=1}^{n}{\Delta\theta_{i} \left(1-PA_y\left(\Delta\theta_i\right)\right)}.
\label{334}
\end{equation}

\vspace{0.7cm}
Putting Equations~\ref{df} and~\ref{334} into Formula~\ref{232} yields:
\vspace{0.7cm}
\begin{equation}
\bar{A_y}\left(\Delta\Theta\right) = \frac{1}{P}\frac{\sum_{i=1}^{n}{\Delta\theta_{i} \left(1+PA_y\left(\Delta\theta_i\right)\right)}-\sum_{i=1}^{n}{\Delta\theta_{i} \left(1-PA_y\left(\Delta\theta_i\right)\right)}}{\sum_{i=1}^{n}{\Delta\theta_{i} \left(1+PA_y\left(\Delta\theta_i\right)\right)}+\sum_{i=1}^{n}{\Delta\theta_{i} \left(1-PA_y\left(\Delta\theta_i\right)\right)}}, 
\label{dupa677}
\end{equation}

\vspace{0.7cm}
which after reductions leads to:
\vspace{0.7cm}

\begin{equation}
\bar{A_y}\left(\Delta\Theta\right) = \frac{\sum_{i=1}^{n}{\Delta\theta_{i} A_y\left(\Delta\theta_{i}\right)}}{\sum_{i=1}^{n}{\Delta\theta_{i}}} = \frac{\Delta\theta_{i} \sum_{i=1}^{n}{A_y\left(\Delta\theta_{i}\right)}}{n \Delta\theta_{i}} = \frac{\sum_{i=1}^{n}{A_y\left(\Delta\theta_i\right)}}{n}. 
\label{dupa887}
\end{equation}

\end{proof}

\clearpage
One should mention at this point that the inference presented here 
does not necessarily demands the subranges $\Delta\theta_i$ to be 
of the same width. This is because one can always divide an 
arbitrarily chosen subrange $\Delta\theta_i$ into $m$ infinitesimally small subsubranges,
division for which the prove has just been presented. 

For $n\to \infty$ the relation B.8 becomes:
\begin{equation}
\bar{A_y}\left(\Delta\Theta\right) = \frac{\int_{\theta_1}^{\theta_2}{A_y\left(\theta\right) \ud \theta}}{\int_{\theta_1}^{\theta_2}{\ud \theta}}.
\label{4}
\end{equation}

\newpage
\clearpage
\pagestyle{fancy}
\chapter{Parity invariance and the analysing power} 
\label{parity}

\vspace{3mm}
{\small
It is shown that 
according to the parity invariance rule the 
analysing power in the polarisation plane equals zero, 
hence the scattering yields in the polarisation plane may be used as 
an absolute measure of the luminosity. 
}
\vspace{5mm}

The parity transformation is the reflection of the
system with respect to the origin of the reference frame.
Parity is a multiplicative quantum number, which is 
conserved by the strong and electromagnetic interactions.

\begin{figure}[H]
  \unitlength 1.0cm
        \begin{center}
  \begin{picture}(14.5,4.5)
    \put(1.0,0.0){
      \psfig{figure=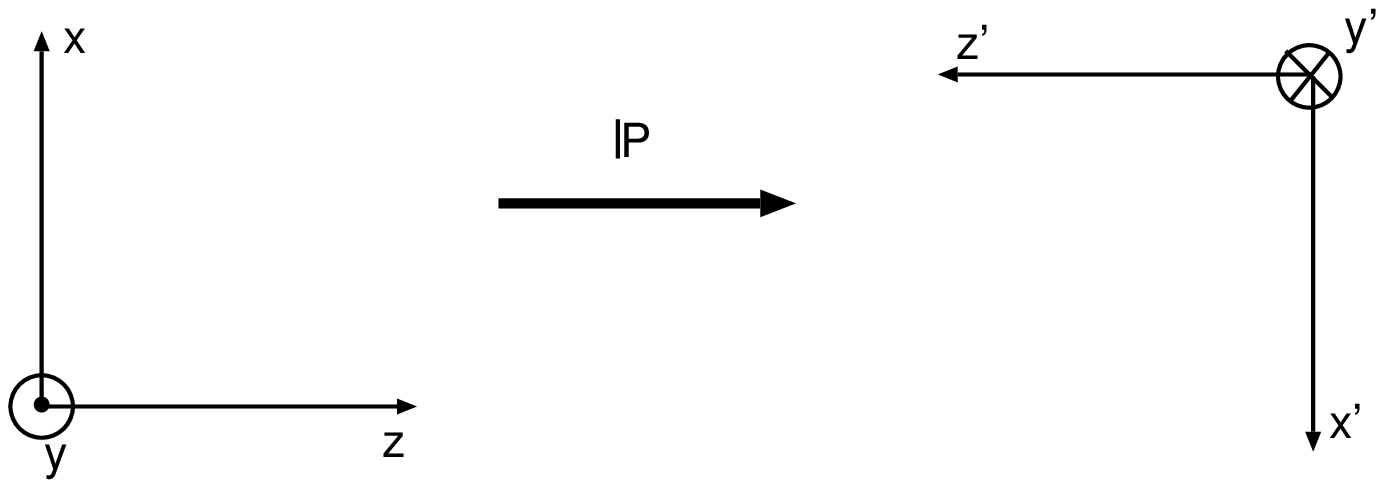,height=4.5cm,angle=0}
    }
  \end{picture}
  \caption{ {\small Parity transformation. 
                 }
 \label{kipszna}
  }
        \end{center}
\end{figure}

Parity transformation may also be represented 
by a threefold reflection of the system in a mirror 
situated in the $y-z$, $x-z$, and $x-y$ planes. 
This operation transforms the right handed (x,y,z) frame into 
the left handed (x',y',z') frame, as depicted in 
Figure~\ref{kipszna}.   

Please note that spin, and as a consequence 
the polarisation vector, is invariant under the parity 
reversal and that these objects transform as 
the {\it pseudovectors}~\footnote{The other 
example of the pseudovector is the angular momentum, which has the same 
transformation properties as a spin.}. Indeed, if we considered the particle 
with its spin along the $y$ axis we would notice that the spin vector changes 
its direction upon reflecting it in the mirrors positioned perpendicularily 
to $x$ and $z$ axis, but does not change the direction when mirrored 
in the plane perpendicular to the $y$ axis. As a consequence 
spin does not change its direction under the parity transformation.

\begin{figure}[H]
  \unitlength 1.0cm
        \begin{center}
  \begin{picture}(14.5,5.5)
    \put(0.2,1.0){
      \psfig{figure=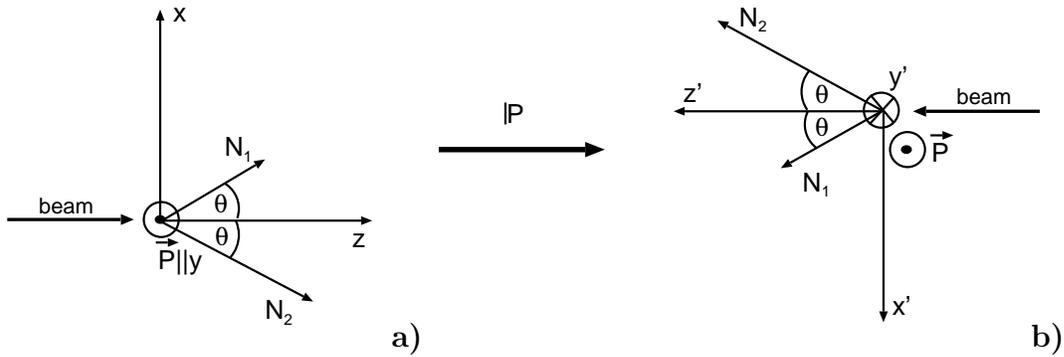,height=4.5cm,angle=0}
    }
	\put(5.5,0.8){{\normalsize {\bf a)}}}
      \put(14.0,0.8){{\normalsize {\bf b)}}}
  \end{picture}
  \caption{ {\small Parity reversal for scattering in the plane 
perpendicular to the polarisation plane.  
                 }
 \label{kipszna2}
  }
        \end{center}
\end{figure}

Now, without loosing generality, let us consider the scattering 
of a particle with spin along the y axis in the $x-z$ plane, 
as depicted in Figure~\ref{kipszna2}.a. 
We shall also assume that there is an asymmetry in the left-right 
scattering in the $x-z$ plane, 
i.e. that the scattering yields $N_1$ and $N_2$ are different.

The parity reversal 
transforms the state presented in Figure~\ref{kipszna2}.a into the
one in Figure~\ref{kipszna2}.b. Please note, that the transformed state of 
Figure~\ref{kipszna}.b is physically identical with the 
initial state from Figure~\ref{kipszna2}.a, as there exist 
an invariant transformation -- which is a rotation by angle $\pi$ around the $y$ axis --
that transforms the final state into initial one. Naturally, 
the (x',y',z') frame remains a left-handed frame, but the physics of the process does
not depend on the choice of the reference frame.

\begin{figure}[H]
  \unitlength 1.0cm
        \begin{center}
  \begin{picture}(14.5,5.5)
    \put(1.5,1.0){
      \psfig{figure=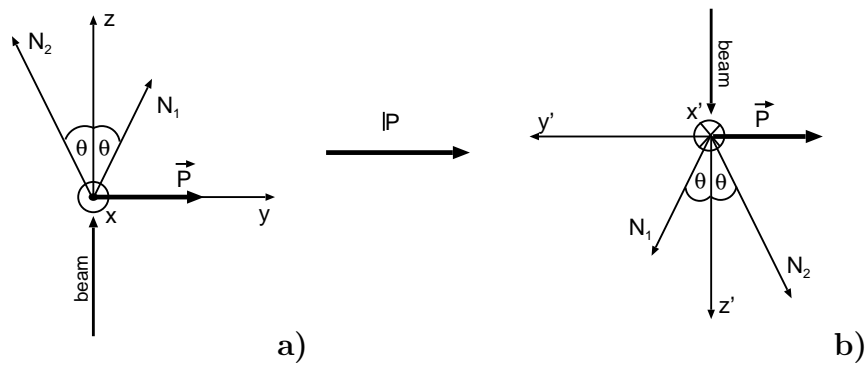,height=4.5cm,angle=0}
    }
	\put(5.2,0.8){{\normalsize {\bf a)}}}
      \put(12.6,0.8){{\normalsize {\bf b)}}}
  \end{picture}
  \caption{ {\small Parity reversal for scattering in the polarisation plane. 
                 }
 \label{kipszna3}
  }
        \end{center}
\end{figure}

Situation is different when we consider the scattering of
a particle with a spin along the $y$ axis in the polarisation plane.
Again we assume that scattering yields $N_1$ and $N_2$ are different. 
The initial state and the state resulting from the parity transformation are
depicted in Figure~\ref{kipszna3}.a and~\ref{kipszna3}.b, respectively.

In this case no invariant transformation exists that could 
retransform the final state into initial one. We note that 
the rotation about the $y$ axis does not work in this case, 
as the directions of N$_1$ and N$_2$ after the rotation are reversed. 
This situation is therefore physically not allowed as it would violate the 
parity invariance rule, which holds for strong interactions.

%

However, the initial and transformed states would be identical 
if $N_1$ and $N_2$ were equal. Therefore, if the studied reaction 
is invariant under the parity transformation it cannot result 
in the asymmetry of yields in the polarisation plane. 
This demands 
the analysing power A$_y$
in the polarisation plane must be equal to zero.

Therefore the scattering in the polarisation plane does 
not depend on the degree of beam polarisation, and  
the scattering yields in the polarisation plane may 
be used as an  absolute measure of the luminosity.    
\pagestyle{fancy}

\end{document}